\documentclass[10pt,review,book]{elsarticle}
\usepackage[titles]{tocloft}
\pdfoutput=1
\usepackage[small]{caption}
\usepackage[bottom]{footmisc}

\usepackage{booktabs}
\usepackage[english]{babel}
\usepackage{amsmath,amssymb,amsbsy,amstext, amsthm, simplewick}
\usepackage{fancyhdr}
 \usepackage{exscale,relsize}

\usepackage{graphicx}
\usepackage{amsfonts}
\usepackage{upgreek}

\usepackage{graphicx}
\usepackage{epstopdf}
\usepackage{amsfonts}
\usepackage{color}
\usepackage{mathrsfs}

\usepackage{graphicx} \usepackage{bm} \usepackage{epsfig}
\usepackage{setspace}
\usepackage{color}
\usepackage{colordvi}
\usepackage{comment}
\usepackage{graphicx}
\biboptions{numbers,sort&compress}
\newcommand{\toUV}{\substack{\longrightarrow \\ {\rm UV}}}
\newcommand{\be}{\begin{equation}}
\newcommand{\ee}{\end{equation}}
\newcommand{\bea}{\begin{eqnarray}}
\newcommand{\eea}{\end{eqnarray}}
\newcommand{\barr}{\begin{array}}
\newcommand{\earr}{\end{array}}
\newcommand{\dl}{\delta}

\newcommand{\bom}{{\boldsymbol{\omega}}}
\newcommand{\bk}{{\boldsymbol{k}}}
\newcommand{\bp}{{\boldsymbol{p}}}
\newcommand{\bq}{{\boldsymbol{q}}}

\newcommand{\bx}{{\boldsymbol{x}}}
\newcommand{\by}{{\boldsymbol{y}}}
\newcommand{\br}{{\boldsymbol{r}}}
\newcommand{\bl}{{\boldsymbol{l}}}
\newcommand{\bs}{{\boldsymbol{s}}}
\newcommand{\bd}{{\boldsymbol{d}}}
\newcommand{\bv}{{\boldsymbol{v}}}
\newcommand{\bz}{{\boldsymbol{z}}}
\newcommand{\bw}{{\boldsymbol{w}}}
\newcommand{\bA}{{\boldsymbol{A}}}
\newcommand{\bS}{{\boldsymbol{S}}}
\newcommand{\ba}{{\boldsymbol{a}}}
\newcommand{\bL}{{\boldsymbol{L}}}
\def\beq{\begin{equation}}
\def\eeq{\end{equation}}
\def\nn{\nonumber}
\allowdisplaybreaks[1]

\newcommand{\pd}{\partial}
\newcommand{\bit}{\begin{itemize}}
\newcommand{\eit}{\end{itemize}}
\newcommand{\im}{\item}

\def\calo{{\cal O}}
\def\Mp{M_{\rm Pl}}

\usepackage{chngcntr}
\counterwithin{equation}{section}

\usepackage[linktocpage=true]{hyperref}
\hypersetup{
colorlinks=true,
citecolor=darkblue,
linkcolor=reddish,
urlcolor=darkblue,
pdfauthor={},
pdftitle={},
pdfsubject={}
}

\journal{Physics Reports}

\begin{document}

\setlength{\skip\footins}{5mm}
\setlength{\footnotesep}{\baselineskip}
\setlength{\jot}{1.5ex}
\interfootnotelinepenalty=10000

\hypersetup{pageanchor=false}
\begin{titlepage}

\begin{center}
{\fontsize{20}{28}\selectfont  \sffamily \bfseries The Effective Field Theorist's Approach\\[0.3cm] to Gravitational Dynamics}
\\[1cm]
{\fontsize{12}{28}\selectfont  Rafael A. Porto}
\\[0.2cm]
 { \it ICTP South American Institute for Fundamental Research,\\ Instituto de F\'isica Te\'orica - Universidade Estadual Paulista}\\ Rua Dr. Bento Teobaldo Ferraz 271, 01140-070 S\~ao Paulo, SP Brazil\\




\end{center}
\begin{figure}[h!]
\centering
\includegraphics[width=0.35\textwidth]{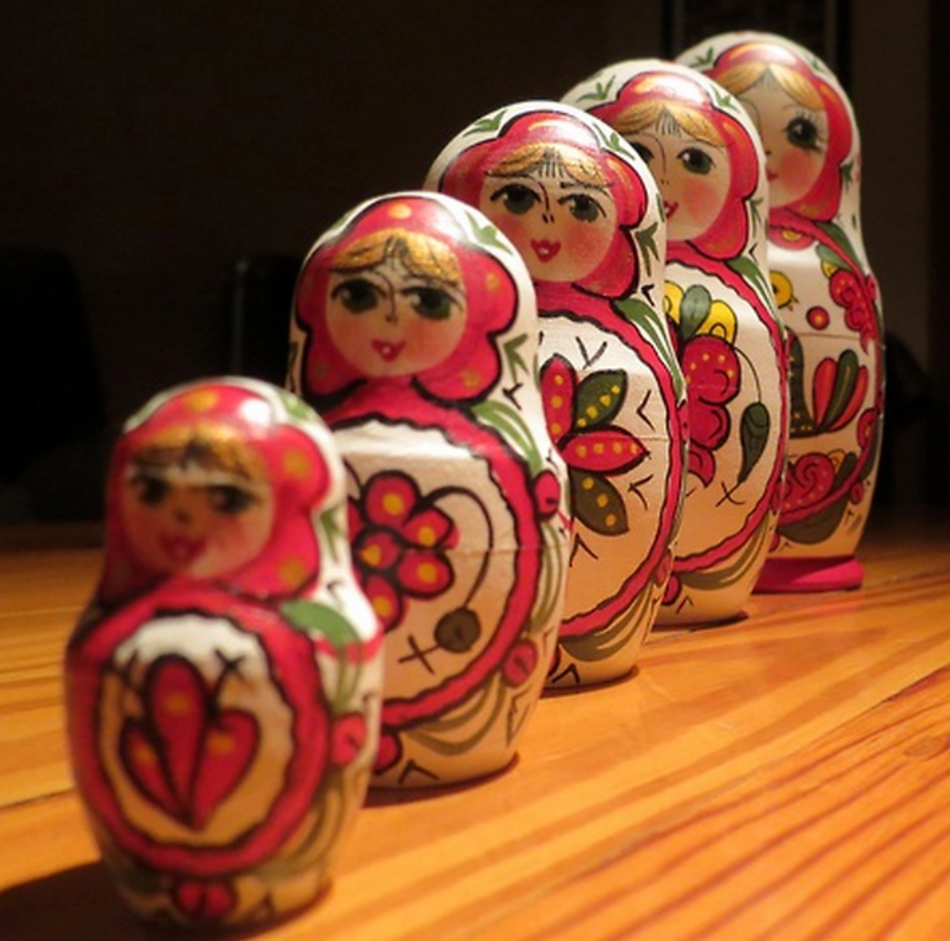}
\end{figure}

\vspace{0.7cm}
\hrule 
\begin{center}\textbf{Abstract}:\end{center} 
We review the effective field theory (EFT) approach to gravitational dynamics. We focus on extended objects in long-wavelength backgrounds and gravitational wave emission from spinning binary systems. We conclude with an introduction to EFT methods for the study of cosmological large scale structures.

 \vspace{0.5cm}
\hrule 
\end{titlepage}
\hypersetup{pageanchor=true}
\thispagestyle{empty}
\newpage
\pagenumbering{roman}

\begin{spacing}{1.19}
\tableofcontents
\end{spacing}
	
\addtocontents{toc}{~\hfill\textbf{Page}\par}
\newpage
\section*{Notation \& Conventions}
\phantomsection
\bit
\im Throughout this review we work in $\hbar=c=1$ units, unless otherwise noted.  

\im We use a reduced Planck mass: $\Mp^{-2} \equiv 32\pi G_N$, with $G_N$ Newton's constant.

\im We use Einstein's summation over repeated indices. Greek and Latin indices ranges are as usual, i.e. $\mu,\nu,\ldots=0,1,2,3,$ and $i,j\ldots = 1,2,3$.
\im Sometimes we use $x^0=t$ to simplify notation, and dots to denote time derivatives, e.g. $\dot E = \frac{dE}{dt}$.

\im We denote ${\bf 3}$-vectors in boldface, e.g. ${\bx}, {\by},\cdots$, and put hats on unit vectors, e.g.  $\hat {\bx}$. To avoid cluttering expressions, sometimes we omit the greek indices when evaluating spacetime (scalar, vector, tensor,$\cdots$) functions, e.g. $f^{\alpha\beta\cdots}(x,y,\cdots) \equiv f^{\alpha\beta\cdots}(x^\mu, y^\mu,\cdots)$. 

\im The Minkowski metric is given by $\eta_{\alpha\beta} \equiv {\rm diag}(+,-,-,-)$. The full metric tensor is denoted as usual, $g_{\mu\nu}(x)$, and often use the standard notation:  $v^2 \equiv v^\mu v_\mu \equiv g_{\mu\nu} v^\mu v^\nu$. 

\im We use $(\alpha\,\beta\,\cdots)$ and $\{\alpha\,\beta\,\cdots\}$ to symmetrize and anti-symmetrize indices,\\ e.g. $A^{(\alpha\,\beta)} \equiv \frac{1}{2} \left(A^{\alpha\beta}+ A^{\beta\alpha}\right)$, $A^{\{\alpha\,\beta\}} \equiv \frac{1}{2} \left(A^{\alpha\beta}- A^{\beta\alpha}\right)$, etc.

\im We use $\tau$ for the proper time: $d\tau^2 = g_{\alpha\beta} dx^\alpha dx^\beta$. 

\im  We write spacetime velocities as $u^\mu(\sigma) \equiv \frac{dx^\mu}{d\sigma}$, with $\sigma$ an affine parameter (sometimes the proper time), and $v^\mu (t) \equiv \frac{dx^\mu}{d t}$, when the choice $\sigma=t$ is made. 

\im The partial derivatives are denoted by $\partial_\mu \equiv \frac{\partial}{\partial x^{\mu}}$\,, and sometimes we simplify notation by repeating indices, e.g. $\partial_{ij} \equiv \partial_i\partial_j$. At times we also use the standard notation $g_{\alpha\beta,\mu} \equiv \partial_\mu g_{\alpha\beta}$. \im For the covariant derivative we use $\nabla_\mu$, and $\frac{D}{D\tau} \equiv u^\gamma \nabla_\gamma$ for the covariant time-derivative. 

\im We follow the convention $R_{\mu\nu}(x) = \partial_\alpha \Gamma^\alpha_{\mu\nu}(x) - \partial_\nu \Gamma^\alpha_{\alpha\mu}(x) +\cdots$, for the Ricci tensor. 

\im A locally-flat frame is described by a vierbein, $e^\mu_a(x)$, such that 
$e^\mu_a(x) e^\nu_b(x) g_{\mu\nu}(x) = \eta_{ab}$, $e^\mu_a(x) e^\nu_b(x) \eta^{ab} = g^{\mu\nu}(x)$, with $a=0,1,2,3$. A co-moving locally-flat frame is denoted as $e^\mu_A$, with $e^\mu_0 = u^\mu$.

\im We use the shortened notation $\int_{\bp, \cdots , \bq} \equiv \int \frac{d^3{\bp}}{(2\pi)^3} \cdots \frac{d^3{\bq}}{(2\pi)^3}$  and $\int_{p_0} \equiv \int \frac{dp_0}{2\pi}$. 
 
\im We denote the symmetric-trace-free electric- and magnetic-type multipole moments as $I^L$ and $J^L$, respectively, using the compact notation $L \equiv ({i_1\ldots i_\ell})$. To comply with previous literature sometimes we also employ $Q^L_E$ and $Q^L_B$. We use $Q^L$ for the multipole moments including traces. We also use the shortened notation $x^L \equiv x^{i_1}\cdots x^{i_\ell}$, $x^{ijL-2}\equiv x^i x^j x^{i_1}\cdots x^{i_{\ell-2}}$ etc.,  throughout. 

\im As it is customary in the literature the time average of a quantity is denoted as $\langle X(t)\rangle \equiv \frac{1}{T} \int_{0}^{T} dt X(t)$, whereas we use $\langle X(t) \rangle_S$ for the background expectation value on short-distance modes. (I apologize in advance if this causes confusion.)
\im We use $\langle T\left\{\cdots\right\}\rangle$ for the `time-ordered product', which in our (classical) setting is short-hand for products of the Green's functions.
\im The $n$-th order in the Post-Newtonian expansion is denoted by $n$PN $\equiv {\cal O}(v^{2n})$. 

\eit 
\newpage
 \pagenumbering{arabic}
 \section*{Preamble: Why Effective Field Theory?}
 \phantomsection
   \addcontentsline{toc}{section}{~~~~Preamble: Why Effective Field Theory?}
 \vspace{-0.1cm}
 {\it  `A physical understanding is a completely un-mathematical, imprecise, and inexact thing, but absolutely necessary for a physicist.'} -- Richard P. Feynman.\vskip 2pt \noindent The Feynman Lectures on Physics -- Volume II section 2-1, ``Understanding Physics."\vskip 6pt

One of the main goals of physics is to be able to reduce all observed phenomena down to a set of (mathematical) laws in a unified picture. However, even assuming such a description is ever achieved --or is even possible-- its inherently fundamental character would be of little use in order to describe phenomena at scales other than at the deepest layers. Throughout the years the lack of a `theory of everything' has not stopped physicists from constructing models that fit observations, put bounds on the scale (of {\it new} physics) at which the models may break down, and ultimately make predictions. Our ability to do so is rooted in basic properties of physical laws as faithful descriptions of nature.\vskip 4pt A physical theory which does not attempt to be valid at all scales is often called an `effective field theory' (EFT) or plainly an `effective theory.' A traditional example of an EFT in particle physics is Fermi's theory of weak interactions. However, we do not need to invoke the electroweak scale, since effective descriptions in nature are commonplace. For instance chemistry may be thought of as an effective theory, for it does not require the theory of quarks and leptons to describe chemical reactions, and the latter can be simply fit into a model of electrons interacting via Coulomb forces. The reader may object that, as powerful a framework as EFTs may be, it is nonetheless preferable to have a theoretical description which remains applicable for the largest possible range of scales. However, even in those cases where such a model may exist, for instance when we concentrate on a subset of interactions, it is often we find it hard --or impossible-- with our current techniques to solve for the dynamics in closed analytic form. That is the reason very sophisticated numerical tools are in constant development, for example lattice methods for the strong interaction (QCD) or simulations in structure formation. It is in these situations that finding reliable EFT descriptions is extremely valuable. That is because they allow us to get a grip on the analytic side, often providing a deeper and more systematic understanding of the dynamics. This is particularly useful when the variables in the problem ought to be scanned over a large parameter space, which is usually computationally expensive. The EFT can also be used to cross check with numerical results within the realm of overlapping validity.  Effective theories are thus a simplified, yet remarkably generic, {\it bottom up} approach which provides us with a powerful instrument to describe physics at the scales of interest.\vskip 4pt
One of the key elements in an EFT framework is the {\it decoupling} of short-distance/high-frequency physics from long-distance/low-frequency observables. The effects of the former upon the latter can be described entirely in terms of long-distance/low-frequency degrees of freedom, and {\it local} --in space {\it and} time-- interactions. The price to pay in any effective description is thus a set of unknown coefficients which are obtained from data, or comparison with a more comprehensive theory, when known. This procedure goes by the name of {\it matching}. There are many instances where decoupling is manifest in nature. The most famous\footnote{~Perhaps second only to $E=mc^2$.}\,equation in physics is an example of decoupling: \beq{ \boldsymbol{F}} = m{\boldsymbol{a}}\nn\,.\eeq  
The dynamics of a long-distance observable, namely the position of the center-of-mass of an object, ${\bx}_{\rm cm}(t)$, can be described locally, in space and time, in terms of the forces applied along its trajectory, ${\boldsymbol{F}}[{\bx}_{\rm cm}(t)]$. The parameter in this case, $m$, is the (inertial) {\it mass}. The Newtonian theory does not give us any extra information about the mass of an object. (The sum of the constituents is as far as we get about a collection of particles.) It does, however, provide a mean to {\it measure} masses using some standardized experimental set-up.  Once the mass is known, predictions about the object's motion under different influences can be made, using its {\it universal} character.\,\footnote{~It is still plausible that one day all masses in nature may be derived from first principles, in {\it natural} units. Or say the ratio between the electron's mass and the electroweak scale (Yukawa coupling). In any case, knowledge of the quantum theory of gravity is not required to build bridges.}\vskip 4pt 

It is also often the case that the force itself depends upon parameters that need to be obtained from observation, or a deeper layer, such that additional information is required to solve for the motion, e.g. the electric charge (or the `spring constant'). A celebrated counter-example --provided the equivalence principle holds--  is gravity. The dependence on the internal structure of the body drops out of the equations, and objects in an external gravitational field follow geodesic motion to very good approximation. This is often referred as the {\it effacement theorem}. This does not mean, however, that effective parameters are not required in gravity. For example, take the gravitational field produced by the sun which varies on a scale $r$ much larger than the size of earth, $r_e \ll r$. Since $r_e \neq 0$ our planet thus experiences slightly different pulls. This is the origin of tidal deformations. The response to tidal forces depend on the inner structure of the objects and, because $r_e/r \ll 1$, these can be incorporated in an EFT framework order by order in the ratio of scales. This is ultimately related to the concept of {\it power counting} in EFTs or, in other words, assessing the number of unknown parameters. Simply put, power-counting means identifying terms in a multipole expansion, or generalized dimensional analysis. At the end of the day, gradients of the gravitational field will couple to a series of multipole moments. Hence, the motion of extended bodies in various situations can be obtained once the (background) value of these moments, and the response to an external gravitational field, are known. This is the matching procedure we alluded to before, which relies on comparison with known examples and later on using this information in more complicated settings.\vskip 4pt

In this review, the dynamics of gravitationally interacting extended objects --such as a binary system or cosmological large scale structures-- will be studied within an EFT framework. In our case the matching consists on obtaining multipole moments order by order in the ratio of scales, and ultimately in terms of the dynamics of the constituents.  This will be our {\it leitmotiv} for the construction of an EFT approach to gravitational dynamics. The novel ingredient with respect to more traditional EFTs is the `method of regions' and the distinction, especially for the binary case, between potential and radiation zones. Throughout this review we will assume --based on a strong experimental and theoretical bias-- that general relativity holds at all relevant scales in our problem (as QCD does for the strong force). Nonetheless, the dynamics of extended objects in gravity is challenging, mainly due to non-linearities and the different scales involved in cases of interest. We will attempt to convince the reader that adopting an EFT framework, when possible, greatly simplifies the computations and provides the required intuition for `physical understanding.'

\newpage
\section*{Introduction}
\phantomsection
 \addcontentsline{toc}{section}{~~~~Introduction}

\vspace{-0.1cm}
{\it  An out-of-towner accidentally drives his car into a deep ditch on the side of a country road. Luckily a farmer happened by with his horse named Zoso. The man asked for help. The farmer said Zoso could pull his car out. So he backed Zoso up and hitched him to the man's car bumper. Then he yelled, ``Pull, Lucy, pull." Zoso did not move. Then he yelled, ``Come on, pull Willy." Still, Zoso did not move. Then he yelled really loud, ``Now pull, Timmy, pull hard." Zoso just stood. Then the farmer nonchalantly said, ``Okay, Zoso, pull," and Zoso pulled the car out of the ditch. The man was very appreciative but curious. He asked the farmer why he called his horse by the wrong name three times. The farmer said, ``Oh, Zoso is blind, and if he thought he was the only one pulling he would not even try."} -- Anonymous \vskip 5pt

After (precisely) one hundred years since Einstein's field equations of gravitation were published \cite{grnov}, 
\vspace{-0.3cm}
\begin{equation}
\includegraphics[width=0.4\textwidth]{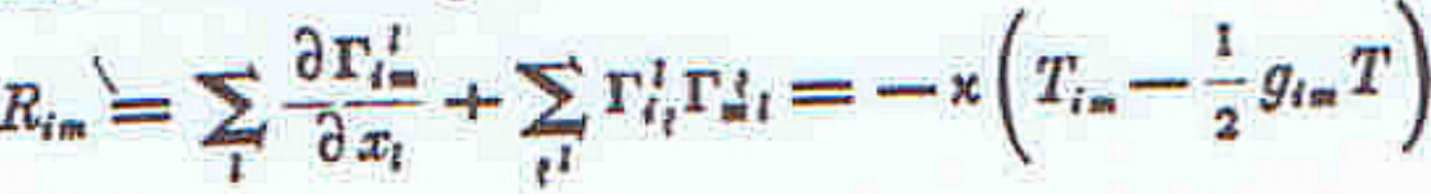}\nn \vspace{-0.42cm}
\end{equation}
Kerr's metric describing neutral, rotating black holes --found almost fifty years later \cite{kerr}-- is among the few known exact solutions in asymptotically flat spacetimes. The~lack of generic solutions to the $N$-body problem in general relativity highlights the importance of analytic and numerical methods as invaluable tools. Exact solutions are difficult to produce primarily due to the non-linear structure of the field equations, but also because of the various disparate scales involved in diverse phenomena of interest. Therefore, similarly to lattice QCD, numerical relativity has matured into a successful area of research. However, unlike the powerful EFTs for QCD, the EFT framework had not been implemented in gravity until recently.\vskip 4pt The EFT approach to the binary inspiral problem was originally proposed by Goldberger and Rothstein in the context of gravitationally bound non-rotating extended objects \cite{nrgr,towers}, and subsequently extended to spinning bodies in \cite{nrgrs,thesis}. The EFT framework was coined non-relativistic general relativity (NRGR), borrowing from similarities with EFTs for heavy quarks in QCD, e.g. NRQCD ~\cite{nrqcd,nrqcd2} and HQET~\cite{hqet0,hqet,hqet2}. While field-theoretic (and diagrammatic) techniques have been applied in gravity in the past, e.g. \cite{Gupta,Kraichnan,Feynmantree,Wsoft2,Wsoft, iwasaki2,iwasaki,hida,duff}, as well as within the Post-Newtonian (PN) expansion, e.g. \cite{oldqft1,todorov3,oldqft2,oldqft3,oldqft4}, NRGR exploits the existence of a separation of scales in the problem which makes it amenable to a novel EFT treatment~\cite{ nrgr,towers, nrgrs, thesis,dis1,dis2, andirad,andirad2,andirad3,chadbr1,andibr1}. The connection with EFTs for the strong interaction does not end with the name-tag, since bound states of heavy quarks interacting via --non-linear-- gluon exchange, and moving non-relativistically, deeply resembles the two-body problem in the PN framework. The main difference is the classical nature of our system whereas QCD is rooted in quantum effects. The classical setting, as we shall see, still shares many of the same computational hurdles, especially dealing with ultraviolet (UV) and infrared (IR) divergences.  \vskip 4pt
  
A combined numerical and analytic approach to the binary problem is of paramount importance in light of the program to observe gravitational waves, directly with the present and next generation of ground- and space-based observatories \cite{Abbott,ligovirgo,ligovirgo2,ligo16,elisa,ET,kagra,agis}, and also through pulsar timing arrays \cite{GWpulsar00,GWpulsar1,GWpulsar11,GWpulsar0,GWpulsar2} (see also \cite{lamst}). The new era of multi-messenger astronomy has began with an outstanding direct detection (from the merger of binary black holes) by Advanced LIGO \cite{ligodetect}.\footnote{\url{https://dcc.ligo.org/LIGO-P150914/public}} Following this remarkable landmark achievement, gravitational wave science will soon~turn into the study of the data to identify the properties of the sources, opening an {\it ear} to the universe which may elucidate fundamental problems in astronomy, astrophysics and cosmology~\cite{cardif,Buoreview}. 
Even though it does not~prevent detection, the~lack of sufficiently accurate templates may hinder parameter estimation and the ability to correctly map the contents of the universe. The need of a faithful template bank has thus driven the development of calculations to high degree of accuracy. While describing the merger requires numerical methods, e.g. \cite{numGR,ninja}, these do not have the power to cover the entire number of cycles during the inspiral phase, and scanning simulations over the binary's parameter space proves costly \cite{formspin3}. Therefore, perturbative (analytic) techniques remain a vital ingredient to tackle the two-body problem in general relativity.\,\footnote{Binary systems are also a natural laboratory to learn about gravity in the strong-field regime. The combination of perturbation techniques with numerical methods has provided novel insight into the structure of non-linear field equations, e.g. \cite{choptuik}, including interesting connections with high energy physics, e.g.~\cite{highnum,cardoso,loustohigh}.}  After an heroic tour de force, the gravitational wave radiation for non-rotating binary systems was completed to 3PN order more than ten years ago \cite{3pn1,3pn2,3pn3} (see e.g. \cite{Blanchet,Buoreview} for an exhaustive list of references). However, until recently most of the computations had been carried out for non-spinning constituents and using traditional methods. Albeit without the precision we envision from future gravitational wave measurements \cite{ligoparam}, recent observations of black hole spins indicate that binary systems may be frequently close to maximally rotating~\cite{McC}. It is then timely and necessary to develop more accurate templates which include spin effects, e.g.~\cite{eobspin,formspin5,formspin4,formspin2,formspin1,formspin0,Mishra:2016whh,Harry:2016ijz}.  \vskip 4pt 

NRGR has been instrumental in describing spinning compact binary systems to 3PN order \cite{eih, nrgrproc, comment, nrgrss,nrgrs2, nrgrso, srad, amps,delphine}. The leading order spin effects were computed many years ago \cite{barker1,barker2,will,kidd}. However, around the time the EFT framework was developed \cite{nrgr,towers,nrgrs,thesis} only spin-orbit terms were known to 2.5PN, corresponding to the next-to-leading order (NLO) \cite{owen,buo1,buo2} (see also \cite{damournloso}). The computations in the EFT approach then triggered a renewed interest in the community, leading to confirmation of the results obtained in \cite{eih, nrgrproc, comment, nrgrss,nrgrs2} for the spin-spin gravitational potentials at NLO (3PN). This was achieved in \cite{Schafer3pn,Schafer3pn2,Hergt:2008jn,HergtEFT,schaferEFT} using the Arnowitt-Deser-Misner (ADM) formalism \cite{steinhoffADM}, and later in \cite{bohennloss} in harmonic gauge \cite{Blanchet}. Moreover, the radiative multipole moments quadratic in the spin, originally computed in \cite{srad} in the EFT approach, have been recently re-derived in~\cite{bohennloss}. (The comparison is pending.) Subsequently, a combined effort has pushed the EFT calculations in the conservative spin sector to higher PN orders, with the computation of the NNLO gravitational spin potentials \cite{levinnlo1,levinnlo2,levinnlo3,equiv4pn}. These results were also obtained with more traditional methods~\cite{hartung, bohennloso,steinhoffnnlo1}, except for finite-size effects, which are incorporated in an EFT framework \cite{nrgrs,nrgrs2}, now generally~adopted. The EFT formalism was also used to compute the leading finite size effects cubic (and quartic) in the spin~\cite{eftvaidya,levis3,marsats3}. The~efficiency of NRGR has been equally demonstrated in the non-spinning case, with the re-derivation of the NNLO (2PN) and  NNNLO (3PN) potentials \cite{nrgr2pn,nrgr3pn}, and partial results to NNNNLO (4PN)~\cite{nrgr4pn}. The latter is in agreement with the (local part of the) complete~4PN Hamiltonian recently achieved in both ADM and harmonic coordinates \cite{4pn,4pndim,4pn1,4pn2,4pnB3,4pnDS}. (Presently, a disagreement between the results in \cite{4pn,4pndim,4pn1,4pn2} and \cite{4pnB3} has not been resolved, see \cite{4pnDS}.) At 4PN order we also find time non-locality in the effective action, due to hereditary effects \cite{4pn2,4pnB3}. This was derived in \cite{andibr1} through the study of radiation-reaction. Moreover, the rich renormalization group structure of NRGR was uncovered in~\cite{andibr1,andirad,andirad3}, naturally incorporating (and resumming) logarithmic contributions to the binding mass/energy, found in \cite{ALTlogx}.\vskip 4pt

Our main goal is therefore to provide an introduction to the EFT approach to gravitational dynamics. We~will not attempt to be fully comprehensive, since  excellent reviews of the standard lore in the two-body problem exist in the literature, with a complete set of references, e.g. \cite{Blanchet,Buoreview, ALT}. Moreover,~a~first course on NRGR can be found in~\cite{nrgrLH}, and in \cite{chadreview} with focus on radiation-reaction (see also \cite{nrgrLH2,riccardocqg,eftgrg20,iragrg}). Nevertheless, we will proceed in a self-contained fashion targeting readers without much familiarity with field-theoretic tools which are at the core of the EFT machinery. This review thus progresses in three parts. In~part~\ref{sec:part1}, we demonstrate how techniques from quantum field theory can be used in classical physics. In sec.~\ref{sec:eff0} we introduce the path integral and the saddle-point approximation. In sec.~\ref{sec:binding} we compute the binding potential perturbatively in a theory of slowly moving `point-like' sources. We~also discuss Wick's theorem and Feynman rules. In sec.~\ref{sec:rad} we introduce the multipole expansion and the long-wavelength effective action. We~derive the total radiated power loss using the optical theorem. We discuss the method of regions and the split between potential and radiation modes in sec.~\ref{methodr}. We summarize part~\ref{sec:part1} in sec.~\ref{sec:sum1}. \vskip 4pt In part~\ref{sec:part2} we describe the tower of EFTs which are needed to analyze the binary inspiral problem. We start with an introduction to the separation of scales. In sec.~\ref{sec:compact} we study the effective point-like description of compact non-rotating extended objects. We also discuss (dimensional) regularization and renormalization as well as the renormalization group flow. This is required to handle the divergences that appear in the point-particle limit. We then introduce the effective action and non-minimal couplings, together with the matching procedure. This fixes the arbitrariness in different renormalization schemes. 
We move onto NRGR in sec.~\ref{sec:nrgr} and study the binary dynamics of two non-rotating compact extended objects, including the binding energy and radiated power loss. We~introduce the long-wavelength effective action and discuss the matching for the radiative multipole moments. We review the rich renormalization structure of the theory, tail effects and the presence of logarithmic corrections. We~also discuss radiation-reaction and the interplay between potential and radiation regions. In~sec.~\ref{sec:spin} we generalize the results from previous sections to the case of spinning compact objects. We~introduce the effective action, non-minimal couplings for rotating bodies and the matching procedure. We~discuss the need of spin supplementarity conditions and the Routhian approach. We then review the computation of the gravitational potentials, and radiative multipole moments, needed to obtain all spin effects in the gravitational wave phase and waveform to 3PN and 2.5PN order, respectively. We summarize the basics of the EFT formalism in sec.~\ref{sec:sum2}. We conclude in part~\ref{sec:part3} with an introduction to EFT methods in cosmology, in particular the Lagrangian-space EFT for large scale structures. We~discuss the pitfalls of standard perturbation theory in sec.~\ref{sec:pitfalls}, the continuum limit of the EFT of extended objects in sec.~\ref{sec:continuum}, and the renormalization procedure and resummation techniques in secs. \ref{sec:LEFTren} and \ref{sec:LEFTres}, respectively. We summarize part~\ref{sec:part3} in sec.~\ref{sec:sum3}.\vskip 4pt

Since NRGR was developed, the EFT formalism has found a variety of applications (besides particle physics). For example, to study the gravitational self-force in the extreme mass ratio limit, in vacuum \cite{chad1,chad2,chad3,largeN} and non-vacuum \cite{Zimmerman} spacetimes; the thermodynamics of caged black holes \cite{cbh,andi2,kolcbh}; gravitational radiation in $d>4$ dimensions \cite{efthighd,efthighd2,efthighd3}; the radiation-reaction force in electrodynamics \cite{chadqed,replyqed}; constraints on modifications of general relativity \cite{eftmodg}; the $N$-body problem \cite{nbodychu};  Casimir forces \cite{iracmt,cmteft,cmteft2}; vortex-sound interactions \cite{vortexE} and dissipation \cite{disfluid,disfluid2} in fluid dynamics \cite{nicofluid1,nicofluid2} (see also \cite{DTSON,HongL}); and for the early universe \cite{eftinfdis,eftinfdis2,eftinf,startingN}. More recently, EFT tools have been applied to the evolution of large scale structures \cite{eftfluid,eftlss,left}, which we also cover in this review. At the end of this journey we expect the reader to appreciate how the EFT framework can be applied across length-scales and disciplines.\,\footnote{~\url{http://www.ictp-saifr.org/?page_id=9163}}

\newpage
\part{The Quantum Field Theorist's Approach to {\it Classical} Dynamics}\label{sec:part1}
We start our review introducing some of the basic elements of the EFT construction while studying the classical limit in quantum field theory. The appearance of $\hbar$ in intermediate steps solely demonstrates the range of validity of the formalism, which naturally incorporates quantum effects. In our context, however, $\hbar$ plays the role of a conversion factor which drops out of the final answer at the end of the day. That is the case because --within the saddle point approximation-- it enters as an overall multiplicative constant in front of the action. Therefore, unless otherwise noted, we work in $\hbar=1$ units. The reader may be puzzled about the soon-to-appear UV divergences in purely classical computations. As we shall see in detail, these arise from a point-like approximation for the sources. There is no need to invoke quantum mechanics to introduce a cutoff to our ignorance on the short-distance dynamics! (In addition, IR divergences are also present. For the case of gravity, these will be related to the radiation problem in part~\ref{sec:part2}.) We restrict our analysis to scalar fields in $d=4$ spacetime dimensions. We study a linear static case first and then allow for time-dependence and non-linearities. At the end of the chapter we overview the method of regions, and the separation between potential and radiation modes. We will return to these concepts later on when we review the binary problem in part~\ref{sec:part2}. We draw (heavily) from Coleman \cite{coleman},  Rothstein \cite{iraeft} and Zee \cite{zee}, which we recommend emphatically for more details, also~e.g.~\cite{eftjoe,georgi,eftanesh,eftkaplan}.

\section{Path-Integral}\label{sec:eff0}
In classical mechanics the dynamics of the system is encoded in the action principle. In quantum field theory the action is the main actor of the path-integral, which is defined as a functional integral:
\beq
\label{zj}
Z[J] \equiv \int D\phi~e^{i S[\phi,J]}.
\eeq  
Here $S[\phi,J]$ represents the action for a set of fields, $\phi(x)$, coupled to external sources, $J(x)$. For macroscopic/classical objects, the factor of $S[\phi,J] \gg 1$ leads to rapid oscillatory behavior. The path-integral is then dominated by the `saddle-point'
\beq
\label{zjphj}
Z[J] \simeq e^{i S[\phi=\phi_J,J]},
\eeq
where $\phi_J(x)$ is a solution that minimizes the action,
\beq
\label{eom1}
\left.\frac{\delta S[\phi,J]}{\delta \phi(x)}\right|_{\phi \to \phi_J} = 0\, .
\eeq
It is useful also to introduce \beq \label{wj} W[J] \equiv -i  \log Z[J]\,.\eeq
Notice in the classical limit we have $W[J] \to S[\phi_J,J]$.\vskip 4pt

For simplicity, let us concentrate on a single massless scalar field. The action reads,
\beq
S[\phi,J] =  \int d^4x ~\left(-\frac{1}{2} \phi (x)\partial^2\phi (x) -V(\phi) + J(x)\phi(x)\right)\,.
\eeq
If we furthermore turn off self-interactions, $V(\phi)=0$, a solution to the field equations from \eqref{eom1} can be written as
\beq
\label{phiJ}
\phi_J(x) = \phi_{J=0}(x) + i \int d^4  y \Delta_{\rm F}(x -  y) J( y)\,,
\eeq
where $\phi_{J=0}(x)$ solves the Klein-Gordon equation (with $J(x)=0$). We will only keep the source term in what follows. The `propagator,' or Green's function, is given by
\beq
\label{eq:DFprop}
\Delta_{\rm F}(x-y) \equiv \int_{\bp} \int_{p_0} \frac{i}{p_0^2-{\bp}^2 + i\epsilon}e^{-ip_0(x_0-y_0)} e^{i{\bp}\cdot({\bx}-{\by})}\,. 
\eeq
The $i\epsilon$ is a choice of boundary condition, also known as Feynman's prescription. Notice it only matters when the momenta goes `on-shell,' $p_0^2=\bp^2$. We will return to this point in sec.~\ref{sec:rad}.\vskip 4pt From \eqref{phiJ} we obtain
\beq
\label{Wfree}  
S[\phi=\phi_J,J] \to \frac{i}{2} \int d^4x d^4 y J(x) \Delta_{\rm F}(x-y) J(y) = W[J]\,.
\eeq
We will show later on how to include non-linearities. The functional $W[J]$ will be the most important object in the development of a --classical-- EFT approach.\vskip 4pt Even though we are using the path integral to define $Z[J]$, the latter is simply obtained by inserting the solution to the classical field equations back into the action. It is nonetheless useful to retain the functional integral, and moments thereof, as a compact way to organize the perturbative expansion. In~particular, from (in Euclidean space, $x^0 \to -ix_4$) 
\beq
\int dx_1\cdots dx_n {\rm exp}\left(-\frac{1}{2}\sum_{i,j} x_i A_{ij} x_j + \sum_i J_ix_i\right) = \frac{(2\pi)^{n/2}}{\sqrt{\det A}} {\rm exp}\left(\frac{1}{2}\sum_{i,j} J_i (A^{-1})_{ij}J_j\right)\,,
\eeq
 it is easy to show we obtain the previous results, after identifying the matrix $A$ with the Laplacian operator in four dimensions.\,\footnote{~In this language the $i\epsilon$-prescription is forced upon us to avoid poles in the complex plane after analytic continuation from Euclidean to Minkowskian space. This leads to Feynman's propagator. Moreover, the determinant does not play a role in the classical limit.} The expression in \eqref{Wfree} is exact for scalar fields coupled only to external sources. That will not be the case for a self-interacting theory.\vskip 4pt  Notice we can read off the propagator from the functional derivative (with normalization~$Z[0]=1$)
\beq
\Delta_{\rm F}(x-y) = -i \left.\frac{\delta^2 W[J]}{\delta J(x)\delta J(y)}~\right|_{J=0} =(-i)^2\left.\frac{\delta^2 Z[J]}{\delta J(x)\delta J(y)}~\right|_{J=0}\,. 
\label{torder}
\eeq
This will be useful later on to set up the perturbative approach once non-linearities are included.

\section{Binding Potential}\label{sec:binding}

\subsection{Static Sources}\label{sec:static}
\vskip 4pt
Let us consider first static point-like sources, e.g.
\beq \label{j12} J({\bx}) = J_1({\bx}) + J_2 ({\bx}) \equiv \frac{1}{M_\phi} \Big[m_1 \delta^3({\bx}-{\bx}_1) + m_2 \delta^3({\bx}-{\bx}_2)\Big]\, , \eeq 
with $M_\phi$ a mass scale related to the strength of the coupling to $\phi$. For off-shell configurations we can ignore the $i\epsilon$ in the propagators altogether. Therefore, the expression in \eqref{Wfree} becomes
\beq
\label{eq:deltapo}
W[J] = \left(\int dt\right) \frac{1}{2} \int d^3{\bx}d^3{\bx}' J({\bx}) J({\bx'})  \int_{\bp,p_0}  \frac{-1}{p^2_0-{\bp}^2} \delta(p_0) e^{i{\bp}\cdot({\bx}-{\bx}')}\,.
\eeq
From $W[J]$ we can identify the binding potential, $V[J]$, via\,\footnote{~Recall $W[J] = S[\phi_J,J]$, which then serves as an (effective) action. Moreover, the path integral is related to the time-evolution operator, $U = \text{exp} \left(-i\int dt \, V \right)$, which we can also associate with the binding potential \cite{iraeft,zee}.}  
\beq
\label{eq:bindE0}
W[J] \to -\int_{t_{\rm in}}^{t_{\rm out}} dt~V[J] \,,
\eeq
for $t_{\rm out (in)} \to +\infty (-\infty)$. Hence, using
\beq
\int_{\bp} \frac{1}{{\bp}^2} e^{-i{\bp}\cdot\br} = \frac{1}{4\pi r}\,,
\eeq
we obtain
\beq
\label{eq:bindE}
W[J] = \left(\int dt\right) \frac{m_1m_2}{4\pi M_\phi^2}\frac{1}{r} \,\,\to\,\, V[J] = - \frac{m_1m_2}{4\pi M_\phi^2}\frac{1}{r}\, ,
\eeq
with $\br \equiv {\bx}_1-{\bx}_2$. We recognize in $V[J]$ the --Coulomb-like-- binding energy for point-like sources interacting via a massless scalar field. Apart from this term, we also find `self-energy' contributions. These occur from products of the sources at the same point,  \beq \int d^3{\bx} d^3{\by}~ \delta^3(\bx-\bx_1(t))\Delta_{\rm F}({\bx}-{\by},t) \delta^3(\by-\bx_1(t)) = \Delta_{\rm F}(0,t) \propto \int_{\bp} \frac{1}{{\bp}^2} \, . \eeq 
Introducing a UV cutoff for the divergent integral, it is easy to show that its contribution can be absorbed into the mass coupling(s) in \eqref{j12}. This is often referred as adding a `counter-term.' We may instead use dimensional regularization (dim. reg.), which sets to zero scale-less integrals as the one above (provided the IR singularities are properly handled \cite{iraeft}). In dim. reg. we can therefore completely ignore these terms. We will discuss regularization in more detail later on in sec.~\ref{sec:regular0} of part~\ref{sec:part2}~.

\subsection{Time-Dependent Sources}\label{sec:slow}
\vskip 4pt
It is straightforward to generalize the previous results to the case of time-dependent sources, 
\beq
\label{jtx}
J(t,{\bx}) = \sum_{a=1,2} \frac{m_a}{M_\phi} \delta^3({\bx}-{\bx}_a(t))\, .
\eeq
\begin{figure}[t!]
\centering
\includegraphics[width=0.4\textwidth]{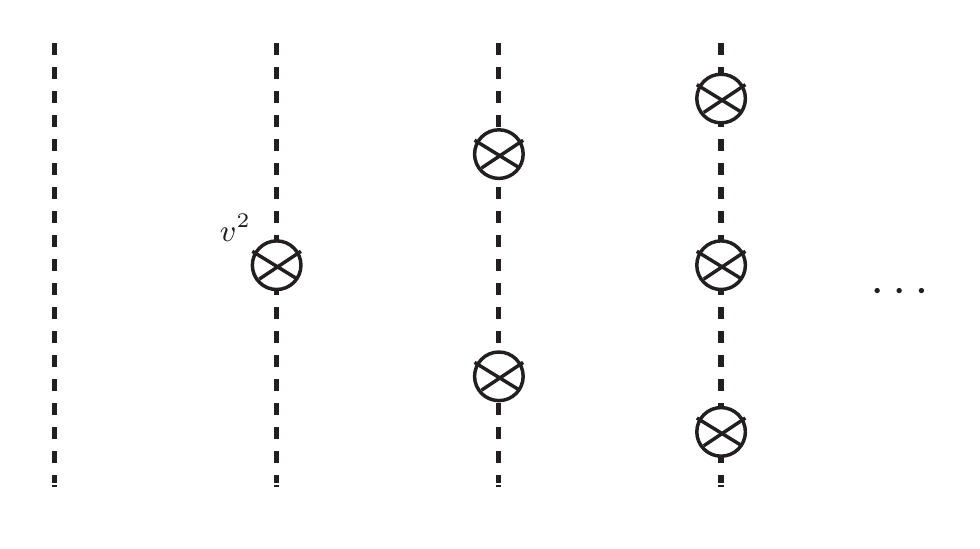}
\caption{The dashed line represents the static propagator, with $p_0=0$. In the subsequent diagrams each cross represents an insertion of a factor of $p_0^2/\bp^2$, see \eqref{expP}.}
\label{fig1}
\end{figure}
The difference is the time integrals, which no longer lead to $\delta(p_0)$ as in \eqref{eq:deltapo}. However, we can still use quasi-instantaneous interactions, provided the sources move {\it slowly}\,: $|\bv_a| \equiv |\dot \bx_a| \ll 1$. Hence, we may expand the (off-shell) Green's function in powers of~$p_0/|\bp|$,
\beq
\label{expP}
\frac{1}{p_0^2-{\bp}^2} \simeq -\frac{1}{{\bp}^2} \left(1 + \frac{p_0^2}{{\bp}^2} +\cdots\right)\,.
\eeq
This expansion is shown diagrammatically in Fig.~\ref{fig1}. The leading term corresponds to
\beq
\label{inverselaw}
W_{(0)} [J] = \frac{m_1m_2}{4\pi M_\phi^2} \int \frac{dt}{|\br (t)|}\,,
\eeq
as we just computed, and the first correction reads
\begin{align}
\label{Sv2}
W_{(v^2)} [J] &=  \frac{m_1 m_2}{M_\phi^2} \int dt dt' \int_{\bp,p_0}  \frac{p_0^2}{{\bp}^4} e^{-ip_0(t-t')}e^{i{\bp}\cdot({\bx}_1(t)-{\bx}_2(t'))} 
= \frac{m_1 m_2}{M_\phi^2} \int dt ~  {\bv}^i_1(t) {\bv}^j_2(t) \int_{\bp} \frac{{\bp}^i{\bp}^j}{{\bp}^4} e^{i{\bp}\cdot {\br}(t)}
\,,
\end{align}
such that \beq
V_{(v^2)} [J]=  \frac{m_1 m_2}{8\pi M_\phi^2} \frac{1}{|\br(t)|^3} \Big[ |\br(t)|^2\left({\bv}_1(t)\cdot  {\bv}_2(t)\right)  -\left(\bv_1(t) \cdot \br(t)\right)\left({\bv}_2(t)\cdot \br(t)\right)\Big]\, , \nn 
\eeq
where we used
\beq
 \int_{\bp} \frac{{\bp}^i {\bp}^j}{{\bp}^4} e^{-i {\bp}\cdot{\br}} = \frac{1}{8\pi r^3} \left( r^2\delta_{ij} - {\br}^i {\br}^j\right).
\eeq
The higher order corrections are computed in a similar fashion. Notice that the Taylor series in \eqref{expP} is performed inside the integral. The validity of this expansion relies on the `method of regions,' which we review in sec.~\ref{methodr}.\,\footnote{~Intuitively, when computing the binding energy we do not expect to hit singularities in the propagators, which --due to unitarity-- are related to on-shell modes and radiation.}

\subsection{Non-Linearities}\label{sec:nonlinear1}
\vskip 4pt
Once we add self-interactions the field equations become difficult to solve in closed analytic form, and we must rely on perturbative techniques in a small coupling expansion or numerical methods. Here we discuss the former, and for illustrative purposes we consider a cubic potential, $V(\phi) = \lambda \phi^3$.\vskip 4pt  
It~turns out adding a cubic term into the action leads to IR divergences for a massless field, e.g.~\cite{nrgrLH2}. Nonetheless, we proceed to study this term and introduce regulators to tame the singular integrals. We~will assume $\lambda$ is {\it small} in a sense we will later specify. The field equations are
\beq
\partial^2 \phi (x) = J(x) - 3\lambda \phi^2(x)\,.
\eeq
The idea is to solve for $\phi_J$ in powers of $\lambda$, with the ansatz 
\beq
\label{ansatz}
\phi_J (x) = \phi^{\lambda=0}_J (x)+ \phi^{\lambda}_J (x)+ \cdots+ \phi^{\lambda^n}_J (x)+\cdots\, .
\eeq
For simplicity, in what follows we assume static sources, as in \eqref{j12}. Then, at first order in $\lambda$ we have
\beq
\partial^2 \phi^{\lambda}_J ({\bx}) = J_\lambda(\bx)\, , ~\text{with}~~J_\lambda(\bx) \equiv -3\lambda \left(\phi^{\lambda=0}_J\right)^2(\bx)\,.
\eeq
To solve for $\phi^\lambda_J$ we use the Green's function, obtaining
\beq
\label{phiJtx}
\phi^\lambda_J({\bx}) = 3i \lambda \int d^3 \by d^3\bz d^3\bw ~\Delta_{\rm F}(\bx - \by) \Delta_{\rm F}(\by-\bz) \Delta_{\rm F}(\by-\bw) J (\bz) J(\bw)\, . 
\eeq
For instance, we find contributions sourced from particle~1 (in Fourier space)
\beq
\label{phij3n}
\phi^\lambda_J({\bk}) =  -3 \lambda \frac{m_1^2}{M_\phi^2} \frac{e^{i{\bk}\cdot{\bx}_1}}{{\bk}^2} \int_{\bq} \frac{1}{({\bk}+{\bq})^2{\bq}^2}  +\cdots =
-3\lambda \frac{m_1^2}{8M_\phi^2} \frac{e^{i{\bk}\cdot{\bx}_1}}{{|\bk|}^3} + \cdots\, , 
\eeq
where we used
\beq
\int_\bq \frac{1}{{\bq}^2({\bk}+{\bq})^2} = \frac{1}{8|{\bk}|}.
\eeq
To compute $W_{(\lambda)}[J]$ we plug $\phi^\lambda_J$ into the action. For example, we find
\beq
\label{halfjphi}
-\frac{1}{2} \phi^\lambda_J \partial^2 \phi^{\lambda=0}_J \to -\frac{1}{2} J \phi^\lambda_J\,,
\eeq
which together with the source term, and using \eqref{eq:bindE0}, gives 
\beq
\label{jphi3}
V_{(\lambda)}[J] = - \frac{1}{2} \int d^3{\bx}  J({\bx})\phi^\lambda_J(\bx) + \cdots = 3\lambda {m_1^2 m_2 \over 64 \pi^2 M_\phi^3} \log \left(\mu r \right) + 1\leftrightarrow 2\cdots\,.
\eeq
Here $\mu$ is introduced as an IR regulator, for instance a scalar mass. The logarithmic potential produces a long-range force scaling as $1/r$ \cite{nrgrLH2}. The procedure continues to all orders in $\lambda$.  Extending these manipulations to the case of non-static sources proceeds as before, see Fig.~\ref{fig1}. \vskip 4pt Notice we will once again run into divergences. For instance, there is a contribution given by the cubic potential evaluated on the unperturbed solution, $\lambda \left(\phi^{\lambda=0}_J\right)^3$. This produces a divergent integral, which represents the self-energy in the scalar field produced by a point-like object,
\beq
\label{selfphi}
\lambda \frac{m^3}{M_\phi^3} \int d^3{\bx}~\frac{1}{|{\bx}_1-{\bx}|^3} \propto \int \frac{dr}{r} =  \log \Lambda/\mu\,.
\eeq
Note in addition to the IR regulator we introduced a short-distance cutoff $\Lambda^{-1}$. As we mentioned, the dependence on the UV cutoff can be absorbed into the couplings of the theory, whereas the IR singularities cancel out after all the long-distance effects are properly incorporated~\cite{jackiw,nrgrLH2}. (Alternatively, we can use dim. reg., although in this case one needs to be careful regarding the appearance of IR poles, e.g. \cite{iraeft}.) \vskip 4pt

Before concluding let us add a few comments which may also help understand some features that appear later on in the gravitational setting. The force which derives from the logarithmic potential in (\ref{jphi3}) could be present in nature, as a fifth force. It can then be contrasted against the gravitational force produced from the cubic coupling in Einstein's theory (see sec.~\ref{sec:EIH}). For comparable masses, we have
\beq
\frac{F_{\lambda \phi^3}}{F_{\rm grav}} \sim \frac{\lambda m^3/(\Mp^3 r) } {m^3/(\Mp^4r^3)} \sim \lambda \Mp r^2,\,
\eeq
(after identifying $M_\phi \to \Mp$). This implies that in order to have a well defined perturbative expansion in the scalar case, and also comply with precision tests in the solar system, we need $\lambda \lesssim 1/(\Mp r^2)$. This turns out to be a rather tiny coupling, as it was argued in \cite{nrgrLH2}, suggesting (universal) non-derivatively coupled scalar long-range interactions are highly constrained in nature. The main difference with the cubic coupling in gravity is two extra derivatives, dictated by the equivalence principle. The derivatives help remediate the IR singularities, while introducing a more prominent UV behavior. As we shall see, at the same time this allows us to set up a well defined derivative expansion in powers of $m/(\Mp^2r)$, plus the addition of counter-terms to absorb  UV divergences. We elaborate on the regularization/renormalization procedures in part~\ref{sec:part2}. \vskip 4pt
At this point it is clear that a diagrammatic approach would greatly simplify the account of all possible terms contributing to $W[J]$. This requires the use of Wick's theorem, which we introduce next.

\subsection{Wick's Theorem}\label{app:Wick}
\vskip 4pt

A compact way to organize our computations is to use the path-integral approach, in which case the main obstacle comes from integrals of the sort 
\beq
\label{pathga}
Z[J] = \int dx_1\cdots dx_n ~{\rm exp}\left(-\sum_{i,j} x_i A_{ij} x_j - \lambda \sum_{ijk} B_{ijk} x_i x_j x_k  + \cdots + \sum_i J_ix_i\right).
\eeq
While closed analytic expressions are not known, it is straightforward to develop a perturbative approach to solve for $Z[J]$. Let us first look at a simple example \cite{zee}. Let us consider a $d=1$ model, with
\beq
\label{lx3}
z[J,\lambda] = \int dx e^{-\frac{a}{2}x^2-\lambda x^3 + Jx}\,,
\eeq
which resembles \eqref{pathga}. In what follows we drag the $\lambda$-dependence explicitly for illustration purposes.\vskip 4pt 

Let us start with just the source term.  In this toy example the `propagator' follows from expanding $z[J,\lambda=0]$ to second order in $J$ and evaluating at $J=0$ ($z_0\equiv z[J=0,\lambda=0]$)
\beq
\langle x^2 \rangle = \frac{1}{z_0} \left.\frac{\delta^2 z[J,\lambda=0]}{\delta J^2}\right|_{J=0} = \frac{ \int dx~x^2~ e^{-\frac{a}{2} x^2}}{ \int dx~ e^{-\frac{a}{2} x^2}} = a^{-1}\,.
\eeq
The expressions for the higher moments can be easily obtained by further differentiating with respect to~$J$, 
\beq
\langle x^{2n}\rangle = \frac{ \int dx~x^{2n}~e^{-\frac{a}{2} x^2}}{ \int dx ~e^{-\frac{a}{2} x^2}} = a^{-n} (2n-1)!!\,,
\eeq
where $(2n-1)!!= (2n-1)\times (2n-3)\times \cdots \times 5 \times 3 \times 1$.  To remember this we simply imagine $2n$ distinct points connected in pairs \cite{zee}. Wick's theorem is then represented as follows, say for the four-point function,
\begin{align}
\langle x^4 \rangle &= \langle x_1 x_2 x_3 x_4 \rangle = \langle x_1 x_2 \rangle \langle x_3 x_4 \rangle +  \langle x_1 x_3 \rangle \langle x_2 x_4 \rangle +  \langle x_1 x_4 \rangle \langle x_3 x_4 \rangle  =  \frac{1}{a^2} (4-1)\times(4-3), 
\end{align}
and each one of these terms is called a `Wick contraction'. In general we will have field variables evaluated at different spacetime points, and each factor of $1/a$ is replaced by a propagator (with normalization~$Z[0]=1$)
\begin{align}
\label{tproduct}
\langle T \left\{\phi(x_1)\phi(x_2)\right\}\rangle_{\lambda=0}  \equiv \int D\phi\,\phi(x_1)\phi(x_2)\,e^{i S[\phi, \lambda=0]} = (-i)^2\left.\frac{\delta^2 Z[J,\lambda=0]}{\delta J(x_1)\delta J(x_2)}\right|_{J=0} = \Delta_{\rm F}(x_1-x_2)\,,
\end{align}
where $\left\langle T\left\{\cdots\right\}\right\rangle$ stands for `time-ordering.'\,\footnote{~In quantum mechanics the order matters, since operators in general do not commute if causally connected (in the Heisenberg picture). This raises the question of which order to choose for Wick's theorem. The time-ordered product, defined as $T\left\{\phi(x_1) \phi(x_2)\right\} \equiv \phi(x_1) \phi(x_2)\theta(t_1-t_2) + \phi(x_2) \phi(x_1)\theta(t_2-t_1)$, is the correct answer. This is ultimately related to the choice of $i\epsilon$-prescription and Feynman's propagator. Let us emphasize, however, we do not need to invoke quantum mechanics to write Wick's theorem, since it is nothing but a consequence of computing moments of a Gaussian integral.} We will use the notation $\left\langle T\left\{\cdots\right\}\right\rangle$ throughout this review to denote different Wick contractions. For the case of the two-point function, it is defined by the right-hand side in \eqref{tproduct}, extended to all orders in the perturbation theory. The reader should keep in mind that, in general, it is shorthand for products of Feynman propagators. These $n$-point functions will be our building blocks for the perturbative expansion. For example,
\begin{align}
&\big\langle T \big\{\phi(x_1)\phi(x_2)\phi(x_3)\phi(x_4) \big\}\big\rangle_{\lambda=0} = \int D\phi~\phi(x_1)\phi(x_2)\phi(x_3)\phi(x_4)~e^{i S[\phi,\lambda=0]} =  \\ & \Delta_{\rm F}(x_1-x_2) \Delta_{\rm F}(x_3-x_4) + \Delta_{\rm F}(x_1-x_3) \Delta_{\rm F}(x_2-x_4) + \Delta_{\rm F}(x_1-x_4) \Delta_{\rm F}(x_2-x_4)\, .\nn
\end{align}

Let us now return to the one-dimensional case and turn on the cubic interaction, $\lambda x^3$. The trick is to expand $z[J,\lambda]$ around the linear theory, that is
\beq
z[J,\lambda] = \sum_n \frac{1}{n!} \left. \frac{\partial^n z[J,\lambda]}{\partial\lambda^n}\right|_{\lambda=0}\lambda^n\,,
\eeq
and using \eqref{lx3}, we have (after redefining $z[J,\lambda] \to z[J,\lambda]/z_0$)
\beq
z[J,\lambda] = \sum_n \frac{1}{n!} \lambda^n \langle(-x)^{3n}\rangle_{(J,\lambda=0)}\, .
\eeq
We now expand in powers of the source. The final product reads:
\beq
z[J,\lambda] = \sum_{n,l} \frac{1}{n!}\frac{1}{l!} \left. \frac{\partial^n \partial^l z[J,\lambda]}{\partial\lambda^n\partial J^l}\right|_{\lambda=0,J=0} J^l\lambda^n =  \sum_{n,l} (-1)^{n} J^l\lambda^n \langle x^{3n+l}\rangle_{(J=0,\lambda=0)}. 
 \eeq
For the case of fields and in Minkowski space (that is putting the $i$'s back in), we have 
\begin{align}
\label{zljall}
&Z[J,\lambda] = \sum_{n,l} \frac{(-i\lambda)^n}{n!} \frac{i^l}{l!} \int  d^4x_1 \cdots d^4x_l J(x_1)\cdots J(x_l) \times  \\ &  
\left\langle T  \left\{ \frac{}{} \phi(x_1)\cdots \phi(x_l)
 \left(\int d^4y_1  \phi^3(y_1)\cdots \int d^4y_n \phi^3(y_n)\right)\right\} \right\rangle_{(J=0,\lambda=0)}\nn\, . 
\end{align}
To compute each one of these moments of a Gaussian integral we make intensive use of Wick's theorem. 

\subsection{Feynman Diagrams and Rules}\label{Frules}
\vskip 4pt

We are now in a position to develop a full fledged diagrammatic approach. As in Fig.~\ref{fig1}, we use dashed lines to represent the static propagator. We notice that propagators are either integrated against sources or contracted with each leg of the $\lambda \phi^3$ self-interaction. We also have an overall spacetime integral which guarantees momentum is conserved at each vertex. The Feynman rules are summarized below in Fig.~\ref{fig2PR}
\begin{figure}[h!]
\centering
\includegraphics[width=0.7\textwidth]{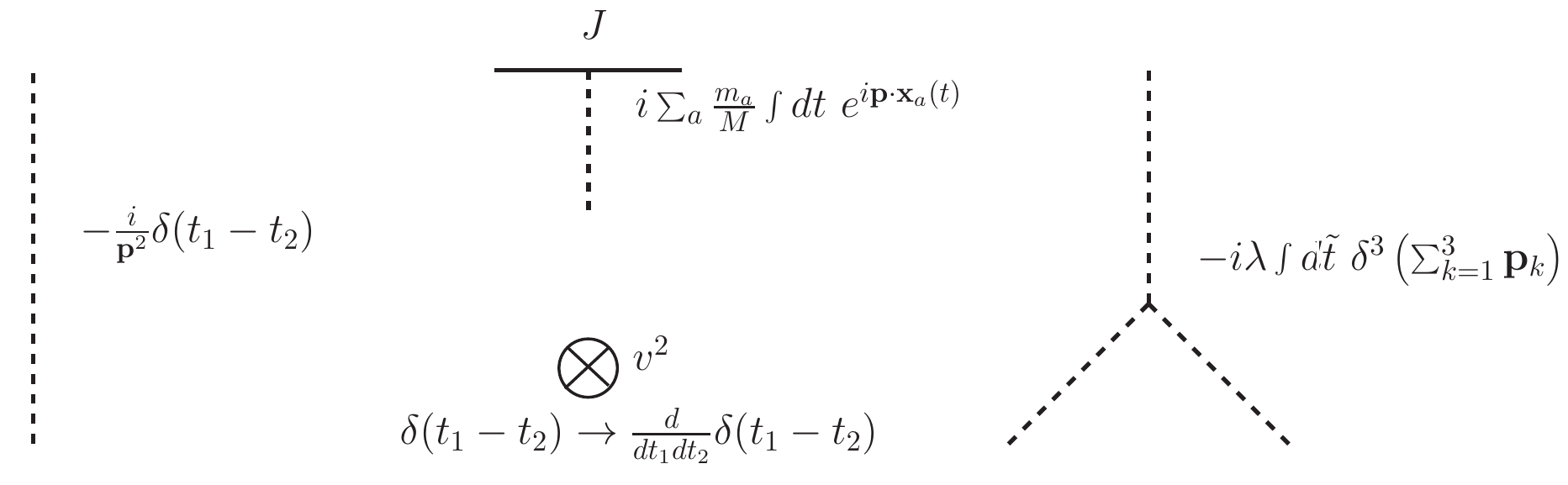}
\caption{Feynman rules. The solid line represents point-like external sources which {\it do not} propagate. It is depicted in this fashion for historical reasons, e.g. \cite{nrgr}.}
\label{fig2PR}
\end{figure}
\bit
\im {\it Propagator}: Include a factor of $\frac{-i}{{\bp}^2} \delta(t_1-t_2)$ for each dashed line connecting two points in the diagram.
\im {\it Non-instantaneity}: Replace the factor of $\delta(t_1-t_2)$ by $\frac{d^2}{dt_1dt_2} \delta(t_1-t_2)$ for one of the (two) propagator(s) connected by a cross. 
\im {\it Point-like sources}: Include a factor of $i \sum_{a=1,2} \frac{m_a}{M_\phi} \int dt~ e^{i{\bp}\cdot {\bx}_a(t)}$, or $i \sum_{a=1,2}\frac{m_a}{M_\phi} \int dt~\delta({\bx}-{\bx}_a(t))$ in coordinate space, for each propagator ending in a source. 
\im {\it Vertex}: Include a factor of $-i \lambda \int d \tilde t~ \delta^3\left(\sum_{i=1}^{i=3}{\bp}_k\right)$ for each $\phi^3$-vertex. This guarantees
conservation of momenta, with ${\bp}_i$ incoming. 
\eit

 At this point one may ask: What are the diagrams needed to compute $Z[J]$, and which for $W[J]$? Moreover, what type of diagrams contribute in the classical limit? As we show next, these questions turn out to be related. 
 
 \subsection{Tree-level and Connected}\label{treeL}
\vskip 4pt
 
 Let us return to the linear theory for which we know the exact result. Let us start considering the case of static sources. Then, only one-scalar diagrams are needed and, furthermore, it is straightforward to show the sum exponentiates, \beq
 \nn
\includegraphics[width=0.67\textwidth]{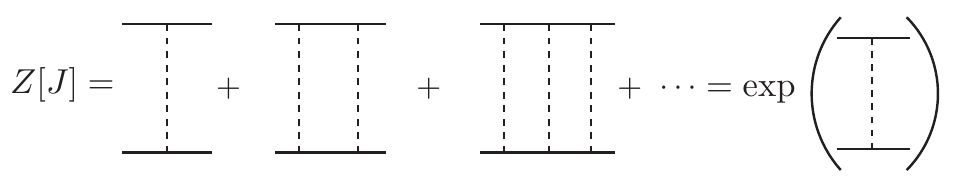}
\eeq
with
\beq
\label{sumz}
W[J] = - i \log Z[J] = \frac{1}{2}\left(\int dt\right)\int d^3\bx d^3\by~J(\bx)\left(i\Delta_{\rm F}(\bx-\by)\right) J(\by)\, .
\eeq
Recall the solid lines represent external sources which do not propagate. Therefore the computation of $Z[J]$ is simply a combinatorial problem. In a theory without self-interactions, $W[J]$ is thus given solely in terms of the one-scalar exchange. This is also what we found in sec.~\ref{sec:static} in the static limit. Furthermore, it is evident that adding non-instantaneous corrections does not modify this result. The expression for $W[J]$ is given in \eqref{Wfree}. Then, we conclude that while a series of diagrams enter in $Z[J]$, only {\it tree-level connected} diagrams contribute~to~$W[J]$. In our context,  a `connected' diagram cannot be split into pieces if we cut off the source `worldines'. On the other hand, `tree-level' means there are no diagrams with closed `loops' after the worldlines are removed. (The loops are responsible for quantum effects.) \vskip 4pt

In the linear theory these statements are easy to show, since only one-scalar exchanges contribute to $Z[J]$ when $\lambda=0$. However, this turns out to be true also after switching on self-interactions \cite{zee}. Hence, in layman terms, to compute $W[J]$ all we need is the sum of tree-level connected diagrams. As an example, let us calculate $W[J]$ to order ${\cal O}(\lambda)$ for static configurations. This results from the three-scalar vertex coupled to three sources, 
\begin{align}
W_{\lambda}[J] &= -i\frac{\lambda m_1^2 m_2}{2 M_\phi^3} \int dt_1 dt_2 dt_3 dt_y d^3{\by} \left\langle T \Big\{ \frac{}{} \phi(t_1,{\bx}_1)\phi(t_2,{\bx}_2)\phi(t_3,{\bx}_1) \phi^3(t_{y},{\by})\Big\} \right\rangle \, ,\label{zlj3a}
\end{align}
plus mirror image. Some of the Wick contractions produce divergences which are removed by counter-terms, and we obtain 
\begin{align}
\label{zlj3n}
V_{\lambda}[J] =3\frac{\lambda m_1^2 m_2}{2 M_\phi^3}  \int_{\bk,\bq}  \frac{e^{i{\bk}\cdot ({\bx}_1-{\bx}_2)}}{{\bq}^2{\bk}^2({\bq}+{\bk})^2}  = 3\frac{\lambda m_1^2 m_2}{16  M_\phi^3}   \int_\bk \frac{e^{i{\bk}\cdot ({\bx}_1-{\bx}_2)}}{|{\bk}|^3} + 1\leftrightarrow 2 \,.
\end{align}
This result agrees with \eqref{jphi3}.

\subsection{Eikonal Approximation}\label{sec:eik}
\vskip 4pt

The relation between $Z[J]$ and $W[J]$ (inserting $\hbar$ for illustrative purposes)
\beq
\label{eq:zjsum}
Z[J] = \sum_{n}^{\infty} \frac{\left\{\frac{i}{\hbar}W[J]\right\}^n}{n!}\,\,,
\eeq
entails a series of terms which build up the exponential in \eqref{wj}. In the classical limit we expect this sum to involve a large number of terms. Indeed, from \eqref{zjphj} and using Stirling's approximation, $e^N \simeq  \frac{N^{N+1/2}}{N!}$ for $N \gg 1$, we notice $Z[J]$ is dominated by~\cite{sgrav,erice} \beq n \simeq N \equiv S[\phi_J]/ \hbar \gg 1\,.\eeq We may then interpret the series of $N$ one-scalar exchanges as building blocks for the classical field, $\phi_J(t,{\bx})$. While the typical momentum exchanged on each {\it kick} is of order $|{\bq}| \sim \frac{\hbar}{r}$, the total momentum {\it transferred} due to the force induced by the binding potential is much (much) bigger, \beq \Delta {\bp} \simeq N \frac{\hbar}{r} \simeq mv \to {|{\bq}| \over |\Delta {\bp}|} \sim {\hbar \over L} \ll 1\,,\eeq with $L = mvr$ the angular momentum. Consequently,
\beq
S[\phi_J] \simeq \int \Delta {\bp} \,d {\bx} \simeq L \gg \hbar\,.
\eeq 
Therefore, in our effective theory macroscopical objects may be treated as `non-propagating' sources \cite{nrgr}. This is tantamount of working in the so called {\it eikonal approximation}. In this approximation the full propagator for a particle described by a field, e.g. a heavy fermion of mass $m_\psi$, is expanded as
\beq
\label{eq:eikap}
\frac{i}{\not{p}-m_\psi+i\epsilon} = \frac{i}{v\cdot p+i\epsilon}\left(1+{\not{v} \over 2}\right) + {\cal O}\left(\frac{1}{m_\psi}\right)\,,
\eeq
where ${\not p} \equiv \gamma_\mu p^\mu$, with $\gamma_\mu$ the Dirac matrices and $\left(1+{\not{v} \over 2}\right)$ is a positive-energy projector, e.g. \cite{zee}. In~coordinate space, and using 
\beq \theta(x) =  \frac{i}{2\pi} \int dt \frac{e^{-ixt}}{t+i\epsilon}\,,\eeq
for a representation of the step-function, the propagator in the rest frame of the particle takes the form
\beq
\label{propx}
\Delta_X(x-y) = \int_\bp e^{i{\bp}\cdot({\bx}-{\by})} \int_{p_0} \frac{i}{p_0+i\epsilon} e^{-ip_0(x^0-y^0)} = \delta^3({\bx}-{\by}) \theta(x^0-y^0)\, .
\eeq
The function $Z[J]$ then follows by computing a series of ladder (and `crossed ladder') diagrams, 
\beq
\includegraphics[width=0.48\textwidth]{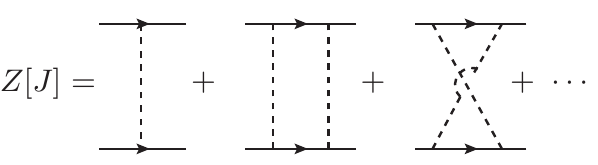}
\eeq
were the external sources are now replaced by a `propagating' field, given by \eqref{propx}. Notice that since the field now `propagates' (described by the arrow on the solid line), we have to include the required crossed ladder diagrams, which are not present for non-propagating sources. 

It is now easy to show that the heavy fermion line(s) behave like 
\beq
\delta^3({\bx}-{\by}) \theta(x^0-y^0) + \delta^3({\by}-{\bx}) \theta(y^0-x^0) = \delta^3({\bx}-{\by})\times \mathbb{I}\,.
\eeq
Therefore, the particle {\it effectively} serves as a localized source which remains unaltered by the interaction with field modes of very soft momenta. In this respect our EFT approach \cite{nrgr} shares common ground with effective theories for heavy quarks in QCD, e.g. HQET \cite{hqet0,hqet,hqet2}. However, while in HQET quantum effects are important, and parameterized in powers of $1/m_\psi$, in NRGR we never depart from the classical regime. Let us do some numerology. For solar-mass black holes with typical velocities of a percent of the speed of light, $v\sim 0.01$, and of the order of hundreds of Km's apart (also known as the LIGO band), we have \beq L = (10^{30}{\rm Kg}) \times (10^5{\rm m}) \times (10^6{\rm m/s}) \simeq 10^{41} {\rm Kg~m^2/s} \to \frac{\hbar}{L} \simeq 10^{-75} \ll \cdots \ll 1\,.\eeq
There is one more point worth mentioning. In various gravitational scenarios, e.g. scattering processes, it is often the case perturbative solutions are sought for in powers of the center-of-mass energy over Planck's mass, $E/\Mp$, e.g. \cite{eftg1,eftg2}. For astrophysical objects, however, we have $E \sim m \gg \Mp$. Therefore this is not a useful parameter in the classical realm. In the PN expansion, we find that each term in the effective action is indeed rather large, of order $Lv^n$, but otherwise organized in powers of $v$. 

\section{Radiated Power Loss}\label{sec:rad}

For the computation of the binding potential the choice of boundary conditions is innocuous. That is the case because we work with quasi-instantaneous modes, for which $p_0 \ll |{\bp}|$. However, once we allow the particles to move, they will accelerate under the influence of the binding forces. Hence, they will lose energy via (scalar) radiation. When the scalar field can be emitted on-shell, $p_0^2 = {\bp}^2$, the $i\epsilon$-prescription becomes important. This is particularly relevant if we want to derive the value of the scalar field in the radiation zone, for which
we ought to impose --causality preserving-- `retarded' boundary conditions, in contrast to the Feynman propagator we have advocated thus far. However, Feynman's prescription --which follows from the path integral-- encodes information about the total radiated power, through the optical theorem. We review here both the computation using retarded boundary conditions and the equivalent procedure using Feynman's propagator. We start with the standard approach.

\subsection{Retarded Boundary Conditions}\label{retard}
\vskip 4pt 
The retarded propagator is given by,
\beq
\label{propret}
\Delta_{\rm ret}(x-y) = \int_{\bp} \int_{p_0} \frac{i}{{(p_0+i\epsilon)}^2-{\bp}^2}e^{-ip_0(x^0-y^0)} e^{i{\bp}\cdot({\bx}-{\by})}\,.
\eeq
Notice, unlike the Feynman propagator, we only have poles in the lower-half complex-plane thus enforcing the required causal condition. In coordinate space we find
\beq
\label{propretx}
i\Delta_{\rm ret}(x-y) =  \frac{1}{2\pi}\theta(x^0-y^0) \delta\left((x-y)^\mu(x-y)_\mu\right)\,.
\eeq

It is easy to see that, for point-like sources, the solution to (\ref{phiJ}) with retarded boundary conditions takes the usual form (i.e. the Lienard-Weichert potential)
\begin{align}
\label{phiJdel}
\phi^{\rm ret}_J(x) 
&= \sum_{a=1,2} \frac{m_a}{4\pi M_\phi} \int^{x^0}_{-\infty} dt~\frac{1}{|{\bx}-{\bx}_a(t)|}\delta\left(x^0-t-|{\bx}-{\bx}_a(t)|\right)  \\
&= \sum_{a=1,2} \frac{m_a}{4\pi M_\phi} \left[\frac{1}{|{\bx}-{\bx}_a(t)|- {\bv}_a(t)\cdot ({\bx}-{\bx}_a(t))}\right]_{\rm ret}\,, \nn
\end{align}
where `ret' stands for the replacement $t=x^0-{\boldsymbol{R}}(t)$, with ${\boldsymbol{R}}(t) \equiv {\bx}-{\bx}_a(t)$. Using Noether's theorem we can then compute the momentum density, ${\boldsymbol {P}}_i (x)= \partial_i \phi(x) \partial_0 \phi(x)$. From \eqref{phiJdel}, we have
\begin{align}
\partial_i \phi^{\rm ret}_J(x) &= \sum_{a=1,2} \frac{m_a}{4\pi M_\phi}\left[ \frac{\hat {\boldsymbol{R}}-{\bv}_a}{{\boldsymbol{R}}^2(1-{\bv}_a\cdot\hat {\boldsymbol{R}})^2} + \frac{\hat{\boldsymbol{R}}}{(1-{\bv}_a\cdot\hat{\boldsymbol{R}})}\left(\frac{\dot{\bv}_a\cdot{\boldsymbol{R}}-{\bv}_a\cdot\hat{\boldsymbol{R}}+{\bv}_a^2}{{\boldsymbol{R}}^2(1-{\bv}_a\cdot\hat{\boldsymbol{R}})^2}\right)\right]_{\rm ret}\,, \\
\partial_0 \phi^{\rm ret}_J (x)&= \sum_{a=1,2} \frac{m_a}{4 \pi M_\phi}\left[  \frac{\dot{\bv}_a\cdot{\boldsymbol{R}}-{\bv}_a\cdot\hat{\boldsymbol{R}}+{\bv}_a^2}{{\boldsymbol{R}}^2(1-{\bv}_a\cdot\hat{\boldsymbol{R}})^3}\right]_{\rm ret}.
\end{align}
In principle the system is simple enough we can obtain an exact result.  However, for the sake of comparison, we will assume $|{\bv}| \ll 1$ and expand in small velocities. Needless to say, we are just after Larmor's formula for scalar radiation. At leading order in the non-relativistic limit we have (for the radiative part)
 \beq
\hat {\boldsymbol{R}}_i\cdot \partial_i \phi^{\rm ret}_J = \partial_0 \phi^{\rm ret}_J = \sum_{a=1,2} \frac{m_a}{4\pi M_\phi} \left[\frac{\ddot {\bx}_a\cdot {\boldsymbol{R}}}{{\boldsymbol{R}}^2}\right]_{\rm ret}.
\eeq
Hence, we obtain for the total radiated power loss,
\bea
\label{pac}
\frac{dP_{LO}}{d\Omega} &=& \frac{1}{16 \pi^2 M_\phi^2} \sum_{a\neq b} m_a m_b  \left(\ddot {\bx}_a\cdot \hat {\boldsymbol{R}}\right) \left(\ddot {\bx}_b\cdot \hat {\boldsymbol{R}}\right)\,, \\
\label{eq:Ploret}
P_{LO} &=&  \frac{1}{12 \pi M_\phi^2} \left\langle \left(\sum_{a=1,2} m_a \ddot {\bx}_a\right) \cdot \left(\sum_{b=1,2} m_b \ddot {\bx}_b\right)\right\rangle  = \frac{(m_1+m_2)^2}{12 \pi M_\phi^2} \left\langle {\ba}_{\rm cm}^2\right\rangle\, .
\eea

\subsection{Optical Theorem}\label{sec:opt}
\vskip 4pt
The Feynman propagator in coordinate space is given by,
\beq
i\Delta_{\rm F}(x-y) = -i\frac{1}{(2\pi)^2} \frac{1}{(x-y)^\mu(x-y)_\mu+i\epsilon}\,.
\eeq 
Notice it has support {\it outside} the light-cone. However, using
\beq
\label{imieps}
 {\rm Im}\left\{\frac{1}{x+i\epsilon}\right\} = -\pi \delta(x)\,,
\eeq
we find 
\beq
\label{imFeyn}
{\rm Re}\,i\Delta_{\rm F}(x-y) = \frac{1}{4\pi} \delta\left((x-y)^\mu(x-y)_\mu\right)\,,
\eeq
which does propagate along the light-cone, but includes (half of) both retarded and advanced contributions.\vskip 4pt

Nevertheless, Feynman's prescription turns out to be the correct choice, also dictated by the path integral, as long as we concentrate on certain observables. In~particular those which do not require explicit knowledge of the radiation field, such as the total radiated power loss. This is due to the type of boundary conditions we impose for computing $W[J]$, namely `in-out' vacuum to vacuum amplitudes. These are time symmetric, and therefore we do not allow for outgoing radiation! However, the power loss is encoded in (the imaginary part of) $W[J]$ through the optical theorem, as we discuss in what follows.\vskip 4pt
Let us first split the effective action in terms of the real and imaginary part,
\beq
\label{imE}
Z[J] = e^{iW[J]} \to e^{i\int^{t_{\rm out}}_{t_{\rm in}} dt~{\rm Re}\,E [J]} \times e^{-\int^{t_{\rm out}}_{t_{\rm in}} dt~{\rm Im}\,E[J]}\, ,
\eeq
with $t_{\rm in(out)}$ (infinitely) long times. Then, Re\,$E[J]$ accounts for the binding energy we studied before, i.e. Re\,$E[J]= K[J]-V[J]$, including in principle also a kinetic term. Whereas the imaginary part gives, 
\beq
\label{imW}
\frac{1}{T} \,{\rm Im}\, W[J] = \langle {\rm Im}\, E[J]\rangle \to \frac{1}{2} \int \frac{d^2\Gamma}{dEd\Omega}  dE d\Omega\,,
\eeq
for $t_{\rm out}-t_{\rm in} = T \to \infty$, with $d\Gamma$ the differential rate of radiation or `decay width.' Multiplying by the energy of the emitted (massless) scalars, and integrating over energy and solid angle, we have
\beq
\label{totpower}
P \equiv \int dP = \int Ed\Gamma =  \int E \frac{d^2\Gamma}{d\Omega dE} dE d\Omega\, .
\eeq
Here is where the $i\epsilon$-prescription turns out to be crucial, since we need to take the scalar field `on-shell.' \vskip 4pt

It is useful to introduce a mixed Fourier space representation, where 
\beq
\label{mixed}
J(t,{\bp}) \equiv \int d^3{\bx} J(t,{\bx})e^{-i{\bp}\cdot{\bx}} =  \frac{1}{M_\phi}\sum_a m_a e^{-i{\bp}\cdot {\bx}_a(t)}\,.
\eeq
Hence, from the expression in \eqref{Wfree} and using \eqref{imieps}, we find
\bea
\label{eq:318}
{\rm Im}\, W[J]& =& \frac{1}{2} \int dt dt'\int_{p_0,\bp} \frac{1}{2|\bp|} \left(\delta(p_0-|{\bp}|)+ \delta(p_0+|{\bp}|)\right)e^{-ip_0(t-t')}J(t,{\bp}) J(t',-{\bp})\\
&=& \int dt dt'~ \int_\bp \frac{1}{2|{\bp}|}J(t,{\bp}) J(t',-{\bp})e^{i|{\bp}|(t-t')}\,,\nn
\eea
such that both retarded and advanced (positive and negative energy) contributions add up. From here, and using \eqref{imW}, we get for the total radiated power
\beq
\label{powercl}
\frac{d^2P}{d|{\bp}|d\Omega}  =  \frac{1}{T} \frac{|{\bp}|^2}{16\pi^3M_\phi^2} \left|  \sum_{a=1,2} \int dt_a m_a e^{i|{\bp}|t_a}e^{-i{\bp}\cdot{\bx}_a(t_a)}\right|^2\,.
\eeq 
We can then expand in powers of ${\bp}\cdot{\bx}_a \sim v$, obtaining a series of terms
\begin{align}
P= \frac{1}{T} \int_\bp \frac{1}{2M_\phi^2}\left|\sum_{a=1,2} \int dt_a m_a e^{i|{\bp}|t_a} \left(1+{\bp}\cdot{\bx}_a(t_a)+\frac{1}{2}({\bp}\cdot{\bx}_a(t_a))^2+\cdots\right)\right|^2 = P_{(0)} + P_{(1)}+\cdots .
\end{align}
The first one, $P_{(0)}$, vanishes. The next term reads
\begin{align}
P_{(1)} &= 
\frac{1}{T} \int_\bp \frac{1}{2M_\phi^2} \sum_{a\neq b} \int dt_a d t_b m_a m_b e^{i|{\bp}|(t_a-t_b)} {\bp}^i{\bp}^j {\bx}^i_b(t_b){\bx}^j_a(t_a)  \\
&= \frac{1}{T}  \int_\bp \frac{1}{2M_\phi^2} {\bp}^i{\bp}^j \left(\sum_{b=1,2} m_b \tilde {\bx}^i_b(|{\bp}|)\right)\left(\sum_{a=1,2} m_a\tilde {\bx}^j_a(-|{\bp}|)\right) \nn\, .
\end{align}
Performing the angular integral we get, 
\begin{align}
\label{p1xx}
P_{(1)}&=\frac{1}{12\pi M_\phi^2} \frac{1}{T} \int_0^{\infty} \frac{d|{\bp}|}{\pi} |{\bp}|^4\left(\sum_{b=1,2} m_b \tilde {\bx}^i_b(|{\bp}|)\right)\left(\sum_{a=1,2} m_a \tilde {\bx}^i_a(-|{\bp}|)\right) \\ &= \frac{1}{12\pi M_\phi^2}  \left\langle \left(\sum_{b=1,2} m_b \ddot {\bx}_b(t)\right)\cdot \left(\sum_{a=1,2} m_a \ddot {\bx}_a(t)\right)\right\rangle \,. \nn \end{align}
in complete agreement with \eqref{pj1}.\vskip 4pt
Intuitively, we may think of the optical theorem as follows. For in-out boundary conditions, we have a system emitting radiation which is later absorbed. The process is then `cut' --compute (twice) the imaginary part-- which may be represented as sending the radiation `backwards' in time,
\beq
\includegraphics[width=0.45\textwidth]{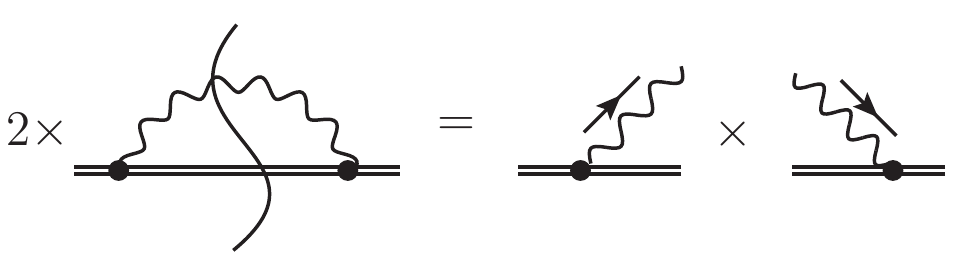}\nn
\eeq
(The double line are the non-propagating constituents of the source, separated by a distance much shorter than the scale of radiation.) Feynman's prescription is thus the backbone of the optical theorem.\,\footnote{~While the optical theorem and unitarity are key ingredients in quantum mechanics, Feynman's propagator can be introduced without evidence of a quantum world. (Feynman himself was originally inspired by particles moving forward and backwards in time \cite{Feynman}.) We could indeed picture a $19^{\rm \small th}$ physicist formulating the optical theorem in a classical setting, introducing an $i\epsilon$-prescription without reference to anti-particles.} (The procedure is more subtle if we attempt to account for the radiation-reaction effects on the dynamics due to the emitted radiation, {\it aka} the self-force. See part~\ref{sec:part2}.)\vskip 4pt Notice the expression in \eqref{eq:318}, and \eqref{powercl}, has the form of the square of an on-shell amplitude~\cite{nrgrLH}
\bea
i{\cal A}(p_0=|\bp|, \bp) &=&  i \sum_{a=1,2} \frac{m_a}{M_\phi} \int dt_a e^{i|{\bp}|t_a}e^{-i{\bp}\cdot{\bx}_a(t_a)}\,\\
\label{Pamp}
P &=& \frac{1}{T} \int_\bp \frac{1}{2|\bp|} |{\bp}| |{\cal A}|^2\,.
\eea
The factor of $(2|{\bp}|)^{-1}$ is part of the `phase space' for a massless scalar \cite{zee}. We will use this expression later on to compute the radiated power for the binary system.

\subsection{Multipole Expansion}\label{sec:mult}
\vskip 4pt

The result in  (\ref{powercl}) is exact, however, we will not be able to produce a similar analytic answer once we add self-interactions as are present in gravity. We ought to develop a perturbative approach to the radiation problem, and the key idea is the multipole expansion. As in electromagnetism, if the particles' positions and velocities do not vary too rapidly, we have a typical wavelength of the radiation given by $\lambda_{\rm rad} \sim  r/v \gg r$. Then, we can treat the combined system as a single localized source endowed with a series of time-dependent multipole moments, with an effective action 
\beq
\label{mult}
S_{\rm eff}^{\rm rad} = \frac{1}{M_\phi}\int dt \left( J_{(0)}(t) \phi(t,{\bx}_{\rm cm}) + J_{(1)}^i (t) \partial_i \phi(t,{\bx}_{\rm cm})+ \frac{1}{2}J^{ij}_{(2)}(t) \partial_i\partial_j \phi(t,{\bx}_{\rm cm}) + \cdots\right)\,,
\eeq
where $\bx_{\rm cm}$ is the center-of-mass of the bound state. (The overall factor of $1/M_\phi$ is inserted for convenience.) To compute the time-dependent couplings, $J_{(n)}^{i_1\cdots i_n}(t)$, we need to develop a matching procedure. This is achieved by inserting the multipole expansion,
\beq
\phi(t,{\bx}) =  \phi(t,{\bx}_{\rm cm}) + ({\bx}-{\bx}_{\rm cm})^i \partial_i \phi(t,{\bx}_{\rm cm}) +\frac{1}{2} ({\bx}-{\bx}_{\rm cm})^i{(\bx}-{\bx}_{\rm cm})^j \partial_i\partial_j \phi(t,{\bx}_{\rm cm}) +\cdots ,
\eeq
into the $J\phi$ interaction in the original action. Let us choose ${\bx}_{\rm cm} = 0$ for simplicity, then
\beq
\label{jbphi}
\int dt d^3{\bx} ~J(t,{\bx})\phi(t,{\bx}) \to  \int dt\left\{\left(\int  d^3{\bx}~J(t,{\bx})\right) \phi(t,0) + \left( \int d^3{\bx}~J(t,{\bx}){\bx}^i \right) \partial_i \phi(t,0)\right\}+ \cdots\, .
\eeq
We can now read off the multipole moments term by term,
\beq
J_{(n)}^{i_1\cdots i_n}(t)= M_\phi\int d^3{\bx}~J(t,{\bx}){\bx}^{i_1}\cdots {\bx}^{i_n}~ \label{jmul}.
\eeq
Notice that in mixed Fourier space we have
\beq
\label{mixed2}
J(t,{\bk}) = \int d^3{\bx} e^{-i{\bk}\cdot{\bx}} J(t,{\bx}) = \sum_n \frac{(-i)^n}{n!} \left(\int d^3{\bx}~J(t,{\bx}) {\bx}_i\cdots{\bx}_{i_n}\right) {\bk}^i\cdots{\bk}^{i_n}\, ,
\eeq
and we can identify the coefficients in this expansion with the $J_{i\cdots i_n}^{(n)}(t)$ factors in \eqref{jmul} \cite{andirad}. For example, 
\beq
\label{jmat}
J_{(0)} = m_1+m_2;~ {\boldsymbol {J}}_{(1)} = m_1{\bx}_1+m_2{\bx}_2;~\rm etc.
\eeq 
The total radiated power loss now follows as before. For instance, the dipole term contributes to the gravitational amplitude as 
\beq
\label{leadingamp}
i{\cal A}_{(1)}(p) = i \frac{1}{M_\phi} {\bp}\cdot{\boldsymbol{ J}}_{(1)}(t)\,,
\eeq
such that the power reads, in agreement with \eqref{eq:Ploret},
\begin{align}
\label{pj1}
P_{(1)} &= \frac{1}{T}\frac{1}{M_\phi^2} \left\langle \int_\bp {\bp}^i{\bp}^j \boldsymbol{ J}^i_{(1)}(t)\boldsymbol{ J}^j_{(1)}(t)\right\rangle =  \frac{1}{12\pi M_\phi^2}  \left\langle \ddot {\boldsymbol{ J}}_{(1)}  \cdot\ddot{ \boldsymbol{ J}}_{(1)}\right\rangle \\ &=\frac{1}{12\pi M_\phi^2}  \left\langle \left(\sum_{b=1,2} m_b \ddot {\bx}_b(t)\right)\cdot \left(\sum_{a=1,2} m_a \ddot {\bx}_a(t)\right)\right\rangle   \nn\,.\end{align} 
The procedure continues to all orders. To organize the computations it is useful to decompose the multipoles into irreducible symmetric-trace-free (STF) parts~\cite{thorne,bala}
\beq
J^{i_1\cdots i_n} =  \sum_{k=1}^{k=[n/2]} \frac{ n! (2n-4k+1)!!}{(n-2k)!(2n-2k+1)!!(2k)!!}~ \delta^{(i_1i_2}\cdots \delta^{i_{2k-1}i_{2k}} J_{\rm STF}^{i_{2k+1}\cdots i_n)\ell_1\ell_1 \cdots \ell_k\ell_k}\, ,
\eeq
with $[n/2]$ the largest integer $\leq n/2$. (Notice some of the indices on the right-hand side are traced over, such that the STF condition applies to the free indices.) We then plug the above expression into \eqref{mult} and use the wave equation to replace $\partial_i^2\phi \to \partial^2_0\phi$, which we then integrate by parts to produce time derivatives of the multipole moments. At the end we obtain~\cite{andirad2}
\beq
\label{smult}
S_{\rm eff}^{\rm rad} = \frac{1}{M_\phi} \sum_\ell  \int dt \frac{1}{\ell !} {\cal J}^L(t) \partial_L \phi(t,0),
\eeq
with $L = (i_1\cdots i_\ell)$, and
\beq
{\cal J}^L (t)= \sum_k \frac{ (2\ell+1)!!}{(2\ell+2k+1)!!(2k)!!}\int d^3{\bx} ~\partial_0^{2k} J(t,{\bx}) |{\bx}|^{2k} \bx^L_{\rm STF}\,.
\eeq
Using this general expression we can compute the flux to all orders. Generalizing the result in \eqref{leadingamp},
\beq
i{\cal A}_{(\ell)}(p) = i \frac{1}{M_\phi}\frac{(-1)^\ell}{\ell!} {\cal J}^Lp^L\,,
\eeq
the total radiated power loss becomes
\beq
\label{equ:totalpower}
P =\frac{1}{4\pi M_\phi^2} \sum_\ell \frac{1}{\ell! (2\ell+1)!!} \left\langle\left( \frac{d^{\ell+1}{\cal J}^L(t)}{ dt^{\ell+1}}\right)^2\right\rangle\,.
\eeq
We return to these manipulations in more detail in section \ref{sec:radgrav} of part \ref{sec:part2}\,. 

\section{Method of Regions}\label{methodr}

So far we have separated the computations of the binding potential and radiated power. Even though we may compute the conservative dynamics ignoring the radiation field (up until we incorporate back-reaction effects), once non-linearities are turned on the radiation modes will inevitably couple to the gravitational potentials which hold the binary together. This takes us to the `method of regions' \cite{iraeft}.\vskip 4pt

For slowly moving objects, $v\ll 1$, the wavelength of the radiation is much longer than their separation, $\lambda_{\rm rad}\sim r/v \gg r$. In such scenario we can split the interaction into different field modes, each one representing the dynamics of potential forces and radiation effects, separately. For this purpose we decompose the scalar field into --non-overlapping-- {\it regions}. We have a short-distance component, or potential region, and also a long-distance radiation mode. In total, we write the scalar field as \beq \label{eq:split1} \phi(t,{\bx}) = \underbrace{\Phi(t,{\bx})}_{\rm potential}+\underbrace{\bar\phi(t,{\bx})}_{\rm radiation}\, ,
\eeq 
with each term scaling as follows:
\begin{align}
\partial_0 \Phi (t,\bx) \sim \frac{v}{r}\, \Phi(t,\bx), ~ \partial_i \Phi (t,\bx)\sim \frac{1}{r} \,\Phi(t,\bx) \,,\,\,\,\, \partial_\mu \bar\phi(t,\bx) \sim \frac{v}{r}\, \bar\phi(t,\bx)\, .  \label{dpot}
\end{align}
For a static source we only find potential modes, which do not have a temporal component, and simply $|{\bp}| \sim 1/r$. Departures from instantaneity are then parameterized in powers of $p_0 \sim v/r \ll 1/r$ for slowly moving sources. On the other hand, the radiation mode is on-shell, with typical momentum of order $(p_0,|\bp|) \sim (v/r,v/r)$. The latter are the only modes that can appear as asymptotic states in the EFT, whereas the potential modes must be solved for, or in the jargon of particle physics: {\it integrated out}. For~the case of a linear theory these two regions decouple, in which case the splitting is trivial and we may compute the binding energy and power emitted separately. In general we have `mode-coupling' and the method of regions allows us to isolate the conservative and dissipative contributions to the effective action.\,\footnote{~
Let us stress an important point. As we discussed in sec.~\ref{retard}, to solve for the radiation field sourced by $J(x)$, $\bar\phi_J(x)$, we need to impose the correct (retarded) boundary conditions. In a more technical sense, what we need is to use the so called `in-in' formalism \cite{inin1,inin2}, where (in practice) $\Delta_{\rm F}(x)$ will be ultimately replaced by $\Delta_{\rm ret}(x)$ \cite{chadbr1,chadbr2,chadprl,chadprl2,chadreview}. However, as a mathematical procedure, the choice of boundary conditions is innocuous to read off the relevant multipole moments which contribute to radiation. The in-in formalism will play a key role to incorporate back-reaction effects. See part \ref{sec:part2}.}

\begin{figure}[t!]
\centering
\includegraphics[width=0.4\textwidth]{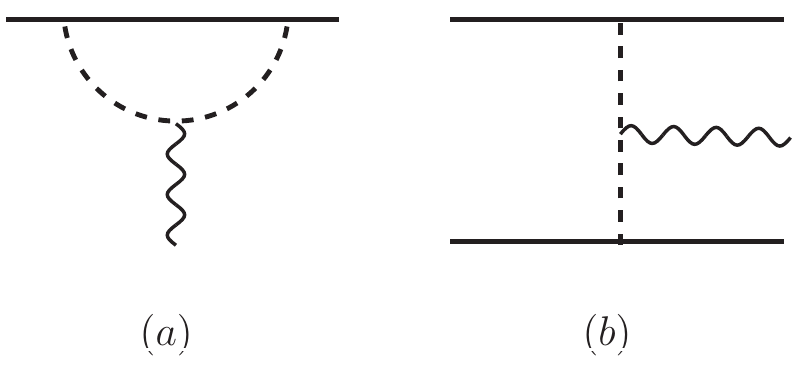}
\caption{Feynman diagrams representing the contributions in \eqref{ja2} (left) and \eqref{ja12} (right). The wavy line is the radiation field.}
\label{figa2}
\end{figure}

\vskip 4pt It is important to distinguish different type of non-linearities induced by short-distance potential modes coupled to the radiation. In our toy model we can have for instance a non-linear interaction of the form $\Phi^2\bar\phi$, but also $\bar\phi^2\Phi$. The first contributes directly to the source multipole moments, whereas the latter is responsible for the scattering off the potential induced by the binary itself.\,\footnote{~This is the so called `tail effect,' e.g. \cite{tail1,tail2,tail3,tail3n,tail4,tail5}, which leads to the introduction of {\it radiative} multipoles. See sec.~\ref{sec:tail} for more details\,. Moreover, we also find self-interactions of the radiation field. These produce a `memory effect,' which we will not cover in this review, see e.g. \cite{memory,tail3,thorneM,favata}.} As an example, let us study the contribution from the $\Phi^2\bar\phi$ coupling to the source multipoles. At linear order in $\lambda$ the total source becomes
 \beq
\label{jradlam}
J^\lambda_{\rm src}(t,\bx) \equiv  J_{\rm pp}(t,\bx) - 3\lambda\left(\Phi^{\lambda=0}_{J_{\rm pp}}(t,\bx)\right)^2, 
\eeq
where $J_{\rm pp}(t,\bx)$ is the `point-particle' contribution, given by the localized point-like expressions we have used thus far in the linear theory, e.g. \eqref{jtx}, and
\beq
\label{phi2st}
\Phi^{\lambda=0}_{J_{\rm pp}}(t,{\bx}) =\frac{1}{(4\pi M_\phi)^2}\left(\frac{m_1}{|{\bx}-{\bx}_1(t)|}+\frac{m_2}{|{\bx}-{\bx}_2(t)|}\right).
\eeq
After performing a partial Fourier transform we face two type of contributions, shown in Fig.~\ref{figa2}. Let us concentrate first on the square of each individual term, 
\begin{align}
J^\lambda_{{\rm src},\ref{figa2}(a)}(t,{\bk})
= -3\lambda \frac{m_{1}^2}{M_\phi^2} e^{-i{\bk}\cdot{\bx}_{1}} \int_{\bq} \frac{e^{-i{\bq}\cdot {\bx}_{1}(t)}}{{\bq}^2}\frac{e^{i{\bq}\cdot{\bx}_{1}(t)}}{({\bk}+{\bq})^2} = -3\lambda \frac{m_{1}^2}{8M_\phi^2} \frac{e^{-i{\bk}\cdot{\bx}_{1}(t)}}{|{\bk}|}\, . \label{ja2}
\end{align}
Notice that this expression diverges in the long-wavelength limit, ${\bk} \to 0$, and therefore does not admit a multipole expansion. However, this contribution is unphysical since it is associated with a divergent self-energy term. (This is again a signal of the IR divergences that plague our toy model.) Indeed, we already encountered a similar term in \eqref{phij3n}. The main difference is the regions involved.

Previously, the typical $|{\bk}|$ was given by $1/r$, which is the scale associated with the (off-shell) binding force. In \eqref{ja2}, on the other hand, the momentum of the {\it external} leg is now on-shell and soft(er). 
We~incorporate these contributions by expanding the Green's function in ${\bk}$, prior to performing the integral. In other words, we not only multipole expand the coupling between the radiation field and the point-like sources, we must also expand the propagators \cite{iraben}, namely
\beq
\label{eq:iraben}
\frac{1}{({\bq}+{\bk})^2} = \frac{1}{{\bq}^2+2{\bk}\cdot {\bq} + {\bk}^2} = \frac{1}{{\bq}^2}\left(1-2\frac{{\bk}\cdot {\bq}}{{\bq}^2}+\cdots\right)\,.
\eeq
After implementing the multipole expansion each term scales homogeneously in the expansion parameter, and at the same time we are guaranteed the integrals that contribute to the expansion in $\bk$ of \eqref{ja2} vanish, upon using dim. reg. \cite{iraben}. Hence, the diagram in Fig~\ref{figa2}$(a)$ can be consistently set to zero altogether.\vskip 4pt After Fig~\ref{figa2}$(a)$ is discarded, we end up with the contribution from Fig~\ref{figa2}$(b)$,
\begin{align}
\label{radjab}
J^\lambda_{{\rm src},\ref{figa2}(b)}(t,{\bk}) = -6\lambda \frac{m_1m_2}{M_\phi^2} e^{-i{\bk}\cdot{\bx}_1} \int_\bq \frac{e^{i{\bq}\cdot ({\bx}_1(t)-{\bx}_2(t))}}{{\bq}^2({\bk}+{\bq})^2}\,, 
\end{align}
which contrary to the previous one has a well defined expansion for $|\bk| \ll |\bq|$,
\beq
\label{ja12}
J^{\lambda}_{{\rm src},\ref{figa2}(b)}(t,{\bk}) = -6\lambda \frac{m_1m_2}{M_\phi^2}\left( (1-i{\bk}\cdot{\bx}_1) \int_\bq \frac{e^{-i{\bq}\cdot ({\bx}_1(t)-{\bx}_2(t))}}{{\bq}^4} - 2\int_{\bq} \frac{{\bq}\cdot{\bk}}{{\bq}^6}e^{-i{\bq}\cdot ({\bx}_1(t)-{\bx}_2(t))} \right)+{\cal O}(\bk^2)\, .
\eeq
From here, and using \eqref{mixed2}, we obtain 
\beq
\left(J^\lambda_{{\rm src},\ref{figa2}(b)}\right)_{(0)} = 3\lambda \frac{m_1m_2}{4\pi M_\phi}|{\br}(t)|\, ,~~\label{lamdipole}
\left({\boldsymbol {J}}^\lambda_{{\rm src},\ref{figa2}(b)}\right)_{(1)} = 3 \lambda \frac{m_1m_2}{8\pi M_\phi}|{\br}(t)| ({\bx}_1+{\bx}_2)\, .
\eeq
It is instructive to re-derive these results directly from Wick's theorem at the level of the path-integral. The task is to  compute the one-point function integrating out the potential modes,
\begin{align}
\label{expbp}
\bar\phi_{J_{\rm src}}(t,\bx) &=\int D\bar\phi~\bar\phi(t,{\bx})~{\rm exp}\left[iS_{\rm free}[\bar\phi]+i \int J_{\rm pp} \bar\phi\right] \\
& \times \int D\Phi~{\rm exp}\left[iS_{\rm free}[\Phi] +i\int \left\{J_{\rm pp}\Phi-\lambda(\bar\phi+\Phi)^3\right\}\right] \nn\,.
\end{align}
At leading order in $\lambda$ we have 
\begin{align}
\label{phi2j1}
 {\bar\phi}_{J_{\rm src}}^\lambda(t,\bx) &=  \lambda \frac{3i}{2!} \int dt_y d^3{\by} \int D\bar\phi~ e^{iS_{\rm free}[\bar\phi]}~\bar\phi (t,{\bx}) \bar\phi(t_y,{\by}) \left\{
\int dt' d^3{\bx}' dt'' d^3{\bx}'' J_{\rm pp}(t',{\bx}')J_{\rm pp}(t'',{\bx}'') \times  \nn\right.\\&\left.\times \int D\Phi~\Phi^2(t_y,{\by})\Phi(t',{\bx}')\Phi(t'',{\bx}'')~e^{iS_{\rm free}[\Phi]}\right\}\,, 
\end{align}
and plugging in the (point-like) source term, we get
\beq
{\bar\phi}_{J_{\rm src}}^\lambda (t,\bx) = i \int dt_y d^3{\by} \Delta_F(t-t_y,{\bx}-{\by})J^\lambda_{\rm src}(t_y,{\by})\, ,
\eeq
with 
\beq
J^\lambda_{\rm src}(t_y,{\by}) =6\lambda\frac{m_1m_2}{M_\phi^2} \int dt_1 dt_2   \Delta_F(t_y-t_1,{\by}-{\bx}_1(t_1))\Delta_F(t_y-t_2,{\by}-{\bx}_2(t_2))\, .
\eeq
After regularizing away pure divergences, we find
\begin{align}
J^\lambda_{\rm src}({\bk}) = -6\lambda\frac{m_1m_2}{M_\phi^2}e^{-i{\bk}\cdot{\bx}_1(t)}\int dt \int_\bq \frac{e^{i{\bq}\cdot({\bx}_1(t)-{\bx}_2(t))}}{{\bq}^2({\bq}+{\bk})^2}\, + \cdots\,, \label{phi2j2}
\end{align}
for static sources. This is the same result we obtained in \eqref{radjab}.

\section{Summary of Part~\ref{sec:part1}}\label{sec:sum1}

The basic ingredients of the EFT approach to compute (scalar) emission for classical binary systems can be summarized as follows:

\bit

\im Decompose the field into potential and radiation {\it regions} coupled to point-like sources.

\im Compute ${\rm Re}\,W[J]$ by integrating out the potential modes ignoring radiation. Departures from instantaneity are incorporated by expanding the propagators. The particles' equations of motion follow from the binding potential (after including a kinetic energy). Discard scale-less divergent integrals. 

\im The long-wavelength radiation theory takes the form of a multipole expansion. Read off the multipole moments by expanding the coupling to the (point-like) sources and integrating out the potential modes

\im Compute the energy loss by integrating out the radiation modes using the optical theorem, relating ${\rm Im}\,W[J]$ to the total radiated power. The latter is given in terms of (squares of) derivatives of the STF multipole moments.

\im Use the equations of motion to compute the time variation of the multipoles, thus obtaining the total power loss in terms of the dynamics of the constituents of the binary system.

\im Proceed systematically in powers of the self-coupling $(\lambda)$ and velocity $(v)$ to all orders. 

\eit

\newpage

\part{The Bird's Eye View on the Binary Inspiral Problem}\label{sec:part2}

We now apply the ideas we developed in part~\ref{sec:part1} to the binary inspiral problem. In general relativity we will find a few extra ingredients for the construction of an EFT framework. In particular, the non-linear interaction between potential and radiation modes, and the existence of renormalization group trajectories for the (Wilson) coefficients of the EFT. Moreover, black holes and neutron stars in binary systems may be rapidly rotating, which requires introducing the concept of spin. As we shall see, the EFT formalism is sufficiently rich to incorporate all of these aspects of the two-body problem in gravity. We start out our journey identifying the hierarchy of scales.

\section*{One Scale at a Time}
\phantomsection
\addcontentsline{toc}{section}{~~~~One Scale at a Time}

The dynamics of the binary problem can be separated into different parts~\cite{Fock}:
\bit
\im {\it The internal zone:} This is the scale of finite size effects. For compact neutron stars or black holes we have $r_s \simeq 2G_Nm$.

\im {\it The near (or potential) zone:} The intermediate region is the {\it orbit scale}, $r$, given by the typical separation between the constituents of the binary.

\im {\it The far (or radiation) zone:} This is the scale of gravitational waves, emitted with typical wavelength $\lambda_{\rm rad} \sim r/v$.  For the radiation problem, the orbit scale becomes also part of the {\it internal} zone.
\eit

For the case of Post-Newtonian sources we have $v \ll 1$, and therefore there is hierarchy of scales,  \beq \lambda^{-1}_{\rm rad} \ll  r^{-1} \ll r_s^{-1}\,.\eeq Here is where the EFT machinery comes to play, since EFTs are tailor made to naturally handle several scales in a tower-like fashion. As we discuss in detail, we proceed in steps:

\bit

\im  {\it Parameterizing our ignorance.} Before we set course we must deal with degrees of freedom at the scale $r_s$. The basic idea is the inclusion of new terms in the point-particle action --beyond minimal coupling-- to account for finite size effects. In gravity, extended objects have a very rich structure, which includes tidal as well as dissipative effects.  For probes that vary on distances of order $|\bk|^{-1}$, the incidence of finite size effects enters in powers of $|\bk| r_s$. 

\im  {\it Conservative dynamics}.  At the orbit scale we have (off-shell) potential modes. In the small velocity approximation we treat the interaction as `instantaneous,' and time derivatives as a perturbation. We~obtain the equations of motion from the binding potential energy. The potential modes, varying on a scale $|{\bk}| \simeq 1/r$, are the probes for the internal structure of the bodies. Hence, the finite size effects for compact objects scale as powers of $r_s/r \simeq v^2$, from the virial theorem.

\im  {\it Radiated power}.   At the scale of radiation the binary can be treated as a point-like source endowed with a series of ($\ell \geq 2$) multipole moments, $(I^L,J^L)$, scaling as $I^L \sim M r^\ell$. The expansion parameter is given by $|\bk| r$, with $|\bk|^{-1} \sim \lambda_{\rm rad}$. The change in the (averaged) binding mass/energy of the system, and the gravitational wave amplitude, are obtained in terms of derivatives of these multipoles as a function of the dynamical variables of the constituents (like positions, velocities and spins).

\im {\it Hereditary effects}. The interaction of the emitted gravitational wave with the static potential, sourced by the binary as a whole, produces the tail effect (or scattering off the geometry). The tails contribute to the radiated power loss and are responsible for the renormalization group evolution in the effective theory. There is also a memory effect, where time-dependent contributions arise from non-linear self-interactions in the radiation field. 

\im {\it Radiation-reaction}. The emission of gravitational waves back-reacts on the motion of the binary system, producing the so called radiation-reaction force. The non-linear couplings involved in the study of back-reaction effects entail a subtle interplay between different regions, featuring both conservative and dissipation contributions. These can be incorporated in the EFT framework (extended to the in-in formalism) by using appropriated (causal) retarded Green's functions. 
\eit

Interferometers, such as Advanced LIGO \cite{ligovirgo,ligovirgo2,ligodetect}, are particularly sensitive to the phase of the signal,  \beq \phi(t) \sim \int dt~\Omega(t)\,,\eeq with $\Omega(t)$ the frequency of a circular orbit (corresponding to half the frequency of the gravitational wave).\,\footnote{~For earth-based detectors, there is a (seismic) low-frequency cutoff. Therefore, interferometers such as Advanced LIGO \cite{ligovirgo,ligovirgo2,ligodetect}, are only sensitive to the last few minutes of the inspiral/merger/ring-down phases. Hence, also due to gravitational wave emission, the orbit has circularized by the time the signal enters the detector's band, e.g.~\cite{Blanchet,Buoreview}.} This can be computed using the adiabatic approximation. The energy balance equation $\langle\dot E\rangle~=~-~P$, where $P$ is the integrated power loss, then allows us to solve for $\Omega(t)$. Using this information we can construct restricted waveforms which are accurate to a given PN order in the phase, but at leading order in the amplitude. To compute the gravitational waveform we solve the instantaneous values of the metric field in terms of the multipole moments. Having the amplitude corrections leads to fully accurate templates.\vskip 4pt At each step we work systematically order by order in the ratio of scales using the power counting rules of the EFT. As we shall see, in principle there is no formal obstacle to calculating to any order beyond the Newtonian approximation. In the EFT approach we map complex integrals into the computation of Feynman diagrams, use textbook regularization tools, and naturally handle spin degrees of freedom at the level of the action. These techniques allow for a very natural systematization and (physical) visualization of the computation that can be automatized almost entirely using {\it Mathematica} code.\,\footnote{~\url{http://feyncalc.org/}}
In the ensuing sections we will sharpen the distinction between the relevant physical regions, elaborate on the necessary steps to construct the effective description at each scale, and perform some pedagogical computations. 
 \newpage
\section{Worldline Effective Theory}\label{sec:compact}
We start by reviewing the effective approach to describe gravitationally interacting extended objects in gravity. As we shall see, the appearance of UV divergences is directly linked to the use of point-like sources, and are naturally handled in the EFT framework through counter-terms and renormalization.

\subsection{Point-like Sources}
\vskip 4pt
Say we are interested in the (static) gravitational field produced by an isolated compact object, for simplicity, a non-rotating black hole of mass $m$. We solve for the metric, $g_{\mu\nu}$, using the Einstein-Hilbert action,
\beq
\label{actR}
S_{\rm EH} = -2 \Mp^2 \int \sqrt{g(x)} R(x) d^4x\, .
\eeq
Of course, we find the Schwarzschild(-Droste) solution after some appropriate coordinate (gauge) choice~\cite{Schwar1,Zeegr}. In most cases of interest, however, a close analytic answer may not be feasible. We must therefore set up a perturbative scheme (or resort to numerical methods). We then start by splitting the metric field~as \beq \label{split3} g_{\mu\nu} = \eta_{\mu\nu} + \frac{h_{\mu\nu}}{\Mp}\,.\eeq
Hence, inserting \eqref{split3} into \eqref{actR} we have (schematically)
\beq
\label{eq:seh}
S_{\rm EH} =  \int \, h \partial^2 h + \frac{1}{\Mp} h h \partial\partial h + \cdots\,,
\eeq
such that we get an equation for $h_{\mu\nu}$ which we may solve iteratively in powers of $r_s/r \ll 1$. This is often referred as the Post-Minkowskian expansion, e.g.~\cite{damourPM}. Since we restrict ourselves to values of $r$ greater than~$r_s$, there is yet an important missing ingredient. Namely, we need to incorporate the imprint of degrees of  freedom at $r < r_s$ as a matching/boundary condition. Moreover, in our worldline description we take the limit $r_s \to 0$. In the effective theory the black hole will be thus described by a localized  source, with
\beq
\label{tm}
-\frac{1}{2\Mp} \int d^4x\,T^{\mu\nu}_{\rm pp} (x) h_{\mu\nu}(x) \to -\frac{m}{2\Mp} \int dt~ h_{00}(t,0) + \cdots \,,
\eeq
as an extra term added to the action in \eqref{actR}. Here, the $T^{\mu\nu}_{\rm pp}$ is the stress-energy tensor for a `point-particle,'  given by
\beq
\label{tpp}
T^{\mu\nu}_{\rm pp} (x) =  m \int d\sigma ~ \frac{u^\mu u^\nu}{\sqrt{u^2}} \frac{\delta^4(x-x(\sigma))}{\sqrt{g(x)}} + \cdots\,.
\eeq 
For the purpose of computing the metric for an isolated object, the linearized expression in \eqref{tm} (evaluated at the origin and at rest) is sufficient.\vskip 4pt As it is well known, distributional sources are problematic in general relativity \cite{geroch} (recall black holes are Ricci-flat solutions \cite{Zeegr}). The point-like approximation leads to divergences due to the non-linear structure of the field equations. As we shall see, these will be removed by counter-terms proportional to higher order derivatives of the metric, encoded in the ellipsis in \eqref{tpp}.  At the same time, these extra terms incorporate finite size effects for extended objects. Due to Birkhoff's theorem, the new coefficients (and non-linear worldline couplings) do not contribute to the computation of the --classical-- one-point function (for non-rotating bodies). Therefore, we postpone their appearance until sec.~\ref{sec:effect}.
\vskip 4pt 
In addition, Einstein's theory enjoys a coordinate invariance (in a flat background)  
\beq
\label{gsym}
x^\mu \to x^\mu + \epsilon^\mu(x)\,,~~ h_{\mu\nu} \to h_{\mu\nu}+ \epsilon_{\mu,\nu} + \epsilon_{\nu,\mu}\, .
\eeq
This will be handled by a `gauge fixing' term. For example, the harmonic gauge, $\Gamma_\mu=0$, with
\beq
\label{gamma}
\Gamma_\mu = \partial_\alpha h^\alpha_\mu - \frac{1}{2}\partial_\mu h^\alpha_\alpha\, .
\eeq
At the level of the action (and path integral), we add 
\beq
\label{eq:Sgf1}
S_{\rm GF} = \int d^4 x~\Gamma_\mu \Gamma^\mu\, .
\eeq
This allows us to find a unique solution to the (Feynman) propagator
\beq
\label{deltaf}
\Delta_{{\rm F}\,\alpha\beta\mu\nu} = P_{\alpha\beta\mu\nu}\, \Delta_{\rm F}(t-t',{\bx}-{\bx}')\,,
\eeq
where  
\beq
\label{pabmn}
P_{\alpha\beta\mu\nu} = \frac{1}{2} \left(\eta_{\alpha\mu}\eta_{\nu\beta}+\eta_{\alpha\nu}\eta_{\mu\beta}-\eta_{\alpha\beta}\eta_{\mu\nu}\right)\,.
\eeq
With these tools we can simply derive, for instance, the Newtonian potential, $V_N$. At linear order in $G_N$, for a static source described by \eqref{tm}, we get 
\beq
 \frac{h_{\mu\nu}}{\Mp}(t,{\bx}) 
= P_{\mu\nu00}\frac{-im}{2 \Mp^2} \int dt'  e^{-ip_0 t'} \int_{\bp,p_0}  \frac{i\, e^{i\left(p_0t-{\bp}\cdot {\bx}\right)}}{p_0^2-{\bp}^2+i\epsilon} = -(2\eta_{\mu0}\eta_{\nu0}-\eta_{\mu\nu})\frac{2G_Nm}{r}\, .
\label{hmn}
\eeq
Notice the $dt'$ integral removed the dependence on $p_0$. That is the reason the $i\epsilon$-prescription plays no role for (quasi-)\,instantaneous computations. The expression in \eqref{hmn} reproduces the familiar result, $g_{00} = 1 + 2V_N$, at leading order in $G_N$. Since we work in harmonic coordinates we also have $g_{ij} = -\delta_{ij}(1-2V_N)$. 

\subsection{Non-Linearities}\label{sec:nonlinear2}
\vskip 4pt

At higher orders in $G_N$ we encounter non-linearities from the Einstein-Hilbert action as well as from the coupling to the source.  The worldline non-linearities (see \eqref{tpp}) will be important later on for the binary problem. However, they do not contribute (at the classical level) to the~metric of an isolated object, and the expression in \eqref{tm} suffices (see Fig~\ref{ydiag}). Unlike the scalar case, the non-linear self-interactions in Einstein's theory involve derivatives, see \eqref{eq:seh}, which add extra factors of $k^2$ to the vertices. This ameliorates the IR divergences we found before, but exacerbates the UV problem. As we shall see, UV poles will be absorbed into counter-terms. This is the starting point for the construction of the effective action in~sec.~\ref{sec:effect}.\vskip 4pt A convenient way to incorporate gravitational non-linearities is to introduce the pseudo stress-energy tensor, ${\cal T}^{\alpha\beta}(x)$, via
\beq
\label{onph}
\frac{h_{\mu\nu}}{\Mp} (k_0,{\bk})= -\frac{i}{2\Mp^2} P_{\mu\nu\alpha\beta} \frac{i}{k_0^2-{\bk}^2} {\cal T}^{\alpha\beta}(k_0,{\bk})\, ,
\eeq
which we may compute to all orders in $G_N$. At leading order we have, see \eqref{tpp},
\beq
{\cal T}^{\mu\nu}_{(1)}(k_0,{\bk}) = m (2\pi) \delta(k_0) e^{-i{\bk}\cdot{\bx}(\tau)} u^\mu u^\nu\, ,~~{\rm etc}.
\eeq
To obtain the $n$-th order contribution, ${\cal T}_{(n)}^{\alpha\beta}$, it is  convenient to work with the background field method~\cite{backgrDW,backgrV,backgrA}. We first proceed to split the metric perturbation into two components,
\beq
\label{splitcalH}
h_{\mu\nu} = \underbrace{{\cal H}_{\mu\nu}}_{\rm background} + \underbrace{H_{\mu\nu}}_{\rm perturbation},
\eeq
and then integrate out $H_{\mu\nu}$ in a non-trivial background.\,\footnote{~The reader should not confuse ${\cal H}_{\mu\nu}$ at this stage with the radiation field. This is nothing but a trick to integrate out the perturbation adding a gauge fixing term which is formally invariant under transformations of the background metric. In the static limit there is no radiation whatsoever. We will, nonetheless, use a similar machinery to study radiation later on.}  
The main difference is the gauge fixing term, which is promoted from \eqref{gamma} to
\beq
\label{gambf}
\Gamma^{(\cal H)}_\mu = \nabla^{(\cal H)}_\alpha H^\alpha_\mu - \frac{1}{2}\nabla^{(\cal H)}_\mu H^\alpha_\alpha\,,
\eeq
with $\nabla^{(\cal H)}$ the covariant derivative compatible with the ${\cal H}_{\mu\nu}$ metric, that is then used to raise and lower indices. We must also re-write the volume factor in \eqref{eq:Sgf1} in terms of the background metric. In this manner we ensure that the resulting path integral
\begin{align}
\label{eq:seff}
e^{i S_{\rm eff}[{\cal H}_{\mu\nu}]} \equiv e^{-\frac{i}{2\Mp}  \int T_{\rm pp}^{\mu\nu}(x) {\cal H}_{\mu\nu}(x)} \int &DH_{\mu\nu} \exp\Big[ iS_{\rm EH}[{\cal H}_{\mu\nu} + H_{\mu\nu}] \\ &+ iS^{({\cal H})}_{\rm GF}[H_{\mu\nu}] - \frac{i}{2\Mp}\int T_{\rm pp}^{\mu\nu}(x) H_{\mu\nu}(x)\Big]\nn
\end{align}
remains invariant under \eqref{gsym}, applied to ${\cal H}_{\mu\nu}$. In this fashion we are guaranteed the associated current, ${\cal T}^{\mu\nu}(x)$, obeys the Ward identity
\beq
\label{dTzero}
\partial_\alpha {{\cal T}}^{\alpha\beta}(x) =0\, .
\eeq
This is nothing but the well-known conservation of the (pseudo) stress-energy tensor, including the {\it self-energy} in the gravitational field \cite{landau}. For example, the ${\cal O}(G_N^2)$ contribution can be straightforwardly computed, see Fig.~\ref{ydiag}(a). The result reads \cite{nrgr}
\begin{align}
\label{tabk2}
{{\cal T}}^{\alpha\beta}_{(2),\ref{ydiag}(a)}(k) &= (2\pi)\delta(k_0)\frac{m^2}{32 \Mp^2} \left[-7(\eta^{\alpha\beta}k^2 -k^\alpha k^\beta)+k^2v^\alpha v^\beta\right]\int_{\bq} \frac{1}{{\bq}^2({\bq}^2+{\bk}^2)}\\
&=  (2\pi)\delta(k_0)\frac{m^2}{16\Mp^2|{\bk}|} \left[-\frac{7}{32}(\eta^{\alpha\beta}k^2 -k^\alpha k^\beta)+\frac{1}{32}k^2 v^\alpha v^\beta\right]\nn ,
\end{align}
with $v^\mu=(1,0)$. Notice that it is indeed conserved: $k_\mu {{\cal T}}^{\mu\nu}(k)=0$. From here, and \eqref{onph}, we get \cite{nrgr}
\begin{align}
g_{00} &= 1-\frac{2G_Nm}{r} + 2\left(\frac{G_Nm}{r}\right)^2 + \cdots\,, \\
g_{ij} &= -\delta_{ij}\left[1 + \frac{2G_Nm}{r} + 5\left(\frac{G_Nm}{r}\right)^2 + \cdots \right].
\end{align}
Not surprisingly one recovers Schwarzschild's solution in harmonic coordinates as a series expansion in powers of~$r_s/r$. (This result was previously derived in \cite{Feynmantree,duff} using similar techniques.) 

\subsection{Regularization} \label{sec:regular0}

\subsubsection{Power-law Divergences}
\vskip 4pt

At this point the reader may wonder about divergences, similar to those we encountered in our toy model in part~\ref{sec:part1}\,. In fact, we cheated a little because the computation of ${\cal T}^{\mu\nu}_{(2)}(k)$ does entail regularizing a divergence, which arises from a term equivalent to a self-energy diagram (see below).  The temporal components do not contribute for a static source, and we end up with an integral of the sort
\beq
A^{ij} ({\bk}) \equiv \int_{\bq} \frac{{\bq}^i{\bq}^j}{{\bq}^2({\bk}+{\bq})^2}\,.
\eeq 
The divergence occurs when we take the trace, $A^{ii}$, in which case the numerator cancels out one of the propagators. Diagrammatically, the singular part of the diagram in Fig.~\ref{ydiag}(a) appears when the $\partial^2$ from the cubic vertex hits the propagator. The resulting integral is then equivalent to a self-energy contribution to the mass. All such diagrams are set to zero in dim. reg.,  since they correspond to power-law divergences:
\beq
\nn
\includegraphics[width=0.25\textwidth]{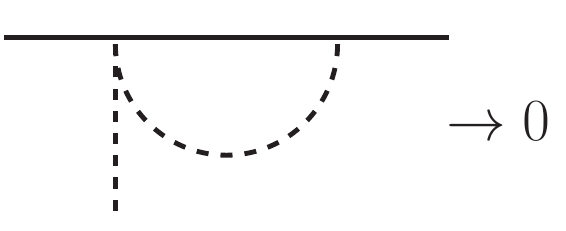}
\eeq
\vskip 4pt
The full integral $A^{ij}$, however, does not vanish identically. To compute its contribution we write it as
\beq
A^{ij}(\bk) = A({\bk}) {\bk}^i{\bk}^j + {\bk}^2 B({\bk})\delta^{ij}\, ,
\eeq
and solve for $A(\bk)$ and $B(\bk)$, using
\begin{align}
A^{ii} (\bk) &= \int_{\bq} \frac{1}{({\bk}+{\bq})^2}\, ,\\
{\bk}^i {\bk}^j A^{ij}(\bk) &= \int_{\bq} \frac{({\bk}\cdot {\bq})^2}{{\bq}^2({\bk}+{\bq})^2} = \frac{1}{4} \int_{\bq} \frac{\left(({\bk}+ {\bq})^2-{\bk}^2-{\bq}^2\right)^2}{{\bq}^2({\bk}+{\bq})^2}\, .
\end{align}
The first equation is the trace, which we set to zero, yielding $A(\bk) = -3B(\bk)$. In the second integral all the terms from the numerator lead to similar divergences, except for one,
\beq
\frac{1}{4} \int_\bq \frac{{\bk}^4}{{\bq}^2({\bk}+{\bq})^2} = \frac{{\bk}^4}{4}\times \frac{1}{8|\bk|} = \frac{|{\bk}|^3}{32}\, ,
\eeq
such that
\beq
A(\bk) + B(\bk) = \frac{1}{32|{\bk}|} \to B(\bk)= -\frac{1}{64|{\bk}|},~A(\bk)= \frac{3}{64|{\bk}|}\,.
\eeq
Combining the two we arrive at the regularized answer:
\beq
\int_\bq \frac{{\bq}^i{\bq}^j}{{\bq}^2({\bk}+{\bq})^2}
 = \frac{1}{64|{\bk}|} \left(3{\bk}^i{\bk}^j-{\bk}^2\delta^{ij}\right)\,.
\eeq

\subsubsection{Logarithmic Divergences}\label{sec:renorm}
\vskip 4pt
\begin{figure}[t!]
\centering
\includegraphics[width=0.7\textwidth]{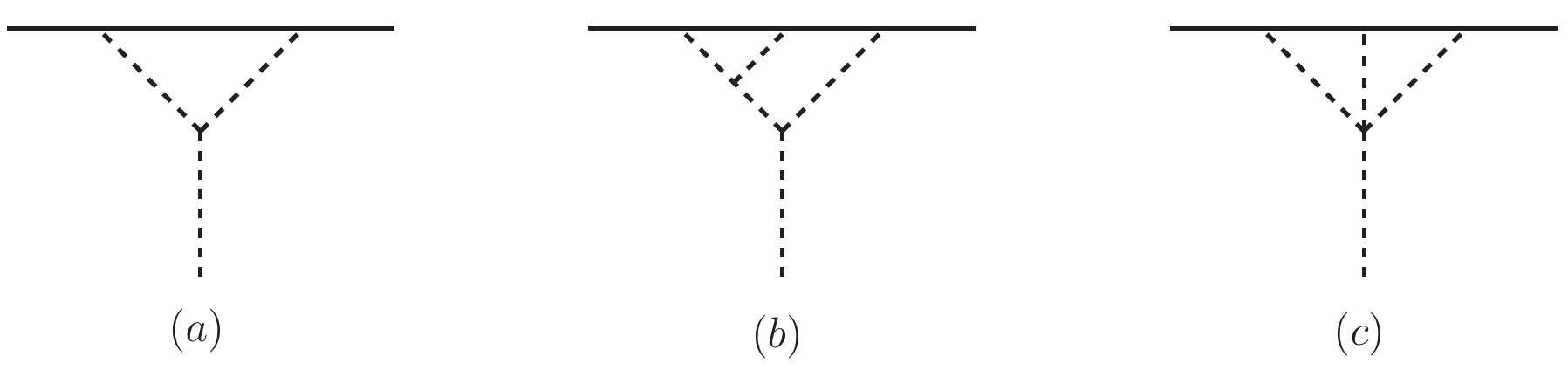}
\caption{Feynman diagrams which contribute to the one-point function to ${\cal O}(G_N^3)$ \cite{nrgr}.}
\label{ydiag}
\end{figure}
So far we have used the words `dim.~reg.' essentially to set to zero scale-less integrals. When logarithmic divergences are present, however, we need to dig a little deeper into regularization in $d$ spacetime dimensions (see \cite{jadasch} for another implementation of dim. reg. within the PN framework).\vskip 4pt

The calculation of ${{\cal T}}^{\alpha\beta}_{(3)}(k)$ to ${\cal O}(G_N^3)$ can be found in \cite{nrgr}. The diagrams which contribute are shown in Fig. \ref{ydiag}$(b$-$c)$, and the result reads \cite{nrgr}
 \beq
 {{\cal T}}^{\mu\nu}_{(3)}(k) = \frac{1}{2\Mp}\left(\frac{m}{2\Mp}\right)^3 (2\pi)\delta(k^0) \left[ a_1(k^\mu k^\nu - \eta^{\mu\nu}k^2) + a_2 k^2 v^\mu v^\nu\right] I_0({\bk})\, ,
 \eeq
 with $a_1,a_2$ some numerical factors. The relevant scalar integral, $I_0({\bk})$, is given by
 \beq
 \label{eq:scalarI0}
I_0({\bk})=\int \frac{d^{d-1}{\bq}}{(2\pi)^{d-1}}\frac{d^{d-1}{\bp}}{(2\pi)^{d-1}} \frac{1}{{\bq}^2{\bp}^2({\bq}+{\bp}+{\bk})^2}\,,
\eeq
written in $d$-dimensions. After analytic continuation in $d$, we find
\beq
\label{eq:dI0}
I_0({\bk}) = \frac{\sqrt{\pi}}{(4\pi)^{d-1}} \frac{\Gamma[4-d]\Gamma[d-3]\Gamma[(d-3)/2]^2}{\Gamma[d/2-1]\Gamma[3/2(d-3)]}\left(\frac{{\bk}^2}{2}\right)^{d-4}\,.
\eeq
This integral contains a logarithmic divergence when $d \to 4$, hidden in the poles of the $\Gamma[z]$ function, 
\beq
\Gamma[z]=\int_0^\infty dt ~e^{-t}t^{z-1}\,.
\eeq
Introducing an expansion parameter: $\epsilon \equiv 4-d$, with $\epsilon>0$, and expanding around $\epsilon \simeq 0$, we have
\beq
\Gamma[-n+\epsilon] = \frac{(-1)^n}{n!}\left(\frac{1}{\epsilon} +\psi(n+1)-\frac{\epsilon}{2}\left(\psi^{(1)}(n+1)-\psi(n+1)^2-\frac{\pi^2}{3}\right) + {\cal O}(\epsilon^2)\right)\, ,
\eeq
where $n$ is a (positive) integer, $\psi[x] =\frac{d}{dx}\log\Gamma[x]$, and $\psi^{(1)}[x]$ its first derivative. For $n=0$ we get
\beq
\Gamma[\epsilon] = \frac{1}{\epsilon} -\gamma_E + {\cal O}(\epsilon)\, ,
\eeq
with $\gamma_E = - \psi(1) \simeq 0.5772$ the Euler-Mascheroni constant. Using this formalism, and expanding  \eqref{eq:dI0} in~$\epsilon$, we arrive at \beq
\label{eq:scalarI}
I_0({\bk})= \frac{1}{32\pi^2} \left[ \frac{1}{\epsilon} + \log(4\pi)-\gamma_E + 3 - \log \frac{{\bk}^2}{\mu^2}\right] + {\cal O}(\epsilon)\,.
\eeq
The logarithm comes from expanding
\beq
\left({\bk}^2\right)^{-\epsilon} = \exp\left[-\epsilon  \log {\bk}^2\right] = 1- \epsilon \log {\bk}^2+ {\cal O}(\epsilon^2)\,,
\eeq
whereas the $\log \mu^2$ follows in a similar fashion.\,\footnote{ ~The scaling dimensions are shifted in $d=4-\epsilon$, e.g.  we have $\Mp \to \mu^{-\epsilon/2} \Mp$, so that  $G_N \to G_N\left(1 + \frac{1}{2}\epsilon \log \mu^2 +\cdots\right)$.} Notice that the regularized answer for $I_0(\bk)$ in \eqref{eq:scalarI} depends on $\mu$. This is usually referred as a `subtraction' or `renormalization' scale. 
\subsection{Renormalization}\label{sec:renCVCR}

For pedagogical purposes, in what follows we ignore the fact that the coefficients we will soon introduce to deal with divergences ultimately drop out of observable quantities. The methodology we are about to discuss applies to all the parameters needed to renormalized the theory, see sec.~\ref{sec:effect}. We discuss in sec.~\ref{sec:RenQM} a more relevant example in the radiation theory.

\subsubsection{Counter-Terms}
\vskip 4pt
Although the regularized answer for $I_0(\bk)$ is finite when $d<4$, we still have a pole as $\epsilon \to 0$. This means we need to remove the divergence by adding a `counter-term.'  A first guess would be to absorb the divergence into the mass term $\propto m\, \delta(k\cdot v)$. However, the singularity we find is proportional to,\,\footnote{~Notice dim. reg. preserves spacetime symmetries which are otherwise explicitly broken by introducing a cutoff.}
\beq
\label{t3eps}
\left({\cal T}^{\mu\nu}_{(3)}\right)_{\text{pole}} = \frac{1}{\epsilon}\,\delta(k\cdot v) \left[c_1 (\eta^{\mu\nu}k^2 - k^\mu k^\nu) + c_2 k^2 v^\mu v^\nu \right]\,,
\eeq
with some pre-factors $c_{1,2}$. Hence, in order to make the answer finite in the $\epsilon\to 0$ limit we need to add counter-terms which, by direct inspection, must take the form
\beq
\label{t3ct}
{\cal T}^{\mu\nu}_{\rm ct} (k) = \delta(k\cdot v) \left[C^{\rm ct}_R ~(\eta^{\mu\nu}k^2 - k^\mu k^\nu) + C^{\rm ct}_V ~ \frac{k^2}{2} v^\mu v^\nu \right]\, .
\eeq

The trick is to choose $C^{\rm ct}_{R(V)}$ to cancel the poles. There is not a priori a preferred way to subtract divergences away. Therefore, we must also introduce a {\it renormalization} scheme. Several such schemes are available and a convenient one is the `minimal subtraction bar' ($\overline{\rm MS}$) scheme, see e.g. \cite{iraeft}. In $\overline{\rm MS}$, not only the $\tfrac{1}{\epsilon}$-pole is removed but also some extra constants. Then, splitting the coefficient into a counter-term and a renormalized piece, $C_{R(V)} = C^{\rm ct}_{R(V)} + C^{\rm ren}_{R(V)}$, we have  
\beq
C^{\rm ct}_R =  -\frac{c_1}{\bar\epsilon}\,,~~\frac{1}{2} C^{\rm ct}_V =  -\frac{c_2}{\bar\epsilon}\, , \label{crven}
\eeq
where
\beq
\label{eq:bareps}
\frac{1}{\bar\epsilon} \equiv \frac{1}{\epsilon} -\gamma_E +\log(4\pi)\,,
\eeq
and $c_{1,2}$ are adjusted to cancel the divergences. After these are removed the remaining finite piece, encoded in $C^{\rm ren}_{R(V)}$,  must be obtained by a matching procedure. For example, by comparison between the full and effective theory within the overlapping realm of validity, see sec.~\ref{smatch}. 

As we will soon discuss these type of coefficients may depend on the internal structure of the objects. However, for reasons that will be clear later on, the terms proportional to $C_{R(V)}$ can be consistently removed altogether \cite{nrgr,damourcr1,damourcr2}. This is a consequence of Birkhoff's theorem (or Gauss' law), see appendix~\ref{app:field-redef}. This will not be the case for all the type of divergences we may encounter and subsequent renormalized parameters. We use these terms in what follows for illustrative purposes.

\subsubsection{Renormalization Group Flow}
\vskip 4pt

After the divergences are removed by the addition of $C^{\rm ct}_R, C^{\rm ct}_V$, the final answer becomes \cite{nrgr}
\begin{align}
\label{tab3kmu}
{{\cal T}}^{\alpha\beta}_{(3)}(k) &= -4\pi \delta(k^0) \left[ (\eta^{\alpha\beta} k^2 - k^\alpha k^\beta)\left\{ C^{\rm ren}_R+ \frac{G_N^2m^3}{12} \left(2 - \log \frac{{\bk}^2}{\mu^2}\right)\right\} \right. \\ &+ \left. v^\alpha v^\beta k^2 \left\{ \frac{1}{2} C^{\rm ren}_V - \frac{G_N^2m^3}{6} \left(\frac{75}{32} - \log \frac{{\bk}^2}{\mu^2}\right)\right\}\right] \nn\, ,
\end{align}
 in the $\overline{\rm MS}$ scheme. Notice there is still an arbitrary scale $\mu$. This, of course, would be physically unacceptable. The resolution is simple and it relies on introducing $\mu$-dependence in the renormalized coefficients, $C^{\rm ren}_{R,V}(\mu)$, such that the overall result is independent of any particular value of $\mu$. A natural choice though may be such as to cancel the logarithm in ${\cal T}^{\mu\nu}_{(3)}(k)$. Namely, by setting $\mu^2=\bk^2$ we obtain
\begin{align}
{{\cal T}}^{00}_{(3)}(k) &= 4\pi \delta(k^0) {\bk}^2 \left[  \left\{ C_R^{\rm ren}(|{\bk}|) + \frac{G_N^2m^3}{6}\right\} + \left\{ \frac{1}{2}C_V^{\rm ren} (|{\bk}|)- \frac{25G_N^2m^3}{64} \right\}\right] \\
{{\cal T}}^{ii}_{(3)}(k) &= -8\pi \delta(k^0) {\bk}^2 \left[C_R^{\rm ren}(|{\bk}|) + \frac{G_N^2m^3}{6}\right] \label{tiicr}\, . 
\end{align}
This choice thus trades the explicit logarithmic dependence into the renormalized coefficients.\vskip 4pt Continuing with this exercise let us solve for $C^{\rm ren}_{R(V)}(|{\bk}|)$ in terms of ${\cal T}^{00}_{(3)}(k)$ and ${\cal T}^{ii}_{(3)}(k)$ and plug it back into (\ref{tab3kmu}), but now evaluated at a different scale. For example, for the trace we find
\beq
\frac{{{\cal T}}^{ii}(\tilde {\bk})}{\tilde {\bk}^2} = \frac{{\cal T}^{ii}({\bk})}{{\bk}^2} + \frac{G_N^2m^3}{3} \log \frac{\tilde {\bk}^2}{{\bk}^2}\,,
\eeq
yielding a relationship between the stress-energy tensor at different values of momenta, without reference to $C_R$. The above expression can also be encoded in terms of the renormalized coefficients. For example, using \eqref{tiicr},
\beq
\label{crrg}
C_R^{\rm ren}(|\tilde {\bk}|) = C_R^{\rm ren}(|{\bk}|) - \frac{G_N^2m^3}{6} \log \frac{|\tilde {\bk}|}{|{\bk}|}\, .
\eeq
Along similar lines, we have 
\beq
\label{cvrg}
C_V^{\rm ren}(|\tilde {\bk}|) = C_V^{\rm ren}(|{\bk}|) + \frac{2G_N^2m^3}{3} \log \frac{|\tilde {\bk}|}{|{\bk}|}\, .
\eeq
From here we obtain the renormalization group equations,
\beq
\label{rgflow0}
|{\bk}|\frac{d}{d|{\bk}|}  C^{\rm ren}_{R}(|{\bk}|) = -\frac{G_N^2m^3}{6}\,,~~|{\bk}|\frac{d}{d|{\bk}|}  C^{\rm ren}_{V}(|{\bk}|) = \frac{2G_N^2m^3}{3}\, ,
\eeq
which in this particular case are rather trivial. See sec.~\ref{sec:RenQM} for another example.\vskip 4pt

Notice that even if a renormalized coefficients vanishes at a given $\bk$, these equations imply that they will reappear at another value with a typical size given by the right-hand side of \eqref{rgflow0}. This is a generic feature. Once we know the renormalized parameters at $\mu \simeq |{\bk}|$, the renormalization group flow dictates how they change as we probe the system across different scales. The above analysis also tells us that the dependence on $\mu$ in the renormalized coefficients is nothing but a trick to keep track --and resum-- logarithms (or any other non-analytic structure). The reader may object we still need initial conditions for the renormalization group equations. This is the matching procedure we discuss in sec.~\ref{smatch}.  We introduce first the effective action, and the necessary coefficients to encapsulate finite size effects.

\subsection{Effective Action}\label{sec:effect} 

When an extended object is placed in an external long-wavelength gravitational field, the center-of-mass of the body will trace a worldline which, to first approximation, corresponds to geodesic motion. However, upon closer inspection, we will observe deviations as a result of tidal effects. In what follows we incorporate finite size effects in a point-particle (effective) action approach. This also provides us with a compact way to organize the necessary counter-terms required due to the use of point-like sources. 

\subsubsection{Decoupling}
\vskip 4pt

Let us consider a compact body, of size $r_s$, in an external gravitational field varying on scales much larger than $r_s$. As in the method of regions in sec.~\ref{methodr}, we start by splitting the metric  as
\beq
g_{\mu\nu} \,\, \,=  \underbrace{g^S_{\mu\nu}}_{\rm short\hbox{-}distance}+ \underbrace{g^L_{\mu\nu}}_{\rm long\hbox{-}distance}\, \hspace{-0.3cm},
\eeq
where $g_{\mu\nu}^{L(S)}$ represent the long- and short-distance modes, compared with $r_s$. For the case of neutron stars, the short-distance physics includes the geometry, $g^S_{\mu\nu}$, and also the positions, $x_p^\mu(\sigma_p)$, of all the approximately $10^{40}/{\rm m}^3$ constituents of the star. Moreover, we must also incorporate other fields which participate in the dynamics on short scales. Hence, analytically solving for the equations of motion of each single particle is an unsurmountable task. However, since we are interested in the motion in long-wavelength backgrounds, we may rely on an effective description which depends solely on long-distance degrees of freedom, e.g. the collective motion of the body.\vskip 4pt To construct the effective theory we thus resort {\it only} to the long-wavelength metric field, $g^L_{\mu\nu}$, as well as the center-of-mass of the compact object,~$x^\mu_{\rm cm}(\sigma)$. The effective action, $S_{\rm eff}\left[x_{\rm cm},g^L_{\mu\nu}\right]$, follows by integrating out the short-distance modes in the saddle-point approximation. For instance, for a neutron star we have
\begin{align}
\label{zspp}
\exp &\Big\{i S_{\rm EH}\left[g^L_{\mu\nu}(x)\right]+ iS_{\rm eff}\left[x_{\rm cm}^\alpha(\sigma),g^L_{\mu\nu}(x)\right]\Big\}= \\
\int  D  g^S_{\mu\nu}(x) &D \delta x^\alpha_p(\sigma_p) \exp\Big\{i S_{\rm EH}\left[g_{\mu\nu}(x)\right]+ iS_{\rm int}\left[x^\alpha_p(\sigma_p), g^S_{\mu\nu}(x)\right]\Big\}\, , \nn
\end{align}
where $S_{\rm \rm int}[x^\alpha_p, g^S_{\mu\nu}]$ describes the dynamics of the internal degrees of freedom. The latter includes not only the short-distance metric, $g^S_{\mu\nu}$, but also the displacement of the constituents of the neutron star measured with respect to the center-of-mass, $\delta x^\alpha_p \equiv x^\alpha_p -x_{\rm cm}^\alpha$.\vskip 4pt  In most cases the exact form of the effective action is unknown. However, we have at hand the symmetries of the long-distance physical system, namely diffeomorphism and reparameterization invariance. Therefore, we write an effective action using all possible --local-- terms respecting such symmetries. For~objects which are spherically symmetric in isolation, we have \cite{nrgr,nrgrLH}
\begin{align}
\label{acrcv}
&S_{\rm eff}\left[x_{\rm cm},g^L_{\mu\nu}\right] =  \int d^4x~d\sigma~\delta^4(x-x_{\rm cm}(\sigma)) \Big( -m~ \sqrt{ g^L_{\mu\nu}(x) u_{\rm cm}^\mu(\sigma) u_{\rm cm}^\nu(\sigma)}\\ &+ C_R \int R^L \left[g^L_{\mu\nu}\right]~\sqrt{ g^L_{\mu\nu}(x) u_{\rm cm}^\mu(\sigma) u_{\rm cm}^\nu(\sigma)}  + C_V \int R^L_{\mu\nu}\left[g^L_{\mu\nu}\right] \frac{u_{\rm cm}^\mu(\sigma) u_{\rm cm}^\nu(\sigma)}{\sqrt{ g^L_{\mu\nu}(x) u_{\rm cm}^\mu(\sigma) u_{\rm cm}^\nu(\sigma)}} +\cdots \Big) \, .\nn
\end{align}
The extra terms beyond the geodesic approximation (or minimal coupling) are organized in powers of derivatives of the long-wavelength perturbation, $g^L_{\mu\nu}$. After the symmetries are incorporated the remaining freedom enters in a series of parameters, often called `Wilson coefficients,' which carry information about the inner structure of the bodies. For (non-rotating) black holes, they are determined solely in terms of~$r_s$ and Newton's constant, e.g. $m=r_s/(2G_N)$. On the other hand, for a neutron star, they depend on the equation of state, making these coefficients suitable to test different models, e.g. \cite{poissonlove,damourlove,tidaltan0,tidaltan,tidalflan,binitidal,iloveq,iloveq2,steinchak2,steinchak,eosNS2,eosNS,eosNS3,binitidalspin}.
 \vskip 4pt 
The fact that all the information about the short-distance dynamics is now encoded in a series of coefficients is one of the pillars of the EFT framework. This is often referred as {\it decoupling} \cite{decoupling}. Decoupling of short-distance (and short-time) physics implies that their influence can be encoded in (derivatively coupled) interactions which look {\it local} on scales $|\bk|^{-1} \gg r_s$ (and frequencies $\omega \ll r_s^{-1}$), and parameterized in powers of $|{\bk}| r_s \ll 1$ (and $\omega r_s \ll 1$). The Wilson coefficients are universal, which means once they are obtained --via a matching procedure we discuss momentarily-- they can be used in different situations. The effective action can be then implemented broadly, for any situation where extended bodies are perturbed by long-distance probes. This allows us to study complicated dynamical configurations, e.g. compact bodies moving in the background field produced by a companion and emitting gravitational waves.
 
\subsubsection{Finite Size Effects: Background and Response}
\vskip 4pt

The reader will notice the two terms proportional to $C_{R(V)}$ in \eqref{acrcv} are the same we introduced before as counter-terms through the stress-energy tensor, e.g.~\eqref{t3ct}. As we mentioned, the $C_{R(V)}$ coefficients can be consistently set to zero as a consequence of Birkhoff's theorem. This will be a general feature in the EFT framework where, at a given order, terms that vanish by lower order equations of motion can be systematically eliminated~\cite{iraeft}.\,\footnote{~As a consequence of Einstein's equations, the terms proportional to the Ricci tensor do not propagate long-distance effects, and can be thus removed from the effective theory. See appendix \ref{app:field-redef}. We drop the `cm' and $L$ labels from now on.}\vskip 4pt  After we discard terms proportional to the Ricci tensor, we re-write the action in terms of the electric, $E_{\mu\nu}$, and magnetic, $B_{\mu\nu}$, components of the Weyl tensor $C_{\mu\alpha\nu\beta}$, 
\begin{align}
\label{eq:defE}
E_{\mu\nu} = C_{\mu\alpha\nu\beta} u^\alpha u^\beta\,,~~ B_{\mu\nu} = \frac{1}{2} \epsilon_{\mu\alpha\beta\sigma} {C^{\alpha\beta}}_{\nu\rho} u^\sigma u^\rho\, ,
\end{align}
which obey $g_{\alpha\beta} E^{\alpha\beta}= g_{\alpha\beta} B^{\alpha\beta}  =0$ and $E_{\alpha\beta}u^\beta=B_{\alpha\beta}u^\beta=0$. \vskip 4pt

Reparameterization invariance is incorporated by using the proper time, $\sigma \to \tau$, so that $u^2=1$. Moreover, we introduce a locally-flat co-moving frame, $e^\mu_A(x)$, with $e^\mu_0 = u^\mu$, such that $E_{AB}(x) = e^\mu_A(x) e^\nu_B(x) E_{\mu\nu}(x)$, and similarly for the magnetic term.  Notice that in this frame $E_{0A} = B_{0A}=0$. Therefore only space-like components, i.e. $E_{ij}$, are relevant. We then write the following term in the effective action at leading order in derivatives 
\beq
\label{sqe}
S_{Q_E} =  \frac{1}{2} \int d\tau~d^4x~\delta^4(x-x(\tau))~ Q_E^{ij}(\tau) E_{ij}(x) \, ,\eeq
and similarly with $E_{ij} \to B_{ij}$. Here the $Q^{E(B)}_{ij}(\tau)$ are the trace-free electric (and magnetic) quadrupole moments. Notice that the $(ij)$ indices transform under $SO(3)$, and therefore can be raised and lowered using the Euclidean metric. While the gravitational interaction is local,  these quadrupole moments may depend on time. We then split them into different components. There is a {\it background} piece which accounts for the short-distance contributions in the absence of external perturbations, and there is a {\it response} to long-wavelength probes, i.e.
\beq
\label{qsr}
Q^{E(B)}_{i\ldots i_n} = \underbrace{\left\langle Q^{E(B)}_{i\ldots i_n}\right\rangle_S}_{\rm background} + \underbrace{\left(Q^{E(B)}_{i\ldots i_n}\right)_{\cal R}}_{\rm response}\, .
\eeq
The expectation value, $\langle \cdots \rangle_S$, is  computed in the background of the short modes. For example, for the (trace-free) quadrupole of an object which is spherically symmetric in isolation, we have $\left\langle Q_E^{ij} \right\rangle_S = 0$. There is no contribution linear in the metric from \eqref{sqe} in such case. (This would not be the case for objects with a permanent quadrupole moment, or rotating bodies. For the latter $\left\langle Q_E^{ij} \right\rangle_S$ is proportional to the spin (squared) of the compact object. See part~\ref{sec:spin} for more details.)\vskip 4pt 

The second term in \eqref{qsr} represents the induced multipole moment in the presence of a long-wavelength perturbation. The $E_{ij}$ serves as a source for the response of the electric-type quadrupole.  Then, using linear response theory, we write 
\beq
\label{responseret}
\left(Q_E^{ij}\right)_{\cal R}(\tau) =  \frac{1}{2} \int d\tau' d^4 x~\delta^4(x-x(\tau'))\Big(i G_{\rm ret}^{ij,kl}(\tau,\tau')\Big) E_{kl}(x) + \cdots\, ,
\eeq
with the retarded Green's function,
\beq
\label{eq:responseret1}
G^{\rm ret}_{ij,kl}(\tau,\tau') = \left\langle \left[Q_E^{ij}(\tau),Q_E^{kl}(\tau')\right] \right\rangle  \theta(\tau-\tau')\, ,
\eeq
and similarly for the magnetic-type multipole moments.\vskip 4pt If we ignore absorption (which is related to the imaginary part, see sec.~\ref{sec:abs}) we can concentrate on the real part of the response. We parameterize the latter in terms of the symmetries of the theory. In our case, the $SO(3)$ invariance --for non-rotating bodies-- which is manifest in the locally-flat frame.\,\footnote{~This is reminiscent of the so called AdS/CFT correspondency, where isometries of the spacetime metric are mapped into global symmetries in the field theory side \cite{juan1,juan2,juan3}.} 
Therefore, we have (in frequency space)
\beq
\label{responseret2}
{\rm Re}\left(iG^{\rm ret}_{ij,kl}(\omega)\right) =Q_{ijkl} \,\,{\rm Re}f(\omega)\, ,
\eeq
with the projector onto symmetric and traceless two-index (spatial) tensors give by
\beq
\label{eq:qijkl}
Q_{ijkl} = \left( \delta_{ik}\delta_{jl} + \delta_{il}\delta_{jk} - \tfrac{2}{3} \delta_{ij}\delta_{kl}\right)\,.
\eeq
Furthermore, since we are interested in long-wavelength perturbations, the function ${\rm Re}f(\omega)$ can be expanded in powers of $\omega r_s \ll 1$. Using the (time-reversal) symmetry of the real part of the response ($\omega \to -\omega$),
\beq
{\rm Re}\,f(\omega) = C_E + C_{\ddot E}\, r_s^2\omega^2 + {\cal O}\left(\omega^4 r^4_s\right),
\eeq
and we find 
\beq
\left(Q_E^{ij}\right)_{\cal R} =  \int d^4x~ \delta^4(x-x(\tau)) \left(C_E E_{ij}(x) + C_{\ddot E} \, r_s^2 \frac{d^2}{d\tau^2} E_{ij}(x) + \cdots \right)\, .
\eeq
The $C_E,C_{\ddot E}$, etc., are (time-independent) Wilson coefficients. If we now plug this result back into \eqref{sqe}, we obtain (similarly with $E \to B$)\,\footnote{~Technically speaking we are not allowed to plug the expression for the response into the action. However, because the Wilson coefficients are fixed by matching, it is easy to show we get the same equations up to a trivial re-scaling. Notice we also absorbed the factor of $r_s^2$ into $C_{\ddot E}$ and integrated by parts. Moreover, we expressed the result in a covariant manner, by using $E_{ij}E^{ij} \to E_{\mu\nu}E^{\mu\nu}$ and $\frac{d}{d\tau} \to u^\alpha \nabla_\alpha$, for a generic frame.}
\beq
\label{Se2}
S_{Q_E}  \to  \frac{1}{2} \int d\tau \int d^4x~ \delta^4(x-x(\tau)) \Big[ C_E \, E_{\mu\nu}(x) E^{\mu\nu}(x) + C_{\ddot E}\, u^\alpha(\tau) \nabla_\alpha E_{\mu\nu}(x)\, u^\alpha(\tau) \nabla_\alpha E^{\mu\nu}(x)  +\cdots \Big]\, ,\eeq
for the first corrections describing finite size effects for extended (non-rotating) bodies. Notice, since we chose the proper time, the reparameterization invariance in \eqref{Se2} is manifest. (This is true also for the factors of velocity within $E_{\mu\nu}$ and $B_{\mu\nu}$, see \eqref{eq:defE}.) It may be re-written in terms of any affine parameter,~$\sigma$, introducing factors of $\sqrt{u^2}$, as in \eqref{acrcv}. The~procedure continues by adding diffeomorphism invariant high(er) derivative terms. As we discuss momentarily, the scaling of the leading term in \eqref{Se2} is connected with the effacement of internal structure in the two-body problem, see sec.~\ref{sec:eff}.

\subsection{Matching: Black Holes vs. Neutron Stars}\label{smatch}
\vskip 4pt  
In order to extract the Wilson coefficients, e.g. $C_{E(B)}$, we need either experimental data, or a (mathematical) procedure by which computations in the full and effective theory can be compared within the overlapping realm of validity. This is the matching we alluded before, e.g. \cite{iraeft}, and for the latter option it may or may not involve observable quantities. In principle the number of Wilson coefficients is infinite. However,  terms in the action scale with a definite power of the expansion parameter(s) in the theory, and therefore we may truncate the series at a desired level of accuracy.\vskip 4pt For concreteness, let us perform the matching for the metric produced by a generic extended compact object, of size $r_s$, at distances $r >r_s$. (We will see later on that the leading order Wilson coefficient for spinning bodies may be obtained in this fashion.) This is in general a complicated function of $(r,r_s)$ and, for the case of a neutron star, also some dimensionful parameters which describe the internal structure. For illustrative purposes, and simplicity, let us consider instead a non-rotating black hole at rest, such that the only relevant scale is given by $r_s$. Moreover, let us concentrate on a scalar observable, say the trace $h^\alpha_\alpha$, and for static configurations. The following steps can be easily generalized to other situations.\vskip 4pt From dimensional analysis we have (in Fourier space and ignoring an overall $\delta(k_0)$) 
\beq \frac{h^\alpha_\alpha(\bk)}{\Mp} = \frac{1}{\bk^2} f( |{\bk}|r_s)\,,\eeq 
which we can compute both in full fledged general relativity as well as in the effective theory, for $z\equiv |{\bk}|r_s \ll 1$. The full answer may be written as,
 \beq
 \label{ffull}
f_{\rm gr}(z)  = P(z) + Q(z)\log(z)\,,
 \eeq
with $P(z),Q(z)$ some analytic functions. For simplicity, we have assumed the non-analytic term is governed by a logarithm, but in general we may have a more complicated structure. On the other hand we have the EFT computation. By construction, the low-energy degrees of freedom are always kept as propagating modes in the effective theory. Therefore, the EFT reproduces the long-distance physics, including the non-analytic behavior in the IR limit $|{\bk}| \to 0$. In this example, this means we will find~\cite{nrgr}
\beq
 \label{feft}
f_{\rm eft}(|{\bk}|,\mu) = \left[{\bar P}(|{\bk}|,C_i(\mu)) + {\bar Q}(|{\bk}|,C_i(\mu))\log {|{\bk}| \over{\mu}}\right]\,,
 \eeq
in the EFT side, with $\bar P$ and $\bar Q$ some polynomials.\,\footnote{Note the logarithm in \eqref{feft} is associated with a UV divergence in the EFT. The long-distance non-analytic behavior in the full theory is thus linked to UV singularities in the EFT, due to the $r_s \to 0$ approximation. This allows us to resum logarithms using renormalization group techniques. See sec.~\ref{sec:interplay} for an explicit example.} The $\mu$ dependence enters through dim. reg., see sec.~\ref{sec:renorm}.
Notice we have no explicit reference to $r_s$. This is not surprising, since in the effective theory we have taken the limit $r_s\to 0$, at the expenses of introducing UV divergences. As we discussed, these are removed by the introduction of counter-terms and finite size effects are encapsulated in the (renormalized) $C_i(\mu)$ coefficients.\vskip 4pt To complete the matching we decide on a level of accuracy, e.g. ${\cal O}(z^N)$, and equate both results, \beq  f_{\rm eft}(|{\bk}|,\mu\simeq r^{-1}_s) = f_{\rm gr} (|{\bk}|r_s \ll 1) + {\cal O}\left(z^{N+1}\right)\label{eq:fgrfeft}\,,\eeq
such that,
\begin{align}
\bar Q({|\bk|},C_i(\mu\simeq 1/r_s)) &\simeq \sum_{\ell=1}^{\ell= N} q_\ell (|{\bk}|r_s)^\ell,~~\bar P({|\bk|},C_i(\mu \simeq 1/r_s)) \simeq \sum_{\ell=1}^{\ell= N} p_\ell (|{\bk}|r_s)^\ell\,,
\end{align}
with  $q_\ell,p_\ell$ the coefficients in the Taylor expansion of $P(z), Q(z)$ for small $z$. From here we solve for the $C_i(\mu \simeq 1/r_s)$ as a function of $r_s$ by isolating the relevant powers of $z$. The choice $\mu \to \mu_{\rm UV} \simeq 1/r_s$ for the UV matching scale in \eqref{eq:fgrfeft} prevents the existence of large logarithms, which may jeopardize the perturbative expansion. After the Wilson coefficients are extracted at that scale, the renormalization group flow allows us to {\it run} to long(er) distances, $r_s^{-1} \simeq \mu_{\rm UV} \leq \mu \to \mu_{\rm IR} \simeq r^{-1}$. (We will analyze a specific example in sec.~\ref{sec:RenQM}, for which the relevant scales are given by: $\mu_{\rm IR} \to \lambda^{-1}_{\rm rad}$ and $\mu_{\rm UV} \to r^{-1}$.)\newpage
The reader may object that, since terms quadratic in the metric perturbation do not contribute to the one-point function, the above procedure does not allow us to extract $C_{E(B)}$ for an isolated object in a flat background. Moreover, Birkhoff's theorem precludes parameters, other than the mass, to enter in the metric outside of a spherically symmetric, non-rotating, stationary object. (This is, after all, the reason we discarded the $C_{R(V)}$ terms from the effective action.) In spite of this, the previous steps are generic and therefore also apply for different matching conditions. For example, instead of a flat background we may place a (non-rotating) body in an external {\it electric} field, ${\cal E}_{\mu\nu} \neq 0$, varying on scales larger than $r_s$. In such case the finite size term scales linearly in the perturbations (defined with respect to the curved background) 
\beq
C_E \int {\cal E}_{\mu\nu} E^{\mu\nu}[h_{\alpha\beta}]\, ,
\eeq 
and our previous analysis follows, allowing us to extract the value of the $C_E$ coefficient. Moreover, we can also consider a boosted black hole in the presence of an external magnetic component, ${\cal B}_{\mu\nu}$, such that $C_B$ also enters in the one-point function.\vskip 4pt

Let us stress an important point. While matching may entail fixing a gauge (or a coordinate system) in general relativity, the final result for the Wilson coefficients is diffeomorphism invariant. That is the case because they appear multiplying gauge invariant terms in the effective action. Moreover, once they are obtained, by whichever mean, there will be no left-over reference to either a particular gauge or the type of (long-wavelength) probe used in the matching procedure. For instance, to read off the $C_{E(B)}$ Wilson coefficients we could have equally considered gravitational wave scattering. The cross section in the EFT framework can be easily computed, and it scales as \cite{nrgrLH}
\beq
\sigma_{\rm eft} (\omega) = \cdots + \frac{C^2_{E(B)}}{\Mp^4}\omega^8 +\cdots\, .
\eeq
On the other hand, in the fully relativistic computation we expect, say for the case of a black hole,
\beq
\sigma^{\rm bh}_{\rm gr} (\omega) = r_s^2 f(r_s\omega)\,,
\eeq
with $f(z)$ an analytic function. (As we discussed, non-analytic terms cancel out in the matching and therefore we do not include them here for simplicity.) If we then expand in powers of $\omega r_s \ll 1$, we obtain
\beq
\sigma^{\rm bh}_{\rm gr} (\omega) \simeq  r_s^2 \,(\cdots + (r_s\omega)^8+\cdots )\,,
\eeq
and comparing both expressions (up to a numerical factor) we find
\beq
C_{E(B)} \sim \Mp^2 r_s^5 \label{rs5}\,. 
\eeq
Unlike the (off-shell) static case, the Wilson coefficients entered here in an on-shell scattering amplitude. The result, however, is universal. We can then use the expression in \eqref{rs5} to derive the value of the one-point function in a curved background, or vice versa. As we shall discuss later on,  the scaling in \eqref{rs5} does not hold for black holes which, somewhat unexpectedly, turn out to have vanishing $C_{E(B)}$'s, see sec.~\ref{sec:eff} and sec. \ref{sec:conclusion}. That is not the case for neutron stars.

\newpage
\section{Non-Relativistic General Relativity}\label{sec:nrgr}

After integrating out the short(est) distance scale, $r_s$, we then collect two compact extended objects and form a binary system. 
The steps toward solving for the dynamics resemble our treatment of the scalar toy model in part~\ref{sec:part1}\,; up to gauge fixing issues, tensor structure and more complicated non-linear interactions. We will also find a rich renormalization group structure for the parameters of the theory. 

\subsection{Potential and Radiation Regions}

The bound state adds two extra regions to the problem. First of all, we have modes varying on scales of the order of the separation of the bodies, $r \sim r_s/v^2$. Furthermore, there is the radiation scale, $\lambda_{\rm rad} \sim r/v$. As we mentioned, when~$v \ll 1$, we have  \beq \lambda_{\rm rad}^{-1} \ll r^{-1} \ll r_s^{-1}\,.\eeq Following the method of regions in sec.~\ref{methodr} we split the metric perturbation into two classes,
\beq
h_{\mu\nu} = \underbrace{H_{\mu\nu}}_{\rm potential}+\underbrace{\bar h_{\mu\nu}}_{\rm radiation}  \, ,
\eeq
 potential and radiation modes, according to the scaling rules
\beq (k_0,\bk)_{\rm pot} \sim (v/r,1/r)\, ,~~(k_0,\bk)_{\rm rad} \sim (v/r,v/r)\,.\eeq 
The latter represents the on-shell propagating degrees of freedom, whereas the former are the off-shell modes which mediate the binding forces between the constituents of the binary. The potential modes do not appear as external states in the EFT, therefore the $H_{\mu\nu}$ field must be integrated out in a background of radiation modes. The non-linear coupling between potential and radiation complicates the calculation of the power loss. However, the fact that the EFT framework separates the relevant scales, one at a time, greatly simplifies the computations as we show next in several steps. 

\subsection{Binding Potential} 
The binding potential is obtained by integrating out the potential modes, ($a=1,2$)
\beq
\label{pathW}
 e^{i W[{\bx}_a]} =  \int  DH_{\mu\nu}  \exp\Big\{i S_{\rm EH}[H_{\mu\nu}]+ iS^{\rm pp}_{\rm eff}[{\bx}_a(t), H_{\mu\nu}]+ iS_{\rm GF}[H_{\mu\nu}]\Big\}\, ,
\eeq 
while ignoring the radiation field.\,\footnote{We will extend this procedure later on to incorporate (conservative) contributions from radiation-reaction. See~sec.~\ref{sec:self}.} We use a gauge fixing term, as in \eqref{eq:Sgf1}, and the coordinate time as affine parameter. The point-particle action, $S_{\rm eff}^{\rm pp}[\bx_a,H_{\mu\nu}]$, includes a series of terms, as in \eqref{Se2}. At~zero-th order in $G_N$ \beq W_{(0)}[{\bx}_a] \to S^{\rm pp}_{\rm eff}[{\bx}_a,\eta_{\mu\nu}] \equiv \int dt ~K[\bx_a]\,,\eeq representing the kinetic part of the effective action, which we must also expand in powers of $v$. Next we compute $W[\bx_a]$, by solving for $H_{\mu\nu}$ perturbatively and plug it back into the action.

\subsubsection{Quasi-Instantaneous Modes}
\vskip 4pt 
The potential modes are off-shell, therefore they only contributes to the real part of $W[\bx_a]$,  
 \beq 
 \label{eq:SKV2}
 {\rm Re}\,W[\bx_a] = \int dt\, \left(K[\bx_a] - \,V[\bx_a]\right)\, ,\eeq	 
from which we read off the binding potential, $V[\bx_a]$. To integrate out $H_{\mu\nu}$ we use the `quasi-instantaneous' Green's function 
\begin{align}
\label{prpH}
\big\langle T\big\{H_{\mu\nu}(t_1,{\bx}_1)H_{\alpha\beta}(t_2,{\bx}_2)\big\}\big\rangle &= -i P_{\mu\nu\alpha\beta}~\left[\delta(t_1-t_2) \int_\bk \frac{1}{{\bk}^2} e^{i{\bk}\cdot({\bx}_1-{\bx}_2)}\right. \\ &+ \left. \frac{d}{dt_1dt_2} \delta(t_1-t_2) \int_\bk \frac{1}{{\bk}^4} e^{i{\bk}\cdot({\bx}_1-{\bx}_2)} + \cdots \right]\,. \nn
\end{align}
As before, we expanded the denominator in \eqref{eq:DFprop} in powers of $k_0/|\bk| \ll 1$, for slowly moving sources. The propagator and velocity corrections will be represented as in Fig.~\ref{fig1}.\vskip 4pt  The leading order term in~(\ref{prpH}) gives us the scaling laws
 \beq
 \label{Hscal}
 \big\langle T\big\{H_{\mu\nu}(t,{\bk})H_{\alpha\beta}(0,{\bq})\big\}\big\rangle \sim \frac{1}{\bk^2} \delta(t) \delta^3({\bk} + {\bq})~ \to ~ [H_{\mu\nu}(t,{\bk})] \sim r^2 \sqrt{v}\,,
\eeq
where we used $t \sim r/v$, $|\bk| \sim 1/r$. We also have (with $L=mvr$)
\beq
\frac{m^2}{\Mp^2} \sim L v ~\to~ [H_{\mu\nu}(t,{\bx})] \sim \Mp \frac{v^2}{\sqrt L}\,.
\eeq
We show in sec.~\ref{sec:eff} that first non-trivial Wilson coefficient in the point-particle action enters at ${\cal O}(v^{10})$ for non-spinning objects. Until then the mass coupling,
\beq
S_{\rm eff}^{\rm pp}[\bx_a,H_{\mu\nu}] = -\frac{m}{2\Mp} \int dt \frac{v^\mu v^\nu}{\sqrt{v^\alpha v_\alpha}} H_{\mu\nu} +\cdots \,
\label{tpphmn},
\eeq 
with $v^\mu =(1,{\bv}^i)$, suffices. Note the leading order term scales as 
\beq
\label{loh00}
\left[\frac{m}{2\Mp} \int H_{00} dt \right] \sim  \frac{m v^2r}{\sqrt{L}v} \sim \sqrt{L}\,.
\eeq
We can use this worldline coupling to calculate $W[\bx_a]$ at leading order in $G_N$. Indeed, this is identical to the computation in sec.~\ref{sec:binding}, and following similar steps we reproduce the Newtonian potential
\beq
\label{W1}
V_{ N}[{\bx}_a] =  -\frac{m_1m_2}{32 \pi \Mp^2} \frac{1}{|{\bx}_1-{\bx}_2|} =  -\frac{G_Nm_1m_2}{r}\,,
\eeq
as in (\ref{eq:bindE}) (after identifying $M \to 2\Mp$ and including a factor of $P_{0000}=1/2$). 

\subsubsection{Einstein-Infeld-Hoffmann Lagrangian}\label{sec:EIH}
\vskip 4pt

\begin{figure}[t!]
\centering
\includegraphics[width=0.7\textwidth]{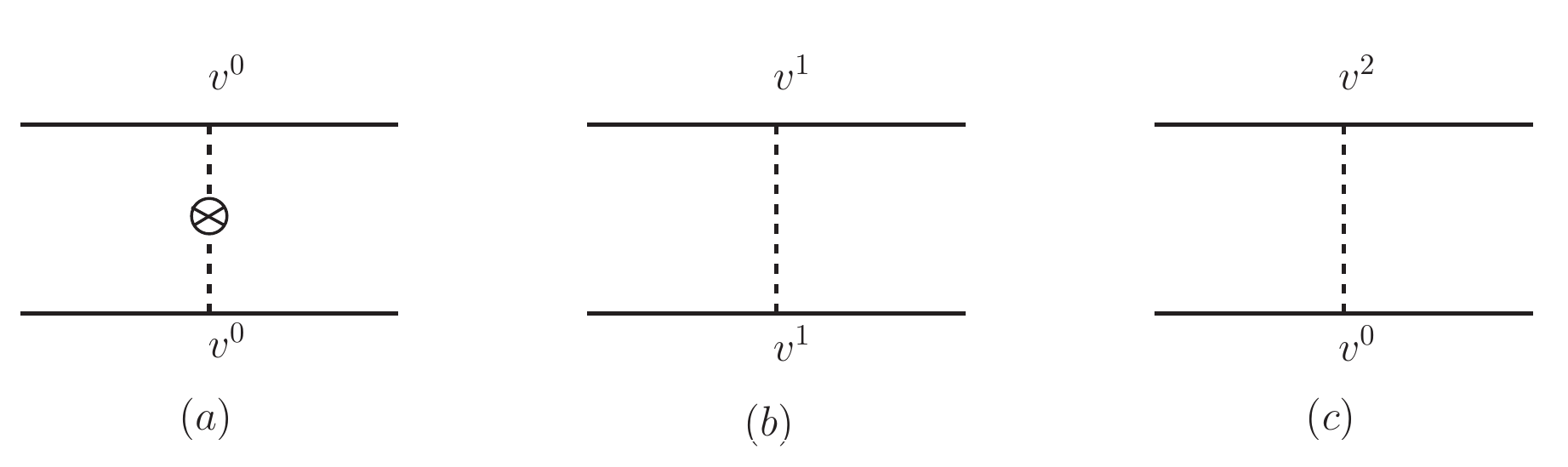}
\caption{ The first diagram accounts for an insertion of $p_0^2$ in \eqref{prpH}. The subsequent diagrams include velocity corrections from the point-particle action in \eqref{tpphmn}.}
\label{fig1EIH}
\end{figure}

\begin{figure}[t!]
\centering
\includegraphics[width=0.7\textwidth]{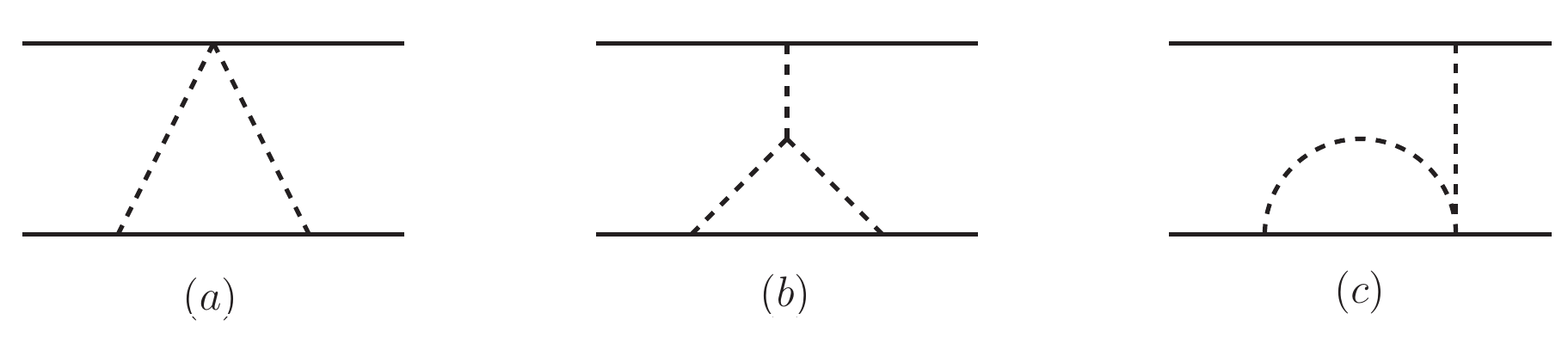}
\caption{ Non-linear (static) contributions from the point-particle source (left), see \eqref{scalseg},  and Einstein-Hilbert action plus gauge fixing term (middle), see \eqref{h00c}. The figure on the right represents a divergent self-energy contribution, which can be removed using dim. reg.}
\label{fig2EIH}
\end{figure}

The first correction to the Newtonian dynamics enters at order $L v^2$, or 1PN. The relevant couplings are summarized in appendix~\ref{app:Fey}. 
The Feynman diagrams are depicted in Figs.~\ref{fig1EIH} and \ref{fig2EIH}\,. 
\vskip 4pt
For the velocity corrections in Fig.~\ref{fig1EIH} we obtain, 
\begin{align}
\label{w05pn}
V_{\ref{fig1EIH}(a)}[\bx_a] &= -\frac{1}{2} \frac{G_Nm_1m_2}{r}\left(\bv_1\cdot\bv_2 - \frac{(\bv_1\cdot \br)( \bv_2\cdot \br)}{r^2}\right) \,,\\
V_{\ref{fig1EIH}(b)}[\bx_a] &= 
4 \frac{G_Nm_1m_2}{r} {\bv}_1\cdot{\bv}_2\,,\\
 V_{\ref{fig1EIH}(c)}[\bx_a] &= -\frac{3}{2}  \frac{G_Nm_1m_2}{r} {\bv}_1^2\,.
\end{align}
The remaining contributions are due to non-linear couplings, shown in Fig.~\ref{fig2EIH}.  For example, for the non-linearity stemming off the worldline we have
\beq
\label{scalseg}
\left[ \frac{m}{8\Mp^2} \int  H_{00}^2 dt \right] \sim m \frac{v^4}{L} \frac{r}{v} \sim v^2\, .
\eeq
This term thus enters at 1PN through the {\it seagull} diagram in Fig.~{\ref{fig2EIH}(a)},
\bea
&& {\rm Re}\,W_{{\ref{fig2EIH}(a)}}[{\bx}_a] =   \frac{1}{2!} \left(\frac{-i m_2}{2\Mp}\right)^2 \frac{im_1}{8\Mp^2}\int dt_1dt_2 d\tilde t_2 \left\langle T\left \{ H_{00}(t_2,{\bx}_2(t_2))H_{00}(\tilde t_2,{\bx}_2(\tilde t_2))H_{00}^2(t_1,{\bx}_1(t_1))\right\}\right \rangle \nn \\
&=& -\frac{m_1^2m_2}{32\Mp^4}\int dt_1dt_2 d\tilde t_2\, \delta(t_1-t_2)\delta(t_1-\tilde t_2) (-iP_{0000})^2 \int_{\bp,\bq} \frac{e^{i{\bp}\cdot ({\bx}_2(t_2)-{\bx}_1(t_1))}}{{\bp}^2}\frac{ e^{i{\bq}\cdot ({\bx}_2(\tilde t_2)-{\bx}_1(t_1))}}{{\bq}^2}\, ,
\eea
and we get \cite{nrgr}
\beq
V_{\ref{fig2EIH}(a)}[{\bx}_a] = - \frac{m_1^2m_2}{(4\pi)^2 128 \Mp^4} \frac{1}{|{\bx}_1(t)-{\bx}_2(t)|^2} = - \frac{G_N^2 m_1^2m_2}{2r^2}.
\eeq
There is one more diagram, Fig. \ref{fig2EIH}$(b)$, for which we need the three-graviton vertex~\cite{eftg1,eftg2,nrgr}, 
\begin{align}
\label{h00c}
& \big\langle T \big\{ H_{00}(t_1,{\bp}_1) H_{00}(t_2,{\bp}_2) H_{00}(t_3,{\bp}_3) \big\} \big\rangle = \\ 
\frac{i}{4 \Mp} \delta (t_1-t_2)\delta (t_1&- t_3) \left(\frac{-i}{{\bp}^2_1}\right)\left(\frac{-i}{{\bp}^2_2}\right)\left(\frac{-i}{{\bp}^2_3}\right)\left( {\bp}_1^2+{\bp}_2^2+{\bp}_3^2\right)\delta^3({\bp}_1+{\bp}_2+{\bp}_3)\nn\, .
\end{align}
To show that the three-graviton contributes at 1PN we use the power counting rule in \eqref{Hscal} \cite{nrgr}
\beq
\left [ \frac{1}{\Mp} \int dt~\delta^3({\bk}) {\bk}^2 \Big(d^3{\bk} H_{\mu\nu}(t,{\bk})\Big)^3\right] \sim \frac{1}{\Mp}\frac{r}{v} r^3 \frac{1}{r^2}\frac{\Mp^3v^6}{{\sqrt L}^3} \sim \frac{v^2}{\sqrt{L}}\,,
\eeq
such that Fig.~\ref{fig2EIH}$(b)$ scales as $(\sqrt{L})^3 v^2/\sqrt{L} \sim Lv^2$.\vskip 4pt
When we compute this diagram we run into divergences. This occurs when the ${\bp}_i^2$ part of the vertex in the numerator cancels the corresponding propagator, thus shrinking the corresponding line to a contact interaction. This produces a self-energy contribution, shown in Fig.~\ref{fig2EIH}$(c)$, which diverges. These are regularized using dim. reg. and set to zero as we discussed previously in sec.~\ref{sec:regular0}. The only term that survives is the one which does not involve a scale-less integral, and we obtain \cite{nrgr}
\beq
V_{\ref{fig2EIH}(b)}[{\bx}_a]  =  \frac{G_N^2 m_1^2m_2}{r^2}\, .
\eeq
Garnering the pieces we arrive at the Einstein-Infeld-Hoffmann Lagrangian~\cite{EIH1pn}
\beq
\label{w1eih}
L_{\rm EIH} = {1\over 8}\sum_{a=1,2} m_a {\bv}^4_a + {G m_1 m_2\over 2 r}\left[3 ({\bv}^2_1+ {\bv}^2_2) - 7({\bv}_1\cdot {\bv}_2)  - {({\bv}_1\cdot  {\br})({\bv}_2\cdot {\br})\over r^2} \right]- {G_N^2 m_1 m_2 (m_1+m_2)\over 2  r^2}
 \, .
\eeq

\subsubsection{Higher PN Orders: Kaluza-Klein Decomposition}\label{sec:KK}
\vskip 4pt

\begin{figure}[t!]
\centering
\includegraphics[width=0.50\textwidth]{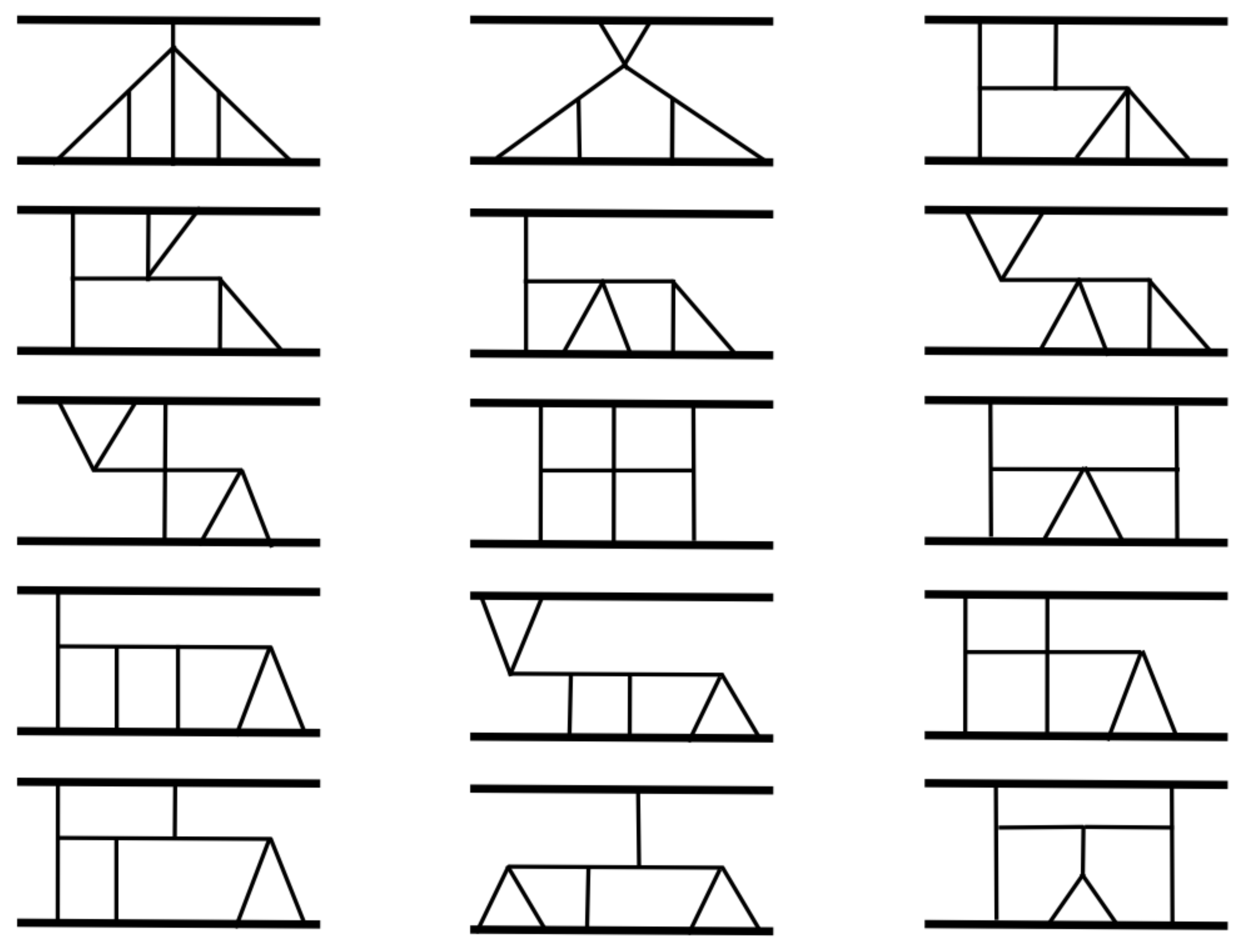}
\caption{Some of the topologies which contribute to 4PN order, see \cite{riccardocqg}. Each line represents scalar, vector and tensor exchanges, and the vertices are determined according to the Feynman rules following from the action.}\label{fig:4pn}
\end{figure}

To compute the Einstein-Infeld-Hoffmann Lagrangian we used the three-graviton vertex. However, only the $H_{00}$ component matters at this order. This suggests that a spacetime decomposition into scalar, vector and  $(3)$-metric may simplify the number of diagrams and Feynman rules at each given order.  A~useful split was then introduced in \cite{smolkin1}, see also \cite{KK2}, consisting on the following `Kaluza-Klein' decomposition:
\beq
\label{eq:kk}
g_{\mu\nu} dx^\mu dx^\nu= e^{2\varphi}\left(dx^0- \bA_i d\bx^i\right)^2- e^{-2\varphi} \gamma_{ij} d\bx^id\bx^j\, . 
\eeq 
(Note this also resembles the ADM formalism \cite{adm}.) The Einstein-Hilbert action takes the form \cite{smolkin1}
\beq
S_{\rm EH} = -2\Mp^2 \int d^4x \sqrt{\gamma}\left[ R^{(3)}[\gamma] + 2\gamma^{ij} \partial_i\varphi\partial_j \varphi + e^{4\varphi} F_{ij} F^{ij}\right]\,,
\label{kkaction}
\eeq
where $F_{ij} \equiv \partial_i \bA_j - \partial_j \bA_i$. The couplings to point-like sources may be obtained from \eqref{eq:kk}, in terms of $(\varphi,\bA_i,\gamma_{ij})$. Notice that $\varphi$ always couples to the vector and tensor perturbations, and therefore there is no scalar cubic coupling. This means the 1PN effective action may be obtained without the diagram shown in Fig.~\ref{fig2EIH}$(b)$~\cite{smolkin1}. The split in \eqref{eq:kk} also proves to be helpful at high PN orders in the conservative sector, with the computation of the 2PN \cite{nrgr2pn} and 3PN \cite{nrgr3pn} potentials respectively, and significant progress toward 4PN order in \cite{nrgr4pn}. A~partial list of topologies are shown in Fig~\ref{fig:4pn} \cite{riccardocqg}. Unfortunately, since physical modes are encoded in~$\gamma_{ij}$, it does not provide much of an advantage in the radiation sector.

\subsubsection{Effacement Theorem for Non-Rotating Objects}\label{sec:eff}
\vskip 4pt

At some point in the PN expansion finite size effects will start to play a role. This requires adding extra terms in the effective action, as we discussed in sec.~\ref{sec:effect}. If we ignore spin, finite size effects first enter through the $C_{E(B)}$ coefficients. Notice that these terms entail four powers of derivatives. For example, for the electric-type term we have
\beq 
 \frac{C_E}{\Mp^2}\int (\partial_i \partial_j H_{00})^2 dt + \cdots\, .
\eeq
We are now in a position to prove the effacement theorem, e.g.~\cite{DamourEN}, which establishes the following:  Finite size corrections  for non-rotating compact binaries enter at $\calo(v^{10})$. This is nothing but a consequence of a simple observation. For a gravitationally bound system $H_{00}/\Mp \sim G_Nm/r$, and derivatives add extra factors of $1/r$. The correction induced by finite size effects thus scales as  
\beq
C_E \frac{G^2_N m^2}{r^6}\frac{r}{v} \sim \left(\frac{r_s}{r}\right)^6 \frac{L}{v^2} \sim L v^{10}\,,
\eeq
where we used $C_E \sim r_s^5/G_N$, following the matching procedure in sec.~\ref{smatch}, see e.g. \eqref{rs5}. This immediately tells us it enters at 5PN order in the binary's dynamics, as advertised. The magnetic-type term requires factors of the velocity and does not couple to $H_{00}$. Therefore, it contributes at higher PN orders.\vskip 4pt

It turns out the (renormalized) $C_{E(B)}$ vanish for black holes in four spacetime dimensions \cite{smolkinlove,poissonlove,damourlove}. This was shown in \cite{smolkinlove} (for the electric-type) by explicitly performing the matching for the one-point function in a curved background as we discussed in sec.~\ref{smatch}, and applies to all the electric and magnetic $\ell$-pole moments. On the other hand, terms involving higher time derivatives, e.g. $\dot E_{ij} \dot E^{ij}$ in \eqref{Se2}, are present for black holes \cite{steinchak2,binitidal}, which means we still need to augment the point-particle action beyond minimal coupling for non-stationary configurations. For neutron stars the $C_{E(B)}$ coefficients are finite and related to the so called `Love numbers,' which have interesting phenomenology, e.g. \cite{poissonlove,damourlove,tidaltan0,tidaltan,tidalflan,binitidal,iloveq,iloveq2,steinchak2,steinchak,eosNS2,eosNS,eosNS3,binitidalspin}. 
The vanishing of Love numbers for black holes in $d=4$ implies they are not only the most compact objects in nature, also the most rigid. We return to ponder on these issues in sec.~\ref{sec:conclusion}.

\subsection{Gravitational Wave Emission}\label{sec:radgrav}

\subsubsection{Radiation Effective Action}
\vskip 4pt 

We now turn on the radiation field, $\bar h_{\mu\nu}$, and write down a long-wavelength effective theory. Like~before, we integrate out the potential modes, but this time in a non-trivial background,
\beq g_{\mu\nu} = \bar g_{\mu\nu} + H_{\mu\nu}/\Mp\,,~~\bar g_{\mu\nu} = \eta_{\mu\nu} + \bar h_{\mu\nu}/\Mp\,.\eeq 
The effective action for the radiation theory, $S^{\rm rad}_{\rm eff}[\bx_a, \bar g_{\mu\nu}]$, is thus obtain from
\begin{align}
\label{zspp2}
e^{i W[\bx_a]} &= \int  D \bar h_{\mu\nu}  \exp\Big\{i S_{\rm EH}[\bar g_{\mu\nu}]+ iS^{\rm rad}_{\rm eff}[\bx_a, \bar g_{\mu\nu}] + i S_{\rm GF}[\bar h_{\mu\nu}]\Big\}\,  \\
&= \int  D H_{\mu\nu} D \bar h_{\mu\nu}   \exp\Big\{i S_{\rm EH}[g_{\mu\nu}]+ iS^{\rm pp}_{\rm eff}[\bx_a, g_{\mu\nu}] +i S^{(\bar h)}_{\rm GF}[H_{\mu\nu}] \Big\}\, , \nn
\end{align}
and subsequently radiation modes are integrated out to compute $W[\bx_a]$. The latter contains (through the optical theorem) information about the total radiated power. The implementation of the background field gauge-fixing term, $S^{(\bar h)}_{\rm GF}[H_{\mu\nu}]$, is similar to sec.~\ref{sec:nonlinear2} (see below).\vskip 4pt

Following the discussion in sec.~\ref{sec:effect}, we use the symmetries of the long-distance theory to construct the effective action, which becomes~\cite{andirad,andirad2}\,\footnote{~We remind the reader $L =(i_i\cdots i_\ell)$, $L-2 = (i_1\cdots i_{\ell-2})$,~etc. We omitted the coupling to the angular momentum, which we study in detail in sec.~\ref{sec:spin}.}
\beq
\label{eq:lag}
 S_{\rm eff}^{\rm rad}[\bx_a, \bar h_{\mu\nu}]  =  \int dt \sqrt{\bar g_{00}} \left[ -M(t) +  \sum_{\ell=2} \left( \frac{1}{\ell!} I^L(t) \nabla_{L-2} E_{i_{\ell-1}i_\ell} - \frac{2\ell}{(2\ell+1)!}J^L(t) \nabla_{L-2} B_{i_{\ell-1}i_\ell}\right)\right]\, .
\eeq 
The binary system is thus replaced by a point-like object endowed with a series of multipole moments. \vskip 4pt We~use the Weyl tensor since terms proportional to the Ricci tensor can be removed via field redefinitions, see appendix~\ref{app:field-redef}.  We place the center-of-mass of the binary at the origin, and at rest with respect to distant observers, such that $\sqrt{\bar g_{00}} dt= d\tau$ is the proper time. The first term, $M$, represents the binding mass/energy of the binary system.  The electric- and magnetic-type ($\ell\geq 2$) multipole moments, $(I^L,J^L)$ are $SO(3)$ symmetric and traceless tensors \cite{thorne,bala}.  Along the worldline, latin indices are defined with respect to a (co-moving) locally-flat vierbein, $e^\mu_A(x)$, with $e^\mu_0 = v^\mu$, such that ${\bar g}^{\mu\nu} = e^\mu_0 e^\nu_0 - \delta^{ij} e^\mu_i e^\nu_j\,,$ and $\nabla_i \equiv e^\mu_i \nabla_\mu$, where $\nabla_\mu$ is the derivative compatible with $\bar g_{\mu\nu}$.\vskip 4pt 

As we discussed in sec.~\ref{sec:effect}, we will decompose the multipole moments into background and response, see \eqref{qsr}. However, for the radiation problem only the background terms, $\left(\langle I^L\rangle_S ,\langle J^L\rangle_S\right)$, will be relevant. Higher powers of the metric field can be incorporated by studying the response of the multipole moments to external perturbations. These extra terms, e.g. $\int E^{ij}E_{ij}$ as in \eqref{Se2}, are important when the binary is embedded in an external gravitational field. In part~\ref{sec:part3} we study an example where the response terms are present. (For the sake of notation, we drop the $\langle\cdots\rangle_S$ in what follows.)

\subsubsection{Radiated Power}
\vskip 4pt
To calculate the total radiated power we use the optical theorem, see sec.~\ref{sec:opt}. We perform the $D\bar h_{\mu\nu}$ integral in \eqref{zspp2} using Feynman's boundary conditions and the same gauge fixing term as in \eqref{eq:Sgf1}, evaluated on a flat Minkowski background. Then, taking (twice) the imaginary part of $W[\bx_a]$, we get
\beq
\includegraphics[width=0.45\textwidth]{optical.pdf}\nn
\eeq
and for the total power,
\begin{equation}
P =   {G_N\over T} \int^\infty_0 \frac{d\omega}{\pi}  \left[ {\omega^6\over 5} \left|I^{ij}(\omega)\right|^2 + {16\over 45} \omega^6\left|J^{ij}(\omega)\right|^2+ {\omega^8\over 189}\left|I^{ijk}(\omega)\right|^2 + \cdots\right] \label{eq:power}\, ,
\end{equation}
This expression can be written in compact form in terms of time averages \cite{andirad2}
\bea
P =G_N \sum^{\infty}_{\ell=2} ~ \frac{(\ell+1)(\ell+2)}{\ell(\ell-1) \ell! (2\ell+1)!!} \left\langle\left( \frac{d^{\ell+1}I^L}{ dt^{\ell+1}}\right)^2\right\rangle + \frac{4~\ell(\ell+2)}{(\ell-1)(\ell+1)! (2\ell+1)!!} \left\langle\left( \frac{d^{\ell+1}J^L}{ dt^{\ell+1}}\right)^2\right\rangle\, ,\label{eq:power3}
\eea
to all order in the multipole expansion. Needless to say \eqref{eq:power3} agrees with the standard result, e.g. \cite{Maggiore}. 
As in sec.~\ref{sec:opt}, it is useful to introduce the gravitational wave amplitude (for polarization $h= \pm 2$)
\begin{equation}
\label{eq:bubbleA}
 i \mathcal A_h(\omega,{\bk}) =  \parbox{21.5mm}{\includegraphics{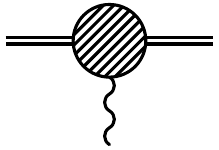}}
\end{equation}
which can be computed, from \eqref{eq:lag}, as a derivative expansion
\begin{align}
i \mathcal A_h(\omega,{\bk}) & =  \parbox{23mm}{\includegraphics{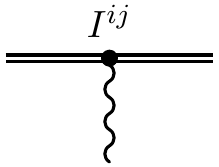}} + \hspace*{2pt} \parbox{23mm}{\includegraphics{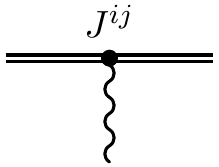}} + \hspace*{2pt} \parbox{23mm}{\includegraphics{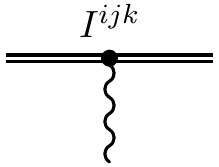}} + \cdots \notag \\
                 & = {i\over  4\Mp} \epsilon^*_{ij}({\bk}, h)  \left[\omega^2 I^{ij}(\omega) + {4\over 3} \omega \, {\bk}^l \epsilon^{ikl}J^{jk}(\omega) -{i\over 3} \omega^2 {\bk}^l I^{ijl}(\omega) + \cdots \right] \label{eq:ampliIIJ}\,\, , 
\end{align}
with $\omega=|\bk|$, and $\epsilon_{ij}({\bk},h=\pm 2)$ the polarization tensor. In terms of ${\cal A}_h(\omega,{\bk})$ we have (see \eqref{Pamp})
\begin{equation}
d\Gamma_h({\bk}) = {1\over T} {d^3 {\bk}\over (2\pi)^3 2|{\bk}|} |{\cal A}_h(|\bk|,{\bk})|^2 ~\to~ P~\big|_{h=\pm 2} = \int_\bk |{\bk}| d\Gamma_h({\bk})\,.
\end{equation} 
From the expansion in \eqref{eq:ampliIIJ} we recover \eqref{eq:power} after summing over polarizations \cite{andirad}.
\subsubsection{Matching: Multipole Moments}\label{sec:gravradM}
\vskip 4pt

To compute observable quantities from the above general procedure we need to be able to extract the multipole moments of \eqref{eq:lag} in terms of short-distance degrees of freedom. To this purpose, we start with the coupling between the short-distance modes and the radiation field \cite{andirad,andirad2,andirad3},
\beq
\label{eq:tadpole}
 -{1\over 2 \Mp}\int d^4 x~ {{\cal T}}^{\mu\nu}(x) {\bar h}_{\mu\nu}(x)\,,
\eeq
to leading order in $G_N$ in the far zone. In this expression, the ${{\cal T}}^{\mu\nu}(x)$ is the pseudo stress-energy tensor which includes all of the (near region) non-linear gravitational contributions from the Einstein-Hilbert action and worldline couplings, without yet incorporating non-linear effects in the wave's propagation (see sec.~\ref{sec:tail} and \ref{sec:RenQM}). The matching procedure now follows similar steps as in sec.~\ref{sec:mult}. We first expand the radiation mode in a Taylor series. For simplicity, let us assume the center-of-mass of the binary system, defined as
\begin{equation}
\label{eq:734}
{\bx}^i_{\rm cm} \equiv \int d^3{\bx}~{\cal T}^{00}(t,{\bx})~{\bx}^i\,,
\end{equation}
is at the origin and undisturbed by external sources, $\dot\bx_{\rm cm}=0$. Hence, the multipole expansion of the field becomes, 
\begin{equation}
{\bar h}_{\mu\nu}(x) = \sum_{n=0} {1\over n!} {\bx}^{i_1} \cdots {\bx}^{i_n} \partial_{i_1} \cdots \partial_{i_n} {\bar h}_{\mu\nu}(t,0)\,,
\end{equation}  
which we then plug back into (\ref{eq:tadpole}). Afterwards we decompose the moments of the stress-energy tensor into irreducible representations of $SO(3)$. We may concentrate on transverse-traceless physical modes, for which $\partial^2 \bar h_{ij}=0$. The gauge invariance is guaranteed by construction \cite{andirad2}.\vskip 4pt The decomposition of moments of ${\cal T}^{ij}(t,\bx)$ is standard in the literature, e.g.~\cite{Maggiore,Blanchet}. For illustrative purposes we recall some of the manipulations at low orders. At leading order we have,\begin{equation}
\label{eq:magg}
\int d^3 {\bx}~{\cal T}^{ij} (t,{\bx})= {1\over 2} \,  \int d^3{\bx}~\partial^2_0{\cal T}^{00} (t,{\bx}){\bx}^i {\bx}^j\, ,
\end{equation}
such that (integrating by parts)
\beq
\label{eq:tijh}
-\frac{1}{2} \int dt\left[\int d^3 {\bx} {\cal T}^{ij}(t,{\bx})\right] {\bar h}_{ij}(t,0) \to -\frac{1}{2} \left(\int d^3{\bx}~{\cal T}^{00} (t,{\bx}){\bx}^i {\bx}^j\right)\frac{1}{2} \partial_0^2 h_{ij}(t,0)\,.
\eeq 
Then, using $E_{ij} = -\frac{1}{2} \partial_0^2 {\bar h}_{ij}$ for on-shell physical modes at linear order in the perturbations, we find
\beq
\label{eq:lolij}
I^{ij}(t,{\bx}) = \int d^3{\bx}~{\cal T}^{00} (t,{\bx})\left[{\bx}^i {\bx}^j\right]_{\rm TF} + \cdots\,.
\eeq 
At next order in derivatives we encounter $\int d^3\bx {\cal T}^{ij} \bx^k$, which can be decomposed as ${\bf 2}\otimes {\bf 1} = {\bf 3} \oplus {\bf 2}$. (We~ignore traces which do not contribute to the radiative multipoles.) The ${\bf 3}$ corresponds to the electric octupole moment, whereas the ${\bf 2}$ contributes to the magnetic quadrupole. The latter reads
\beq
\left[\int d^3\bx\, {\cal T}^{ij} (t,{\bx}) \bx^k\right]_{\bf 2} \partial_k {\bar h}_{ij}(t,0) = -\frac{4}{3} J^{ij}(t,{\bx}) B_{ij}(t,0)\, ,
\eeq
where, 
\beq
J^{ij}(t,{\bx}) = -\frac{1}{2}\int d^3\bx \Big( \epsilon^{ikl} \left[{\cal T}^{0k}(t,{\bx})\bx^j\bx^l\right]_{\rm TF} + i \leftrightarrow j \Big) +\cdots\,.
\eeq
Continuing with the derivative expansion we find the moment $\int d^3\bx {\cal T}^{ij} \bx^k \bx^l$, which can be decomposed as $\bf 2 \otimes \big(1\otimes 1\big) = {\bf 4} \oplus {\bf 3} \oplus {\bf 2}$, plus traces \cite{andirad2}. The ${\bf 4}$ contributes to the electric sixteen-pole, the ${\bf 3}$ to the magnetic octupole, whereas the ${\bf 2}$ modifies the expression in \eqref{eq:lolij}, and we get \cite{andirad2}
\beq
\label{eq:lolij2}
I^{ij} (t,{\bx})= \int d^3{\bx}~\left({\cal T}^{00}(t,{\bx}) +{\cal T}^{kk} (t,{\bx}) -\frac{4}{3}\dot{\cal T}^{0k}(t,{\bx})\bx^k + \frac{11}{42} \ddot{\cal T}^{00}(t,{\bx})\bx^2 \right) \left[{\bx}^i {\bx}^j\right]_{\rm TF}  + \cdots\,.
\eeq
After extensive use of the Ward identity, integration by parts and the wave equation, we arrive at
\cite{andirad2}
\bea
\label{eq:ILJL}
I^L &=&\sum_{p=0}^{p=\infty} \frac{(2\ell+1)!!}{(2p)!!(2\ell+2p+1)!!}\left(1+\frac{8p(\ell+p+1)}{(\ell+1)(\ell+2)}\right)\left[\int d^3\bx \partial_0^{2p}{\cal T}^{00}(t,\bx){\bx}^{2p} \bx^L\right]_{\rm STF}\\
&+& \frac{(2\ell+1)!!}{(2p)!!(2\ell+2p+1)!!}\left(1+\frac{4p}{(\ell+1)(\ell+2)}\right)\left[\int d^3\bx \partial_0^{2p}{\cal T}^{kk}(t,\bx){\bx}^{2p} \bx^L\right]_{\rm STF}\nn\\
&-&\frac{(2\ell+1)!!}{(2p)!!(2\ell+2p+1)!!}\left(\frac{4}{\ell+1}\right)\left(1+\frac{2p}{(\ell+2)}\right)\left[\int d^3\bx \partial_0^{2p+1}{\cal T}^{0m}(t,\bx){\bx}^{2p} \bx^{mL}\right]_{\rm STF}\nn\\
&+& \frac{(2\ell+1)!!}{(2p)!!(2\ell+2p+1)!!}\left(\frac{2}{(\ell+1)(\ell+2)}\right)\left[\int d^3\bx \partial_0^{2p+2}{\cal T}^{mn}(t,\bx){\bx}^{2p} \bx^{mnL}\right]_{\rm STF}\nn\\
J^L &=& \frac{(2\ell+1)!!}{(2p)!!(2\ell+2p+1)!!}\left(1+\frac{2p}{(\ell+2)}\right)\left[\int d^3\bx \epsilon^{k_\ell mn} \partial_0^{2p}{\cal T}^{0m}(t,\bx){\bx}^{2p} \bx^{nL-1}\right]_{\rm STF}\\
&-& \frac{(2\ell+1)!!}{(2p)!!(2\ell+2p+1)!!}\left(\frac{1}{(\ell+2)}\right)\left[\int d^3\bx \epsilon^{k_\ell mr} \partial_0^{2p+1}{\cal T}^{mn}(t,\bx){\bx}^{2p} \bx^{nrL-1}\right]_{\rm STF}\,,\nn
\eea
for the $\ell$-pole moments, where the symmetric trace-free (STF) parts are constructed as, e.g. \cite{thorne},
\beq
Q^{i_1\cdots i_\ell} =  \sum_{k=1}^{k=[\ell/2]} \frac{ \ell! (2\ell-4k+1)!!}{(\ell-2k)!(2\ell-2k+1)!!(2k)!!} \delta^{(i_1i_2}\cdots \delta^{i_{2k-1}i_{2k}} I^{i_{2k+1}\cdots i_\ell)\ell_1\ell_1 \cdots \ell_k\ell_k}\, .
\eeq
To obtain the moments entering in \eqref{eq:ILJL} it is useful to work in mixed Fourier space, and expand in $\bk\cdot \bx \sim \bv$,
\begin{equation}
\label{mixedTk}
{\cal T}^{\mu\nu}(t,{\bk}) = \sum_{\ell=0}^\infty {(-i)^\ell\over \ell!} \left(\int d^3 {\bx}~{\cal T}^{\mu\nu}(t,{\bx}) \, {\bx}^{i_1}\cdots {\bx}^{i_\ell}\right) {\bk}_{i_1} \cdots {\bk}_{i_\ell}\,.
\end{equation}
The final step consists on calculating ${\cal T}^{\mu\nu}(t,{\bk})$, which may be read off from the gravitational amplitude in~\eqref{eq:bubbleA},
\vspace{-0.2cm} 
\beq 
\label{eq:tadpole2}
  i \mathcal A_h(\omega,\bk) = -\frac{i}{2\Mp}  \epsilon^*_{ij}(\bk, h) {\cal T}^{ij}(\omega,\bk)\,.  
\eeq

This amplitude includes, in addition to contributions from worldline couplings, also from the binding potential energy. When we solve for $H_{\mu\nu}$, we must guarantee the preservation of the long-distance symmetries of general relativity. This is achieved in the background field method, see sec.~\ref{sec:nonlinear2}, using 
\beq
\label{gambf2}
S^{(\bar h)}_{\rm GF}[H_{\mu\nu}] = \int d^4 x~ \sqrt{\bar g(x)}~\Gamma^{(\bar h)}_\mu(x) \Gamma^{(\bar h)\mu}(x)\, 
\eeq
in \eqref{zspp2}, with
\beq
\Gamma^{(\bar h)}_\mu = \nabla^{(\bar h)}_\alpha H^\alpha_\mu - \frac{1}{2}\nabla^{(\bar h)}_\mu H^\alpha_\alpha\,. \,\,
\eeq
The gauge-fixing term now contributes to the potential-radiation coupling, ${\bar h} HH$, and Feynman rules. It~also enforces the Ward identity, $\partial_\mu {\cal T}^{\mu\nu}(x)=0$, for the pseudo stress-tensor in \eqref{eq:tadpole} (as a consequence of \eqref{gsym}), such that unphysical modes do not radiate.\footnote{Notice that, up to this point, we allowed for (non-linear) couplings between potential and radiation modes, but only to linear order in~$\bar h_{\mu\nu}$. Hence, strictly speaking, the ${\cal T}^{\mu\nu}(x)$ in \eqref{eq:tadpole} would be {\it covariantly} conserved: $\nabla^{(\bar h)}_\mu {\cal T}^{\mu\nu}(x)=0$, had we turned on non-linearities in the far zone. We study these tail effects later on in~secs.~\ref{sec:tail} and~\ref{sec:RenQM}. As we shall see, the conserved (in the usual sense) pseudo stress-tensor, which we will continue denoting as ${\cal T}^{\mu\nu}(x)$, will be modified to incorporate non-linear gravitational interactions in the wave's propagation.}   
\subsubsection{Power Counting}
\vskip 4pt
At leading order we may ignore non-linear gravitational effects. Then, see \eqref{tpp},
\beq \label{tmnpp3} {\cal T}^{\mu\nu} (t,\bk) \to T^{\mu\nu}_{\rm pp} (t,\bk) = \sum_{a=1,2} m_a \frac{v^\mu_a(t) v^\nu_a(t)}{\sqrt{1-\bv^2_a(t)}} e^{-i \bk\cdot \bx_a(t)} + \cdots\,.
\eeq
The ellipsis include finite size effects, which will be important at higher orders. From this expression, and introducing
\beq 
\label{eq:KTmn}
K_\ell^{\mu\nu} \equiv \int d^3\bx~{\cal T}^{\mu\nu}(t,\bx) \bx^L\, ,
\eeq 
 we find the following scaling laws at leading order (for non-rotating objects)
\beq 
\label{eq:Kij00}
 K_\ell^{00} \sim M r^\ell  \, ,~~
 K^{0i}_\ell \sim M r^\ell v \, ,~~
 K^{ij}_\ell \sim M r^\ell v^2  \, .
\eeq
Since derivatives of the radiation field scale as $\partial_\mu \sim \lambda_{\rm rad}^{-1}$, the effective action in \eqref{eq:lag} becomes an expansion in powers of $r/\lambda_{\rm rad} \sim v$, as expected. The leading order multipoles are given by~\cite{andirad}
\bea
I^{ij}   &=&  \sum_a  m_a \left[{\bx}^i_a {\bx}^j_a\right]_{\rm TF} + {\cal O}(v^2)\label{eq:quadlo}\, ,\\
I^{ijk}   &= & \sum_a m_a \left[{\bx}_a^i {\bx}_a^j {\bx}_a^k\right]_{\rm TF} + {\cal O}(v^2) \label{eq:octo}\, , \\ 
J^{ij} & = &  \sum_a m_a \left[({\bx}_a\times {\bv}_a)^i {\bx}^j_a \right]_{\rm STF}+ {\cal O}(v^3) \label{eq:cquad}\, .
\eea
Let as add a few comments regarding the multipole expansion. Notice the first two terms in \eqref{eq:lolij2}. The trace, ${\cal T}^{kk}(x)$, originated at second order in $\bx\cdot \partial \ll 1$, whereas the temporal component, ${\cal T}^{00}(x)$, appeared at leading order, e.g.~\eqref{eq:tijh}.\,\footnote{~In general, the electric-type $\ell$-pole moment at order $(\bx\cdot\partial)^n$ receives contributions from the expansion of $\bar h_{ij}$ at orders $(n+\ell, n+\ell-1, n+\ell-2)$, whereas the $(n+\ell+1,n+\ell, n+\ell-1)$ contribute to the magnetic-type~\cite{andirad2}.} It turns out, in the velocity expansion for non-rotating bodies, the trace is suppressed by a factor of $v^2$ with respect to the $00$-component displayed in \eqref{eq:quadlo}. However, this is not the case at ${\cal O}(G_N)$, see \eqref{eq:Kij00} and \eqref{eq:egt00tij} below. The decomposition into irreducible representations is thus essential for setting up the power counting, and isolate the relevant contributions to a given order. 

\subsubsection{Power Loss to Next-to-Leading Order}\label{sec:NLOPR}
\vskip 4pt

\begin{figure}[t!]
\centering
\includegraphics[width=0.75\textwidth]{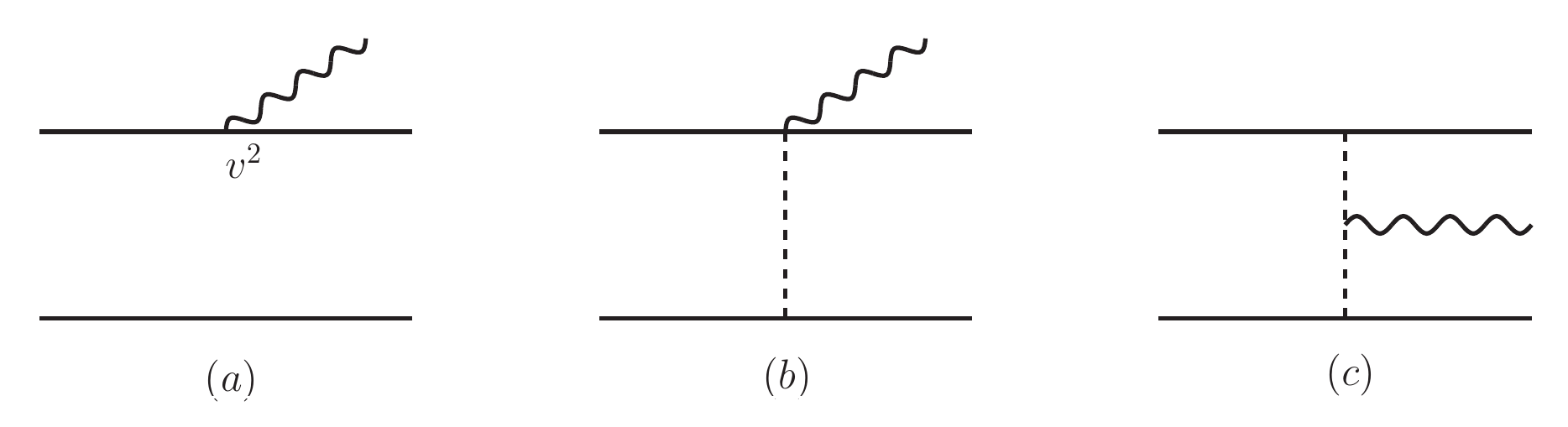}
\caption{Feynman diagrams which contribute to the one-point function, $\bar h_{\mu\nu}(x)$, at NLO. The dashed lines are potential modes sourced by the particles, while the wavy line represents the radiation field.}
\label{fig:tmn}
\end{figure}

We are now all set to compute the radiated power to NLO. We need, the factors of velocity from \eqref{tmnpp3} shown in Fig.~\ref{fig:tmn}$(a)$, as well as non-linear couplings from Figs.~\ref{fig:tmn}$(b)$ and \ref{fig:tmn}$(c)$. We only require~NLO corrections for the mass quadrupole, while the current quadrupole and octupole are needed to leading~order. Therefore, we are after the combination ${\cal T}^{00}(t,\bk)+{\cal T}^{kk}(t,\bk)$ in \eqref{eq:lolij2}. Computing the gravitational amplitude we extract (the diagram in \ref{fig:tmn}(b) does not contribute to the trace at this order),
\begin{eqnarray}
{\cal T}^{00}_{\ref{fig:tmn}(a)}(t,\bk)+{\cal T}^{kk}_{\ref{fig:tmn}(a)}(t,\bk) &=& \frac{3}{2} \sum_a m_a  {\bv}^2_a e^{-i {\bk}\cdot {\bx}_a}\,,\\
{\cal T}^{00}_{\ref{fig:tmn}(b)}(t,\bk) &=& {i\over 4 \Mp^2} \sum_{a\neq b} m_a m_b \int_{\bq} e^{i{\bq}\cdot {\bx}_{ab}}{- i\over 2 {\bq}^2} e^{-i {\bk}\cdot {\bx}_a} = \sum_{a \neq b} {G_N m_a m_b\over|{\bx}_{a}-{\bx}_b|} e^{-i {\bk}\cdot {\bx}_a}\,,\\
{\cal T}^{00}_{\ref{fig:tmn}(c)}(t,\bk) +
{\cal T}^{kk}_{\ref{fig:tmn}(c)}(t,\bk)   &=& {1\over 2!} \frac{- i}{2 \Mp} \sum_{a\neq b}{m_a m_b} \int_{\bq} e^{i{\bq}\cdot {\bx}_{ab} } {- i\over 2 {\bq}^2} {- i\over 2 ({\bq} + {\bk})^2} {4 i\over \Mp} \left(\bq^2 +  \bq \cdot  \bk\right) e^{-i{\bk}\cdot {\bx}_b} \nonumber \\
&=& - 2 \sum_{a\neq b} {G_N m_a m_b\over|{\bx}_a-{\bx}_b|} e^{-i {\bk}\cdot {\bx}_a}\,. \label{eq:egt00tij}
\end{eqnarray}
We then multipole expand these expressions in powers of ${|\bk|/|\bq|}$ \cite{iraben}, and read off the relevant moments from \eqref{mixedTk}. Hence, combining the terms and expanding to ${\cal O}(\bk^2)$, we obtain 
\begin{eqnarray}
\nonumber
I^{ij} &=& \sum_a m_a \left(1 + {3\over 2} {\bv}^2_a -  \sum_b {G_N m_b\over |{\bx}_a -{\bx}_b|}\right) \left[{\bx}^i_a {\bx}^j_a\right]_{\rm TF} + {11\over 42}\sum_a m_a {d^2\over dt^2} \left({\bx}_a^2 \left[{\bx}^i_a {\bx}^j_a\right]_{\rm TF}\right)\\
& & {}-{4\over 3}\sum_a m_a {d\over dt} \Big({\bx_a}\cdot {\bv}_a \left[{\bx}^i_a {\bx}^j_a\right]_{\rm TF}\Big) + {\cal O}(v^4) \label{eq:equadv2}\,.
\end{eqnarray}
It is customary to assume gravitational wave emission has circularized the orbit by the time it enters the detector's band, such as LIGO, which is sensitive only to the final few minutes of the inspiral~\cite{Blanchet,Buoreview}. If~we denote the orbital angular frequency by $\Omega$, and work in the center-of-mass frame, we obtain~\cite{andirad}
\begin{align}
I^{ij}(t)& =  \mu \left\{1 -\left(\frac{1}{42} + \frac{39}{42} \nu \right) x \right\} \left[{\br}^i {\br}^j \right]_{\rm TF} + \frac{11}{21} \mu {\br}^2 \left(1 - 3 \nu\right) \left[{\bv}^i {\bv}^j \right]_{\rm TF}  \label{eq:QuadCirc},\\
J^{ij}(t) & =  \mu \sqrt{1 - 4 \nu} \left[({\br} \times {\bv})^i {\br}^j \right]_{\rm STF} \label{eq:CQuadCirc},\\
I^{ijk}(t) & = \mu \sqrt{1 - 4 \nu} \left[{\br}^i {\br}^j {\br}^k \right]_{\rm TF} \label{eq:OctoCirc},
\end{align}
where we defined $\bv \equiv \dot\br$ and used the equations of motion which follow from the 1PN Lagrangian in \eqref{w1eih}. 

It~is~standard to write the results in terms of the following mass ratios: 
\beq\label{eq:munu} \mu_m\equiv{m_1 m_2}/{M}\,, \,\,\,\, \nu_m \equiv \mu_m/M\,,\eeq 
as well as the parameter \beq \label{eq:xOm} x \equiv (G_N M \Omega)^{2/3}~,\eeq
which scales as $x \sim v^2$ in the PN expansion,  e.g. \cite{Blanchet}. In terms of these variables, we obtain for the multipole moments needed to NLO\,\footnote{~The factors of $\delta(0)$ appear from squaring the multipoles, e.g. $\delta(\omega-2\Omega)\delta(\omega-2\Omega) = \delta(0) \delta(\omega-2\Omega)$. We also discarded terms proportional to $\delta(\omega)$ which do not contribute to radiation. Note, since in \eqref{eq:power} we integrate only over positive frequencies, we also omitted terms such as e.g. $\delta(\omega + 2\Omega)$, which only amount to factors of $2$ already incorporated. We will identify $2\pi \delta(0) = \underset{\omega \to 0}{\lim} \int dt e^{i\omega t}\to T$, in order to compute the average radiated power.}
\begin{eqnarray}
\label{eq:IdeltaO}
\left| I^{ij}(\omega)\right|^2 &=&\frac{2\pi\delta(0)}{G_N^2} {\pi \nu_m^2  x^5  \over 2\Omega^6}  \left[1+ \left(-{107\over 21}+  {55\over 21}\nu_m\right)x\right]  \delta(\omega-2\Omega)\,, \\ 
\left| J^{ij}(\omega)\right|^2 &=& \frac{2\pi\delta(0)}{G_N^2} {\pi  \nu_m^2  x^6\over  2\Omega^6}(1-4\nu_m)\,\delta(\omega-\Omega)\,, \nn\\ 
\left| I^{ijk}(\omega)\right|^2 &=& \frac{2\pi\delta(0)}{G_N^2} {\pi  \nu_m^2  x^6\over 4 \Omega^8} (1-4\nu_m) \left[ \delta(\omega-3\Omega) +  {3\over 5}\delta(\omega-\Omega)\right].\nn
\end{eqnarray}
From here we conclude that at leading order emission is quadrupolar, with $\omega_{\rm rad} \simeq 2\Omega$, and the total radiated power is given by
\beq
P_{\rm LO} =  \frac{32}{5} \frac{\nu_m^2}{G_N} x^5 \sim {\cal O}(v^{10})\label{eq:powerLO}\, .
\eeq
Moreover, plugging the expressions in \eqref{eq:IdeltaO} into \eqref{eq:power}, we find at NLO the well-known 1PN  result, e.g.~\cite{Blanchet},
\begin{equation}
{P_{\rm NLO} \over P_{\rm LO}} =1 - \left(\frac{1247}{336} + \frac{35}{12} \nu_m \right) x\,.
\end{equation}
These steps continue to all orders, modulo non-linear effects in the radiation zone we discuss in sec.~\ref{sec:tail}.

\subsubsection{Gravitational Waveform}\label{sec:amplitude}
\vskip 4pt

The previous analysis allows us to compute multipole moments describing the binary system. These multipoles can be used to compute the total radiated power as well as the gravitational waveforms. The latter can be obtained as follows. The physical degrees of freedom are transverse-traceless (TT), therefore to construct the waveform we need to project away the unphysical modes,
\begin{equation}
 \bar h_{ij}^{TT}(x) \equiv  \frac{1}{\Mp} \Lambda_{ij,kl} \,\bar h_{kl}(x)\,,
\end{equation}
with the projector,
\begin{equation}
 \Lambda_{ij,kl}  = \left(\delta_{ik} - {\hat \br}_i {\hat \br}_k\right) \left(\delta_{jl} - {\hat \br}_j {\hat \br}_l\right) - \frac{1}{2} \left(\delta_{ij} - {\hat \br}_i {\hat \br}_j\right) \left(\delta_{kl} - {\hat \br}_k {\hat \br}_l\right) \, .
 \end{equation}
The on-shell gravitational waveform for a given source, described by a stress-energy tensor ${\cal T}^{ij}(x)$, can be then obtained from the linearized Einstein's equations in the radiation zone, 
\beq
\label{htt2}
  \bar h_{ij}^{TT}(t,{\bx}) = -\frac{4 G_N}{r} \Lambda_{ij,kl} \int_{-\infty}^{+\infty} \frac{d\omega}{2\pi} {\cal T}^{kl}(\omega,\omega {{\hat \br}}) e^{i\omega t_{\rm ret}} + \calo\left(1/r^2\right)\,,
\eeq 
with $t_{\rm ret}$ the retarded time, see sec.~\ref{retard}.\,\footnote{~Formally speaking the expression in~(\ref{htt2}) is only valid for sources with compact support. See the next sub-section.}\vskip 4pt Following similar steps as in sec.~\ref{sec:gravradM} we can implement a multipole expansion for the source. Hence, in terms of the multipole moments we find \cite{andirad2}
\beq
\label{eq:wavett}
\bar h_{ij}^{TT}(t,\bx)  = - \frac{4 G_N}{r} \Lambda_{ij,k_{\ell-1} k_\ell} \! \sum_{\ell = 2}^{\infty}\Bigg[  \frac{1}{\ell!} \, \partial_0^{\ell} I^{L} (t_{\text{ret}}) \, {\hat \br}^{L-2}- \!  \frac{2 \ell}{(\ell+1)!} \,\epsilon^{mn(k_\ell}  \partial_0^{\ell} J^{k_{\ell-1})nL-2} (t_{\text{ret}}) \, {\hat \br}^{mL-2} \Bigg], 
\eeq
in agreement (up to an overall sign convention) with the known result, e.g. \cite{Blanchet}.  This expression also follows directly from \eqref{eq:lag} by computing the one-point function with retarded boundary conditions. The waveform in \eqref{eq:wavett} can be used to re-derive the total power, e.g. \eqref{eq:power}, from
\begin{equation}
P = \frac{1}{32 \pi G_N} \int r^2 d\Omega \left\langle\dot {\bar h}_{ij}^{TT}(t,\bx) \dot {\bar h}_{ij}^{TT}(t,\bx)\right\rangle \, .
\end{equation}
\subsection{Tail Effects} \label{sec:tail}
\vskip 4pt
The binary system sources the ambient geometry in which the waves propagate. The metric (far away from the binary) is that of a Schwarzschild (Kerr) spacetime with a total mass/energy $M$. Hence, in addition to contributions from the worldline couplings and quasi-instantaneous gravitational potentials, we now need to include also the gravitational interactions with the background geometry in the far zone. At~leading order in $G_N$ we have a gravitational potential, varying on scales of order $|\bk| \simeq \lambda^{-1}_{\rm rad} \ll 1/r $, given by the Newtonian approximation,
\beq \label{tailphi} \Phi_{\rm binary}(\bk) = -\frac{G_N M}{{\bk}^2} + {\cal O}(G_N^2) \,.\eeq
As we discuss in what follows this is the origin of the tail effect, e.g. \cite{tail1,tail2,tail3,tail3n,tail4,tail5,logx1,ALTlogx,4pn2,4pnB3,4pnDS}. From \eqref{tailphi} we notice the gravitational tails can be parameterized in powers of the ratio $\eta\equiv {\cal R}_s/\lambda_{\rm rad}$, with ${\cal R}_s \equiv 2 G_N M$ the `gravitational radius' of the system \cite{andirad}.  For Post-Newtonian sources we have  $\eta \sim v^3$, and the tail first enters at 1.5PN  order beyond the leading effects for binary inspirals.\vskip 4pt

As we shall see, the tail effects lead to the introduction of so called {\it radiative} moments, as opposite to {\it source} multipoles which are computed as in sec.~\ref{sec:gravradM}. The expression for the total power remains as in~\eqref{eq:power}, but written in terms of radiative multipoles. In the following we exercise an abuse of notation, and retain ${\cal T}^{\mu\nu}(x)$ for the (pseudo) stress-energy tensor. The latter now incorporates both the gravitational non-linearities from the potentials in the near zone as well as in the radiation region. It is guaranteed to obey the Ward identity provided we use the background field method, as in sec. \ref{sec:nonlinear2}, to include non-linear interactions in the far zone after the short-distance potential modes are integrated out (see~sec.~\ref{sec:RenQM}). 

\subsubsection{Radiative Multipole Moments}
\vskip 4pt 
The contribution to the total power from the tail effect may be obtained from the optical theorem. For~example, at leading order~in~$\eta$ in the radiation zone, we have  %
\vspace{-0.1cm}
\beq
\nn
\includegraphics[width=0.45\textwidth]{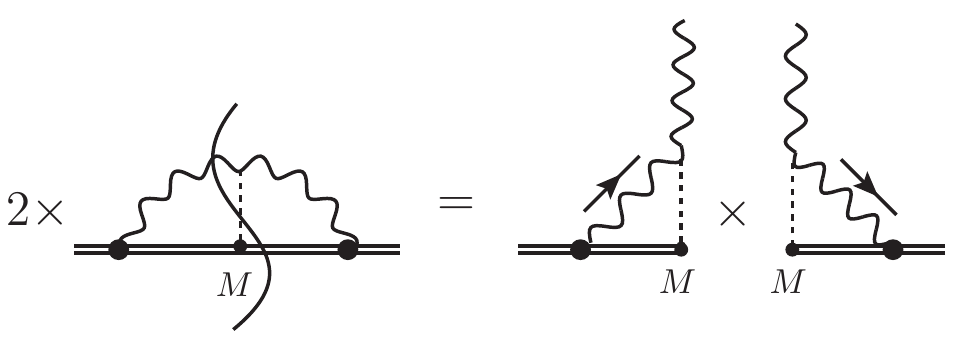}
\vspace{-0.2cm}
\eeq
for each $(I^L, J^L)$ multipole emission. The interaction between the outgoing radiation and the binary's potential is represented by the $M$ insertion. Higher orders in $\eta$ corresponds to extra factors of \eqref{tailphi}, and diagrammatically to extra ladders. In general, the tail contribution to the gravitational amplitude can be written as 
\beq
i \mathcal A_{\rm tail}(\omega,{\bk}) =  \frac{i}{4\Mp} \epsilon_{ij}^\star(\bk) \left[\bk^2 I_{\rm rad}^{ij}(\omega) + \cdots\right]\, ,
\eeq
which defines the radiative multipole moments \cite{Blanchet}. From this definition it is straightforward to show that the form of the total power loss remains the same, as in \eqref{eq:power3}, but computed in terms of radiative multipoles.\vskip 4pt As an example, let us calculate the tail contribution to the radiative quadrupole moment. At leading order in $\eta$ we have~\cite{tail1,tail2,tail3,andirad}
\begin{align}
 \label{eq:andir}
i \mathcal A^{(1)}_{\rm tail}({\bk}) &=  \parbox{21mm}{\includegraphics[width=0.09\textwidth]{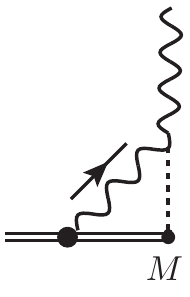}} 
\end{align}
\beq
= i {\cal A}^{(0)}(\omega,\bk) \times \left(i G_N M |\omega| \right) \left[- {(\omega+ i\epsilon)^2\over \pi \mu^2} e^{\gamma_E}\right]^{(d-4)/2} 
 \times\left[{2\over d-4} - {11\over 6} + (d-4)\left({\pi^2\over 8} + {203\over 72}\right)+\cdots\right]\, ,\nn
\eeq
 where dim. reg. was used to regularize an IR-divergent integral \cite{andirad}.\,\footnote{~The factor of $(\omega+i\epsilon)^2$ is due to the choice of retarded boundary condition, whereas in \cite{andirad} we find $\omega^2+i\epsilon$, from Feynman's prescription. The difference is irrelevant for the total power, but affects the amplitude~\cite{amps}.} There is a pole at $d=4$, and once again we introduced the (arbitrary) scale $\mu$. The $i{\cal A}^{(0)}$, given by
 \beq
  i{\cal A}^{(0)}(\omega,\bk) = \frac{i}{4\Mp} \epsilon_{ij}^\star(\bk) \,\omega^2 I_{\rm src}^{ij}(\omega)\,,
  \eeq
is the amplitude computed in terms of the source quadrupole. Expanding around $\epsilon_{\rm IR}=d-4 \simeq 0$, we obtain\,\footnote{~\label{foot22} Note IR singularities are regularized with $\epsilon_{\rm IR} \equiv (d-4)_{\rm IR} >0$, whereas for UV divergences we use $\epsilon_{\rm UV} =-(d-4)_{\rm UV} > 0$.}
 \beq
\label{eq:Itail}
I^{ij}_{\rm rad}(\omega) = I^{ij}_{\rm src}(\omega) \left\{1+G_N M \omega\left( {\rm sign}(\omega)\pi + i\left[\frac{2}{\epsilon_{\rm IR}}+\log {\omega^2\over \pi\mu^2}+\gamma_E -{11\over 6}\right]\right)\right\}\,.
\eeq  
For the power we need the total amplitude, ${\cal A} = {\cal A}^{(0)} + {\cal A}^{(1)}$, (note only the real part in \eqref{eq:Itail} contributes)
\begin{equation}
\label{fintail}
\left|{{\cal A}\over {\cal A}^{(0)}}\right|^2 = 1 + 2 \mbox{Re} {{\cal A}^{(1)} \over {\cal A}^{(0)}} + \cdots = 1 + 2\pi G_N M |\omega| + {\cal O}(\eta^2)\,.
\end{equation}
From this expression we can immediately compute the tail correction to the radiated power, obtaining up to NLO~\cite{andirad}
\beq
\label{eq:powerTA}
\frac{P_{\rm tail}}{P_{\rm LO}} = 4\pi x^{3/2}\left( 1 - \frac{1}{4} \left[ \frac{8191}{692} + \frac{583}{24} \nu_m\right]x\right)\,,
\eeq
where we used the source quadrupole to NLO in \eqref{eq:IdeltaO} (also the 1PN correction to $M$, see \eqref{eq:Ea} below). The formula in \eqref{fintail} was derived from the quadrupolar emission, but in fact it is universal. Indeed, each tail introduces extra factors of $2\pi G_N M |\omega|$. This is related to the cancelation of the associated IR poles, which we discuss~next.
 
\subsubsection{Infrared Behavior: Sommerfeld Enhancement \& Phase Shift}\label{sec:IR}
\vskip 4pt
The previous computation entailed the regularization of an IR divergence.  The reason is the presence of a long-range $1/r$ potential in which the massless gravitational waves propagate. (This is analogous to what occurs in Coulomb scattering in quantum mechanics \cite{Wsoft}.) However, the IR pole did not contribute to the total radiated power in \eqref{fintail}. This turns out to be the case to all orders in $G_N$, as expected.\vskip 4pt

At~${\cal O}(\eta^n)$ one find the amplitude factorizes \cite{andirad} 
\begin{equation}
{\cal A}^{(n)}_{\rm tail} \rightarrow  {\cal A}^{(0)} \times {1\over n!} \left[-32\pi (G_N M {\bk}^2)\int {d^{d-1}{\bq}\over (2\pi)^{d-1}} {1\over {\bq}^2} {1\over 2 {\bk}\cdot {\bq}}\right]^n\,.
\end{equation}
Then, using
\begin{equation}
\int {d^{d-1}{\bq}\over (2\pi)^{d-1}} {1\over {\bq}^2} {1\over 2 {\bk}\cdot {\bq}} = -{i\over 16\pi |{\bk}|} {1\over \epsilon_{\rm IR}} \, ,
\end{equation}
we notice the IR poles wind up exponentiating into an overall phase in the amplitude, i.e. \cite{andirad}
\beq
\sum_{n=0}^\infty  {\cal A}^{(n)}_{\rm tail} = {\cal A}^{(0)} \times \exp\left(i\frac{2 G_N M \omega}{\epsilon_{\rm IR}}\right)\,,
\eeq
and cancel out in the total power. In this respect, tail effects are reminiscent of the factorization of soft terms in electrodynamics and gravity~\cite{Wsoft,Muzi}.\vskip 4pt 

While the harmful divergences drop out, there are still finite terms (factors of $\pi$) associated with these poles, e.g. \cite{Kpi1,Kpi2}. This is the origin of the Sommerfeld enhancement at low frequencies  
\begin{equation}
{\cal A}(\omega) \rightarrow  \frac{4 \pi G_N M \omega}{1 - \exp\left(- 4 \pi G_N M  \omega \right)} {\cal A}(\omega)\,.
\end{equation}
The alert reader will immediately notice that the IR singularity may still remain in the gravitational waveform, as an infinite overall phase. (The reason can be traced back to the form of the radiative multipole moments which now enter in \eqref{eq:wavett}.) However, because the divergence is linear in $\omega$, it is easy to see we can absorb the IR pole into a time redefinition, 
\beq
\label{bare}
 t_{\rm b} \to  t_{\rm ren} - 2 G_N M/\epsilon_{\rm IR}\,.
\eeq
The time $t_{\rm ren}$ represents the (renormalized) variable used by the experimentalists, and $t_{\rm b}$ an unobserved {\it bare} coordinate time. This shift is associated with a choice of time origin to start tracking the gravitational wave signal, which is of course arbitrary.\vskip 4pt The tail effect in the mass and current quadrupole contribution to the waveform then reads (we remove the `ret' in $t$ for notational simplicity)
\begin{align}
\label{eq:httIM}
\left(h_{ij}^{TT}\right)_{I\mbox{-}\rm tail}(t_{\rm ren},\bx) &= -\frac{2G_N}{r}\Lambda_{ij,kl}\int_{-\infty}^{+\infty}\frac{d\omega}{2\pi}\left[-\omega^2 e^{i\omega t_{\rm ren}(\mu)} e^{i\theta_{\rm tail}(\omega,\mu)}\left(1+ G_N M |\omega|\pi\right)\right] I^{ij}_{{\rm src}}(\omega),\\
\label{eq:httJM}
\left(h_{ij}^{TT}\right)_{J\mbox{-}{\rm tail}}(t_{\rm ren},\bx) &= -\frac{2G_N}{r}\Lambda_{ij,kl}\int_{-\infty}^{+\infty}\frac{d\omega}{2\pi}\left[-\omega^2 e^{i\omega  t_{\rm ren}(\mu)} e^{i\varphi_{\rm tail}(\omega,\mu)}\left(1+ G_N M |\omega|\pi\right)\right] J^{ij}_{{\rm src}}(\omega)\,,
\end{align}
with \cite{tail1,tail2,tail3,tail4,tail5,amps} 
\beq
\label{theta12}
\theta_{\rm tail}(\omega,\mu) \equiv G_N M\omega\left(\log {\omega^2\over \pi\mu^2}+\gamma_E-\frac{11}{6}\right)\,,~~\varphi_{\rm tail}(\omega,\mu) \equiv G_N M\omega\left(\log {\omega^2\over \pi\mu^2}+\gamma_E -{7\over 3}\right)\,.
\eeq
The $\mu$ dependence in the renormalized time ensures the final expression is independent of any particular choice. A shift in $\mu$ is absorbed into a concurrent time shift, \beq t_{\rm ren} (\mu_2) \to t_{\rm ren} (\mu_1) + 2G_N M \log (\mu_2/\mu_1)\, .\eeq 
A convenient choice is given by $\mu = \omega_s$, the seismic cutoff frequency for earth-based interferometers \cite{blan2}, which is then related to a reference frequency scale at $``t_{\rm ren}=0"$, the time at which the signal enters the detector's frequency band. The onward time evolution is what corresponds to a measurable quantity. Once a choice for $\mu$ is made, the above expressions dictate how to correlate the predicted gravitational wave evolution with experimental observations. Notice that the constants associated with the logarithms in \eqref{theta12} are arbitrary. However, the difference is not, since we are free to choose $\mu$ only once. The $\gamma_E$'s are unphysical, but the rational numbers are non-universal, as can be seen from \eqref{theta12}, and the difference is thus measurable.
 
\subsection{Renormalization} \label{sec:RenQM}
\subsubsection{Quadrupole Moment}\label{sec:Qren}
\vskip 4pt

While one can show that the cancelation of IR singularities occurs to all orders, UV divergences may still be present. That is the case, for instance, when computing the `tail-of-the-tail' at ${\cal O}(\eta^2)$ \cite{tailtail1,tailtail2}.\,\footnote{\label{foot20}~When IR and UV divergences are simultaneously present one has to carefully isolate the relevant regions. In particular, one cannot simply set to zero scale-less integrals. Since we have, e.g. \cite{iraeft},
\beq
\int \frac{d^dq}{(2\pi)^4} \frac{1}{q^4} =-\frac{i}{4\pi^2}\left( \frac{1}{(d-4)_{\rm UV}} - \frac{1}{(d-4)_{\rm IR}}\right)\, \nn,
\eeq
the IR singularity may be needed to cancel other poles, leaving behind a UV divergence in need of a counter-term.}
A detailed calculation using dim. reg. yields  ($d=4-\epsilon_{\rm UV}$) \cite{andirad}
\bea
\label{eq:amp2}
\left|{{\cal A}\over {\cal A}^{(0)}}\right|^2  &=& 1 + 2 \pi G_N M |\omega|  + \left (G_N M |\omega|\right)^2 \left[ {214\over 105}\left( {1\over \epsilon_{\rm UV}} -\gamma_E - \log {\omega^2\over \pi\mu^2}\right) + {4\pi^2\over 3} + {634913\over 44100}\right]\,.\eea
The UV pole in \eqref{eq:amp2} must be subtracted away. This is similar to what we did in sec.~\ref{sec:renorm}, except that now the coefficients of the effective action in \eqref{eq:lag} are time (frequency) dependent. Hence, we split the quadrupole moment into a renormalized piece and a counter-term. The latter is given by,  in the $\overline{\rm MS}$ scheme~\cite{andirad},
\beq
\label{eq:Ict}
I_{\rm ct}^{ij}(\omega) \equiv Z_{\rm ct}(\omega) I_{\rm src}^{ij}(\omega) =  - {107\over 105} \left(G_N M \omega\right)^2  \left[{1\over \epsilon_{\rm UV}} -\gamma_E +\log 4\pi\right]\times I_{\rm src}^{ij}(\omega)\,,
\eeq
where $I_{\rm src}^{ij}(\omega)$ is the source quadrupole moment.\vskip 4pt 
 
 From \eqref{eq:Ict} we can read off the renormalization group flow equation\,\footnote{In $d=4-\epsilon$ dimension we have for the bare quadrupole,
\beq
I^{ij}_{\rm bare}(\omega) = \mu^{-2\epsilon} \frac{Z_{\rm ct}(\omega)}{\epsilon} I_{\rm src}^{ij}(\omega) + I_{\rm ren}^{ij}(\omega,\mu)\,.\nn
\eeq
Taking $\mu \frac{d}{d\mu}$ on both sides we get to this order,
\beq
0= -2\epsilon \,  \, \frac{Z(\omega)}{\epsilon}  I_{(0)}^{ij}(\omega) + \mu \frac{d}{d\mu}I_{\rm ren}^{ij}(\omega,\mu) \to \mu \frac{d}{d\mu}I_{\rm ren}^{ij}(\omega,\mu)  = 2 Z(\omega) I_{\rm ren}^{ij}(\omega)\nn\,.
\eeq
}  
\beq
\label{eq:rgI}
\mu {d\over d\mu} I_{\rm ren}^{ij}(\omega,\mu) = - {214\over 105} (G_N M\omega)^2 I_{\rm ren}^{ij}(\omega,\mu)\, ,
\eeq
which leads to \cite{andirad}
\begin{equation}
\label{eq:RGsoln}
I_{\rm ren}^{ij}(\omega,\mu) = \left[{\mu\over \mu_0}\right]^{- {214\over 105} (G_N M\omega)^2} I_{\rm ren}^{ij}(\omega,\mu_0)\,.
\end{equation}
Then, the renormalized amplitude becomes
\beq
\label{amptail}
\left|{{\cal A}(\omega) \over {\cal A}^{(0)}(\omega,\mu)}\right|^2  = 1 + 2 \pi\left (G_N M \omega \right)  + \left (G_N M \omega \right)^2 \left[ -{214\over 105} \log {\omega^2\over 4\mu^2} + {4\pi^2\over 3} + {634913\over 44100}\right]  + {\cal O}(\eta^3)\, ,
\eeq
where ${\cal A}^{(0)}(\omega,\mu)$ is now written in terms of $I_{\rm ren}^{ij}(\omega,\mu)$ such that the total expression is $\mu$-independent. 
It~is convenient to choose $\mu\simeq\omega$, where $\omega \sim v/r$ is the typical gravitational wave frequency, while $\mu_0$ is the short-distance scale at which we perform the matching for the quadrupole moment, i.e. $\mu_0 \simeq r^{-1}$.  The renormalization group equation is then used to sum the series of  so called `leading logarithms,' (using $\mu/\mu_0 \sim v$)~\cite{andirad}
\begin{align}
\label{eq:logv}
\left|
{{\cal A}(\omega)\over {\cal A}^{(0)}(\omega,\mu_0)}\right|^2_{\log v} = 1 &- {428\over 105} (G_N M \omega)^2 \log v + {91592\over 11025} (G_N M \omega)^4  (\log v)^2  \\ &- {39201376\over 3472875}  (G_N M \omega)^6(\log v)^3+\cdots. \nn
\end{align}
An explicit computation in \cite{22pn}, for a test particle on a circular orbit around a Schwarzschild back hole, agrees with the expression in \eqref{eq:logv} for the leading logarithms up to order $(G_N M \omega)^{14}(\log v)^7$. In our case, however, these never become too large since they come accompanied by a factor of $\eta^2 \simeq (G_N M \omega)^2$. In~the PN regime we have $\eta \sim v^3$, such that $\eta^2 \log (\omega r) \sim v^6\log v$ remains small throughout the inspiral. \vskip 4pt

While the full series of logarithms might not be phenomenologically relevant, it does contain important information about the dynamics. The coefficients in the series of terms are universal, and a direct consequence of the renormalization group flow. Their structure may be totally random or a consequence of a deeper structure in gravitational dynamics. We will return to these issues later on in sec.~\ref{sec:conclusion}. 

\subsubsection{Binding Mass/Energy}\label{sec:Mren}
\vskip 4pt

\begin{figure}[t!]
\centering
\includegraphics[width=0.6\textwidth]{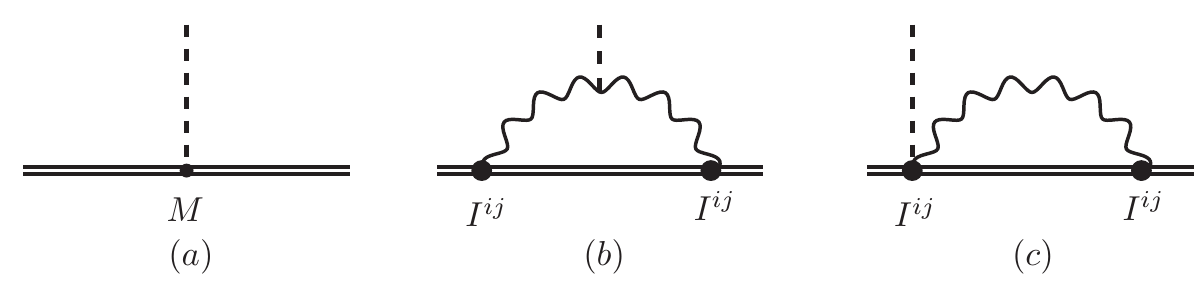}
\caption{The diagrams which contribute to ${\cal T}^{00}(x)$, up to non-linearities at ${\cal O}(\eta)$. The wavy lines are radiation modes, whereas the dashed line represents the (non-radiative) $00$-mode.}
\label{fig:Mren1}
\end{figure}

\begin{figure}[t!]
\centering
\includegraphics[width=0.6\textwidth]{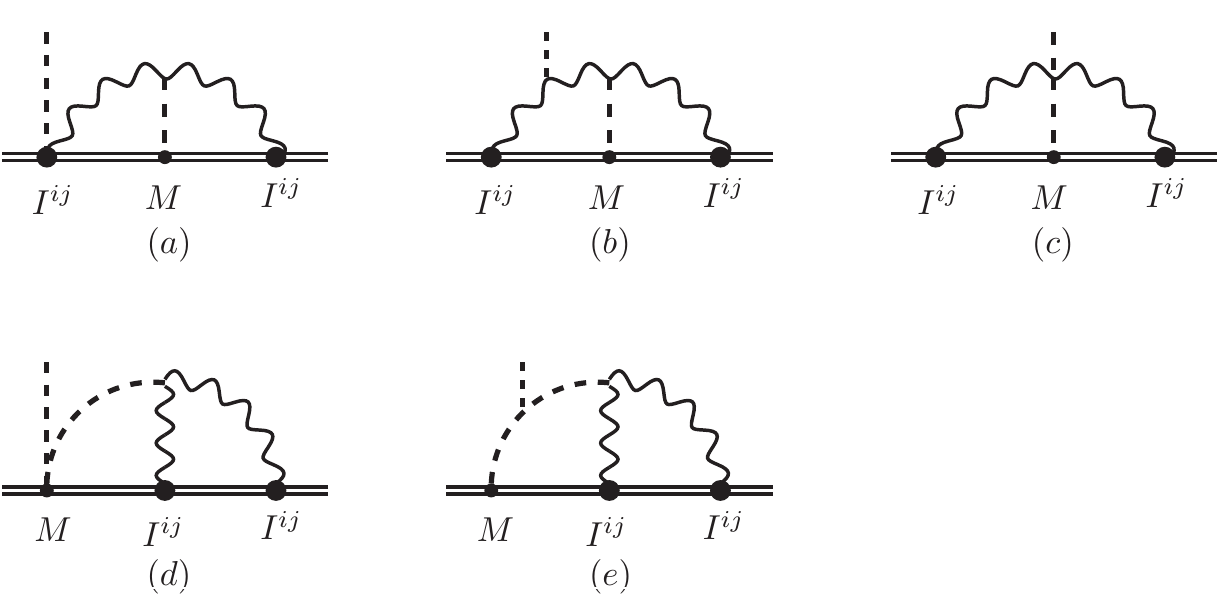}
\caption{The diagrams which contribute to the renormalized energy/mass at ${\cal O}(\eta^2)$. At this order, the factor of $M$ represents the leading order binding mass/energy of the system, since the time (and scale) dependence enters at higher orders.} 
\label{fig:Mren2}
\end{figure}

\begin{figure}[t!]
\centering
\includegraphics[width=0.25\textwidth]{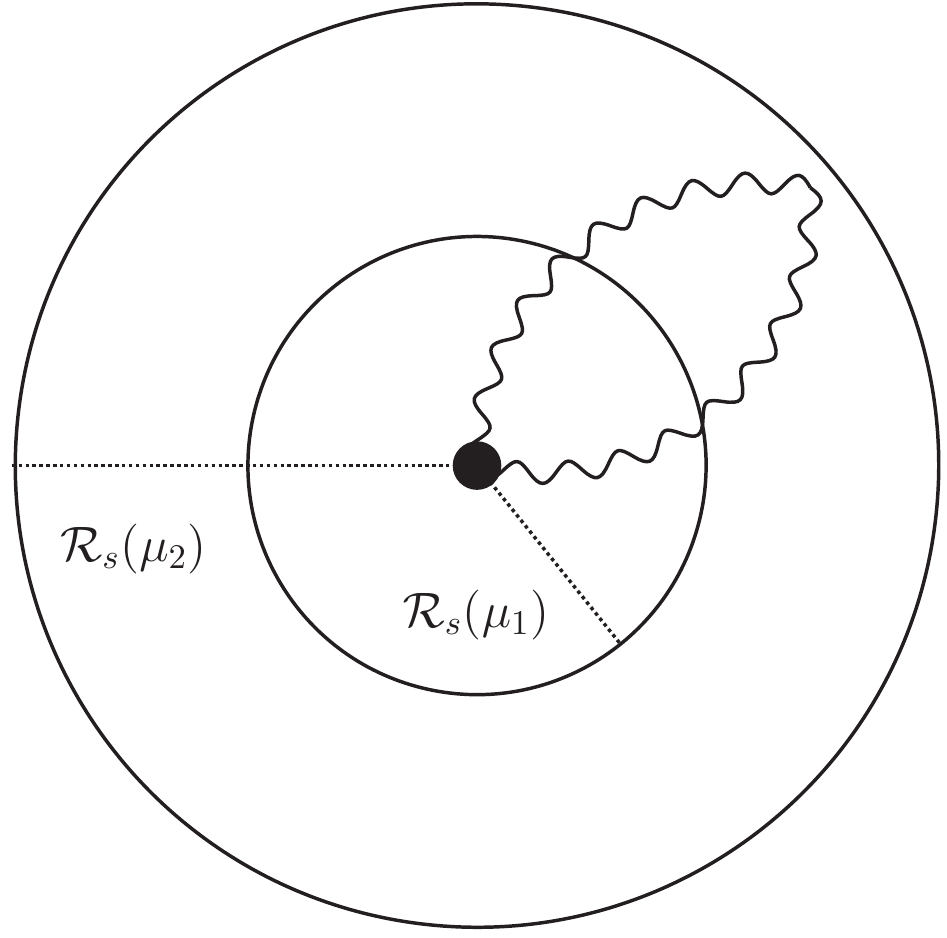}
\caption{At the short-distance scale, $\mu_0$, we have ${\cal R}_s(\mu_0)\sim r$ (shown as a black disc). As $\mu$ flows to long(er) distances, $\mu_2 \ll \mu_1 \ll \mu_0$, the gravitational size of the system grows, $r \ll {\cal R}_s(\mu_1) \ll {\cal R}_s(\mu_2)$, together with the (renormalized) gravitational mass/energy: $M(\mu) \equiv {\cal R}_s(\mu)/(2G_N)$. Similarly for the multipole moments. }
\label{fig:Mren3}
\end{figure}

The binding mass/energy, $M$, which requires computing ${\cal T}^{00}(t,\bk)$, is also renormalized by gravitational non-linear interactions \cite{andirad3}. As we show here, this leads to a logarithmic correction to the binding energy at 4PN order \cite{ALTlogx}.\vskip 4pt 
At leading order in the radiation zone we simply have the contribution from the one-point function, shown in Fig.~\ref{fig:Mren1}$(a)$, 
\beq 
\label{eq:T00a}
{\cal T}_{\ref{fig:Mren1}(a)}^{00}(t,\bk) = \langle M(t)\rangle\,.\eeq  Furthermore, to 1PN order for the sources we can also perform the matching, see Figs.~\ref{fig:tmn}$(a$-$c)$, and obtain $M(t) = \sum_a  E_a +{\cal O}(v^4)\,$, where (similarly with $ 1 \leftrightarrow 2$)
\beq
\label{eq:Ea}
E_1 = m_1\left[1 + {1\over 2} {\bv}^2_1 -{1\over 2}  {G_N m_2\over |{\bx}_1 -{\bx}_2|}\right] \,.
\eeq
The conservation of ${\cal T}^{\mu0}(x)$ thus enforces $ \langle\dot M(t)\rangle=0$. This is the case because (see \eqref{eq:734})
\bea
\dot\bx^i_{\rm cm} = \int d^3\bx ~\partial_0 {\cal T}^{00} (t,\bx) \bx^i &=& \int d^3\bx ~{\cal T}^{0i} (t,\bx)=0 ~\to~ {\cal T}^{0i}(t,\bx) \propto \partial^i\delta^3(\bx)\, \nn\\
 \int d^3\bx~\partial_\mu{\cal T}^{\mu0}(x)&=&  0 ~\to~  \langle \dot M (t)\rangle \propto \int \frac{d^3\bx}{|\bx|} \to 0\, , \label{eq:consM}
\eea 
where we used $\partial^2\delta^3(\bx) \propto 1/|{\bx}|$, and dim. reg. to set to zero the divergent integral. Notice that, had we used a cutoff in \eqref{eq:consM} we would have to absorb a time-dependent (power-law) divergent term into $M$. This suggests the mass/energy can be itself time-dependent. 
\vskip 4pt
The non-linear gravitational effects in the radiation zone are shown in Fig.\,\ref{fig:Mren1}$(b$-$c)$, to first~order~in~$\eta$. The stress-energy tensor acquires an extra piece \cite{LucALT,andirad3} (the $(n)$ indicates the number of time-derivatives)
\beq
{\cal T}_{\ref{fig:Mren1}(a\mbox{-}c)}^{00}(t,\bk) =  \left\langle M(t) + \frac{G_N}{5} \int_{-\infty}^t dt \, I^{(5)}_{ij}  I^{(1)}_{ij}+\cdots\right\rangle \, .
\eeq
The Ward identity gives us the quadrupole formula for the power loss, e.g. \eqref{eq:power}. We then obtain an expected result, with the change in the total mass/energy of the binary encoded in the radiated power.
\vskip 4pt
At higher orders we will encounter new type of non-linear effects. The relevant diagrams are in Fig.~\ref{fig:Mren2}. Isolating the logarithmic contribution the result reads  \cite{andirad3}
\beq
{\cal T}^{00}_{\ref{fig:Mren2}(a\mbox{-}e)}(\omega,\bk) = i\frac{32\pi^2}{10} G_N^2  \langle M\rangle  I_0(\omega)\int \frac{d\omega_1}{2\pi} \frac{d\omega_2}{2\pi} I_{ij}(\omega_1) I_{ij}(\omega_2) \left( \omega_1^5\omega_2 + \frac{1}{2}\omega^3_1\omega_2^3 -\omega_1^4\omega_2^2+ 1\leftrightarrow 2\right)\,.
\eeq
Not surprisingly, the scalar integral,  $I_0(\omega)$, is the same we found in sec.~\ref{sec:renorm}, see~\eqref{eq:scalarI0}. Using dim. reg. we have,
\beq
I_0(\omega)= \frac{1}{32\pi^2} \left[ \frac{1}{\epsilon_{\rm UV}} -\gamma_E + \log 4\pi+3 - \log{\frac{\omega^2}{\mu^2}}\right] + {\cal O}(\epsilon)\,.
\eeq
The divergence is removed by adding a counter-term. In this case it can be absorbed into a `tadpole' 
\beq 
\label{eq:tadpoleM}
-\frac{i}{2} \int \frac{d\omega}{2\pi}\,M_{\rm ct} (\omega) \bar h_{00}(\omega,0)\,, 
\eeq
which at this order enters through Fig.~\ref{fig:Mren1}(a). The counter-term is given by \beq
M_{\rm ct}(\omega) =\frac{1}{\bar\epsilon_{\rm UV}}  \frac{2 G_N^2}{5}\langle M\rangle \int \frac{d\omega_1}{2\pi} \frac{d\omega_2}{2\pi} I_{ij}(\omega_1) I_{ij}(\omega_2) \left( \omega_1^5\omega_2 + \frac{1}{2}\omega^3_1\omega_2^3 -\omega_1^4\omega_2^2+ 1\leftrightarrow 2\right)\delta(\omega-\omega_1-\omega_2)\,,
\eeq
where the $\bar\epsilon_{\rm UV}$ includes the extra constants which are removed in the $\overline{\rm MS}$ scheme, see \eqref{eq:bareps}.
This leads to a renormalization group equation for the renormalized binding mass/energy \cite{andirad3}
\begin{align}
 \mu {d\over d\mu}\log M_{\rm ren}(t,\mu) & = - \frac{2 G_N^2}{5}\left(2 I^{(5)}_{ij}(t) I^{(1)}_{ij}(t) - 2 I^{(4)}_{ij}(t) I^{(2)}_{ij} (t)+ I^{(3)}_{ij}(t) I^{(3)}_{ij}(t)\right)\, , \label{eq:RGmass}
\end{align}
or, after performing the time averaging, \cite{andirad3}
\beq
\label{eq:RGmass0}
\mu \frac{d}{d\mu} \log \langle M_{\rm ren}(t,\mu)\rangle = -2 G_N^2 \left\langle ~ I^{ij(3)}_{\rm ren} (t,\mu)I^{ij(3)}_{\rm ren}(t,\mu)\right\rangle \, .\eeq
The dependence in $\mu$ on the right-hand side is due to the renormalization of the quadrupole, see \eqref{eq:RGsoln}. An interpretation of the running of the binding mass/energy is depicted in Fig.~\ref{fig:Mren3}. The solution to \eqref{eq:RGmass} reads \cite{andirad3} (we drop the `ren' tag below to simplify notation)
\beq
\frac{\langle M(t, \mu)\rangle}{  \langle M(t, \mu_0)\rangle}=  {\rm exp}\left\{ -\frac{105}{214}\left(\frac{\left\langle I_{ij}^{(2)}(t,\mu_0) I_{ij}^{(2)}(t,\mu_0)\right\rangle-\left\langle I^{(2)}_{ij}(t,\mu)I^{(2)}_{ij}(t,\mu)\right\rangle}{ \left \langle M^2(t,\mu_0)\right\rangle} \right)\right\}.
\eeq
Then, using (see \eqref{amptail})
\beq
\label{tailexp}
\left|{{\cal A}(\omega) \over {\cal A}^{(0)}(\omega,\mu)}\right|^2  = 1 + \pi {\cal R}_s(\mu)\omega +\cdots\, ,  
\eeq 
with  \beq {\cal R}_s(\mu) \equiv 2 G_N \langle M(t, \mu)\rangle\,,\eeq 
and setting $\mu\simeq \omega$, we find a new series of logarithms~\cite{andirad3}
\begin{align}
&\left|{{\cal A}(\omega) \over {\cal A}^{(0)}(\omega)}\right|^2  = 1 + \pi {\cal R}_s(\mu_0)\omega\left(1 -\frac{1}{2} \frac{\left\langle I^{(3)}_{ij}(t,\mu_0)I^{(3)}_{ij}(t,\mu_0)\right\rangle}{\langle M(t,\mu_0)\rangle^2}{\cal R}^2_s(\mu_0)\log v\,\right. \\
 &+ \left.\frac{107}{420} \frac{\left\langle I^{(4)}_{ij}(t,\mu_0)I^{(4)}_{ij}(t,\mu_0)\right\rangle}{\langle M(t,\mu_0)\rangle^2}{\cal R}^4_s(\mu_0)(\log v )^2 - \frac{11449}{132300}\frac{\left\langle I^{(5)}_{ij}(t,\mu_0)I^{(5)}_{ij}(t,\mu_0)\right\rangle}{\langle M(t,\mu_0)\rangle^2}{\cal R}^6_s(\mu_0)(\log v )^3\right)\,, \nn
\end{align}
after choosing $\mu_0 \simeq 1/r$, such that $\log\mu/\mu_0 \simeq \log v$. These logarithms are present in addition to those in \eqref{eq:logv}, which follow from the running of the quadrupole moment.\vskip 4pt
By inspection, we notice that the logarithmic contribution to the binding mass/energy is {\it conservative}, since it enters as a total time derivative. Therefore, it can be absorbed into a redefinition of the binding energy of the binary system, we shall denote as $E$. In terms of the variables we introduced before, see \eqref{eq:munu} and \eqref{eq:xOm}, we find~\cite{andirad3}
\beq
\label{eq:logEx}
E_{\log} (x) = -\frac{\mu_m}{2} \frac{448}{15}\nu_m x^5 \log x + \cdots \, .
\eeq
In~this expression, the $(\nu_m, \mu_m)$ are defined at the short-distance scale in terms of $M_{\rm ren}(\mu_0 \sim 1/r)$. The~result in \eqref{eq:logEx} is a 4PN correction \cite{ALTlogx}. We return to the origin of this term below.\,\footnote{~Unlike the `conservative' binding mass/energy $M(\mu)$, $E$ is $\mu$-independent. However, by choosing $\mu \simeq \omega$, we can trade the logarithmic dependence into the renormalization group evolution, such that $E = M_{\rm ren}(\mu \simeq \omega)$ at the radiation scale.}

\subsection{Gravitational Radiation-Reaction}\label{sec:self}
\vskip 4pt
In sec.~\ref{sec:opt} we discussed how the optical theorem allows us to compute the total radiated power. Even though we use vacuum `in-out' boundary conditions --without outgoing radiation-- the power follows from Feynman's prescription by calculating (twice) the imaginary part of the effective action, ${\rm Im}\,W[\bx_a]$ (see also the discussion leading to \eqref{eq:power}). This procedure, however, will not account for the back-reaction effects on the dynamics of the binary. The reason is simple, we do not allow for any radiation! This can be immediately seen from the following observation \cite{chadbr1}. Let us integrate out the radiation field in \eqref{zspp2} using the leading order quadrupole coupling, 
\begin{align}
\label{eq:br1}
\hspace{-1cm} i W[\bx_a] \, &=  \, \parbox{21mm}{\includegraphics[width=0.18\textwidth]{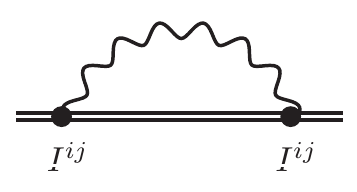}} \nn \end{align}
\vspace{-1cm}
\begin{align}
&= \left( \frac{1}{2} \right) \left( \frac{i}{2 \Mp} \right)^2 \int dt \int dt^\prime I^{ij}(t) \big\langle T\big\{ E_{ij} (t, 0) E_{kl}(t^\prime, 0)\big\} \big\rangle I^{kl}(t^\prime)\,,
\end{align}
and the $i\epsilon$-prescription,
\beq
i W[\bx_a]= - \frac{ i }{ 80 \Mp^2} \int  dt \int dt^\prime \, I^{ij} (t) I^{ij}(t^\prime) \int _{p_0, \bp} e^{i p _0 (t-t^\prime) } \frac{ p_0^4 }{ p_0^2 - \bp^2 + i \epsilon}  \,.
\eeq
Notice that from \eqref{imieps} the Im\,$W[\bx_a]$ readily gives us the total quadrupole radiation, as expected from the optical theorem, e.g. \eqref{eq:power}. On the other hand for the real part we find \cite{chadbr1}
\beq
{\rm Re}\, W[\bx_a] =  -\frac{2\pi G_N}{5} \sum_{n=0}^\infty \frac{ 1}{ n!} \int dt \,I_{ij}(t) I^{(n)}_{ij} (t) \underbrace{\int ds\, s^n \, \left(\,{\rm PV}\,\int_{\bp,p_0} e^{ i p_0s} \frac{p_0^4  }{ p_0^2 - \bp^2 }\right)}_{I(n,4,0)} \,\,\,,
\eeq
where `PV' stands for principal value. Only the $I(5,4,0) = {5! \over 4\pi}$ survives, and we obtain
\beq
{\rm Re}\, W[\bx_a] =  -\frac{G_N}{10} \int dt\,  I_{ij}(t) I^{(5)}_{ij} (t) = -\frac{1}{2} \int V_{\rm BT}(t) dt \,, \hspace{1cm} {(\rm Feynman)} \label{eq:feybr}
\eeq
with, the Burke-Thorne radiation-reaction potential~\cite{thorneBT1,thorneBT2},
\beq
V_{\rm BT} (t) =  \frac{ G_N}{5}\sum_a \bx_a^i(t) \bx_a^j(t) I^{(5)}_{ij}(t)\,.
\eeq
\vskip 4pt At first sight \eqref{eq:feybr} appears to lead to the correct answer. However, integrating by parts we obtain a total time derivative which does not contribute to the dynamics,
\beq
{\rm Re}\,W[\bx_a] \propto \int dt \, \frac{ d}{ dt}\left\{ I^{(2)}_{ij} I^{(2)}_{ij}\right\}\,. \hspace{1cm} {(\rm Feynman)}
\eeq
The resolution relies on incorporating causal propagation in NRGR, or in other words implementing the `in-in' formalism \cite{inin1,inin2}. This was studied in detail in \cite{chadbr1,chadbr2,chadprl,chadprl2}. In a compact notation the in-in formalism entails doubling the numbers of fields, e.g. $\bx_a \to \{\bx^{(1)}_a,\bx^{(2)}_a\}$, also for the metric perturbation, and integrating over a closed-time path. This replaces the action~in~\eqref{zj} by $S \to S_1-S_2$, and similarly for the source term. We point the reader to \cite{chadreview} for a thorough review.\vskip 4pt The computation in \eqref{eq:br1} is then substituted~by
\beq
\label{eqbrplus}
\left( \frac{1}{2} \right) \left( \frac{i}{2 \Mp} \right)^2 \int dt \int dt^\prime I_A^{ij}(t) \big\langle E^A_{ij} (t, 0) E^B_{kl}(t^\prime, 0) \big\rangle I_B^{kl}(t^\prime)\,,
\eeq
where $A,B = \pm$, with $\bx^{(+)}_a = \tfrac{1}{2}(\bx^{(1)}_a+\bx^{(2)}_a)$ and $\bx^{(-)}_a =(\bx^{(1)}_a-\bx^{(2)}_a)$. The propagator in \eqref{eqbrplus},
\beq
\Delta^{AB}(x'-x) = \left(
\begin{array}{cc}
0 &  \Delta_{\rm adv}(x'-x)\\
 \Delta_{\rm ret}(x'-x) & 0
\end{array}
\right)\,,
\label{eq:GreenAB}
\eeq
replaces Feynman's prescription. This guarantees that we enforce the correct boundary conditions \cite{chadreview}. The equations of motion are obtained from the effective action, $W[\bx_a^{(\pm)}]$,
\beq
\label{eq:actin}
\frac{\delta W[\bx_a^{(\pm)}]}{\delta \bx^{(-)}_a}\, \Bigg |_{\bx_a^{(-)}=0}= ~0\,.
\eeq
Hence, keeping only the term which does not vanish after setting $\bx^{(-)}=0$, we get \cite{chadbr1}
\beq
\label{eq:WBT}
W[\bx^{(\pm)}_a] =  -\frac{G_N}{5} \int dt\,  I^{ij}_{(-)}(t) I^{ij(5)}_{(+)} (t) \,, \hspace{1cm} {(\rm Retarded)} 
\eeq
(note this is already a real quantity in the in-in formalism), with 
\beq 
\label{eq:Qminus}
I^{ij}_{(-)} = \sum_a m_a \left(\bx_a^{i(-)} \bx_a^{j(+)} +  \bx_a^{i(+)} \bx_a^{j(-)} - \frac{2}{3}\, \delta^{ij} \bx_a^{(+)}\cdot \bx_a^{(-)}\right)\,.\eeq
Then, using \eqref{eq:actin}, we arrive at the desired result for the radiation-reaction acceleration~\cite{thorneBT1,thorneBT2},
\beq
\label{eq:BTpot}
\left(\ba_a^i\right)_{\rm rr} = - \frac{ 2G_N}{5} \bx_a^j(t) I^{ij(5)}(t)\,. 
\eeq
From here we can also derive the quadrupole radiation formula, as an energy balance,
\beq
\label{eq:balance1}
\langle \dot M(t)\rangle = \sum_a m_a \left\langle \left(\ba_a^i\right)_{\rm rr} \cdot \bv_a(t)\right \rangle = - \frac{G_N}{5}\left\langle  I^{(3)ij}(t) I^{ij(3)}(t)\right\rangle\,,
\eeq
with $M(t)$ the binding mass/energy (the Noether charge in the `conservative' sector).\vskip 4pt The previous steps can be generalized to all $\ell$-order multipoles, 
\vspace{-0.1cm}
\beq
\hspace{-5cm} iW[\bx_a^{(\pm)}]\, = \,  \sum_{\ell \ge 2}~~ \parbox{21mm}{\includegraphics[width=0.4\textwidth]{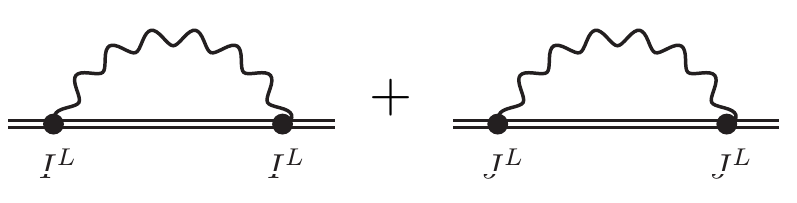}} \label{eq:Wtailall}
\eeq
\vspace{-0.2cm}
\begin{align}
W[\bx_a^{\pm}] = G_N \sum_{\ell \ge 2}\, \frac{(-1)^{\ell+1}(\ell+2)}{(\ell-1)}  \int dt & \left(\frac{2^\ell (\ell + 1)}{\ell (2 \ell + 1)!} I_{(-)}^L(t) I_{(+)}^{L \, (2\ell + 1)}(t)+\frac{2^{\ell+3} \ell}{(2 \ell + 2)!}  J_{(-)}^L(t)  J_{(+)}^{L \, (2\ell + 1)}(t)\right)\,,\nn
\end{align} 
incorporating radiation-reaction effects and total radiated power, at linear order in $G_N$ in the far zone. 

\subsection{Interplay Between Potential and Radiation Regions}\label{sec:interplay}
\vskip 4pt
The gravitational radiation-reaction entails a subtle interplay between near and far zones, in particular when the tails are incorporated. As we shall see, hereditary effects lead to time non-locality as well as both dissipative and conservative terms. The latter includes a UV singularity --intimately connected to an IR divergence in the near region-- which requires renormalization. As a result, we recover the renormalization group evolution for the binding mass/energy and the universal long-distance logarithms~\cite{andibr1}.
\subsubsection{Time Non-Locality}\label{sec:tailNL}
\begin{figure}[t!]
\centerline{{\includegraphics[width=0.29\textwidth]{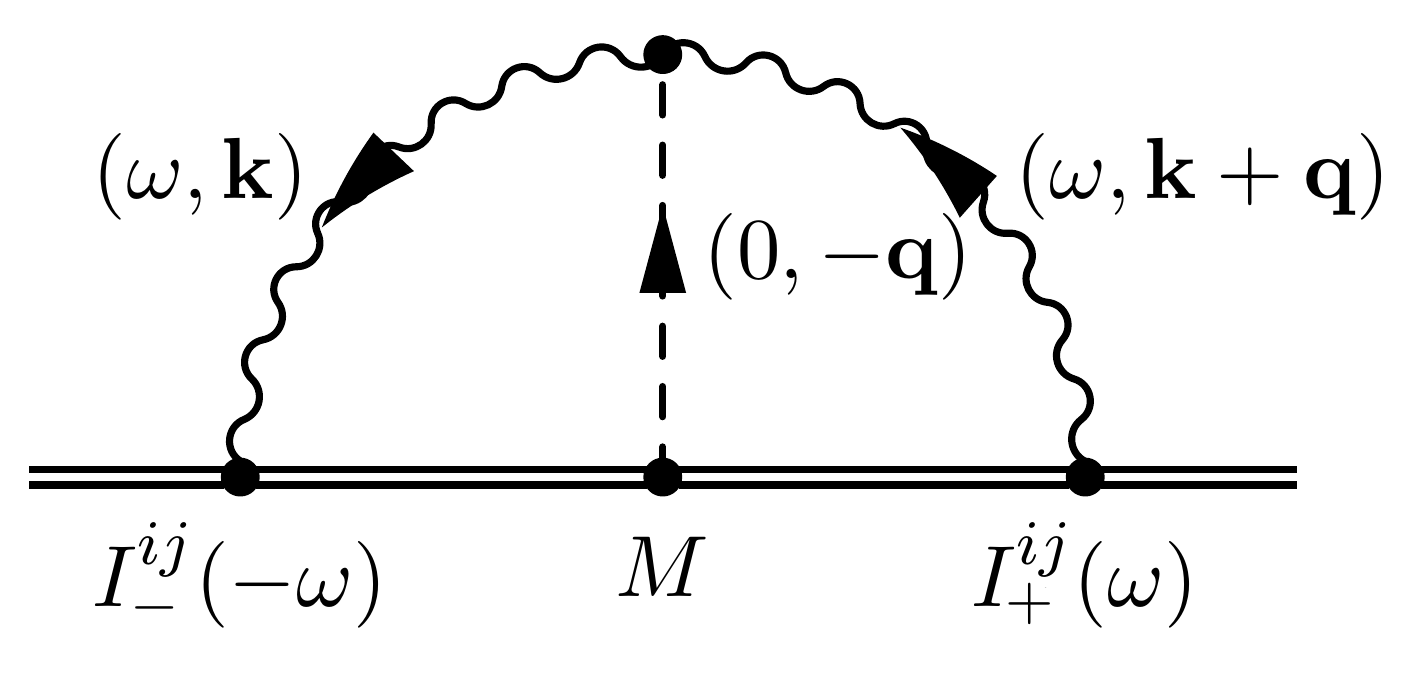}}}
\caption[1]{The tail contribution to gravitational radiation-reaction. The factor of $M$ may be taken as the leading time-averaged value, i.e. $\langle M \rangle$.} 
\label{nltail}
\end{figure}

\vskip 4pt
The contribution from the tail effect to the effective action is shown in Fig.~\ref{nltail}, yielding 
\beq
 i W_{\rm tail}[\bx_a^\pm]  = \int \frac{d \omega}{2 \pi} \int_{\bk, \bq} \langle M\rangle I^{ij}_-(- \omega) I^{ij}_+(\omega)~[{\bar h\bar h\Phi}]~\frac{i}{- \bq^2 } \frac{i}{(\omega + i \epsilon)^2- \bk^2} \frac{i}{(\omega + i \epsilon)^2- (\bk + \bq)^2} \, , \label{eq:tailRR}
\eeq
where $[{\bar h \bar h\Phi}]$ represents the three-graviton coupling between the radiation and background geometry.  After some laborious manipulations, we find \cite{andibr1}
\begin{align}
 i W_{\rm tail}[\bx_a^\pm] = -i \int \frac{d\omega}{2 \pi} \,  \frac{(d-3)\langle M\rangle \, \omega^4 \, I_-^{ij}(-\omega) I_+^{ij}(\omega)}{32 (d-2)^2 (d-1)(d+1)} \Big[(d^2 - 2 d + 3) I_0 - \frac{d(d-2)(d-1)}{d-4} \omega^2 J_0 \Big]\,,
\end{align}
in terms of two integrals,
\begin{align}
 I_0 &= \left( \int_{\bk} \frac{1}{(\omega + i \epsilon)^2 - \bk^2 } \right)^2 = \frac{\left(\Gamma \left[-\frac{d-3}{2}\right]\right)^2}{(4 \pi)^{d-1}} \left[-(\omega + i \epsilon)^2\right]^{d-3}\,, \\
 J_0 & =   \int_{\bk} \frac{1}{(\omega + i \epsilon)^2 - \bk^2 } \int_{\bq} \frac{1}{[(\omega + i \epsilon)^2 - \bq^2]^2 }  =   \, \frac{\Gamma \left(-\frac{d-3}{2}\right) \Gamma \left(-\frac{d-5}{2}\right)}{(4 \pi)^{d-1}} \left[-(\omega + i \epsilon)^2\right]^{d-4} \, .
\end{align}
The result is UV divergent, and expanding around $d = 4$ we have 
\cite{andibr1}
\begin{align}
W_{\rm tail}[\bx_a^\pm] & =   \, \frac{2 G_N^2  \langle M\rangle}{5} \int_{-\infty}^\infty \frac{d\omega}{2 \pi} \,  \omega^6 \, I_-^{ij}(-\omega) I_+^{ij}(\omega) \left[-\frac{1}{(d-4)_{\rm UV}} - \log \frac{\omega^2}{\mu^2} + i \pi \, \text{sign}(\omega) \right] .\label{eq:RRnl}
\end{align} 
After removing the pole by a counter-term (see sec.~\ref{sec:iruvmix}) we arrive at
\begin{align}
W_{\rm tail}[\bx_a^\pm] & = - \frac{2 G_N^2 \langle M\rangle}{5}\int_{-\infty}^\infty \frac{d\omega}{2 \pi} \left(  \omega^6 \, I_-^{ij}(-\omega) I_+^{ij}(\omega) \left[\log \frac{\omega^2}{\mu^2} - i \pi \,\text{sign}(\omega) \right] \right)\,. \label{eq:WtailRR}
\end{align}
We can now Fourier transform from frequency space back to the time domain, obtaining 
\begin{align}
W_{\rm tail}[\bx_a^\pm] = \frac{4 G_N^2  \langle M\rangle}{5} \Big({\rm PV}\int dt \, I_-^{(3)ij}(t) \int_{-\infty}^t dt' \, I_+^{ij(3)}(t') \left[\frac1{t-t'} \right]  +  \int dt \, I_-^{(3)ij}(t) I_+^{ij(3)}(t) \log \mu\Big) \label{finitetail}\,.
\end{align}
(Note that the $\text{sign}(\omega)$ in \eqref{eq:WtailRR} is essential to preserve causality.) The expression in \eqref{finitetail} can also be written as
\beq
W_{\rm tail}[\bx^{\pm}_a] = -\frac{4 G_N^2 \langle M\rangle}{5}\,\,  {\rm PV} \int dt\, I_-^{ij}(t) \int_{-\infty}^t dt' \, I_+^{ij(7)}(t')\log(|{t-t'}|\mu)\,.
\eeq
This is equivalent to the results found in \cite{4pn2,4pnB3,4pnDS} (see also \cite{tail3n}).

\subsubsection{Long-Distance Logarithms}\label{sec:longlog}
\vskip 4pt 
After renormalization the effective action, $W[\bx^{\pm}_a]$, becomes a function of the renormalized potential, 
\begin{align}
\label{eq:WtailRR2}
W[\bx^{\pm}_a] = \int \frac{d\omega}{2\pi} \Big( K[\bx_a^\pm;\omega] - V_{\rm ren}[\bx_a^\pm;\omega,\mu]\Big) - \frac{2 G_N^2 \langle M\rangle}{5}\int \frac{d\omega}{2 \pi}   \omega^6 \, I_-^{ij}(-\omega) I_+^{ij}(\omega) \left[\log \frac{\omega^2}{\mu^2}- i\pi\,\text{sign}(\omega) \right]\,, 
\end{align}
where we also added a kinetic term. This expression leads to a modification of the dynamics which includes both a conservative piece (from the logarithmic term) and a dissipative part (due to the $\text{sign}(\omega)$). The latter is the only term that is not invariant under $\omega \to -\omega$. Therefore, renormalization occurs in the conservative sector while the non-conservative terms are finite at this order~\cite{andibr1}. The $\mu$-independence of the effective action gives us the renormalization group flow for the (renormalized) potential, in standard variables,\,\footnote{~Because of \eqref{eq:actin}, the factor of $4 \log \mu$ in \eqref{eq:WtailRR2} turns into a $2 \log \mu$ when translating from $\bx_a^\pm$ to $\bx_a$.} 
\beq
\mu \frac{\partial}{\partial \mu} \,W[\bx^{\pm}_a]= 0 \to \mu \frac{\partial}{\partial \mu}V_{\rm ren}[\bx_a;\omega,\mu] = \frac{2 G_N^2 \langle M\rangle }{5} \omega^6 I^{ij}(-\omega) I^{ij}(\omega)\,. \label{eq:RGWV}
\eeq
\vskip 4pt
Fourier transforming to the time domain, we have 
\beq
\label{eqVtm}
V_{\rm ren}[\bx_a;t,\mu] = V_{\rm ren}[\bx_a;t,\mu_0] +  \frac{2G_N^2 \langle M\rangle}{5} I^{(3)}_{ij}(t) I^{(3)}_{ij}(t)\log \frac{\mu}{\mu_0}\, .
\eeq
The equations of motion may be then written in the form of an energy balance,
\beq
\label{eq:balance}
\dot M_{\rm ren}(t,\mu)= \sum_a m_a  \left(\ba_{a}\right)_{\rm cons}\cdot \bv_a +\cdots  =   \frac{2 G_N^2 \langle M\rangle}{5} \, I^{ij(1)}(t) \int \frac{d\omega}{2\pi} I^{ij(6)}(\omega) e^{i\omega t} \log \frac{\omega^2}{\mu^2}+\cdots\,.
\eeq
The renormalized binding mass/energy, $M_{\rm ren}(t,\mu)$, includes contributions from both $V_{\rm ren}[\bx_a;t,\mu]$ and $K[\bx_a;t]$, and
\beq
 \left(\ba^j_{a}\right)_{\rm cons} (t,\mu) = \frac{4 G_N^2 \langle M\rangle}{5} \, \bx_a^i(t) \int \frac{d\omega}{2\pi} I^{ij(6)}(\omega) e^{i\omega t} \log \frac{\omega^2}{\mu^2}\,
 \eeq
 is the conservative part of the radiation-reaction acceleration \cite{andibr1}. The ellipsis in \eqref{eq:balance} account for other --non-conservative-- terms, in addition to the one from the tail (see below). These may be obtained from \eqref{eq:Wtailall}, as in e.g.~\eqref{eq:balance1}, or using \eqref{eq:power}.\vskip 4pt We can now average over a circular orbit such that, 
\beq
\label{eq:balance2}
\left\langle\dot M_{\rm ren}(t,\mu)\right\rangle =  -\frac{4 G_N^2 \langle M\rangle}{5} \,\left\langle I^{ij(1)}(t) I^{ij(6)}(t)\right\rangle\, \log\, (\lambda_{\rm rad} \mu)+\cdots\,,
\eeq
where we used $\omega \simeq (2\pi)\lambda_{\rm rad}^{-1}$, see \eqref{eq:IdeltaO}, and absorb the factor of $2\pi$ into $\mu$. Hence, applying the following identity 
\beq
 I^{(1)}_{ij}(t) I^{(6)}_{ij}(t) = \frac{d}{dt} \left( I^{(5)}_{ij}(t) I^{(1)}_{ij}(t) -  I^{(4)}_{ij}(t) I^{(2)}_{ij} (t)+ \tfrac{1}{2} I^{(3)}_{ij}(t) I^{(3)}_{ij}(t)\right)\,, 
 \eeq
on the right-hand side of \eqref{eq:balance2}, we find for the {\it conservative} binding energy:
\beq
\label{eq:Eb}
E \equiv M_{\rm ren}(t,\mu) + \frac{2G_N^2 \langle M\rangle}{5} \left(2I^{(5)}_{ij}(t) I^{(1)}_{ij}(t) -  2I^{(4)}_{ij}(t) I^{(2)}_{ij} (t)+ I^{(3)}_{ij}(t) I^{(3)}_{ij}(t)\right) \, \log\, (\lambda_{\rm rad} \mu)\,.
\eeq
This expressions yields, after choosing $\mu \simeq r^{-1}$, the correction proportional to $\log v$ in \eqref{eq:logEx}. Conversely, at the radiation scale we have \beq E = M_{\rm ren}(t,\mu \simeq  \lambda_{\rm rad}^{-1})\,.\eeq Therefore, the same result follows from studying the renormalization group evolution for the binding mass/energy, 
\beq
 \label{eq:Mrendef}
\mu \frac{d}{d\mu} \langle E\rangle =0 \to \mu \frac{d}{d\mu} \log \left \langle M_{\rm ren}(t,\mu)\right\rangle = -  2G_N^2  \left\langle I^{(3)}_{ij}(t) I^{(3)}_{ij}(t)\right\rangle\,,
\eeq
choosing $\mu \simeq \lambda_{\rm rad}^{-1}$ and $\mu_0 \simeq r^{-1}$ for the matching scale. This agrees with the result in \eqref{eq:RGmass0} \cite{andirad3}.\vskip 4pt

The divergences in~the~computation of the tail effects are due to a $1/r$ long-range force. However, we are deriving the tail contribution to the dynamics of the binary using a long-wavelength EFT in which we shrunk the two-body system to a point. This transforms the IR behavior we encountered for the radiative multipole moments, e.g.~\eqref{eq:Itail}, into the UV singularities we find in radiation-reaction. An IR logarithm ($\log r$) is thus converted into a UV divergence in the limit the separation, $r$, is sent to zero, see Fig.~\ref{fig:Mren3}. This demonstrates a remarkable feature of an EFT framework, which allows us to use the renormalization group machinery to resum logarithms, e.g. \cite{iraeft}, as in \eqref{eqVtm} and~\eqref{eq:Mrendef}. This analysis thus explain the origin of the correction in \eqref{eq:logEx} at 4PN order. Higher order terms may be obtained by including tail-of-the-tails contributions.\vskip 4pt

In~addition, the tail contributes a dissipative term through the $i\pi\,\text{sign}(\omega)$ in \eqref{eq:WtailRR}. Following a similar procedure as in \eqref{eq:balance1}, and \eqref{eq:balance}, we can readily obtain the tail correction to the total radiated power, re-deriving the leading expression in \eqref{eq:powerTA}, i.e. $P_{\rm tail}/P_{\rm LO} = 4\pi x^{3/2}$. As we discussed in sec.~\ref{sec:IR}, this makes manifest the important factors of $\pi$, but now through the study of radiation-reaction effects and, remarkably, without the appearance of the associated IR poles in the phase. See \cite{andibr1} for more details.

\subsubsection{IR/UV Mixing}\label{sec:iruvmix}
\vskip 4pt
In order to arrive at~\eqref{eq:WtailRR} we need to add a counter-term, given by
\begin{equation}
- \int V_{\rm ct}[\bx_a^{\pm}] dt = \,\frac{1}{(d-4)_{\rm UV}} \frac{2 G_N^2 \langle M\rangle}{5}\int\,dt \, I_-^{(3)ij}(t) I_+^{(3)ij}(t)\,.\label{eq:Vpmct}
\end{equation}
The counter-term renormalizes the vacuum amplitude for the conservative sector in the near region, in~contrast to the tadpole in \eqref{eq:tadpoleM}. Notice the pole in \eqref{eq:RRnl} cannot be associated to the UV~behavior in the short-distance theory. The reason is twofold. First of all, UV divergences at the orbit scale are related to finite size effects for extended objects, and renormalized through higher derivative terms in the point-particle worldline action, sec.~\ref{sec:effect}. Moreover, see sec.~\ref{sec:eff}, the leading  finite size effects for (non-rotating) binaries enters at 5PN, whereas the above contributes~at~4PN. Nevertheless, the UV pole in \eqref{eq:RRnl} arises in a point-particle limit, the one in which the binary is shrunk to a point-like source by sending the separation between constituents to zero. (This is shown with a double line in Fig. \ref{nltail}.)  It~turns~out, however, the~{\it ultraviolet} divergence from the tail contribution to radiation-reaction is in fact linked to an~{\it infrared} singularity in the~theory of potentials.\vskip 4pt The presence of IR divergences in the potential region was recently uncovered in~\cite{4pn1,4pn2} in the~ADM formalism, see~also~\cite{4pnB3,4pnDS}. The calculations in \cite{4pn1,4pn2,4pnB3,4pnDS}  demonstrate the existence of an IR pole in the near zone which is intimately related to the UV singularity in \eqref{eq:RRnl} \cite{andibr1}.\,\footnote{~While the renormalization methodology is scheme-dependent, the factor in front of the logarithm in \eqref{eq:RRnl} is physical. (It contributes to the binding/mass~energy.) The same factor is related to the coefficient of the IR/UV poles in dim.~reg.} The IR/UV divergences cancel each other out in the conservative sector. (The same occurs in scale-less integrals, see footnote~\ref{foot20}.) That must be the case since there are no divergences in the full theory computation, which readily features~a~logarithm of the ratio of relevant scales, i.e. $\log \omega r \sim \log v$. These IR/UV singularities appear because of an overlap between potential ($\log \mu r$) and radiation ($\log \omega/\mu$) regions. In~fact,~they are a symptom of double-counting,\footnote{~Something similar occurs in~QCD, see ~e.g.~\cite{ApDiMu,bramb1,bramb2,Hoang}.} which introduces ambiguities. This led to the necessity, for~the~regularization approaches advocated in~\cite{4pn2,4pnB3,4pnDS}, of input beyond the Post-Newtonian framework to complete the knowledge of the conservative dynamics to 4PN order. (For instance semi-analytic self-force calculations,~e.g.~\cite{binigsf}.)\vskip 4pt However, this need not be the case, provided the double-counting is properly addressed. Within an EFT formalism the overlapping is avoided by implementing the `zero-bin~subtraction'~\cite{zerobin}. This~removes the IR divergences in the computation of the potential in NRGR, turning them into UV singularities which can be removed by field-redefinitions to 4PN order. (At higher orders the zero-bin subtraction may be required to obtain the correct renormalization group evolution.) The UV poles from the tails in the radiation zone are handled by counter-terms, as in \eqref{eq:Vpmct}, and the renormalized potential in \eqref{eqVtm} is matched into a (local) IR-safe quantity at the orbit scale. As expected in the EFT framework, the (universal) long-distance logarithms are present in both sides of the matching computation at $\mu \simeq r^{-1}$, once the contribution in the near region from the (long-wavelength) radiation modes is included, see~sec.~\ref{smatch}. Therefore, the results in sec.~\ref{sec:longlog} are not affected by the details of the~matching across regions~\cite{andibr1,zerobinNRGR}. 

\subsection{Absorption}\label{sec:abs}
\vskip 4pt

In sec.~\ref{sec:effect} we constructed an effective action which incorporates the imprint of internal degrees of freedom into the long-distance dynamics. At leading order in derivatives we write\,\footnote{~Note we use a different normalization, see e.g. \eqref{eq:lag}. That is because of the $E \leftrightarrow B$ duality we are about to~use.}  
\beq
\label{sqe1}
S_Q = \int dt\left(Q_E^{ij}(t) E_{ij}  + Q_{B}^{ij} (t) B_{ij}(t)\right)\, .
\eeq
We then decomposed into background and response parts. The real part is responsible for tidal effects, see \eqref{responseret2}. We show here the imaginary part accounts for absorption.\vskip 4pt 

\begin{figure}[t!]
    \centering
    \includegraphics[width=0.21\textwidth]{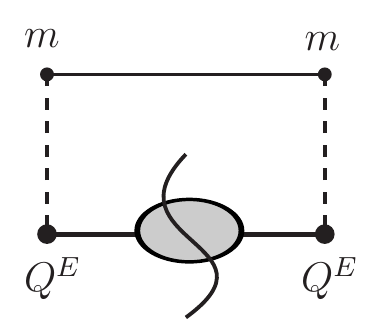}
\caption{Leading order contribution to the absorptive effects in the binary system. The cut represents the imaginary part of the $\langle T(Q^E Q^E)\rangle$ correlator, which account for dissipation into the internal degrees of freedom.}
\label{abspot} 
\end{figure}

Let us consider gravitational wave scattering off a non-rotating black hole. From the optical theorem we can obtain the absorption cross section through the imaginary part of the forward amplitude, with $|{\bk}|=\omega$,~\cite{dis1}
\beq
\nn
\includegraphics[width=0.28\textwidth]{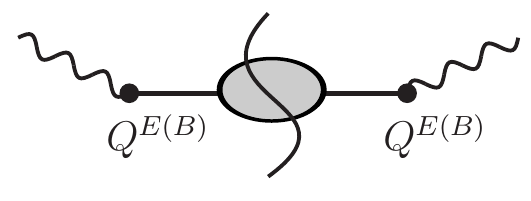}
\eeq
\begin{align}
\label{eq:sigmaeftabs}
\sigma^{\rm eft}_{\rm abs}(\omega) &= \frac{2}{\omega}\times { 1 \over 8  \Mp^2}~\mbox{Im}\, i\,\int dt\, e^{-i\omega t}\left[
\omega^4 \epsilon^*_{ij}(\bk,h)\epsilon_{kl}(\bk,h) \left\langle T\Big(Q_E^{ij}(0) Q_E^{kl}(t)\Big)\right \rangle \right.  \\ &+  \left. \omega^2 ({\bk}\times \epsilon^*(\bk,h))_{ij}  ({\bk}\times \epsilon(\bk,h))_{kl} \left\langle T\Big(Q_B^{ij}(0) Q_B^{kl}(t)\Big)\right\rangle\right]\,.\nn
\end{align}
The $\epsilon_{ij}(\bk,h)$ is the polarization tensor and $({\bk}\times\epsilon)_{ij}={{\bf \epsilon}_{ikl}} {\bk}_k\epsilon_{lj}$. (The $1/8$ comes from expanding an exponential to second order, as in e.g. \eqref{zljall}, and using $E_{ij}~\simeq~-\frac{1}{2} \partial_0^2 h_{ij}$. Similarly for the magnetic~term.)\vskip 4pt Following the steps in sec.~\ref{sec:effect}, see \eqref{responseret} and \eqref{eq:qijkl}, we parameterize the two-point function as
\begin{equation}
\label{eq:g2pt}
2 \,{\rm Im} \left(i\,\int dt \, e^{-i\omega t} \left\langle T \left(Q_E^{ij}(0) Q_E^{kl}(t)\right)\right\rangle\right) = Q_{ijkl} \,{\rm Im}\,\tilde f(\omega)\,.
\end{equation}
From here, and using the invariance under $E_{ij}\rightarrow -B_{ij}$, $B_{ij}\rightarrow E_{ij}$ of the linearized perturbations, we obtain~\cite{dis1,dis2}
\begin{equation}
\label{eqseftabs}
\sigma^{\rm eft}_{\rm abs}(\omega)={\omega^3\over 2 \Mp^2}\mbox{Im}  \tilde f(\omega)\,,
\end{equation}
for each polarization state. On the other hand, the low-frequency absorption cross section for polarized gravitational waves off non-spinning black holes in general relativity is given by, e.g. \cite{page},  
\begin{equation}
\label{eq:gcross}
\sigma^{\rm gr}_{\rm abs}(\omega)={1\over 45} 4\pi r^6_s\omega^4\,,
\end{equation} 
in terms of the Schwarzschild radius, $r_s$. Hence, after matching to \eqref{eqseftabs}, we read off\,\footnote{~The absolute value takes into account the symmetries of the Feynman propagator, which is used in the optical theorem. The imaginary part agrees with the choice of retarded Green's function in \eqref{responseret2} for $\omega>0$, since $\text{Im}f(\omega)= \text{sign}(\omega)\text{Im}\tilde f(\omega)$.} 
\begin{equation}
\label{eq:univImf}
\mbox{Im}\,  \tilde f(\omega) = 16 G^5_N m^6 |\omega|/45\,.  
\end{equation}
From here we find the following scaling for the quadrupole moments, \beq \label{eq:scaleQ}\left\langle Q_{E(B)} Q_{E(B)}\right\rangle \sim G^5_N m^6\omega^2\,.\eeq 
The result in \eqref{eq:univImf} is universal, and therefore can be applied to more complicated dynamical situations, such as the binary system. In such case, the (off-shell) potential modes become the long-distance gravitational perturbation. We can power count the importance of dissipation in the binary's dynamics. In the non-relativistic limit we find, see \eqref{Hscal}, 
\begin{equation}
\label{eq:qev13}
\int dt\,Q_E^{ij}(t) E_{ij}[H]\sim v^{13/2}\,,
\end{equation}
for potential modes (and similarly for the magnetic term). By rotational invariance, $\langle Q_E^{ij}\rangle=0$, hence the leading order absorptive contribution to Im\,$W[\bx_a]$ comes from a box diagram, see Fig. \ref{abspot}, and from the electric-type coupling. 
Moreover, from \eqref{eq:qev13} we conclude that the box diagram contributes at order $Lv^{13}$ for non-spinning black holes \cite{poissonab1}. The calculation is straightforward, yielding
\begin{eqnarray}
iW[\bx_a] &=& {m^2_2 \over 8 \Mp^4} \int dt_1 d{\bar t}_1 dt_2  d{\bar t}_2   \left\langle T \Big(H_{00}(t_2) E_{ij}(t_1)\Big)\right\rangle  \left\langle T\Big( H_{00}({\bar t}_2) E_{kl}({\bar t}_1)\Big)\right\rangle\\
& &{} \times\left \langle T \Big(Q_E^{ij}(t_1) Q_E^{kl}({\bar t}_1)\Big)\right\rangle+ 1\leftrightarrow 2 + \cdots\,.\nn
\end{eqnarray}
Then, using (recall for potential modes the $i\epsilon$-prescription is innocuous)
\begin{equation} 
\langle T H_{00}(t,{\bx}) E_{ij}(0) \rangle = -{i\over 16\pi} \delta(t)\,q_{ij}(t)\,,
\end{equation}
with $q_{ij}(t) = \frac{1}{r^3}\big(\delta_{ij}-3\hat \br_i \hat \br_j\big)$, we arrive at
\cite{dis1}
\begin{equation}
{\rm Im}\,W[\bx_a] = {1\over 2} G_N^2\left(\sum_{a\neq b} m_b^2 \int {d\omega\over 2\pi} \,{\rm Im}\,\tilde f_{(a)}(\omega)\left|q_{ij}(\omega)\right|^2 \right)+\cdots\, .
\end{equation}
From here we obtain the (time-averaged) absorption power loss \cite{dis1}
\begin{equation}
P_{\rm abs} = {16\over 45} G^7_N \left\langle\sum_{a\neq b} m^6_a m_b^2 {\dot q}_{ij} {\dot q}_{ij} \right\rangle\,.
\end{equation}
For example, when we have a hierarchy of masses, $m \ll M$, and on a circular orbit, we get 
\begin{equation} 
P_{\rm abs}={32\over 5} G_N^7 m^6 M^2 \left\langle {{\bv}^2\over r^8}\right\rangle\,,
\end{equation}
in agreement with the result in \cite{poissonab1,tagoshi}. We can now easily generalize the analysis to the case of neutron stars~\cite{dis1}, 
\begin{equation}
\label{eq:general}
{dP_{\rm NS, abs}\over d\omega} = {1\over T} {G_N\over 32\pi^2}\left\langle \sum_{a\neq b} {\sigma^{(a)}_{\rm NS, abs}(\omega)\over\omega^2} m^2_b |q_{ij}(\omega)|^2 \right\rangle\,,
\end{equation}
where $\sigma_{\rm NS, abs}(\omega)$ is the neutron star's cross section for gravitational wave absorption. Notice that unlike the real part of the response, which contributes through Wilson coefficients which are constant, e.g. $C_E$ in \eqref{Se2} (with time and spatial variation associated with the scaling of the higher derivative terms), the dependence on the internal dynamics in the absorption cross section enters as a function of the frequency.\,\footnote{~Dissipative modes are  low-energy (gapless) collective excitations, as in e.g. a fluid \cite{disfluid}.} This offers a much richer spectrum of possibilities to probe the inner structure of neutron stars. We show in sec.~\ref{sec:abspin} these effects may be enhanced for spinning bodies.
\section{Spinning Extended Objects in Gravity}\label{sec:spin}

In what follows we review the EFT formalism for spinning extended objects~\cite{nrgrs,nrgrproc,eih,comment,dis2,nrgrss,nrgrs2,nrgrso,srad,amps}.  As we shall see, the main advantage of an EFT framework is the introduction of an effective action from the onset.\,\footnote{~See \cite{wheel,wheel2} for an interesting alternative approach (based on the coset construction \cite{coset1,coset2}) for slowly rotating bodies. For such systems, however, spin effects are highly suppressed in gravitational wave emission. The approach in \cite{wheel,wheel2} might be useful for other astrophysical processes, although both the action presented here \cite{nrgrs} and in \cite{wheel,wheel2} are essentially equivalent in the Newtonian limit, in which rotation is described in terms of a $\bf 3$-vector and $S^{0i}$ may be~ignored (without the need of supplementarity conditions, see below). The formalisms differ when relativistic corrections are incorporated.}

\subsection{A Twist in Action}
\vskip 4pt

The equations governing the motion of spinning bodies in gravity are known as the Mathisson-Papapetrou equations, and read \cite{Math,Papa,Tulc,dixon}
\beq
 \frac{D p^\mu}{D\tau} = - \frac{1}{2} {R^\mu}_{\nu \alpha\beta} \,u^\nu S^{\alpha\beta},~~~~~  \frac{DS^{\mu\nu}}{D\tau}=p^{\mu}u^{\nu}-u^{\mu}p^{\nu}\label{eq:spin1}.
\eeq
Here $p^\mu$ is the momentum of the particle (which is not given only by $m u^\mu/\sqrt{u^2}$, see below) and $S^{\alpha\beta}=-S^{\beta\alpha}$ is the spin tensor. To derive these equations from an action principle we introduce rotational degrees of freedom in terms of a co-rotating frame, $e_\mu^I$, obeying
\bea 
\label{eq:orthS}
\eta^{IJ}=e_{\mu}^Ie_{\nu}^J\,g^{\mu\nu},\,\, \eta_{IJ} e_{\mu}^I e_{\nu}^J = g_{\mu\nu}\,.
\eea
Notice the $(IJ)$-indices transform under a residual Lorentz invariance. We then define the angular velocity tensor, $\Omega_{\mu\nu}=-\Omega_{\nu\mu}$, which will pave our way to construct the Lagrangian,
\beq
\Omega_{\mu\nu}= e_{I\mu} \frac{De_{\nu}^I}{D\sigma} = e_{I\mu} u^\gamma \nabla_\gamma e_{\nu}^I ~ \to ~
{\dot e}^I_{\mu}= \frac{D e^I_{\mu}}{d\sigma}= -\Omega_{\mu\nu}e^{I\nu}\,.
\label{eq:Omspin}
\eeq
Along the spacetime worldline of the object, $x^\alpha(\sigma)$, the co-rotating frame is described by a time-dependent Lorentz transformation, 
\beq
\label{eq:localLor}
e^I_\mu (x^\alpha(\sigma) )= \Lambda^I_a(\sigma) e^a_\mu(x^\alpha(\sigma)),
\eeq
relative to a locally-flat frame, $e^a_\mu$, also obeying \eqref{eq:orthS} (and as a rotation relative to a co-moving frame,~$e^A_\mu$). The action, including spin, is thus constructed as a function of these variables \cite{regge,israel,nrgrs}
\beq 
S = \int d\sigma~L(x^\alpha,\dot x^\alpha,e^I_\mu,\dot e^I_\mu)\,,
\eeq
following the long-distance symmetries in our problem, i.e. diffeomorphism, reparameterization, and local Lorentz invariance.\vskip 4pt In this formalism the spin tensor, and momentum, are defined as conjugate variables to the velocities, 
\beq
\label{eq:deltaL}
\delta L = - p^\mu \delta u_\mu - \frac{1}{2} S^{\mu\nu} \delta \Omega_{\mu\nu}\, ,
\eeq 
with the overall sign related to the non-relativistic limit (with our set of conventions). Because of the vanishing of the Hamiltonian, we then have, consistent with the symmetries, \cite{regge,nrgrs}
\beq
\label{eq:LSpin}
L = - p^\mu u_\mu - \frac{1}{2} S^{\mu\nu}(\Omega) \Omega_{\mu\nu}\, .
\eeq
We introduce finite size effects momentarily. The Mathisson-Papapetrou equations in \eqref{eq:spin1} follow from \eqref{eq:LSpin}, using \eqref{eq:deltaL}, while enforcing the constraints in \eqref{eq:orthS} \cite{nrgrs}. Moreover, varying the action with respect to the metric we can read off the point-particle stress-energy tensor \cite{nrgrs,israel,dixon},
\beq
T^{\alpha\beta}_{\rm pp}= \int d\sigma\left[ p^{\alpha}u^{\beta}~\frac{\delta^4(x^\mu-x^\mu(\sigma))}{\sqrt{g(x)}}
-\nabla_\mu\left( S^{\mu(\alpha}u^{\beta)}\frac{\delta^4(x^\mu-x^\mu(\sigma))}{\sqrt{g(x)}}\right)\right]\label{eq:TDix}\,.
\eeq
From here, after imposing $\nabla_\alpha T^{\alpha\beta}=0$, we also recover the Mathisson-Papapetrou equations.\vskip 4pt It is convenient to re-write the action in the locally-flat frame. We first introduce the local angular velocity \cite{regge,nrgrs}
\beq
\Omega_L^{ab} = \eta_{IJ} \Lambda^I_a \frac{D \Lambda^J_b}{D\sigma} \,,
\eeq
with $\Lambda_a^I$ in \eqref{eq:localLor}. Hence, using the relationship\,\footnote{The $\omega_\mu^{ab}$ are the Ricci rotation coefficients: $\omega_\mu^{ab} \equiv e^b_\nu \nabla_\mu e^{a\nu}$.} 
\beq
e^a_\mu e^b_\nu \Omega^{\mu\nu} = \Omega_L^{ab} + u^\mu \omega^{ab}_\mu\,, 
\eeq
and defining $S^{ab} \equiv e_\mu^a e^b_\nu  S^{\mu\nu}$, we find 
\beq
\label{eq:Lspin}
L = - p^\mu u_\mu - \frac{1}{2}  S_{ab}\Omega_L^{ab} -\frac{1}{2} \omega^{ab}_\mu S_{ab} u^\mu\,.
\eeq
This expression separates the kinetic part of the action from the spin coupling to the gravitational field. 
\subsubsection{Constraints \& Spin Supplementarity Conditions}
\vskip 4pt
The dynamics from our action leads to a vanishing Hamiltonian, $H=0$, which means evolution is generated by {\it first-class} constraints \cite{regge,teitel}. In our case we have ($S^2 \equiv \tfrac{1}{2} S^{\mu\nu}S_{\mu\nu}$)
\beq
\label{eq:C1} C_1 = p^2-m^2(S^2), \,\,\,  C_2=S^{\mu\nu}S^{*}_{\mu\nu}\,.
\eeq
The  first one, $C_1=0$, encapsulates reparameterization invariance, whereas $C_2=0$ is related to the choice of 
$e^{I=0}_{\mu}$~\cite{nrgrs} (or in other words, the center-of-rotation of a spinning object \cite{kidd,will}). In what follows we fix~\cite{regge,nrgrs}
\beq \label{eq:spinG} x^0(\sigma)=\sigma\,, ~~{\rm and} ~~e_\mu^{I=0} = \frac{p_\mu}{m}\,.\eeq  

In addition, we also notice that an antisymmetric $4\times 4$ matrix, $S^{\mu\nu}$, has three too many degrees of freedom to represent the three necessary angles to describe the rotation of a body.\,\footnote{~This is a common theme in physics, namely, in order to preserve the symmetries of the theory (in this case local Lorentz invariance) we are forced into redundancies, and more variables than physical degrees of freedom are needed, e.g. \cite{zee}.} Hence, the above constraints must be augmented by extra set of conditions. We then implement so called `spin supplementarity conditions' (SSC) to reduce the number of variables.\vskip 4pt The SSC are a set of {\it second-class} constraints \cite{teitel}, $V^\mu = 0$, and the following \cite{regge}
\bea
V_{\rm cov}^\mu &=& S^{\mu\nu} p_\nu  \label{eq:covssc}\,, \hspace*{2.8cm} {\rm (covariant)}\label{eq:sscov}
\,\\
V_{\rm NW}^\mu &=& S^{\mu0}- S^{\mu j}\left(\frac{\tilde \bp^j}{\tilde p_0 +m}\right) \label{eq:nwssc}\,, \hspace*{12pt} {\rm (Newton\mbox{-}Wigner)}\,
\eea 
are often used in the literature. The Newton-Wigner SSC is enforced in the locally-flat frame, with $\tilde p^a \equiv e^a_\mu p^\mu$. Notice the SSC imply $S^{j0} \propto S^{jk}u^k$, which suppresses the temporal component. In the rest frame, rotation is thus simply described in terms of a spin ${\bf 3}$-vector (with $\epsilon_{\mu\nu\alpha\beta}$ the Levi-Civita symbol) 
\beq 
S_\mu = \frac{1}{2m} \epsilon_{\mu\alpha\beta\nu} S^{\alpha\beta} p^\nu  \to \bS_k=\frac{1}{2}\epsilon_{kij}S^{ij}\,.\eeq
We also need the SSC to hold at all times, i.e.~$\frac{D V^\mu}{D\sigma}=0$. For example, for the preservation of the covariant SSC we require
\beq
p^{\alpha}= \frac{1}{p^\mu u_\mu} \left(p^2 u^{\alpha}+\frac{1}{2}R_{\beta\nu\rho\sigma}S^{\alpha\beta}S^{\rho\sigma}u^{\nu}\right)\,.
\end{equation}
This expression can be solved for $p^\alpha$,
\beq
p^{\alpha}= \frac{1}{\sqrt{u^2}}\left(m u^{\alpha}+\frac{1}{2m }R_{\beta\nu\rho\sigma}S^{\alpha\beta}S^{\rho\sigma}u^{\nu} +\cdots \right)\,,\label{eq:up}
\eeq
with $m \equiv m(S^2)$, as in \eqref{eq:C1}. The ellipsis include higher order terms in the spin and curvature. The~relationship between spin and angular velocity can also be derived from here, following~our choice of gauge~\cite{nrgrs}. Using \eqref{eq:spinG}, we find
\beq
\frac{Dp^{\mu}}{D\sigma} =  -\Omega^{\mu\nu}p_\nu = - \frac{1}{2} {R^\mu}_{\nu\alpha\beta} u^\nu S^{\alpha\beta}\,,
\eeq
and, from \eqref{eq:sscov} and \eqref{eq:up}, we get 
\begin{equation}
\Omega_{\mu\nu} = \frac{\sqrt{u^2}}{m} \left( f'(S^2) S_{\mu\nu} + \frac{1}{2} R_{\mu\nu\alpha\beta}S^{\alpha\beta} + \cdots \right)
\label{eq:OmS}\,,
\end{equation}
with $f(S^2) \equiv m^2(S^2)$. This expression was obtained by matching in the limit of a flat background,~where~\cite{regge}
\beq\, \Omega_{\mu\nu}= \,\frac{\sqrt{u^2}}{m} f'(S^2) S_{\mu\nu}\, .\hspace*{1.7cm}({\rm Minkowski})\, \eeq \vskip 4pt

Similar manipulations can be performed in the Newton-Wigner SSC. Although apparently more cumbersome at first, it has the advantage that it leads to a canonical (Dirac) algebra in the reduced phase space under certain circumstances, unlike the covariant SSC (see below). In general, one can try solving say for the covariant SSC explicitly at the level of the action, and arduously work out the transformation to a canonical symplectic structure, e.g. \cite{eliminEFT,equiv4pn}. Since these are second-class constraints, one can also simply implement the SSC through Lagrangian multipliers. The latter are fixed by the preservation of the SSC upon time evolution. We pursue the second route here. 

\subsubsection{Routhian Formalism}
\vskip 4pt
The Lagrangian in \eqref{eq:Lspin} is a function of the angular velocity, $L(x^\alpha,u^\alpha,\Omega^{\alpha\beta})$, without explicit dependence on the angular variables. This invites us to perform a partial Legendre transformation for the latter. This procedure then introduces a Routhian \cite{nrgrss} (see also \cite{yee})
\beq
\label{eq:Routh}
-{\cal R} =  -\frac{1}{2}  S_{ab}\Omega_L^{ab} - L= p^\mu u_\mu + \frac{1}{2} \omega_\mu^{ab} S_{ab} u^\mu \,.
\eeq
The minus sign is for convenience, such that we retain the same rules for the spin-independent part. The dynamics derives from, 
\begin{equation}
\frac{\delta }{\delta x^\mu} \int  {\cal R}~ d\sigma = 0, \;\;\; \frac{d
S^{ab}}{d\sigma} = \{S^{ab},{\cal R}\}\label{eq:eomS}.
\end{equation}
The relevant Poisson brackets are given by
\bea
 \{x^\mu ,S^{ab}\}&=& 0\,,\;\;\; \{{\cal P}^\alpha,S^{ab}\}= 0  \\
\{S^{ab},S^{cd}\} &=& \eta^{ac} S^{bd} +\eta^{bd}S^{ac}-\eta^{ad} S^{bc}-\eta^{bc}S^{ad}\label{eq:algebraS}
\,,\label{eq:palgs} 
\eea
with ${\cal P}_\mu \equiv p_\mu + \frac{1}{2} \omega_\mu^{ab} S_{ab}$, the canonical momentum. It is straightforward to show the Mathisson-Papapetrou equations follow from \eqref{eq:eomS}. For example, using \eqref{eq:algebraS}, the spin equation becomes
\beq
\frac{dS^{ab}}{dt} = \{V , S^{ab}\} = 4\,S^{c\{a}\eta^{b\}d}\frac{\partial}{\partial S^{cd}} V \label{eomV}\,,
\eeq
where the potential is given by $V= - {\cal R}$. The reader may worry about encountering derivatives of the spin tensor, $\dot S^{ab}$, when computing the spin contributions to the gravitational potential. They appear, for instance, through departures from instantaneity in the propagators. These time derivatives may be either integrated by parts, or replaced using the lower order equations of motion prior to applying \eqref{eomV}~ \cite{nrgrso,nrgrs2,nrgrss}. This~simply amounts to a field redefinition, as it is often the case after removing terms proportional to the acceleration from the effective action, e.g. \cite{nrgr2pn,nrgr3pn}. (This procedure has been used for instance to prove the equivalence between the EFT and ADM calculations to 4PN order \cite{equiv4pn}.)\vskip 4pt

We still need to enforce the SSC. For definiteness we work with the covariant version in \eqref{eq:sscov}. We~proceed then to add a set of Lagrangian multipliers, $\lambda_\mu$, to the Routhian, 
\beq
{\cal R}=  -\left( p^\mu u_\mu + \frac{1}{2} \omega_\mu^{ab} S_{ab} u^\mu +  \lambda_\mu V^\mu\right)\,.
\eeq
The $\lambda_\mu$ parameters are fixed by imposing the preservation of the SSC, $\dot V^\mu = 0$, and we obtain~\cite{nrgrss,nrgrs2}
\beq
\label{eq:Routh2}
{\cal R} =  -\left( m \sqrt{u^2} + \frac{1}{2} \omega_\mu^{ab} S_{ab} u^\mu + \frac{1}{2m }R_{\nu\alpha\rho\sigma}S^{\rho\sigma}u^{\nu} S^{\alpha\beta} u_\beta + \cdots\right)\,,
\eeq
after a field redefinition. This expression guarantees the SSC holds upon time evolution. Notice, as a function of the velocity, the SSC involves an infinite series of terms from the relationship between $p^\alpha$ and $u^\alpha$, e.g. \eqref{eq:up}, which also contribute to \eqref{eq:Routh2}. However, $S^{\mu\nu} u_\nu \sim {\cal O}(RS^3)$, and moreover, the extra terms are already coupled to curvature. Therefore, up to effects which are higher order in the spin and curvature (encoded in the ellipsis), we can replace $ p^\mu \to u^\mu$ in the covariant SSC. 

\subsubsection{Feynman Rules \& Power Counting}\label{sec:FeynS}
\vskip 4pt
Armed with an effective action we can compute the gravitational potentials and multipole moments which
contribute to the total radiated power loss, now including spin effects. For that purpose we need to work out first the power counting and identify the relevant Feynman diagrams which contribute to a given order in the PN expansion. To derive the scaling rules we proceed as follows.\vskip 4pt We are interested in compact objects, for which
\beq 
I_{S} \sim m r_s^2 \to S=I_S \Omega \sim I_S \frac{v_{\rm rot}}{r_s}\sim mv_{rot}r_s \leq mr_s\, ,
\eeq
with sub-luminal rotational speed, $v_{\rm rot} \leq 1$. Then, for maximally rotating bodies with $v_{\rm rot}\lesssim 1$, spin scales as 
\beq
\label{eq:slv}
S \sim m r_s = m r \frac{r_s}{r} \sim L v\, ,
\eeq
where $L = m r v$ and we used $r_s/r \sim v^2$. To obtain the Feynman rules, we simply expand the spin-gravity coupling in powers of the metric perturbation. Using
\beq
e_\mu^a =  \delta^a_\mu + \frac{1}{2}\delta^\nu_a \big({h^\nu}_\mu - \frac{1}{4} {h^\nu}_\alpha {h^\alpha}_\mu\big) +\cdots\, ,
\eeq
we find up to quadratic order (see appendix \ref{app:Fey} for a collection of spin couplings)
\beq
L_{\rm spin} = \frac{1}{2\Mp} \delta^\alpha_a \delta^\beta_b h_{\alpha\gamma,\beta}u^{\gamma} S^{ab} + \frac{1}{4\Mp^2}\delta^\beta_a \delta^\gamma_b {h_\gamma}^\lambda \left({1\over 2} h_{\beta\lambda,\mu}+h_{\mu\lambda,\beta}-h_{\mu\beta,\lambda}\right)u^{\mu} S^{ab} +\cdots \,. \label{eq:spingr}
\eeq
When we expand expressions such as \eqref{eq:spingr}, we will no longer distinguish generic indices from those in the locally-flat frame. Everywhere the spin tensor is defined with respect to the $e^a_\mu$ locally-flat vierbein. 
\subsubsection{Finite Size Effects}\label{sec:finiteSS}
\vskip 4pt
To incorporate finite size effects we introduce a series of Wilson coefficients. For the reader's convenience we reproduce some of the expressions from sec.~\ref{sec:effect}. In a locally-flat co-moving frame we have, see \eqref{sqe}, 
\beq
\label{sqe2}
S_{Q_E} = \frac{1}{2} \int dt~Q_E^{ij}(t) E_{ij} \, ,\eeq
at leading order in derivatives. The quadrupole moment is further split as,~see~\eqref{qsr},
\beq
\label{qsr2}
Q^{E}_{ij} = \underbrace{\left\langle Q^{E}_{ij}\right\rangle_S}_{\rm background} + \underbrace{\left(Q^{E}_{ij}\right)_{\cal R}}_{\rm response}\, .
\eeq
The response encodes tidal effects, $\left(Q^{E}_{ij}\right)_{\cal R} \propto E_{ij}$. This time, for rotating bodies, the Love numbers also depend on the spin. For the background piece, unlike the case of non-rotating bodies, spin allows us to write terms which couple linearly to the metric perturbation. The first we find is the quadrupole moment produced by the body's own rotation.  We parameterize the background expectation value as (the factor of $1/m$ is for later convenience)
\beq
\label{vevQES}
\left\langle Q^{E}_{ij}\right\rangle_S = \frac{C_{ES^2}}{2m}  S^{ik}{S_k}^j\, 
\eeq
(up to a trace which do not couple to $E_{ij}$). Plugging this expression into \eqref{sqe2}, we obtain~\cite{nrgrs,eih,nrgrs2}
\beq
\label{eq:sce2}
S_{ES^2} = \frac{C_{ES^2}}{2m} \int d\sigma \int d^4x \delta^4(x-x(\sigma)) \frac{E_{ab}(x)}{\sqrt{u^2}} {S^a}_c S^{cb}\,,
\eeq
where we have written the contribution to the action in a reparameterization invariant manner and in a generic locally-flat frame, see sec.~\eqref{sec:effect}. It is straightforward to generalize the above expression to incorporate higher order spin-dependent multipoles, $ Q_{E(B)}^{i_1\ldots i_\ell}$. The~Feynman rules follow by expanding \eqref{eq:sce2} in powers of $h_{\mu\nu}$~\cite{nrgrs2,srad}. (See appendix~\ref{app:Fey}.)
\vskip 4pt
  
To extract the value of $C_{ES^2}$ we proceed as outlined in sec.~\ref{smatch}. The calculation of the one-point function ({\it aka} the Kerr metric) follows the same steps as in secs.~\ref{sec:nonlinear2} and~ \ref{sec:regular0}. The topologies of the first relevant diagrams are given in Fig~\ref{ydiag} and so on, but this time with spin insertions. In addition, we incorporate the contribution from \eqref{eq:sce2}. At leading order in $G_N$ the $S^2$ dependence in 
\beq
\label{metric}
\left.\frac{H^{00}}{\Mp}\,\right|_{\rm Kerr}= \, S^2 \frac{G_Nm}{r^3}\big( 3(\boldsymbol{j}\cdot \hat \br)^2 -\boldsymbol{j}^2\big)\,,
\eeq
with $|\boldsymbol{j}|=S/m$, is matched into the EFT contribution from the finite size term, with $\bS_i = \tfrac{1}{2}\epsilon_{ijk}S^{jk}$,
\beq
\left.\frac{H^{00}}{\Mp}\,\right|_{\rm EFT} = \,C_{ES^2} \frac{G_N}{m r^3} \big(3(\bS\cdot \hat\br)^2-{\bS}^2\big)\, .
\eeq
Notice the matching is independent of the SSC. Comparing both expressions we conclude that $C^{\rm bh}_{ES^2} =1$, for rotating black holes. The above is nothing but the quadrupole moment of the Kerr spacetime, given by: $Q_E^{\rm bh} = m |\boldsymbol{j}|^2 = S^2/m$. On the other hand, for neutron stars the quadrupole moment will be larger, $C^{\rm ns}_{ES^2} \simeq 4\,\hbox{-}\,8$, depending on the equation of state and internal dynamics, e.g~\cite{poissonNS}.\vskip 4pt  The $C_{ES^2}$ coefficient enters at leading order in the spin sector for maximally rotating bodies (see below). Therefore, it offers an excellent venue to distinguish black holes from neutron stars, as well as testing different models for the latter~\cite{buoce2,bini2,bini3,bini4,binitidalspin,iloveq}. For tidal effects, on the other hand, the leading order effects enter through a term similar to \eqref{Se2} via the spin dependence of the $C_{E(B)}$ parameters. Remarkably, as for the non-rotating case, the $C_{E(B)}$ coefficients vanish for spinning black holes \cite{Poissontidal,Panitidal}. That is not the case for rotating neutron stars, e.g. \cite{Poissontidal,Panitidal,Panitidal2,ericBB}. Because of the scaling of the spin tensor, e.g. \eqref{eq:slv}, tidal effects due to spin are suppressed in the inspiral phase. Tidal disruptions may, however, become important near merger, e.g.~\cite{spinumtid}.\vskip 4pt

Before we proceed, let us add a comment regarding $C_{ES^2}$. As we see from the matching, this coefficient is fixed in terms of the quadrupole moment of an {\it isolated} rotating black hole. Therefore, unlike the $C_E$ coefficient which enters as a response, the $C_{ES^2}$ does not get renormalized. This is expected from Birkhoff's theorem, and the fact the divergences arise from the point-particle limit. In other words, it does not depend on scale, nor receive (classical) logarithmic corrections. Namely, it is a `self-induced' finite size effect, which may be generated, for instance, by implementing a cutoff regulator.

\subsection{Gravitational Spin Potentials}
\subsubsection{Leading Order}
\begin{figure}[t!]
\centering
\includegraphics[width=0.4\textwidth]{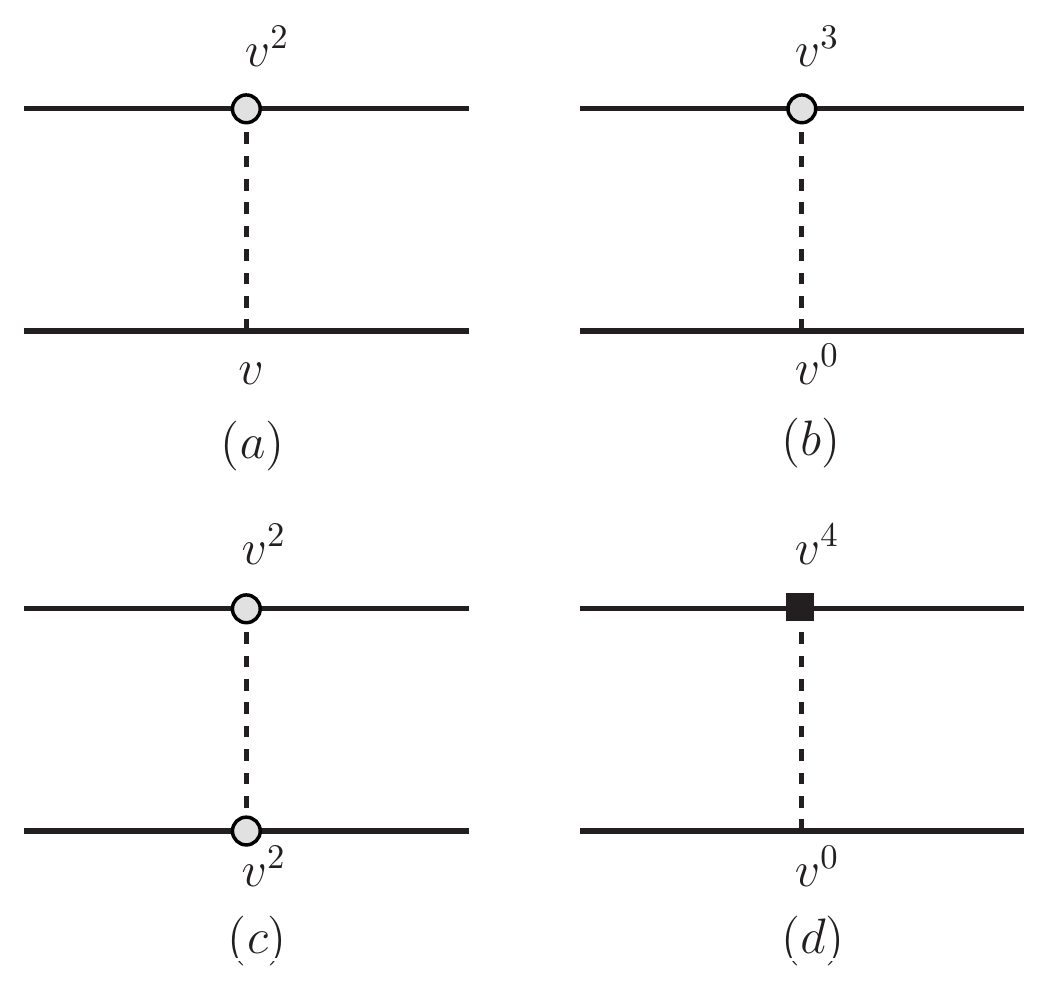}
\caption{Feynman diagrams which contribute to the leading order spin potentials. The gray blob and black square represent spin insertions and finite size effects, respectively. See appendix~\ref{app:Fey}.}
\label{figLOS}
\end{figure}

\vskip 4pt
The leading order spin potentials were obtained many years ago \cite{barker1,barker2}. For us they simply follow from the one-graviton exchange with spin insertions, see Fig.~\ref{figLOS}, yielding \cite{nrgrs} 
\bea
\label{eq:loso}
V_{\rm SO}^{\rm LO} &=& \frac{G_N m_2}{r^3}\br^j \left[ S_1^{j0} + S_1^{jk}\left(\bv^i_1-2\bv_2^k\right)\right] + 1\leftrightarrow 2\,,\\
V_{\rm SS}^{\rm LO} &=& -\frac{G_N}{r^3} \left[\bS_1\cdot\bS_2 - 3(\bS_1\cdot \hat \br) (\bS_2\cdot \hat\br) \right] \label{eq:loss}\,,\\
V_{{\rm S}^2}^{\rm LO} &=& -\frac{G_N}{r^3} \sum_{a=1,2} \frac{C^{(a)}_{ES^2}}{2m_a} \left[ \big({\bS}_a^2-3(\bS_a\cdot \hat\br)^2\big)\right]\,,\label{eq:los2}
\eea
at 1.5PN for spin-orbit and 2PN for spin-spin, respectively.  
For the equations of motion we use \eqref{eq:eomS} prior to imposing the~SSC. For instance for the spin dynamics we get, with ${\bv} = {\bv}_1-{\bv}_2$, 
\begin{equation}
\frac{d{\bS} _1}{dt} = 2\frac{m_2
G_N}{r^2}({\hat \br} \times{\bv} )\times{\bS} _1 +
\frac{m_2 G_N}{r^2}({\bS} _1\times{\hat \br} )\times {\bv}_1 \label{ds1dt}\,.
\end{equation} 
Notice the spin norm is not conserved. The more standard form follows via a spin redefinition~\cite{nrgrs,nrgrso}
\beq
{\tilde {\bS}} _1 = (1-\frac{1}{2}{\tilde \bv} _1^2){\bS} _1 + \frac{1}{2}{\tilde \bv} _1({\tilde \bv} _1\cdot{\bS} _1)\label{pnshift}\,,
\eeq
where ${\tilde \bv}_1$ is the velocity in the locally-flat frame, $\tilde v^a \equiv e^a_\mu v^\mu$, which agrees with the global PN frame at leading order (see sec.~\ref{sec:NW}).  Then, from \eqref{ds1dt} we obtain the spin precession~\cite{will,kidd} 
\beq
\frac{d{\tilde {\bS}} _1}{dt} = \left(2+\frac{3m_2}{2m_1}\right)\frac{\mu_m
G_N}{r^2}({\hat \br} \times{\bv} )\times{\tilde {\bS}} _1\,.  \label{ds1dtn}
\eeq
The alert reader will immediately realize that applying the Newton-Wigner SSC (which reads $S^{j0}=\frac{1}{2}S^{jk}v^k$ at leading order) in the potential of \eqref{eq:loso} we obtain directly the result in \eqref{ds1dtn}. 

\subsubsection{Next-to-Leading Order}
\begin{figure}[t!]
\centering
\includegraphics[width=0.65\textwidth]{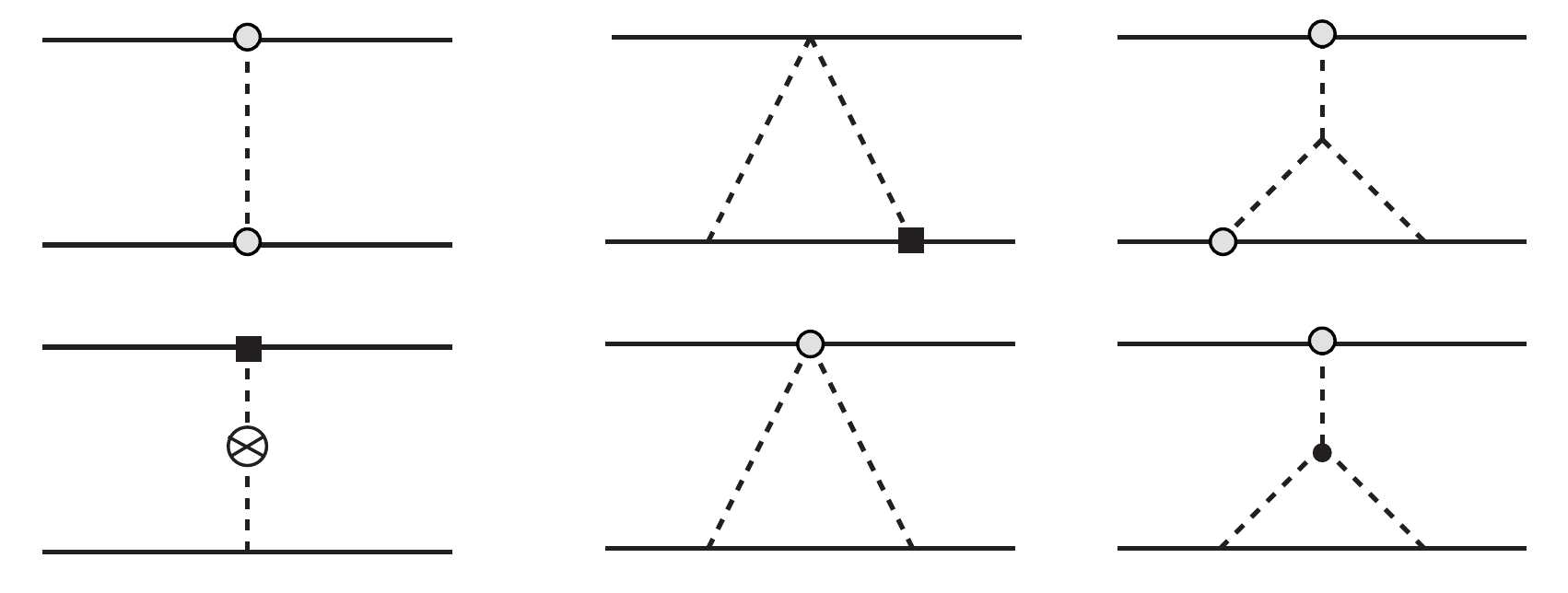}
\caption{Sample of Feynman diagrams which contribute to the NLO spin potentials. We depict a non-instantaneous correction to the vertex by a black dot. See appendix~\ref{app:Fey}.}
\label{fignlS}
\end{figure}

\vskip 4pt
The NLO spin-orbit and spin-spin potentials enter at 2.5PN and 3PN, respectively. The topologies of Feynman diagrams are the same we find in the computation of the 1PN non-spinning Einstein-Infeld-Hoffmann action, see Fig.~\ref{fig2EIH}. For illustration purposes, we display a sampling of the relevant diagrams~in~Fig.~\ref{fignlS}. In total, the spin potentials read~\cite{eih,comment,nrgrproc,nrgrso,nrgrss,nrgrs2}
\bea
\label{eq:nloso}
V^{\rm NLO}_{\rm SO} &=& \frac{G_Nm_2}{r^3}\left[\left\{S^{i0}_1\left(2{\bv}_2^2-2 {\bv}_1\cdot {\bv}_2-\frac{3}{2r^2} ({\bv}_2\cdot {\br})^2-\frac{1}{2} {\ba}_2\cdot {\br}\right)\right.\right.\\ &+&\left.\left.\left(2 {\bv}_1\cdot {\bv}_2+\frac{3({\bv}_2\cdot {\br})^2}{r^2}-2{\bv}_2^2+{\ba}_2\cdot{\br}\right)S_1^{ij}{\bv}_2^j \right. \right. \nn \\  &-&  \left. \left.\left(\frac{3}{2r^2} ({\bv}_2\cdot {\br})^2 +\frac{1}{2} {\ba}_2\cdot{\br}\right)S_1^{ij}{\bv}_1^j+ 2S_1^{ij}{\ba}_2^j{\bv}_2\cdot{\br}+r^2S^{ij}_1\dot {\ba}_2^j \right\}{\br}^i \right.\nn\\ &+& S^{i0}_1\left(({\bv}_1-{\bv}_2)^i {\bv}_2\cdot {\br}-\frac{3}{2}{\ba}_2^i r^2\right)+ S_1^{ij}{\bv}_1^i{\bv}_2^j{\bv}_2\cdot{\br}-r^2S^{ij}_1{\ba}_2^j{\bv}_2^i-\frac{1}{2}r^2S^{ij}_1{\ba}_2^j{\bv}_1^i \Big]\nn \\  &+& \frac{G_N^2m_2}{r^4} {\br}^i\Big[-(m_1+2m_2)S_1^{i0}+\left(m_1-\frac{m_2}{2}\right)S_1^{ij}{\bv}_1^j+\frac{5m_2}{2} S_1^{ij}{\bv}_2^j\Big]+1\leftrightarrow 2\nn\,,\\
\label{eq:nloss}
V^{\rm NLO}_{\rm SS} &=& -\frac{G_N}{r^3}\Big[(\delta^{ij}-3{\hat \br}^i {\hat \br}^j)\left( S^{i0}_1S^{j0}_2+\frac{1}{2}{\bv}_1 \cdot {\bv}_2 
S^{ik}_1S^{jk}_2+\bv_1^m\bv_2^k S^{ik}_1S^{jm}_2- \bv_1^k \bv_2^m S^{ik}_1S^{jm}_2 \right.  \\ 
 &+&   \left.
S^{i0}_1S^{jk}_2(\bv_2^k-\bv_1^k)+S^{ik}_1S^{j0}_2(\bv_1^k-\bv_2^k)\right)+ \frac{1}{2}S_1^{ki}S_2^{kj}\Big( 3 { \bv} _1\cdot{\hat \br}{ \bv} _2\cdot{\hat\br}(\delta^{ij}-5{\hat \br}^i{\hat \br}^j)  \nn\\   &+&   
   3 {\bv} _1\cdot{\hat\br} (\bv_2^j{\hat \br}^i+v_2^i{\hat \br}^j)+3{\bv} _2\cdot{ \hat\br} (\bv_1^j{\hat \br}^i +\bv_1^i{\hat \br}^j) -\bv_1^i\bv_2^j-\bv_2^i\bv_1^j\Big)+(3{\hat \br}^l{\bv}_2 \cdot{\hat \br} -\bv_2^l) S_1^{0k}S_2^{kl}  \nn \\ 
 &+ &  (3{\hat \br}^l{\bv}_1 \cdot{\hat \br} -\bv_1^l) S_2^{0k}S_1^{kl}  \Big]+ \left(\frac{G_N}{r^3}- \frac{3M G_N^2}{r^4}\right) S_1^{ij}S_2^{kj}(\delta^{ik}-3 \hat \br^k {\hat \br}^i)\,,\nn \\
\label{eq:nlos2}
V^{\rm NLO}_{{\rm S}^2}&=&  C^{(1)}_{ES^2}\frac{G_Nm_2}{2m_1r^3}\Big[ S_1^{j0}S_1^{i0}(3{\hat \br}^i{\hat \br}^j-\delta^{ij}) 
-2 S_1^{k0} \left( ({\bv}_1\times{\bS}_1)^k - 3 ({\hat \br}\cdot {\bv}_1)({\hat \br}\times{\bS}_1)^k\right)\Big] 
  \\ &+& C^{(1)}_{ES^2}\frac{G_Nm_2}{2m_1r^3}\Big[ {\bS}_1^2 \left( 6 ({\hat \br}\cdot{\bv}_1)^2 - \frac{15}{2} {\hat \br}\cdot{\bv}_1{\hat \br}\cdot{\bv}_2 + \frac{13}{2}{\bv}_1\cdot{\bv}_2 - \frac{3}{2}{\bv}_2^2 - \frac{7}{2}{\bv}_1^2\right) \nn \\ & +&   ({\bS}_1\cdot {\hat \br})^2 \left ( \frac{9}{2}({\bv}_1^2+{\bv}_2^2)-\frac{21}{2}{\bv}_1\cdot{\bv}_2 - \frac{15}{2} {\hat \br}\cdot{\bv}_1 {\hat \br}\cdot{\bv}_2\right)+ 2{\bv}_1\cdot{\bS}_1{\bv}_1\cdot{\bS}_1\nn \\ & -&   3{\bv}_1\cdot{\bS}_1{\bv}_2\cdot{\bS}_1 
-6 {\hat \br}\cdot{\bv}_1{\hat \br}\cdot{\bS}_1{\bv}_1\cdot{\bS}_1+ 9 {\hat \br}\cdot{\bv}_2{\hat \br}\cdot{\bS}_1{\bv}_1\cdot{\bS}_1+ 3 {\hat \br}\cdot{\bv}_1{\hat \br}\cdot{\bS}_1{\bv}_2\cdot{\bS}_1\Big]\nn \\
&+&  C^{(1)}_{ES^2}\frac{m_2G_N^2}{2r^4}\left(1+\frac{4m_2}{m_1}\right) \left( {\bS}_1^2 - 3({\bS}_1\cdot{\hat \br})^2\right) -  \frac{G_N^2m_2}{r^4}\left({\bS}_1\cdot {\hat \br}\right)^2 \nn \\  &+& S_1^{0l} {\tilde\ba}^l_{1{\rm (SO)}}+ {\bv}_1\times{\bS}_1\cdot {\tilde\ba}_{1{\rm (SO)}} + 1 \leftrightarrow 2\,. \nn
\eea
Notice we also included the contribution from \eqref{eq:Routh2} which enforces the preservation of the covariant SSC. 
This is responsible for the two terms in the last line, where
\beq
\label{covasoloc}
{\tilde \ba}_{1{\rm (SO)}} =  \frac{3m_2G_N}{m_1r^3} \Big[2{\hat\br}~ ({\bv} \times {\bS}_1)\cdot {\hat\br } +  {\hat\br}\cdot{\bv} ({\hat\br} \times {\bS}_1) -{\bv} \times {\bS}_1\Big]\,,
\eeq
is the leading order spin-orbit acceleration in the locally-flat frame, i.e. $\tilde\ba = \dot{\tilde \bv}$. Note, in addition, the spin-orbit potential depends on the acceleration of the bodies. These may be removed by using the leading order equations of motion, e.g. $\ba_2 = \frac{G_Nm1}{r^3}\hat\br$. Collecting all the pieces together the spin potential to 3PN order is thus given by
\beq
V^{\rm 3PN}_{\rm spin} = \eqref{eq:loso}+ \eqref{eq:loss} + \eqref{eq:los2}+ \eqref{eq:nloso} + \eqref{eq:nloss} + \eqref{eq:nlos2}\,.\nn
\eeq
To complete the calculation we must use \eqref{eq:eomS} and afterwards impose the SSC. We apply the covariant SSC, $S^{ab} \tilde v_b=0$ in a locally-flat frame, for which we need expressions in terms of the vierbein, 
\bea
e^0_0(\bx_1) &=& 1 - \frac{G_N m_2}{r} + \cdots\,, \\ 
e^k_0(\bx_1) &=& - 2 \frac{G_N m_2}{r}{\bv}_2^k + \frac{G_N}{r^2} ({\hat \br}\times {\bS}_2)^k+\cdots\,,  \\ 
e^i_j(\bx_1) &=& \delta^i_j \left(1+\frac{G_N m_2}{r}\right) +\cdots\,,
\eea 
such that
\bea
\label{tildv}
\tilde {\bv}^{a=j}_1 &=& {\bv}^i_1 \left(1+\frac{G_Nm_2}{r}\right) - \frac{2G_Nm_2}{r} {\bv}^j_2+\frac{G_N}{r^2} ({\hat \br}\times {\bS}_2)^j\,, \\
\tilde v^{a=0}_1 &=& \left(1-\frac{G_Nm_2}{r}\right) \label{eq:tild0}\,.
\eea
The covariant SSC then has the form (e.g. for particle 1)
\beq
\label{eq:s01ssc}
{\bS}_1^{(0)} =  \left({\bS}_1\times\frac{\tilde {\bv}_1}{\tilde v_1^0}\right)=\left(1+\frac{2G_Nm_2}{r}\right) ({\bS}_1\times{\bv}_1) -2\frac{G_Nm_2}{r}(\bS_1\times \bv_2) + \frac{G_N}{r^2}\left[\bS_1\times({\hat \br} \times {\bS}_2)\right]\,,
\eeq
where we introduced the notation $\bS^{(0)i} \equiv S^{0i}$. The position dynamics now follows from an action principle, whereas the spin equations derive from \eqref{eomV}, using
\bea
\{{\bS}^i,{\bS}^j\} &=& -\epsilon^{ijk}{\bS}^{k}\,,\\
\{{\bS}^i,{\bS}^{(0)j}\} &=& -\epsilon^{ijk}{\bS}^{(0)k}\,. \nn
\label{alsscnew}
\eea  
For example, for the spin-orbit contribution, writing \eqref{eq:loso} and \eqref{eq:nloso} as \beq \label{decomso}V_{\rm SO} = {\bA}_1^{\rm SO} \cdot {\bS}_1 + \bom_1^{\rm SO}\cdot {\bS}_1 + 1\leftrightarrow 2\,,\eeq   
the spin equation of motion becomes
\beq
\label{eq:spind}
\frac{d{\bS}_1}{dt} = \bom^{\rm SO}_1 \times {\bS}_1 + \left(\frac{\tilde {\bv}_1}{\tilde v_1^0}\times{\bS}_1\right)\times {\bA}_1^{\rm SO}\,.
\eeq
This equation does not conserve the spin's norm. However, we can make a spin redefinition which transforms \eqref{eq:spind} into precession form \cite{nrgrso}.  We can perform similar manipulations to derive the full ${\cal O}(\bS_1\bS_2)$ and ${\cal O}(\bS_1\bS_1)$ dynamics. \vskip 4pt

Notice once the SSC is enforced in the locally-flat frame, say for $\bS_1^{(0)}$ in \eqref{eq:s01ssc}, we generate a term which depends not only on $\bS_1$ but also 
$\bS_2$, through the veirbein. This means the spin-orbit potential also contributes to the ${\cal O}(\bS_1\bS_2)$ dynamics. This is one of the subtleties of dealing with spin degrees of freedom. We elaborate briefly on this issue in what follows, see \cite{nrgrs,nrgrss,nrgrs2,nrgrso} for more details.

\subsubsection{Newton-Wigner SSC}\label{sec:NW}
\vskip 4pt
It is straightforward to show that applying the covariant SSC from \eqref{eq:s01ssc} into $V^{\rm 3PN}_{\rm spin}$ produces a potential which requires a non-canonical algebra in order to obtain the correct dynamics. This is already clear at leading order for the spin-orbit potential, and more so at higher orders. Hence, the convenience of having a simple SSC is eclipsed by the necessity of a non-trivial algebraic structure in the reduced phase space. While this is not necessarily a problem, since one can postpone the application of the SSC until the equations of motion are obtained using \eqref{eq:eomS}, it would be perhaps desirable to work at the level of a spin $\bf 3$-vector obeying a canonical algebra. We show here that enforcing instead the Newton-Wigner SSC, while ignoring ${\cal O}(\bS_{1(2)}\bS_{1(2)})$ terms, we obtain a spin potential with a canonical symplectic structure \cite{nrgrso}.\vskip 4pt

The Newton-Wigner SSC in the locally-flat frame is given by, say for particle 1, \cite{regge,nrgrso}
\beq
S^{i0}_{1({\rm NW})} = \frac{{\tilde p}^0_1}{{\tilde p}^0_1+m} S^{ij}_{1({\rm NW})}  \frac{{\tilde {\bp}}^j_1}{{\tilde p}^0_1}\,,
\eeq
which is reparameterization invariant. In terms of the velocities we have
\beq
S^{i0}_{1{(\rm NW)}} = \frac{{\tilde v}^0_1}{{\tilde v}^0_1+\sqrt{{\tilde v}^a_1 {\tilde v}_{1a}}} S^{ij}_{1({\rm NW})}   \frac{{\tilde {\bv}}^j_1}{\tilde v^0_1}\,.
\eeq
Then, choosing $\sigma =t$, we obtain 
\bea
\label{eq:nwssc2}
{\bS}^{(0)}_{1(\rm NW)} &=& \frac{1}{1+\sqrt{1-\left(\frac{{\tilde {\bv}}_1}{\tilde v^0_1}\right)^2}}\left({\bS}_{1(\rm NW)}\times \frac{{\tilde {\bv}}_1}{\tilde v^0_1}\right)\\ &
=& \frac{1}{2} \bS_1 \times \bv_1 + \cdots + \frac{G_N}{2r^2}\left[\bS_1\times({\hat \br} \times {\bS}_2)\right] + \cdots \nn\,.
\eea
We now apply the Newton-Wigner SSC directly in the gravitational potential. At ${\cal O}(\bS_1\bS_2)$ this amounts~to first replacing the LO $\bS^{(0)}_1$ in \eqref{eq:nloss}, and secondly, inserting the term quadratic in spin from \eqref{eq:nwssc2} into the LO spin-orbit result of \eqref{eq:loso} \cite{eih,nrgrproc,comment,nrgrss}. At the end of the day the potential is given to 3PN order by (removing the `NW' tag on the spin)
\bea
V^{\rm NW}_{\rm SS} &=& -\frac{G_N}{2r^3}\left[ {\bS}_1 \cdot {\bS}_2\left({3\over2}{\bv}_1\cdot {\bv}_2-3{\bv}_1\cdot {\hat \br} {\bv}_2\cdot {\hat \br}
-\left({\bv}_1^2+{\bv}_2^2\right)\right)
-{\bS}_1\cdot {\bv}_1{\bS}_2\cdot {\bv}_2  \right. \\ &-&\frac{3}{2}{\bS}_1\cdot {\bv}_2 {\bS}_2\cdot {\bv}_1+
{\bS}_1\cdot {\bv}_2 {\bS}_2\cdot {\bv}_2
+ {\bS}_2\cdot {\bv}_1 {\bS}_1\cdot {\bv}_1+
3{\bS}_1\cdot{\hat \br}{\bS}_2\cdot {\hat \br}
\left({\bv}_1\cdot{\bv}_2+5{\bv}_1\cdot{\hat \br} {\bv}_2\cdot{\hat \br}\right)\nn  \\
&-& 3{\bS}_1\cdot{\bv}_1{\bS}_2\cdot {\hat \br}{\bv}_2\cdot {\hat \br}-
3 {\bS}_2\cdot{\bv}_2{\bS}_1\cdot{\hat \br}{\bv}_1\cdot{\hat \br} + 3({\bv}_2\times{\bS}_1)\cdot{\hat \br}({\bv}_2\times {\bS}_2)\cdot {\hat \br}
\nonumber\\
&+& \left.
3( {\bv}_1\times {\bS}_1)\cdot {\hat \br}( {\bv}_1\times {\bS}_2)\cdot{\hat \br}-
\frac{3}{2}({\bv}_1\times{\bS}_1)\cdot{\hat \br}({\bv}_2\times{\bS}_2)\cdot{\hat \br} - 6({\bv}_1\times{\bS}_2)\cdot{\hat \br}({\bv}_2\times{\bS}_1)\cdot{\hat \br}
\right]  \nonumber\\
&+&  
\frac{G_N^2(m_1+m_2)}{2r^4}\left(5{\bS}_1\cdot{\bS}_2-17{\bS}_1\cdot{\hat \br}{\bS}_2\cdot{\hat \br}\right) - \frac{G_N}{r^3} \left({\bS}_1\cdot {\bS}_2 - 3{\bS}_1\cdot{\hat \br}{\bS}_2\cdot{\hat \br}\right)\,. \nn
\end{eqnarray}
The equations of motion follow from a canonical procedure, e.g. \beq \dot \bS_1 = \frac{\partial V^{\rm NW}_{\rm SS}}{\partial \bS_1}\times \bS_1\,.\eeq
The precession equation is equivalent to the one from \eqref{eq:loss} and \eqref{eq:nloss}, after a spin redefinition \cite{eih,nrgrproc,comment,nrgrss}. On the other hand, using the decomposition in \eqref{decomso} and applying the Newton-Wigner condition from \eqref{eq:nwssc2}, we have
\beq
\label{nwpot}
V^{\rm NW}_{\rm SO}= \left\{ \frac{1}{1+\sqrt{1-\left(\bv^{\rm F}_1\right)^2}}
\left(\bv^{\rm F}_1 \times {\bA}^{\rm SO}_1 \right) + \bom_1^{\rm SO}\right\}\cdot {\bS}_1 + 1\leftrightarrow 2\,,
\eeq
with $ \bv^{\rm F}_1 \equiv \frac{{\tilde {\bv}}_1}{\tilde v^0_1}$. The dynamical equations follow via a canonical procedure, and are equivalent to the derivation from \eqref{eq:loso} and \eqref{eq:nloso} using \eqref{eomV}, after a change of variables \cite{nrgrso}. (The extra term in the square root, from enforcing the SSC in the locally-flat frame, turns out to be essential to match both~results.)\vskip 4pt

The previous manipulations are very suggestive. In fact, we can add a few more words regarding the choice of Newton-Wigner SSC in a locally-flat frame.
The transformation between covariant and Newton-Wigner SSCs in Minkowski space is given by \cite{regge}
\beq
\label{b13}
{\bS}_{\rm NW} = \frac{m}{H_0} {\bS}_{\rm cov} +   \frac{{\bp}\cdot {\bS}_{\rm cov}}{H_0(H_0+m)}~ {\bp}\,, \hspace{2cm} \rm{(Minkowski)}
\eeq
with $H_0= \sqrt{{\bp}^2+m^2}$.~Then, using
\beq
{\bp} = m \left(1+\frac{{\bv}^2}{2}\right) {\bv} + \ldots, ~{\rm and}~~{\bp}^2 = m^2 \left( {\bv}^2 + {\bv}^4+\cdots \right)\,, 
\eeq
we get for the map between SSCs in flat space,
 \beq
 \label{b14}
 {\bS}_{\rm NW} = \left(1 - \frac{{\bv}^2}{2}-\frac{1}{8}{\bv}^4 +\ldots \right){\bS}_{\rm cov} +\frac{1}{2} {\bv} ({\bv}\cdot {\bS}_{\rm cov})\left(1 + \frac{{\bv}^2}{4} +\ldots \right)\,.\eeq
This turns out to be precisely the same transformation we find in general relativity, provided we replace ${\bv} \to {\bv}^{\rm F} \equiv {\tilde {\bv}}/{\tilde v^0}$  \cite{nrgrso}. On behalf of the equivalence principle, it is indeed not surprising that the same redefinition as in Minkowski space still applies, once re-written in a locally-flat frame. Hence, the Newton-Wigner SSC allows us to reduce the number of degrees of freedom at the level of the action while retaining a canonical structure. This result is also consistent with the findings in \cite{buobarau} for the Hamiltonian of a spinning body in a curved background at linear order in the spin. The above manipulations are justified provided we ignore ${\cal O}(\bS_{1(2)}\bS_{1(2)})$ contributions. The latter are due to finite size effects, as well as SSC-preserving terms, which we did not include. A more careful treatment is required in order to achieve a canonical structure when these terms are added (see e.g. \cite{Vines:2016unv}). As we emphasized, we may also simply enforce the SSC through Lagrange multipliers. 

\subsection{Gravitational Wave Emission} 
\vskip 4pt
To include spin effects in the long-wavelength effective theory we proceed as follows. First of all, we must add the contribution from the total angular momentum, $L_{ab}$, to \eqref{eq:lag},
\begin{align}
\label{eq:lag2}
S^{\rm rad}_{\rm eff} =  \int dt \sqrt{\bar g_{00}} \Big[ -M(t) - \frac{1}{2} \omega_0^{ab} L_{ab}(t)
+  \sum_{\ell=2} \left( \frac{1}{\ell!} I^L(t) \nabla_{L-2} E_{i_{\ell-1}i_\ell}-  \frac{2\ell}{(2\ell+1)!}J^L(t) \nabla_{L-2} B_{i_{\ell-1}i_\ell}\right)\Big]\, ,
\end{align}
where we used $v^\mu=\delta^\mu_0$ in the coupling of \eqref{eq:Lspin}. The spin term does not radiate, but it sources the (static) Kerr metric produced by the binary system. Secondly, the multipole moments themselves incorporate spin-dependent contributions from the gravitational couplings. We review next the necessary ingredient to obtain the NLO spin contributions to the gravitational wave phase and amplitude.

\subsubsection{Spin Effects to Third Post-Newtonian Order}
\vskip 4pt

Spin-orbit effects in the energy flux at LO arise both from mass and current quadrupole radiation,~and yield 1.5PN corrections. Beyond LO, additional factor of $v^2$ may enter in different ways. We need the NLO mass and current quadrupole, both without spin and at linear order. Moreover, we need the LO~octupoles. Since the leading tail yields a 1.5PN correction to the flux, we also have to include the tail contribution to the energy flux linear in spin. Spin-spin contributions to the energy flux first appear at 2PN order, from the current quadrupole. The leading mass quadrupole (at 1.5PN) already contributes in the energy flux at 3PN in the spin-spin sector. Therefore, only the current quadrupole is required to NLO (see Table~\ref{tabl_Tscalings}).\vskip 4pt 

Besides, we also have multipole moments at ${\cal O}(\bS_a^2)$. The LO contributions arise from finite size effects in the mass quadrupole (2PN), which we need to NLO. The current quadrupole and mass octupole are needed to LO, along with the non-spinning counter-parts. 
 \begin{table}
\begin{center}
  \begin{tabular}{| r | c | c | c |}
    \hline
    {} & ${\cal O}(\not {\hspace*{-2pt}\bS})$ & ${\cal O}(\bS_a)$ & ${\cal O}(\bS_a^2)$\\ \hline
    $K^{00}_\ell$ & $m r^\ell$ & $m r^\ell v^3$ & $m r^\ell v^4$ \\ \hline
    $K^{0i}_\ell$ & $m r^\ell v$ & $m r^\ell v^2$ & $m r^\ell v^5$ \\ \hline    
    $K^{ij}_\ell$ & $m r^\ell v^2$ & $m r^\ell v^3$ & $m r^\ell v^6$ \\ \hline    
  \end{tabular}
  \caption{Scalings for $K^{\mu \nu}_\ell$, defined in \eqref{eq:KTmn}, valid for $\ell \geq 2$. See the text for a discussion of the necessary ingredients to compute the radiated power to NLO including spin effects.}
  \label{tabl_Tscalings}
\end{center}
 \end{table}
In our EFT there cannot be worldline couplings at ${\cal O}(\bS_1 \bS_2)$. Thus, the latter must involve potential modes. The leading contribution arises from the mass quadrupole moment at 3PN. There is yet another contribution at ${\cal O}(\bS_1 \bS_2)$, which appears once the $S^{i0}$ is replaced through the SSC in the leading order expression linear in spin. This is similar to what we described for the NLO conservative dynamics. Furthermore, spin-dependence in the energy flux may always enter from non-spinning multipole moments, but using the spin-dependent conservative equations~of~motion.\vskip 4pt 

We now follow the same steps as in sec.~\ref{sec:radgrav}. Namely, we compute all Feynman diagrams contributing to ${\cal T}^{\mu\nu}(t,{\bk})$ at the desired order. The topology of the diagrams is the same as in Fig.~\ref{fig:tmn}, but including now also spin insertions from the point-particle sources. A sample of such diagrams is shown in Fig.~\ref{figradnlS} and the various couplings are summarized in appendix~\ref{app:Fey}.\vskip 4pt

Some of these diagrams, e.g. Fig.~\ref{figradnlS}$(a)$, simply consist on radiation modes coupled to the spin on the worldline. For these terms there is no coupling to potential modes, therefore we use the point-particle stress-energy tensor which follows from \eqref{eq:spingr}, 
\beq
\label{eq:set}
{\cal T}_{S,\rm pp}^{\mu\nu} (t,\bx) \to T_{S,\rm pp}^{\mu\nu} (t,\bx) = \frac{1}{2} \sum_{a=1,2} \int dt' ~\partial_\alpha \delta^4(x -x_a(t')) \Big(S_a^{\nu\alpha}(t') v_a^\mu(t')+ S_a^{\mu\alpha}(t') v_a^\nu(t')\Big)\,,
\eeq 
that is the linearized version of \eqref{eq:TDix}. In mixed Fourier space
\begin{align}
 {\cal T}_{S,\rm pp}^{00}(t, \bk) & = \sum_{a=1,2} S_a^{0i} \, i \bk^i e^{-i {\bk} \cdot  \bx_a} \label{eq:WLrad00SA}\\
 {\cal T}_{S,\rm pp}^{0i}(t, \bk) & = \frac{1}{2} \sum_{a=1,2} \left(S_a^{ij} \, i \bk^j + S_a^{0j} \bv_a^i \, i \bk^j + S_a^{0i} \, i \bk \cdot \bv_a - \dot S^{0i} \right) e^{-i {\bk} \cdot  \bx_a} \label{eq:WLrad0iSA}\\
 {\cal T}_{S,\rm pp}^{ij}(t, \bk) & = \frac{1}{2}\sum_{a=1,2} \Big\{\left(S_a^{il} \bv_a^j + S_a^{jl} \bv_a^i \right) i \bk^l + i \bk \cdot \bv_a \left(S_a^{0i} \bv_a^j + S_a^{0j} \bv_a^i \right)  \\
&  {} \hspace*{39pt} - \dot S_a^{0i} \bv_a^j - \dot S_a^{0j} \bv_a^i - S_a^{0i}  \ba_a^j - S_a^{0j} \ba_a^i \Big\} e^{-i {\bk} \cdot \bx_a} \label{eq:WLradijSA}\,.\nn
\end{align}

The remaining contributions represent non-linear gravitational effects which arise from two distinct sources. Firstly, we have radiation stemming off the explicit non-linear terms from the spin couplings (schematically) $SH\bar h$ and $S^2 H\bar h$\,. Secondly, we have the three-graviton vertex, $[HH\bar h]$, which follows from the Einstein-Hilbert action plus a (background gauge) fixing term. The latter contributes in the spin sector by coupling the potential modes to the spin tensors at the worldlines. In this case the radiation is emitted off the binding mass/energy of the system.\vskip 4pt

To illustrate the procedure, let us calculate in somewhat more detail the contribution to the quadrupole moment from the Feynman diagrams in Fig.~\ref{figradnlS}$(b$-$c)$.
\begin{figure}[t!]
\centering
\includegraphics[width=0.70\textwidth]{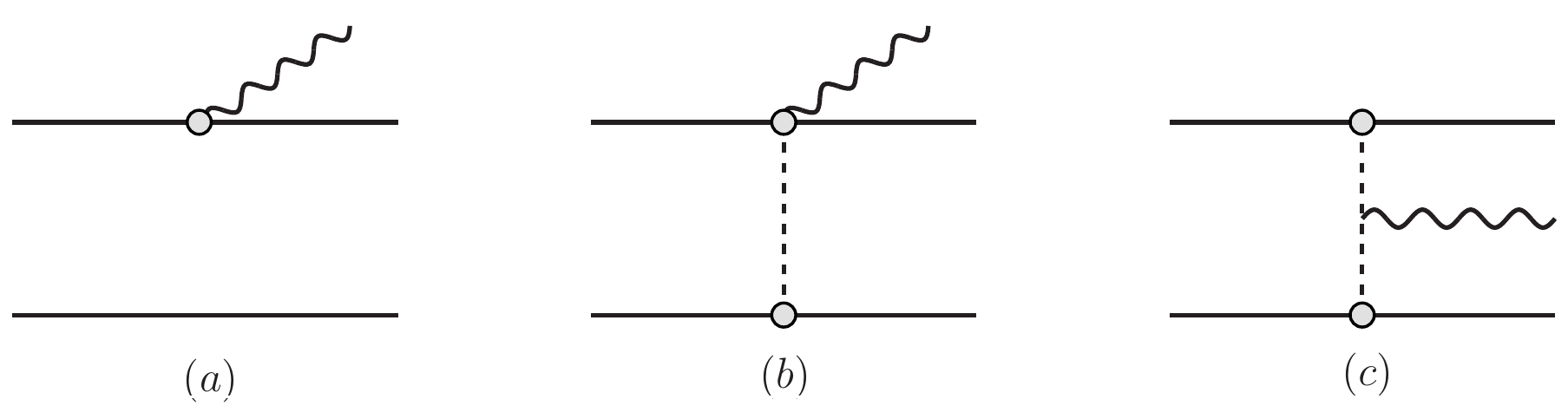}
\caption{Sample of Feynman diagrams which contribute to the NLO multipole moments. In total we need diagrams with spin insertions as well as finite size effects. See~\cite{srad} for more details.}
\label{figradnlS}
\end{figure}
For the diagram in Fig.~\ref{figradnlS}$(b)$ we need the $S H \bar h$ coupling, which requires the two-point function
\beq
\big\langle T\big\{H_{0i}({\bx}_1,t_1)H_{0j,l}({\bx}_2,t_2)\big\}\big\rangle = \frac{1}{2} \delta_{ij} \delta(t_1-t_2)\int_\bq \frac{-i}{ {\bq}^2 }~(i{\bq}^l)~
e^{i {\bq}\cdot ({\bx}_1(t_1)-{\bx}_2(t_2))} = i\delta^{ij}\frac{1}{8\pi} \frac{{\hat\br}^l}{r^2}\,.
\eeq
After summing over mirror images, we obtain
\bea
{\cal T}^{00}_{\ref{figradnlS}(b)}({t,\bk})&=&  \frac{G_N}{r^3}\sum_{a\neq b} \left(({\hat \br}\times {\bS}_b)\times {\bS}_a\right)^l (-i{\bk}^l) e^{-i{\bk}\cdot {\bx}_a}\,,\\
{\cal T}^{ii}_{\ref{figradnlS}(b)}(t,{\bk})&=&-\frac{2G_N}{r^3} \sum_{a\neq b} \left( {\bS}_a\cdot {\bS}_b - 3 {\bS}_a\cdot {\hat\br} {\bS}_b \cdot {\hat \br}\right)  e^{-i{\bk}\cdot {\bx}_a}\,.
\eea
Hence, using \eqref{mixedTk} and \eqref{eq:lolij2}, we find
\beq
 I^{ij}_{\ref{figradnlS}(b)}=  \sum_{a\neq b} \frac{2G_N}{r^3}\left[{\bS}_a\cdot {\br} {\bS}_b^i {\bx}_a^j - {\bS}_a\cdot {\bS}_a {\br}^i{\bx}_a^j -\left( {\bS}_a\cdot {\bS}_b - 3 {\bS}_\cdot {\hat \br} {\bS}_b \cdot {\hat \br}\right) {\bx}_a^i{\bx}_a^j \right]_{\rm STF}.
\eeq 
Let us move now to the diagram in Fig.~\ref{figradnlS}$(c)$. To compute it we need the three-graviton coupling, $[HH\bar h({\bk})]$, combined for the temporal component and spatial trace. Then, after multiplying by the two potential propagators, it reads
\beq
\big\langle T\big\{H_{0i}({\bf p+k})H_{0j}({\bf p})\big[HH\bar h({\bk})\big]\big\} \big\rangle = \frac{i}{4\Mp}[2{\bk}^i{\bk}^j+2({\bf p}^i{\bk}^j-{\bk}^i{\bf p}^j)]\frac{-i}{{\bf p}^2}\frac{-i}{({\bk+p})^2}\,.
\eeq
The open indices must be contracted with the spin couplings on the worldlines, yielding  
\beq
({\cal T}^{00}+{\cal T}^{ii})_{\ref{figradnlS}(c)}(t,{\bk})=\frac{1}{4\Mp^2} \int_\bp  \big[{\bk}^i{\bk}^j+({\bf p}^i{\bk}^j-{\bk}^i{\bf p}^j)\big] \frac{1}{{\bf p}^2}\frac{1}{({\bk+p})^2} S_1^{ik} {\bf p}^k S_2^{jl}(-{\bf p}^l-{\bk}^l) e^{i{\bf p}\cdot{\bx}_1} e^{-i({\bf p+k})\cdot {\bx}_2}.
\eeq
Expanding to to second order in $\bk$, and performing the Fourier integration, we obtain
\beq
({\cal T}^{00}+{\cal T}^{kk})_{\ref{figradnlS}(c)}(t,{\bk})=\frac{4G_N}{r^3} S_1^{im}S_2^{jl}\left(r^2\delta^{ml}-{\br}^m{\br}^l \right) \left(-\frac{{\bk}^i{\bk}^j}{2}\right) + \cdots ,
\eeq
and subsequently
\beq
 I^{ij}_{\ref{figradnlS}(c)}=\sum_{a\neq b} \frac{2G}{r^3} \left[ {\br}^i{\br}^j {\bS}_a\cdot {\bS}_b -2{\bS}_a\cdot {\br} {\bS}_b^i{\br}^j\right]_{\rm STF}.
\eeq

\subsubsection{Multipole Moments}
\vskip 4pt
Garnering all the ingredients we finally arrive at (in the covariant SSC)~\cite{srad}
\bea
\label{Qijtot}
 I^{ij}_{{\bS}_a,{\bS}_a^2,{\bS}_a{\bS}_b}&=& \sum_a \left[\frac{8}{3} ({\bv}_a\times {\bS}_a)^i {\bx}_a^j  -\frac{4}{3} ({\bx}_a\times {\bS}_a)^i{\bv}_a^j-\frac{4}{3}({\bx}_a\times \dot {\bS}_a)^i{\bx}_a^j\right.  \\ 
& -& \left.  \frac{4}{3}\frac{d}{dt} \left\{{\bv}_a \cdot {\bx}_a({\bv}_a\times{\bS}_a)^i{\bx}_a^j\right\}+\frac{1}{7} \frac{d^2}{dt^2}\left\{\frac{1}{3} {\bx}_a\cdot{\bv}_a  ({\bx}_a\times{\bS}_a)^i{\bx}_a^j\right. \right. \nn \\ & +& 
\left. \left. 4 {\bx}_a^2 ({\bv}_a\times{\bS}_a)^i{\bx}_a^j+{\bx}_a^2 ({\bS}_a\times{\bx}_a)^i{\bv}_a^j-\frac{5}{6}({\bv}_a\times{\bS}_a)\cdot {\bx}_a~{\bx}_a^i{\bx}_a^j \right\} \right]_{\rm STF} \nn \\
&+ &  \sum_{a\neq b} \frac{2G_Nm_b}{r^3} \left[ ({\bv}_b\times {\bS}_a)\cdot {\br}({\bx}^i_b{\bx}^j_b-2{\bx}_a^i{\bx}_a^j)+({\bv}_a\times {\bS}_a)\cdot {\br}({\bx}^i_a{\bx}^j_a+{\bx}_b^i{\bx}_b^j)\right. \nn  \\ &+ & \left.  2r^2\left\{ ({\bv}_b\times {\bS}_a)^i ({\bx}_b^j -{\bx}_a^j) +({\br}\times{\bS}_a)^i\left({\bv}_b^j-{\bv}_a^j-\frac{{\bv}_b\cdot {\br}}{r^2}({\bx}_a^j+{\bx}_b^j)\right)\right\} \right]_{\rm STF} \nn \\ &- &   \frac{2}{3} \sum_{a\neq b} \frac{d}{dt} \left[ \frac{G_Nm_b}{r^3} \left\{r^2\left( ({\bx}_b\times {\bS}_a)^i{\bx}_a^j-3({\bx}_a\times {\bS}_a)^i{\bx}_a^j +3 ({\bx}_b\times {\bS}_a)^i{\bx}_b^j-({\bx}_a\times {\bS}_a)^i{\bx}_b^j \right) \right.\right. \nn \\ &- &\left. \left. 2{\br}\cdot{\bx}_b~({\br}\times{\bS}_a)^i({\bx}^j_a+{\bx}_b^j)+ ({\bx}_a\times{\bS}_a)\cdot {\bx}_b ({\bx}_a^i{\bx}_a^j-2{\bx}_b^i{\bx}_b^j)\right\}\right]_{\rm STF} \nn \\
&+ &  \sum_a \frac{C_{ES^2}^{(a)}}{m_a}\left[{\bS}_a^i{\bS}_a^j\left(-1+\frac{13}{42} {\bv}_a^2+\frac{17}{21} {\ba}_a\cdot {\bx}_a\right)+{\bS}_a^2\left(-\frac{11}{21}{\bv}^i_a{\bv}^j_a +\frac{10}{21}  {\ba}_a^i{\bx}_a^j\right)\right. \nn \\ &- & \left.  \frac{8}{21}{\bx}^i_a{\bS}_a^j {\ba}_a\cdot {\bS}_a+\frac{4}{7}{\bv}^i_a{\bS}_a^j{\bS}_a\cdot {\bv}_a-\frac{22}{21}{\ba}_a^i{\bS}_a^j {\bS}_a\cdot {\bx}_a\right]_{\rm STF} \nn \\ & +&  \sum_{a\neq b} \frac{G_N}{2r^3}\left[\frac{C_{ES^2}^{(b)}m_a}{m_b}\left({\bS}_b^2+9({\bS}_b\cdot {\hat \br})^2\right){\bx}_b^i{\bx}_b^j+6\frac{C_{ES^2}^{(b)}m_a}{m_b} r^2 {\bS}_b^i{\bS}_b^j \nonumber \right. \\ & +&  \left.\left(\frac{C_{ES^2}^{(b)}m_a}{m_b}\left(3({\bS}_b\cdot {\hat \br})^2-{\bS}_b^2\right)+12{\bS}_a\cdot {\hat \br} {\bS}_b\cdot {\hat \br}-4{\bS}_a\cdot {\bS}_b \right){\bx}_a^i{\bx}_a^j\right. \nonumber \\ &- & \left.  4 \frac{C_{ES^2}^{(b)}m_a}{m_b} {\bS}_b^2{\bx}_a^i{\bx}_b^j+4\left( 3\frac{C_{ES^2}^{(b)}m_a}{m_b}{\bS}_b\cdot {\br}+2 {\bS}_a\cdot {\br}\right){\bS}_b^i{\bx}_b^j\right]_{\rm STF}\,,\nn \\
\label{Jijtot}
J^{ij}_{{\bS}_a,{\bS}_a^2}&=&  \sum_a \left[ \frac{3}{2}{\bS}_a^i {\bx}_a^j + \frac{C_{ES^2}^{(a)}}{m_a}  ({\bv}_a\times {\bS}_a)^i {\bS}_a^j+
{\bS}_a^i {\bx}_a^j\left(\frac{2}{7} {\bv}_a^2-\frac{5}{7} {\ba}_a\cdot {\bx}_a\right)-
\frac{3}{7} {\bv}^i_a{\bS}_a^j {\bv}_a\cdot {\bx}_a \right.  \\ &+ & \left. 
\frac{11}{28} {\bS}_a^i{\ba}_a^j {\bx}_a^2+\frac{2}{7} {\bS}_a\cdot {\bx}_a {\ba}_a^i{\bx}_a^j+\frac{1}{7} {\bx}_a^i{\bx}_a^j {\ba}_a\cdot {\bS}_a-\frac{3}{7}{\bS}_a\cdot {\bv}_a {\bv}^i_a{\bx}_a^j+\frac{11}{14} {\bS}_a\cdot {\bx}_a {\bv}_a^i{\bv}_a^j\right]_{\rm STF}\,, \nonumber \\
 &+&\sum_{a\neq b}\frac{G_Nm_b}{2r^3}\left[3 {\bS}_a\cdot {\bx}_b ({\bx}_b^i{\bx}_b^j-{\bx}_a^i{\bx}_a^j)+ {\bS}_a\cdot {\bx}_a (2{\bx}_a^i{\bx}_a^j+{\bx}_a^i{\bx}_b^j-3{\bx}_b^i{\bx}_b^j) +{\bS}_a^i{\bx}_a^j ({\bx}_a\cdot {\br}-6r^2) \right]_{\rm STF}\nn \,, \\
  \label{Qoctup}
I^{ijk}_{{\bS}_a, {\bS}_a^2} &=& \sum_a \left[ \frac{9}{2}({\bv}_a\times {\bS}_a)^i{\bx}_a^j {\bx}_a^k -3({\bx}_a\times {\bS}_a)^i{\bv}_a^j {\bx}_a^k -3\frac{C_{ES^2}^{(a)}}{m_a} {\bS}^i_a{\bS}_a^j {\bx}_a^k \right]_{\rm STF}\,,\\
 \label{Joctup} J^{ijk}_{{\bS}_a} &=& 2 \sum_a \left[ {\bx}_a^i {\bx}_a^j {\bS}_a^k\right]_{\rm STF}\,.
 \eea
 Together with the non-spinning counter-parts \cite{andirad,srad}
 \bea
\label{Qijsless}
I^{ij}_{\not {\bS}} &=& \sum_a  m_a \left[ \left( 1+ \frac{3}{2} {\bv}_a^2- \sum_{B} \frac{G_N m_b}{r}\right){\bx}_a^i{\bx}_a^j + \frac{11}{42} \frac{d^2}{dt^2}\left\{{\bx}_a^2 {\bx}_a^i{\bx}_a^j \right\}  
-\frac{4}{3} \frac{d}{dt}\left\{ {\bx}_a\cdot{\bv}_a {\bx}_a^i{\bx}_a^j\right\}\right]_{\rm TF}  \\
J_{\not{\bS}}^{ij} &=&  \sum_a m_a \left(1+\frac{{\bv}_a^2}{2}\right) \left[({\bx}_a\times{\bv}_a)^i{\bx}_a^j\right]_{\rm STF} + \sum_{a\neq b} \frac{G_Nm_am_b}{r} \left[ 2({\bx}_a\times{\bv}_a)^i{\bx}_a^j  \right.  \\ & &\left.  -\frac{11}{4}({\bx}_b\times{\bv}_a)^i{\bx}_b^j-\frac{3}{4}({\bx}_b\times{\bv}_a)^i{\bx}_a^j +({\bx}_a\times{\bv}_a)^i{\bx}_b^j+\frac{7}{4}({\bx}_a\times{\bx}_b)^i{\bv}_a^j\right.  \nn \\ & & \left.  +
\frac{{\bv}_a\cdot{\br}}{4r^2} ({\bx}_a\times{\bx}_b)^i({\bx}_a^j+{\bx}_b^j)\right]_{\rm STF} + \frac{1}{28}\frac{d}{dt}\left[ \sum_a  m_a({\bx}_a \times {\bv}_a)^i (3 {\bx}_a^2 {\bv}_a^j-{\bx}_a\cdot {\bv}_a {\bx}_a^j) \right. \nn \\ & & + \left.  \sum_{a\neq b} \frac{G_Nm_am_b}{2r^3}  {\bx}_a^i ({\bx}_a\times{\bx}_b)^j(6{\bx}_a^2 -7 {\bx}_a \cdot {\bx}_b +7{\bx}_b^2 )\right]_{\rm STF}\,, \nn \\
I^{ijk}_{\not{\bS}} &=& \sum_a m_a \left[ {\bx}_a^i{\bx}_a^j{\bx}_a^k\right]_{\rm TF}\,,\\
J^{ijk}_{\not{\bS}} &=&  \sum_a m_a \left[({\bx}_a\times{\bv}_a)^i{\bx}_a^j{\bx}^k_a\right]_{\rm STF}\,,
\label{octsless}
\eea
these are the source multipoles necessary to include spin effects in the gravitational wave phase to 3PN order. In principle we must also include the contributions from tail effects to the radiative multipoles. This is required only to lowest orders, i.e. for the current quadrupole. The computation is straightforward using the results discussed in sec.~\ref{sec:amplitude} \cite{srad}, see also \cite{tail5}.\vskip 4pt
For the waveform the previous results are not sufficient. That is the case because the amplitude is like taking the `square-root' of the power. To compute it to NLO (2.5PN) we require a few more multipole multipoles, most notably the NLO current octupole (also in the covariant SSC) \cite{amps}
\bea
\label{J3f}
J^{ijk}&=& \sum_a 2 \left[{\bS}_a^i {\bx}_a^j {\bx}_a^k\right]_{\rm STF}+ \sum_a \left[ -\frac{2}{3}\left({\bv}_a^i {\bx}_a^j{\bx}_a^k({\bS}_a\cdot {\bv}_a)  - {\bv}_a^2 {\bS}_a^i {\bx}_a^j {\bx}_a^k-2{\bv}_a^i {\bv}_a^j{\bx}_a^k ({\bS}_a\cdot {\bx}_a)\right.\right. \\ &+& \left. 2({\bx}_a\cdot {\bv}_a){\bS}_a^i {\bv}_a^j{\bx}_a^k  \right) + \frac{1}{6}{\ba}_a^i {\bx}_a^j{\bx}_a^k ({\bS}_a\cdot {\bx}_a)- \frac{5}{6}{\bS}_a^i {\bx}_a^j{\bx}_a^k ({\ba}_a\cdot {\bx}_a) \nn \\ &+& \left.
\frac{2}{9} ({\ba}_a\cdot{\bS}_a){\bx}_a^i{\bx}_a^j{\bx}_a^k + \frac{2}{3} {\bx}_a^2 \left( {\bv}_a^i{\bv}_a^j{\bS}_a^k + {\bx}_a^i{\ba}_a^j{\bS}_a^k\right)\right]_{\rm STF} \nn \\
&+& \sum_{a\neq b} \frac{G_Nm_b}{r^3} \left[\frac{1}{2}({\br}\cdot {\bx}_a-8r^2) {\bS}_a^i {\bx}_a^j {\bx}_a^k + \frac{4}{3} \left({\bx}_a^i {\bx}_a^j {\bx}_a^k-{\bx}_b^i {\bx}_b^j {\bx}_b^k\right){\bS}_a\cdot {\br} - \frac{1}{2} {\bx}_a^i {\bx}_a^j {\br}^k {\bS}_a\cdot {\bx}_a\right]_{\rm STF}\nn,\\
I^{ijkl}&=&\frac{8}{5} \sum_a \left[ 4({\bv}_a\times {\bS}_a)^i {\bx}_a^j {\bx}_a^k {\bx}_a^l - 3({\bx}_a\times {\bS}_a)^i {\bv}_a^j {\bx}_a^k {\bx}_a^l\right]_{\rm STF}\,,\\
J^{ijkl}&=&  \frac{5}{2}\sum_a \left[{\bS}_a^i {\bx}_a^j {\bx}_a^k{\bx}_a^l\right]_{\rm STF}\,,\\
J^{ijklm}&=& 3 \sum_a \left[{\bS}_a^i {\bx}_a^j {\bx}_a^k{\bx}_a^l{\bx}_a^m\right]_{\rm STF}\label{J5f}\,.
\eea

\subsection{Absorption \& Superradiance}\label{sec:abspin}
\vskip 4pt

One of the most amusing new aspects which spin brings to the table is superradiance \cite{zeld,Misner,starobinski,unruh,beke,super}. For the case of gravity,\,\footnote{\,Superradiance was first discovered in the realm of electromagnetism \cite{zeld}, and shows up in many examples \cite{beke} where gravity does not play a role, although rotation (acceleration) is often present.} the mass of a rotating black hole may decrease as a result of gravitational wave scattering, when the following condition is met:
\begin{equation}
\omega-m_\ell\Omega < 0\,.\label{eq:super}
\end{equation}
Here $\omega$ is the frequency of the incoming wave, $m_\ell$ the azimuthal angular momentum with respect to the axis of rotation, and $\Omega$ the  angular velocity. The wave is thus amplified at the cost of the black hole's rotational energy. Our task now is to incorporate this effect into the analysis of sec~\ref{sec:abs}.\vskip 4pt 
 
In the case of non-spinning black holes matching is attained by comparison with the total absorption cross section. However, for the rotating case, superradiance poses a subtlety to the application of the optical theorem. The reason is that what is gained through absorption may be less than what is lost into amplifying the wave. This leads to a negative value for the low-frequency amplitude ($G_N=1$) \cite{page}
\begin{equation}
\Gamma_{s=2,\omega,\ell=s,m_\ell,h}=\frac{16}{225}\frac{A}{\pi} m^4[1+(m_\ell^2-1)a^2_{*}][1+(\frac{m_\ell^2}{4}-1)a^2_{*}]\omega^5(\omega-m_\ell\Omega)\,,\label{astar}
\end{equation}
when \eqref{eq:super} is satisfied. Here $A$ is the area, and $\Omega \simeq {a_{\star} \over 4 m}$ is the rotational angular velocity, with $a_{\star} \equiv \frac{S}{m^2}$. (Also, $h=\pm 2$ is the polarization of the wave, and we display only the $l=s$ dominant mode \cite{page}.)\vskip 4pt 

In~order to avoid dealing with negative `probabilities', we use a basis of polarized spherical waves~\cite{ryan,Handler} 
\begin{equation}
\psi^s_{\ell m_\ell} = Y^s_{\ell m_\ell}(\theta,\phi) \frac{1}{\sqrt{2\pi\omega}}\frac{e^{i(kr-\omega t)}}{r},
\end{equation}
with $Y^s_{\ell m_\ell}(\theta,\phi)$ the spheroidal harmonics \cite{thorne}. Then, we decompose a $h=-2$ plane wave moving in the $\hat {\bz}$ direction in terms of spherical modes. The coefficients are given by  $c^{s=2}_{\ell} = i^\ell \sqrt{2\ell+1}$~\cite{ryan,thorne}. Keeping only the dominant $\ell=2$ mode, the difference between plane and spherical waves becomes a factor of $\sqrt{5}$ for each external state, altogether an overall factor of $5$~\cite{dis2}.\vskip 4pt

The steps are now similar as before. We start by parameterizing the two-point function as\,\footnote{~There is an important difference here with respect to the non-spinning case. For rotating bodies we found a non-zero expectation value $\langle Q_E^{ij}\rangle_{\rm spin} \propto S_{ik}{S^k}_j$,  see \eqref{vevQES}. Therefore, fluctuations must be defined with respect to the background, i.e. $ \delta Q_{ij}^E \equiv Q_E^{ij}-\langle Q_E^{ij}\rangle_{\rm spin}$. For simplicity, in what follows we will continue using the same notation as in sec.~\ref{sec:abs}, but the reader should keep in mind that the correlation functions are computed in terms of $\delta Q_E^{ij}$ \cite{dis2}.}
\cite{dis2}
\begin{equation}
\label{absp2} i \,\int dt e^{-i\omega t} \left\langle T \Big(Q_{E(B)}^{ij}(0)
Q_{E(B)}^{kl}(t)\Big)\right\rangle_{\rm spin} =  \frac{1}{2} S_{ijkl}(\omega) f_s(\omega)\,, 
\end{equation}
where $f_s(\omega)$ is a function of the frequency. The tensor structure is slightly more intricate than before, 
\begin{equation}
S_{ijkl} = \left[\delta_{ik}S_{jl} +\delta_{il}S_{jk}+\delta_{jl}S_{ik}+\delta_{jk}S_{il} \right] (1+\alpha_s {\bS}^2+\cdots ) + \cdots \,. 
\end{equation} 
In this expression $\alpha_s$ is a matching coefficient and the ellipsis account for higher non-linear terms in the spin tensor. These~can be shown to be suppressed at low frequencies \cite{dis2}.\vskip 4pt We now match for the polarized absorption cross section. In the full theory (general relativity) side we obtain
\begin{equation}
\sigma^{\rm gr}_{\rm abs}(\omega) = \frac{4\pi}{45}r_s^5\omega^3(1+3a^2_{\star})a_\star+\cdots \label{astarcr}\,,
\end{equation}
whereas in the EFT (using $\bk = \omega \hat {\bz}$, ${\bS} \cdot \hat {\bz} = a_{\star }G_Nm^2$)
\begin{equation}
\sigma^{\rm eft}_{\rm abs}(\omega) = \frac{\omega^3}{\Mp^2} f_s(\omega) (1+\alpha_s G_N^2m^4a_{\star}^2) G_N m^2 a_{\star} + \cdots\,,
\label{sigeft}
\end{equation}
after using the electric-magnetic duality, as in sec.~\ref{sec:abs}.\,\footnote{~Even though the rotating background explicitly breaks the symmetry, since $\langle Q_{ij}^B\rangle$ vanishes for a Kerr black hole, it is still possible to show that the linearized equations for the perturbations are invariant under the duality transformations~\cite{dis2}.} Inspecting both results we find 
\bea
f_s(\omega) =  \frac{4}{45}G_N^3 m^3\,,~~\alpha_s = \frac{3}{m^4G_N^2}\,. 
\eea
Notice that at low frequencies $f_s$ is independent of $\omega$. Moreover, the imaginary part comes from the contraction between polarization and spin tensors. This means the time dependence is entirely in the spin, as we would expect. That is because it is the rotational energy which is  extracted to amplify the gravitational waves.\vskip 4pt

\begin{figure}[t!]
    \centering
    \includegraphics[width=0.21\textwidth]{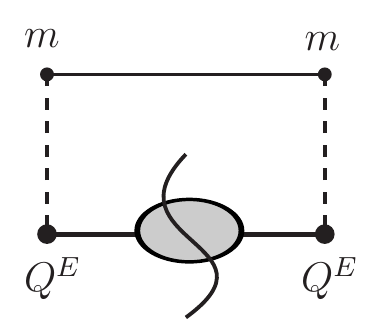}
\caption{Leading order contribution to the absorptive effects due to spin in the binary system. The correlator $\langle T(Q^E Q^E)\rangle$ now includes spin-dependent terms.}
\label{abspot2} 
\end{figure}

Let us power count the contributions from dissipative effects to the binary dynamics, this time including spin effects. From the above expressions we have, in the non-relativistic limit, \cite{dis2}
\begin{equation}
\int dt \left(Q_{ab}^E\right)_{\rm spin} E^{ab}[H] \sim \sqrt{a_{\star}\left(1+3a_{\star}^2\right)} v^5\,.
\end{equation}
This enters at order $L v^{10}$ through the box diagram in Fig.~\ref{abspot}, displayed again in Fig.~\ref{abspot2}, which represents an enhancement of three powers of the velocity parameter with respect to the non-rotating case \cite{tagoshi,poissonab2}. This can be directly associated to the presence of the superradiance effect, allowing us to avoid an extra factor of $\omega$ in the matching,~see~\eqref{astar}.\vskip 4pt Computing the diagram in Fig.~\ref{abspot2} we obtain for the power loss \cite{dis2}\begin{equation}
P^{\rm spin}_{\rm abs}  = -{8\over 5}G_N^6 m_1^2m_2^2\left\langle {\frac{{\bl}\cdot{\boldsymbol{\xi}}}{r^8} }\right\rangle\,, \label{pspinabs}
\end{equation}
where ${\bl} = {\br}\times{\bv}$, ${\bv} =\dot{\br}$,  and \beq \boldsymbol{\xi} \equiv \sum_b m_b^3{\hat \bS}_b \left(a^{(b)}_{\star}+3 (a^{(b)}_{\star})^3\right).\eeq 
 For instance, we find
\begin{equation}
P^{\rm spin}_{\rm abs} = -\epsilon \frac{8}{5} \frac{G_N^6m^5M^2 \omega}{r^6} \left(a_{*}+3 a_{*}^3\right)\,,\label{psab}
\end{equation}
for a rotating test particle ($m,a_\star$) moving on a circular orbit in the equatorial plane of a black hole of mass  $M \gg m$.
Here $\omega$ is the orbital frequency and $\epsilon = \hat \bl \cdot {\hat \bS}$. This is in agreement with the results in~\cite{tagoshi,poissonab2}.\vskip 4pt
As we did for the non-rotating case, we may also extrapolate our results to spinning neutron stars. There are, nonetheless, a few caveats. First of all neutron stars are not maximally rotating, which was assumed in the scaling law for the spin couplings. Therefore, there is likely an extra suppression factor depending on the spin of the neutron star, e.g. \cite{AnderssonR,pulsar}. (We do expect, however, gravitational wave emission from binary pulsars with sizable spins \cite{pulsar2}.) Furthermore, it is unclear whether superradiance is even present for pulsars, and if so, whether it is as prominent as it is for the case of black holes. Therefore, in practice, this is likely to impose an extra suppression factor. It is then reasonable to assume that absorption effects for spinning neutron stars might not be as enhanced as for rotating black~holes.\vskip 4pt

Another simple application of the formalism consists on considering slowly-varying time-dependent backgrounds, described by time-dependent  electric and magnetic components of the Weyl tensor. Following similar steps as before we find~\cite{dis1}
\beq
W[\bx_a]= \frac{i}{2} \int dt_1 dt_2\left\langle T \Big(Q_E^{ij}(t_1) Q_E^{kl}(t_2)\Big)\right\rangle\Big[E^{ij}(t_1) E^{kl}(t_2)+B^{ij}(t_1) B^{kl}(t_2)\Big]+\cdots\,,
\eeq
from which we obtain, for a test particle with parameters $(m,a^\star)$, \cite{dis2}
\begin{equation}
P_{\rm abs} = {8\over 45}a_*(1+3a_*^2) G^4_N m^5\left \langle \left({\dot E}_{ij}\dot E_{il}+  {\dot B}_{ij} \dot B_{jl}\right)s_{jl}\right \rangle \,,
\label{newabs}
\end{equation}
with $s_{ij} = \epsilon_{ijk} \hat\bS_k$. The time derivatives are defined as ${\dot E}^{ij} \equiv e^i_\mu e^j_\nu \,v^\alpha \nabla_\alpha E^{\mu\nu}$, in a locally-flat frame.

\newpage
\section{Summary of Part~\ref{sec:part2}} \label{sec:sum2}
\subsection{A Mamushka of EFTs}
\vskip 2pt
The dynamics of compact objects in a binary system and total emitted power follow from the effective action,
\beq
\nn
W[\bx_a]\equiv -i \log Z [\bx_a]\,  = \underbrace{{\rm Re}\,W[{\bx}_a] }_{\rm binding}\,+\, i\, \underbrace{{\rm Im}\,W[{\bx}_a] }_{\rm radiation}\,.
\eeq 
By extremizing the real part,
\beq
\nn
{\delta \over \delta {\bx}_a(t)}~{\rm Re}\,W[{\bx}_a]  = 0\, ,
\eeq
we obtain the equations of motion for the binary constituents, whereas via the optical theorem,
\beq
\nn
\frac{1}{T}\, {\rm Im}\,W[\bx_a] =  \frac{1}{2}  \int \frac{d^2\Gamma}{dE d\Omega}  dE d\Omega\, ,
\eeq
we get the radiated power, with $dP = E d\Gamma$. The computation of $W[\bx_a]$ proceeds in stages by integrating out --one scale at a time-- the short(er) distance degrees of freedom at the various relevant scales in the problem: $r_s^{-1} \gg r^{-1} \gg \lambda_{\rm rad}^{-1}$, within the saddle-point approximation. Schematically,
\beq
\nn
e^{i W} = \int  D\big[\lambda^{-1}_{\rm rad}\big] D\big[r^{-1}\big] D\big[r^{-1}_s\big]  e^{i S_{\rm full}}\,.
\eeq
At each step we construct a point-like effective action where only modes with $k < \mu$ are kept,
\beq
\nn
\int D\big[\mu \big]\,e^{iS} \to e^{i S_{\rm eff}}\,.
\eeq
The latter is constrained solely by the symmetries of the long(er) distance physics. For the binary problem, 
 \beq
 \nn
 S_{\rm eff} =\int d\tau \left[ -M(\tau) - \frac{1}{2} \omega_\mu^{ab} S_{ab}(\tau) u^\mu(\tau) +   \sum_{\ell=2} \left( \frac{1}{\ell!} I^L(\tau) \nabla_{L-2} E_{i_{\ell-1}i_\ell}- \frac{2\ell}{(2\ell+1)!}J^L(\tau) \nabla_{L-2} B_{i_{\ell-1}i_\ell}\right)\right]\,,
\eeq
which describes a point-like object interacting with a long-wavelength gravitational field. The STF multipole moments are decomposed as 
\beq
\nn
I^L (t) = \underbrace{\left\langle I^L(t)\right\rangle_S}_{\rm background} + \underbrace{I^L_{\cal R}(t)}_{\rm response}\,,
\eeq
and similarly for $J^L(t)$. The response is further separated into (in frequency space)
\beq
\nn
I^L_{\cal R}(\omega) =  \underbrace{{\rm Re}\,I^L_{\cal R}(\omega)}_{\rm tidal\hbox{-}effects}\, +\, i \, \underbrace{ {\rm Im} \, I^L_{\cal R}(\omega)}_{\rm absorption}\,.
\eeq
The point-like limit introduces UV divergences which are regularized in dim. reg.  Power-law divergences are discarded.  After renormalization through counter-terms the functions above depend on scale, $\mu$, and obey renormalization group equations whose boundary conditions are obtained via a matching procedure. 
\begin{figure}[t!]
\centering
\includegraphics[width=0.65\textwidth]{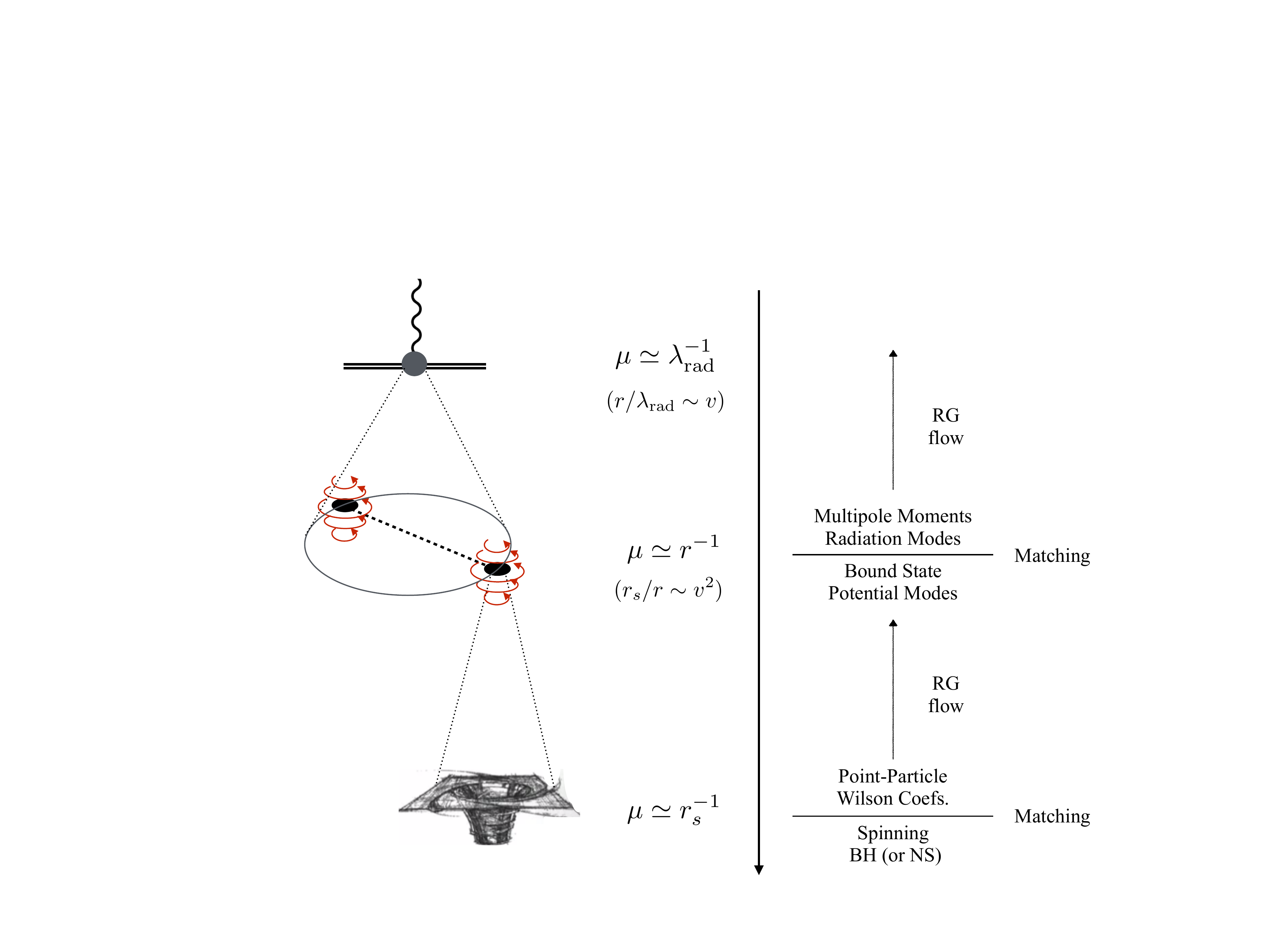}
\caption*{The EFT approach to the binary inspiral problem: One scale at a time.}
\end{figure}
\vspace{-0.7cm}
\bit
\im $\mu \simeq r_s^{-1}$: The $(M,S_{ab})$ are the mass and spin of a black hole or neutron star. The $\left\langle I^L\right\rangle_S$ are proportional to permanent moments, e.g. $\propto  S^2$. For the real part of the response we have (similarly for the magnetic-type) \vspace{-0.3cm}
\beq \nn 
{\rm Re}\,I^L_{\cal R}(\omega) = C_E^{(\ell-2)} \nabla_{L-2} E_{i_{\ell-1}i_\ell}(\omega) + \cdots \,,  \vspace{-0.1cm}
\eeq 
where $C^{(\ell-2)}_{E(B)}$ are a series of Wilson coefficients (the Love numbers). The ellipsis include~higher powers of $\omega$. All of the $C_{E(B)}^{(\ell-2)}$ vanish for black holes in $d=4$. The imaginary part of~the~response incorporates absorption and may be matched via low-frequency gravitational wave scattering.

\im $\mu \simeq r^{-1}$: Matching for the binding mass/energy, total angular momentum, and source multipole moments,  $\left(M, S_{ab},\left\langle I^L\right\rangle_S,\left\langle J^L\right\rangle_S\right)$, is performed in terms of moments of the (pseudo) stress-energy tensor, ${\cal T}^{\mu\nu}(x)$. The latter is obtained after integrating out the potential modes. The background field method ensures~$\partial_\mu{\cal T}^{\mu\nu}(x)=~0$. The~response parts would be relevant in an external~field. 

\im $\lambda_{\rm rad}^{-1} < \mu < r^{-1}$:  We encounter non-linearities induced by gravitational tails which contribute to the one-point function. These are responsible for the {\it radiative} multipole moments, and the --universal-- renormalization group flow in the effective theory, i.e. $M(\mu)$, $I^L(\mu)$. We find logarithmic corrections to the binding energy and gravitational wave amplitude, as well as an IR divergent phase for the latter. The IR pole is absorbed into a time redefinition which does not affect the physical waveforms. 
\eit
These steps can be generalized, using the `in-in' formalism, to include radiation-reaction effects. The~tail contribution entails a subtle interplay between potential and radiation regions, featuring time non-locality and both dissipative and conservative terms. The latter includes a UV divergence which is linked to an IR singularity in the near zone. The IR/UV mixing is a signature of double-counting. This is addressed in the EFT by means of the zero-bin subtraction. The~logarithmic corrections to the binding mass/energy are captured by the renormalization~group evolution.

\subsection{NRGR State of the Art}
\vskip 4pt
The daunting task of producing gravitational wave templates to high level of accuracy has led~to~decades of arduous --yet outstanding-- computations, see e.g. \cite{Blanchet,Buoreview,ALT} for a complete list of references. In~contrast, the EFT approach to the binary inspiral problem was developed some ten years ago \cite{nrgr,towers,nrgrs,thesis}. Since then, the gravitational potentials for non-rotating objects were reproduced to NLO (1PN)~\cite{nrgr}, NNLO (2PN) \cite{nrgr2pn} and NNNLO (3PN) \cite{nrgr3pn}. Most of these computations were carried out using the metric decomposition introduced in~\cite{smolkin1}. Currently the derivation within NRGR of the NNNNLO conservative dynamics at 4PN order is underway~\cite{nrgr4pn}. The (partial) results reported in~\cite{nrgr4pn} are in agreement with the (local part of the) full 4PN Hamiltonian, recently completed in \cite{4pn,4pndim,4pn1,4pn2,4pnB3,4pnDS} within the ADM and harmonic frameworks. (At the moment, the disagreement between these two computations has not been resolved \cite{4pnB3,4pnDS}. This highlights the importance of independent methodologies.)
In the radiation sector for non-spinning bodies, the radiated power loss was re-derived to NLO (1PN) \cite{andirad} and NNLO (2PN) \cite{andiunp}, which is nonetheless still behind the highest PN calculation at NNNLO (3PN), e.g.~\cite{3pn1}.\vskip 4pt

The EFT approach \cite{nrgr,towers,nrgrs,thesis}, on the other hand, has played a key role extending the state of the art~knowledge of the binary's dynamics and emitted power for rotating extended objects. The NLO spin-spin potentials were computed in \cite{eih,nrgrproc,comment,nrgrss,nrgrs2} to 3PN order. These results were also obtained using the ADM~formalism in~\cite{Schafer3pn,Schafer3pn2,schaferEFT,Hergt:2008jn,HergtEFT,steinhoffADM}, and more recently re-derived in \cite{bohennloss} in harmonic gauge \cite{Blanchet}. The NLO spin-orbit potential at 2.5PN was computed in \cite{nrgrso,delphine}, and shown to agree with previous calculations~\cite{buo1,damournloso}. The conservative effects to NLO were reproduced in~\cite{levinloss,levinloso,Levi:2015msa} using the EFT framework of \cite{nrgr,towers,nrgrs,thesis, eih,comment,nrgrproc,nrgrss,nrgrs2,nrgrso}, but applying the decomposition in \cite{smolkin1} together with some technical tools for implementing the~SCC. The radiative multipole moments needed to add spin effects in the gravitational wave phase~to~3PN order appeared in~\cite{srad}, and  in~\cite{amps} for the gravitational waveform to 2.5PN order. The~former were also re-derived in~\cite{bohennloss}, although the comparison is pending. More recently, the EFT formalism has been used to calculate~the NNLO spin-orbit and spin-spin gravitational potentials at 3.5PN and 4PN order, respectively~\cite{levinnlo1,levinnlo2,levinnlo3,equiv4pn}. These results were computed with more traditional methods in \cite{hartung, bohennloso,steinhoffnnlo1}, except for finite-size effects, which are more efficiently handled in an EFT framework \cite{nrgr,towers,nrgrs,thesis, eih,comment,nrgrproc,nrgrso,nrgrss,nrgrs2}, now broadly~adopted. An~effective action approach has also been used to incorporate leading order effects cubic (and quartic) in the spin~\cite{eftvaidya,levis3,marsats3}.\vskip 4pt 

NRGR was extended to the study of gravitational back-reaction in \cite{chadbr1}, developing the in-in formalism in a classical setting \cite{chadprl,chadprl2}. The computation of the radiation-reaction force to 3.5PN order for non-spinning bodies was carried out in \cite{chadbr2}, and shown to agree with previous results \cite{Iyer1,Iyer2,samaya}. In~\cite{andibr1} the non-local (in time) term in the effective action, discussed in \cite{4pn2,4pnB3,4pnDS}, was re-derived.  (This calculation was also investigated in \cite{tailfoffa}.) For more details on radiation-reaction effects (and extreme-mass-ratio inspirals) see \cite{chadreview}. Finally, the presence of logarithmic corrections to the binding potential and mass/energy at 4PN order~\cite{logx1,ALTlogx} was re-discovered in~\cite{andirad3,andibr1}, as a consequence of the renormalization group structure of the effective theory. The issue of IR/UV divergences and double-counting was addressed in \cite{andibr1,zerobinNRGR}. \vskip 4pt

Presently, in addition to assembling all the necessary ingredients to complete the gravitational wave templates to 4PN order and beyond, the main efforts in NRGR have shifted toward more efficient ways to organize the computations, as well as implementing non-perturbative methods. A~line of sight in this direction was advocated in \cite{largeN}, using the ultra-relativistic `large-$N$' expansion (see e.g. VII.4~in~\cite{zee}), and in \cite{iraduff,eftvaidya} using so called on-shell methods, e.g. \cite{BCF,BCFW,zvi1,zvi2,zvi3,onshell1,onshelltree,onshell2,zvi4,JJTASI}. We elaborate on future directions~in~sec.~\ref{sec:conclusion}.

\newpage
\addtocontents{toc}{\protect\newpage}\part{The Effective Theory of Cosmological Large Scale Structures}
\label{sec:part3}
\vskip 4pt
\section*{Introduction and Motivation}
\phantomsection
 \addcontentsline{toc}{section}{~~~~Introduction and Motivation}

The discovery of the present acceleration of the expansion of the universe \cite{lambda1,lambda2} has motivated extremely ambitious observational programs to make very precise measurements of the evolution of large scale structures (LSS), e.g. \cite{euclid,lsst}, with the goal of constraining the nature of {\it dark energy} \cite{darkenergy,Joyce:2014kja}. Moreover, after the outstanding results form the Sloan Digital Sky Survey \cite{Alam:2015mbd,Reid:2015gra,Cuesta:2015mqa}, the Planck satellite~\cite{Ade:2015lrj,Ade:2015ava,Ade:2015xua}, and unprecedented sensitivity to $B$-modes by the BICEP2/Keck experiments \cite{bicep}, future surveys will become leading probes to infer properties about the initial seed of structure formation~\cite{dore,danlss}. For~instance, the current bounds on primordial non-Gaussianity \cite{Ade:2015ava} are still above well-motivated physical thresholds, e.g. \cite{Bmode1,Dansusy,Flauger:2013hra,grgessay,Baumann:2015nta,Arkani-Hamed:2015bza,Mirbabayi:2015hva}. The study of structure in the universe then provides a venue to improve on our present understanding of cosmology and address the origin of both a late and (very plausibly) early phase of accelerated expansion \cite{slavabook,BaMc}. The combination of  vast amounts of new data, together with the fundamental questions it may shed light upon, has therefore reanimated efforts to make precise theoretical predictions for the dynamics of~LSS, e.g. \cite{Baldauf:2016sjb}.\vskip 4pt

Similarly to the binary inspiral problem, numerical methods occupy a prominent place in the study of LSS, e.g. \cite{whitereview,coyote,coyote2,cola,cola2,darksky}. In fact, a universe filled with cold dark matter can, in many respects, be considered a solved problem. Ignoring short- and long-distance relativistic corrections, simulations can in principle run with exquisite control of the numerical outputs. There are, however, a few reasons to adopt instead --when possible-- an analytic framework, or at least a hybrid approach. First of all, numerical simulations including baryonic matter are currently challenging. Secondly, simultaneously scanning over a large space of parameters, and a wide range of scales, becomes computationally expensive the more precision is targeted. (See \cite{springel} for a recent critical assessment of their ultimate feasibility and level of accuracy.) Perhaps more importantly, even for cases with only dark matter particles, an analytical understanding of the dynamics can lead to practical improvements. A well-known example is the reconstruction technique for baryonic acoustic oscillations (BAO)~\cite{Eisenstein:2006nk,Tassev:2012hu,White:2015eaa,Vargas-Magana:2015rqa}. Presently, reconstruction is performed using linear theory or Zel'dovich approximation \cite{Zeldovich}. On the other hand, a more accurate technique will ultimately lead to better bounds on dark energy. There is then still room for improvement by having analytic control over the imprint of non-linear modes at BAO scales. Furthermore, the more modes amenable to an analytic treatment the more accurate the constraint on observables, such as primordial non-Gaussianity.\vskip 4pt

The recent main focus of analytic studies in LSS can be thus broadly classified into two categories: 
\bit
\im The impact of IR/UV fluctuations on modes around the scale of the BAO.
\im The realm of validity of perturbative methods.
\eit
\noindent The influence of soft modes has been studied from different perspectives, most notably using symmetry arguments (equivalence principle) as well as incorporating resummation techniques, see e.g.  \cite{left,Bernardeau:2011vy,Baldauf:2011bh,Sherwin:2012nh,Kehagias:2013yd,Peloso:2013zw,Creminelli:2013mca,Creminelli:2013poa,Creminelli:2013nua,Schmidt:2013gwa,Valageas:2013zda,Kehagias:2013paa,Senatore:2014via,Ben-Dayan:2014hsa,Mirbabayi:2014gda,Baldauf:2015xfa,Baldauf:2015vio}. On the other hand, the development of the EFT of LSS has emerged as a powerful tool to parameterize the influence of UV modes on long-distance observables, and push the validity of perturbation theory toward short(er) scales~\cite{eftfluid,eftlss,left,Hertzberg:2012qn,Pajer:2013jj, Mercolli:2013bsa,Carrasco:2013sva,Carrasco:2013mua,Carroll:2013oxa,Assassi:2014fva,Angulo:2014tfa,Baldauf:2014qfa,Senatore:2014eva,Senatore:2014vja,Lewandowski:2014rca,Mirbabayi:2014zca,McQuinn:2015tva,Angulo:2015eqa,Foreman:2015uva,Assassi:2015jqa, Baldauf:2015tla,Vlah:2015sea,Baldauf:2015zga,Baldauf:2015aha,Foreman:2015,Garny, Vlah:2015zda, Abolhasani:2015mra,Assassi:2015fma,Zaldarriaga:2015jrj,Lewandowski:2015ziq,Bertolini:2015fya}.\vskip 4pt

Perturbation theory for LSS has a long and distinguished history, both Eulerian and Lagrangian space, e.g.~\cite{Zeldovich, Peebles:1980,lssreview}. It has been extremely successful at describing dark matter clustering at linear order in density perturbations, e.g. \cite{lssreview}. However, computations beyond linear theory, or so called `loop corrections,' are not under theoretical control.\,\footnote{~The loop integrals are due to the (iterative) perturbative approach and stochastic nature of the initial conditions \cite{lssreview}, rather than intrinsic quantum effects as in quantum field theory \cite{zee}. To date, computations in Euler space have been carried out to three-loop order~\cite{3loop}.}\,The origin for the failure of standard perturbation theory to describe the correct dynamics can be isolated in a simple example. Namely, an Einstein-de Sitter (EdS) universe\,\footnote{~An EdS universe is a matter dominated spatially flat Friedmann-Lemaitre-Robertson-Walker (FLRW) cosmology.} with power-law initial conditions: $P_L \propto k^n$.  Already at one-loop order the power spectrum is UV divergent, unless the spectral index is bounded: $n<-1$ \cite{lssreview}. (There is also a lower bound, $n > -3$, related to IR divergences.) This is the case even for very soft external momenta, where perturbation theory is expected to work. To produce finite answers thus requires the introduction of an (arbitrary) UV cutoff, the existence of which results in lose of predictability in the standard approach. As we shall see, like in the binary problem, the sensitivity to a UV cutoff signals the need of counter-terms, together with renormalized --finite size-- parameters, leading to a modification of the standard perturbative approach. This is the origin of the EFT approach to the dynamics of LSS \cite{eftfluid,eftlss,left}. For the case of our universe, perturbative computations are not strongly UV-sensitive. Hence, with few exceptions~\cite{Scoccimarro:1996se}, the field has ignored these issues. However, in the era of precision cosmology, the infinite error produced for the simplest of all universes is telling us that the standard approach --while finite-- still does not capture the correct imprint from non-linear modes in the dynamics of LSS, in contrast to the EFT formalism.\vskip 4pt
 
The effective theory is naturally formulated in Lagrangian space, LEFT \cite{left}, as the continuum limit of NRGR \cite{nrgr}. There are, nonetheless, a few crucial differences. In NRGR terms proportional to the Ricci tensor and scalar can be removed by a field redefinition. In the continuum limit, however, different regions overlap, therefore these terms turn out to be important for the consistency of the theory. Moreover, since typical frequencies are of order of the Hubble scale, which is also the IR scale in the problem, there is no decoupling in time \cite{left,Carrasco:2013mua,Carroll:2013oxa}. Hence, while at the level of the effective action locality is manifest, in time and space, the response functions display time non-locality and depend upon previous values over a Hubble period.  For non-linear modes, on the other hand, the typical momentum is $k_{\rm UV} \gtrsim k_{\rm NL} \gg H$. There is then a separation of scales in space, which is amenable to an EFT treatment. At~a~given soft momentum, $k$, the effects from UV physics may be then parameterized in powers~of~$k/k_{\rm NL}$. Finally, the EFT of LSS is stochastic in nature, since we only have a statistical description of the initial conditions, plausibly originating from an early phase of accelerated expansion \cite{slavabook}. This is, after all, how LSS carries information about the early universe. This means we will not only have background values and response functions, we will also need stochastic terms. The latter plays an important role in the matching procedure as well as in the regularization of {\it would-be}~divergences of the standard perturbative approach. In what follows we review the the basic elements and virtues of~LEFT \cite{left}.
\section{Pitfalls of Perturbation Theory}\label{sec:pitfalls}

Lagrangian-space computations present a series of advantages, e.g. \cite{svetlinLPT,LPTVlah,svetlinmild,svetlinmild2}, which also motivated the construction of LEFT. However, the same type of UV divergences in Euler space are found in Lagrangian perturbation theory (LPT). We will use then Lagrangian space as our arena to pinpoint the stumbling blocks in the standard perturbative approach, and also later on for the development of the EFT framework.

\subsection{Lagrangian-space}\label{sec:LS}
\vskip 4pt

In LPT one solves for the displacement of the (dark matter) particles, 
\beq
\bz(\bq,\eta) = \bq + \bs(\bq,\eta), 
\eeq
labeled in terms of a continuum variable, $\bq$. The equations of motion are \cite{lssreview}
\bea
\label{boxed0}
\frac{d^2 {\bz^i}({\bq},\eta)}{d \eta^2} +{\cal H}\frac{d{\bz^i}({\bq},\eta)}{d\eta} &=&  - \partial_i \Phi[{\bz}({\bq},\eta)]\,,\\
\partial^2\Phi({\bx},\eta) &=& \frac{3}{2} {\cal H}^2 \Omega_M \delta({\bx},\eta)\,,\label{boxed02}
\eea
where the density perturbation is given by
\bea
\label{mapqx}
1+\delta({\bx},\eta)= \int  d^3{\bq}~\delta^3( {\bx} - {\bz}({\bq},\eta)) =\left[{\rm det} \left(1+\tfrac{\partial \bs^i}{\partial \bq^j}\right)\right]^{-1}\,.
\eea
Here $\eta$ is the co-moving time, ${\cal H} \equiv a H$, and $\Omega_M \equiv \frac{8\pi G_N}{3H^2} \bar\rho_M(\eta)$, with $\bar\rho_M$ the average matter density.\vskip 4pt The equation for the displacement is solved iteratively. Working in Fourier space (with respect to the $\bq$-variables), we find
\beq
\label{vecsn}
\bs^{(n)}(\bk,\eta) = \frac{iD^n(\eta)}{n!} \int_{\bp_1}\ldots \int_{\bp_n} (2\pi)^3\delta^3\left(\bk_t - \bk\right)\,{\bL}^{(n)}(\bk_1\ldots \bk_n) \delta_0(\bk_1)\ldots \delta_0(\bk_n),
\eeq
with $\delta_0 \equiv \delta(\eta_0)$, $\bk_t = \sum_{i=1}^n \bk_i$, and $D(\eta)$ is the linear growth factor (normalized to $D(\eta_0)=1$). The~expressions for the  
$\bL^{(n)}$ can be found in e.g. \cite{Matsubara:2015ipa,Baldauf:2015tla}. From here we can compute the $n$-point correlation functions, e.g.
\bea
\langle \bs_i(\bk_1) \bs_j(\bk_2)\rangle &=& -(2\pi)^3 \delta^3(\bk_1+\bk_2) C_{ij}(\bk_1,\bk_2)\, ,\\
\langle \bs_i(\bk_1) \bs_j(\bk_2)\bs_l(\bk_3)\rangle &=& +i (2\pi)^3 \delta(\bk_1 + \bk_2 + \bk_3)  C_{ijl}(\bk_1,\bk_2,\bk_3)\, ,
\eea
which entails convoluted integrals with the power spectrum. In a scaling universe, the density power spectrum at linear order is given by, 
\beq
P_L (k,\eta) = A(\eta) k^n \equiv 2\pi^2 \frac{k^n}{k_{\rm NL}^{n+3}(\eta)}\,,
\eeq
where $k=|\bk|$ and $A (\eta) \equiv A (\eta_0) D^2(\eta)$.\vskip 4pt

\begin{figure}[t!]
\centering
\includegraphics[width=0.75\textwidth]{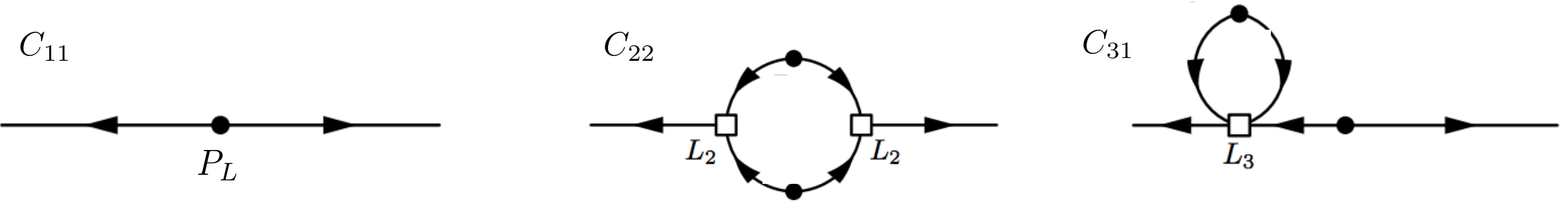}
\caption{Diagrammatic representation in perturbation theory to one-loop order, e.g. \cite{lssreview,Matsubara:2015ipa}.}
\label{lss1loop}
\end{figure}

To one-loop order we find for the two-point function, \cite{Matsubara:2008wx}
\bea
C_{ij}^{(11)}(\bk,-\bk) &=&- \frac{\bk_i \bk_j}{\bk^4} P_L(k)\, , \\
C_{ij}^{(22)}(\bk,-\bk) &=& -\frac{9}{98} \frac{\bk_i \bk_j}{\bk^4} Q_1(k)\, , \\
C_{ij}^{(13)}(\bk,-\bk) &=& C_{ij}^{(31)}(\bk,-\bk)=- \frac{5}{21} \frac{\bk_i \bk_j}{k^4} R_1(k)\, , 
\eea
whereas for the bispectrum, 
\bea
\int_p C_{ijl}^{(112)}(\bk,-\bp,\bp-\bk) &=& \frac{3}{14} \left( -\frac{\bk_i \bk_j \bk_l}{\bk^6} (R_1(k) + 2 R_2(k)) + \delta_{jl}\frac{\bk_i}{\bk^4} R_1(k)\right)\,, \\ 
\int_p C_{ijl}^{(211)}(\bk,-\bp,\bp-\bk) &=& \frac{3}{14} \left(- \frac{\bk_i \bk_j \bk_l}{k^6} (Q_1(k) + 2 Q_2(k)) + \delta_{jl}\frac{\bk_i}{\bk^4} Q_1(k)\right)\,,
\eea
and $C_{ijl}^{(121)}=C_{ijl}^{(112)}$. The diagrammatic representation of these contributions is shown in Fig.~\ref{lss1loop}. The~functions on the right-hand side read\bea
Q_{n}(k) &=& \frac{k^3}{4\pi^2} \int_0^\infty dr\, P_L(kr) \int_{-1}^{1}dx \, P_L(kr) \frac{\tilde Q_n(r,x)}{(1+r^2-2rx)^2}\,, \label{eqQ12}\\
R_{n} (k) &=& P_L(k) \frac{k^3}{4\pi^2} \int_0^\infty dr\, P_L(kr) \tilde R_{n}(r)\, ,
\eea
with $\tilde R_{n}(k), \tilde Q_{n}(k)$ given in e.g. \cite{Matsubara:2008wx}. 
In the UV limit we have $\tilde Q_n(r\to \infty) \propto r^2$, and
 \beq
 \tilde R_1(r \to \infty) \to \frac{16}{15}\,, ~~ \tilde R_2(r \to \infty) \to -\frac{4}{15} \nn \, .
\eeq
Notice $R_n(k)$ depends explicitly on the linear power spectrum, $P_L(k)$, which is still perturbative for $k\ll k_{\rm NL}$. However, there is a UV divergence when $n \geq -1$, which is sensitive to an integral over hard modes. Implementing a cutoff, $\Lambda$, we find
\beq
R^{\Lambda}_1 (k) = \frac{8}{15}  k^2 P_L(k) l_\Lambda^2(\eta)\, , ~~R^{\Lambda}_2(k) = -\frac{1}{4} R^{\Lambda}_1 (k)\ ,  \label{eq:tildeR12}
\eeq
where we introduced\,\footnote{~In \eqref{eq:tildeR12}, the $l^2_\Lambda$ accounts for the highest power of $\Lambda$. For $n>-1$ we will find also lower powers. We will keep only this divergent piece in what follows. Needless to say, the counter-terms can be adjusted to remove all positive powers of $\Lambda$.}
\beq
\label{eq:lLam}
l_\Lambda^2(\eta) \equiv\frac{1}{2\pi^2}\int^\Lambda_0 dp\; P_L(p) = \frac{\Lambda^{n+1}}{k_{\rm NL}^{3+n}(n+1)}\ . 
\eeq
On the other hand, the terms which depend upon $Q_{1(2)}$ lead to divergences which scale as $k^2 \Lambda^{2n-1}$ in the displacement power spectrum, and more importantly {\it are not} proportional to the linear $P_L(k)$. As we shall see, both these divergences will be handled by counter-terms in LEFT. The former through response functions, as {\it tidal effects}, whereas the latter will introduce a novel feature in the theory: stochastic terms. Since there is no dependence on the soft power spectrum in \eqref{eqQ12}, the renormalization through stochastic counter-terms involves `contact' interactions.\,\footnote{~For a scaling universe, instead of a cutoff we may use dim. reg., see e.g. \cite{left,lssreview,Pajer:2013jj}. However, in the real universe the shape of the power spectrum does not allow for a clean analytic implementation.}
\vskip 4pt 

Notice that at one-loop order the above integrals converge in the UV, provided $n<-1$.  Because of the shape of the power spectrum, the same happens for our universe. There are no divergencies in the loop calculations in the real world and the integrals converge simply by the lack of power on short scales. The problem, however, arises when perturbation theory is used to describe short scale fluctuations for which $|\delta(k)|$ has become large. The hard and soft modes couple through non-linear interactions as the loop integral covers all momenta, hence the error on short scales pollutes long-distance observables. The failure of standard perturbation theory cannot be fixed by a resummation approach. This is clearly seen in a simple example.
\subsection{Spherical Model}
\vskip 4pt
Let us consider an in-falling set of particles which starts homogeneous at $t=0$. Then, at time $t=t_c$, collapses to a point, $\bz_c$, and subsequently bounces back at a constant speed. This process is described by 
\beq
\label{eq:sol1}
\bz(\bq,t)=\bq-(\bq-\bz_c)\tfrac{t}{t_c}\  \to  \bs(\bq,t)=-(\bq-\bz_c)\tfrac{t}{t_c}\,,
\eeq 
with the density contrast 
\beq
\label{eq:poled}
1+\delta(\bx,t)=\tfrac{1}{|1-t/t_c|}\, .
\eeq 
The perturbative approximation amounts to finding a solution of the form ($c_0=1$) \beq 1+ \sum_{i=1}^n \delta{}^{(i)}(\bx,t)=\sum_{i=0}^n c_i (t/t_c)^i\,,\label{eq:coll}\eeq 
to $n$-th order. Notice, as $t \to t_c$, more and more terms are needed and the series diverges for $t \geq t_c$. The failure to describe the correct dynamics is due to a pole at $t=t_c$ in the exact solution in \eqref{eq:poled}. In general, the series will not converge beyond a circle in the complex-$t$ plane of radius $|t|=t_c$.\vskip 4pt The above case is somewhat pathological, let us consider instead the time-reversed solution,
\be\label{eq:sol1n}
\bz(q,t)=\bq+(\bq-\bz_c)\tfrac{t}{t_c} \to  1+\delta(\bx,t)=\tfrac{1}{1+t/t_c}\ ,
\ee 
which describes innocuous-looking out-flowing matter with a smooth density perturbation $\delta(\bx,t)$. One~may think this case allows for a perturbative expansion, as in \eqref{eq:coll}.  However, the exact answer still has a pole, now at $t=-t_c$. Hence, the series does not converge beyond $|t|=t_c$ either. This is somewhat counter-intuitive, the dilution process becomes highly non-trivial in the perturbative expansion with more and more terms needed for $t < t_c$. This very simple example underlies the complications of the perturbative approach. We can highlight these issues more explicitly in LPT by studying a spherical  model, e.g. \cite{Peebles:1980,Munshi:1994zb,Sahni:1995rr}.\vskip 4pt Let us consider a spherical shell of radius $r(a) \equiv |\bz(\bq,a)|$, initially at $r_0\equiv |\bz(\bq,0)|=|\bq|$, with $a(\eta)$ the scale factor we may use as a clock. In a matter dominated universe the gravitational potential is given by Poisson's equation in \eqref{boxed02}, $\Phi(r) \propto \frac{r^2}{a} \delta(a)$,\,\footnote{~Since the linear growth factor is given by $D(a) \simeq a$~\cite{lssreview}, the potential is time-independent at leading order.} 
and following mass conservation we have
\beq
\label{eq:densa}
(1+\delta(a))r^3  = r_0^3\,.
\eeq
The equation of motion for the position of the shell can be solved for an initial over-density, $\delta_0>0$, and may be parameterized as \cite{lssreview}
\bea
r(\theta) &=& \frac{3}{10}\frac{r_0}{a(\theta)} \left(1 - \cos\theta\right)\,,\\
\epsilon(\theta) &=& \frac{3}{5}\left(\frac{3}{4} (\theta-\sin\theta)\right)^{2/3}\,, 
\eea
where $\epsilon(\theta) \equiv \delta_0 a(\theta)$.
The density contrast follows from \eqref{eq:densa} 
\beq
1+\delta (\theta)= \frac{9}{2}\frac{\left(\theta-\sin\theta\right)^2}{\left(\cos\theta-1\right)^3}\,\cdot \label{eq:exacth}
\eeq
Like before, the over-density case is singular due to the inevitable collapse of the spherical shell. On the other hand, the under-density solution obtained by replacing $\epsilon \to -\epsilon$ never becomes singular.\footnote{In terms of the parameterization in \eqref{eq:exacth} the under-density solution is found by replacing $\theta \to i\theta$.}  On the contrary, the total density $(1+\delta)$ becomes smaller and smaller with the asymptotic value $\delta \to -1$.\vskip 4pt The~perturbative solution can also be constructed in LPT. For example, using \eqref{eq:densa} up to third order, one finds \cite{Sahni:1995rr,lssreview}
\beq
1+\sum_{i=1}^3 \delta^{(i)} =  \left( 1 - \frac{1}{3}\epsilon - \frac{3}{21} \epsilon^2 - \frac{23}{1701}\epsilon^3 \right)^{-3}\,.
\eeq
It is straightforward to show the perturbative solutions fares well at early times. However, once the density falls under a critical value, $\delta_c \simeq - 0.7$ \cite{Sahni:1995rr}, the accuracy degrades the more LPT orders we add. The linear theory becomes the most accurate approximation to the exact solution. One can show that the situation is even worse, with pathological $\delta^{(2n)}$ LPT terms predicting collapse for the under-density situation!\vskip 4pt In~this case the reason for the failure of LPT is the presence of a singularity in the spherical collapse model, which also enters in the time-reversed dynamics. Since our universe is bound to be significantly more intricate, this simple example then demonstrates that perturbative schemes cannot correctly describe the dynamics when $\delta$ is large, and resummation techniques cannot remediate it. 

\section{Effective Action: Continuum Limit} \label{sec:continuum}

As we discussed in the previous section, computations with \eqref{boxed0}-\eqref{boxed02} require introducing a cutoff in order to keep the perturbative expansion under control. In coordinate space, this entails working with smoothed displacement fields, $\bz_L(\bq,\eta)$, which describe the motion of extended objects interacting with long-wavelength gravitational perturbations. In part~\ref{sec:part2} we studied at length the implementation of an EFT for extended objects. However, because of the continuum limit, a few extra ingredients appear here which we did not encounter before for the binary problem. We review LEFT in what follows.

\subsection{Dark Matter Point-Particles}
\vskip 4pt
Let us consider a set of $q=1\ldots N$ point-like dark matter particles interacting gravitationally, in a FLRW background dominated by a dark matter component (and in principle also with a cosmological constant). The~equations of motion follow by varying the point-particle action,
\beq
S_{\rm pp}^{\rm DM}= -\sum_q m_q \int d^3\bx \,d\eta_q~\delta^3({\bx}-{\bz}_q(\eta_q)) \sqrt{g_{\mu\nu}({\bx},\eta)\dot z^\mu_q(\eta_q)  \dot z^\nu_q(\eta_q)} \ .\label{eq:acDMpp}
\eeq
The $z^\mu_q(\eta_q)$ are the co-moving coordinates for the $q$-particle and dots are taken with respect to $\eta$. In~what follows we set $z^0\equiv \eta(t)$ and $\eta_q = \eta$ for all particles. We now take the continuum limit over the $q$-index,
\beq
\sum_q m_q \to \int  \bar\rho_M(\eta_0)\,a^3(\eta_0)\,d^3{\bq} = \int  \bar\rho_M(\eta)\,a^3(\eta)\,d^3{\bq}\,,
\eeq  
with $\bar\rho_M(\eta) = \bar\rho_M(\eta_0) \frac{a_0^3}{a^3(\eta)}$. Then, the point-particle action becomes
\bea
S_{\rm pp}^{\rm DM}=-\int d^3{\bx} \,d^3{\bq}\, d\eta\,a^3(\eta)\,  \rho({\bz}({\bq},\eta),{\bx})\, \sqrt{g_{\mu\nu}({\bx},\eta)\dot z^\mu({\bq},\eta)  \dot z^\nu({\bq},\eta)}\,,\label{eq:actiondm}
\eea
where we introduced the mass density per unit of $q$-cell,
\beq
\rho({\bz}({\bq},\eta),{\bx})\equiv\bar\rho_{m}(\eta) ~\delta^3({\bx}-{\bz}({\bq},\eta))\,.
\eeq
For latter convenience we also define the density perturbation,
\beq
\delta\rho({\bz}({\bq},\eta),{\bx}) \equiv \bar\rho_M(\eta)\left[\delta^3({\bx}-{\bz}({\bq},\eta))-\delta^3({\bx}-{\bq})\right]\,.
\eeq
In the Newtonian limit we have (in co-moving coordinates)
\beq
g_{\mu\nu} dx^\mu dx^\nu =a^2(\eta)\left[(1+2\Phi) d\eta^2-(1-2\Phi)d\bx^2\right],
\eeq
and also $\dot \bz_q/a(\eta) \ll 1$. The total action, including the gravitational part, turns into,
\beq
\label{action2}
S_{\rm tot}=  \int d^3{\bx} d^3{\bq} d\eta\,a^4(\eta)\left\{\rho({\bz}({\bq},\eta),{\bx}) \left(-1+\frac{1}{2}\left(\frac{d{\bz^i}({\bq},\eta)}{d\eta}\right)^2\right)-\delta\rho({\bz}({\bq},\eta),{\bx})\Phi({\bx},\eta)\right\}+ S_{\rm EH}^{(2)}\, ,
\eeq
where we canceled the tadpole against the linear term in the Einstein-Hilbert action, while retaining the quadratic part, 
\beq\label{EHnewt}
S_{EH}^{(2)}=-4 \Mp^2 \int d^3\bx d\eta\, a^4 \Phi(\bx,\eta)\frac{\pd^2}{a^2}\Phi(\bx,\eta) \ .
\eeq
As expected, $S_{\rm tot}$ starts quadratic in the perturbations. Varying the action we obtain, 
\beq
\pd^2\Phi(x)=4\pi G_N a^2\,\delta\rho(x)=\frac{3}{2}{\cal H}^2(\eta)\Omega_M(\eta) \delta(x)\ ,
\eeq
as in \eqref{boxed02}, where 
\beq
\delta\rho(\bx,\eta)=\int d^3{\bq}\;\delta\rho({\bz}({\bq},\eta),{\bx})\ ,\quad \delta({\bx},\eta)\equiv \frac{\delta\rho({\bx},\eta)}{\bar\rho_M(\eta)}\ .
\eeq
For the particles' trajectory we recover the expression in \eqref{boxed0}, from
\beq
\int d^3\bx\left[ \frac{d}{d\eta}\left(a^4(\eta) \bar\rho_{m}(\eta)  \delta^3({\bx}-{\bz}({\bq},\eta))  \frac{d{ \bz^i}({\bq},\eta)}{d\eta} \right)+ a^4(\eta) \bar\rho_{m}(\eta)  \delta^3({\bx}-{\bz}({\bq},\eta)) \pd_i\Phi \right]=0\ .
\eeq

\subsection{Long-Distance Universe}\label{sec:longdistance}
\vskip 4pt

\subsubsection{Relativistic Theory}
\vskip 4pt
\noindent Following part~\ref{sec:part2}, the continuum limit of the long-distance effective theory is described by~\cite{left}\,\footnote{~To avoid confusion with the `long-wavelength' label we opened the $L$-indices in the multipole moments, see~\eqref{eq:lag}.}
\bea
\label{eq:LEFTac}
S_{\rm LEFT} &=& S_{L}^{\rm DM} - \int a^3 d\eta \,d^3{\bq}\, d^3{\bx}~ \rho_L({\bz}_L({\bq},\eta),{\bx}) \Bigg\{ \frac{1}{2}  {\dot z}_L^\mu(\bq,\eta) \omega_\mu^{ab}(\bx,\eta) L^{ab}({\bq},\eta)\nn \\ &-&
\sum_{\ell=2} \Bigg( \frac{1}{\ell!} I^{i_1\ldots i_\ell}(\bq,\eta) \nabla_{{i_1\ldots i_{\ell-2}}} E_{i_{\ell-1}i_\ell}(\bx,\eta)  -   \frac{2\ell}{(2\ell+1)!}J^{i_1\ldots i_{\ell}}(\bq,\eta) \nabla_{i_1\ldots i_{\ell-2}} B_{i_{\ell-1}i_\ell}(\bx,\eta)\Bigg) \nn \\&-& \sum_{\ell=0} \Bigg(C_0^{i_1\ldots i_\ell}(\bq,\eta) \nabla_L R(\bx,\eta) + C^{i_1\ldots i_\ell j}_v e_j^\nu \nabla_{i_1\ldots i_\ell} R_{\mu\nu}(\bx,\eta) {\dot z}_L^\mu(\bq,\eta)\nn \\ &-&  C^{i_1\ldots i_\ell}_{v^2}({\bq},\eta) \nabla_{i_1\ldots i_\ell} R_{\mu\nu}(\bx,\eta){\dot z}_L^\mu(\bq,\eta) {\dot z}_L^\nu(\bq,\eta)\Bigg) \Bigg\}\,, 
\eea
where $L^{ab}$ is the angular momentum, and we introduced
\beq
\rho_L({\bz}({\bq},\eta),{\bx})\equiv \bar\rho_M(\eta) ~\delta^3({\bx}-{\bz}_L({\bq},\eta))\,.
\eeq
The terms proportional to curvature (and derivatives thereof) are evaluated with the long-wavelength metric field,  $g_{\mu\nu}^L$. Moreover, the $(I^L,J^L)$ are the symmetric trace-free multipole moments, while the $C^L$'s couple to the traces of the curvature tensor. The first term in \eqref{eq:LEFTac} includes the binding mass/energy, and is given by
\bea
\label{eq:actiondmL}
S_L^{\rm DM}=-\int a^3 d^3{\bx} d^3{\bq} d\eta\,\rho_E(\bz_L(\bq,\eta),\bx)  \sqrt{g_{\mu\nu}^L(\bx,\eta)\dot z_L^\mu({\bq},\eta)  \dot z_L^\nu({\bq},\eta)}\,. \nn
\eea
This is the same expression as in \eqref{action2} with the replacement: $\bz \to \bz_L$, $g_{\mu\nu} \to g_{\mu\nu}^L$, and  
\beq
\rho_E(\bz_L(\bq,\eta),\bx) \equiv \bar\rho_M( \eta) \left[ K(\bq,\eta) + V_S(\bq, \bx,\eta)\right] \delta^3 (\bx - \bz_L(\bq,\eta))\,.
\eeq
Note $\rho_E$ includes a kinetic term, $K(\bq,\eta)$, as well as a potential, $V_S(\bq,\bx,\eta)$, which is a function of both the internal binding potential of a given smoothed object, through the $\bq$-level, and the potential induced by the other (plausibly overlapping) regions, through the $\bx$-dependence. The total action also features the Einstein-Hilbert term.\vskip 4pt

Since particles interact only gravitationally we do not have multi-particle vertices. Provided there is a separation of scales with respect to the cutoff of the EFT, we can expand such vertices, and write
 \beq
 \label{contact}
S_{\rm LEFT} \supset \int d\eta \int d^3\bq\;  {\cal L}(\pd_{q^j}\bz_L^i\pd_{q^j}\bz_L^i,\; \pd_{q^iq^l}\bz_L^j\pd_{q^i q^l}\bz_L^j,\ldots)\ .
 \eeq
These interactions describe different UV models, and may be used, for example, to incorporate baryons. We will not discuss these terms here. 

\subsubsection{Newtonian Limit}
\vskip 4pt
We now take the Newtonian limit of our relativistic theory. This is relevant for the study of structure formation (up until we reach the Hubble scale in the IR, where relativistic effects may be important). Hence, we only include the coupling to the long-wavelength potential, $\Phi_L$, and ignore velocities and spin. As a consequence many of the terms in \eqref{eq:LEFTac} combine, and we arrive at
\bea
 \label{effnewt}
 S_{\rm LEFT} &\simeq& \int  d \eta d^3{\bx} d^3{\bq}\,a^4(\eta)  \Bigg\{-\rho_E({\bz}_L({\bq},\eta),{\bx})- \delta\rho_L({\bz}_L({\bq},\eta),{\bx})\,\Phi_L({\bx},\eta)\\ &+&\rho_L({\bz}_L({\bq},\eta),{\bx})\Bigg[ \frac{1}{2}\left(\frac{d{\bz}_L({\bq},\eta)}{d\eta}\right)^2 + \sum_{\ell=2} \frac{1}{\ell!} I^{i_1\ldots i_{\ell}}(\bq,\eta) \partial_{{i_1\ldots i_{\ell-2}}} E^\Phi_{i_{\ell-1}i_\ell}(\bx,\eta)\nn\\ &+&\sum_{\ell=0} C^{i_1\ldots i_\ell} (\bq,\eta)\partial_{{i_1\ldots i_\ell}} R^\Phi(\bx,\eta)\Bigg]\Bigg\} + \cdots\,,\nn
\eea
where we absorbed all contributions from the Ricci tensor into $C^L(\bq,\eta)$, and introduced
\bea
E^\Phi_{ij} &\equiv& - \left(\partial_i\partial_j-\frac{1}{3}\delta_{ij}\partial^2\right) \Phi_L\,,~~ R^\Phi \equiv -\partial^2 \Phi_L\,,\\
\delta\rho_L({\bz_L}({\bq},\eta),{\bx}) &\equiv& \bar\rho_M(\eta)\left[\delta^3({\bx}-{\bz_L}({\bq},\eta))-\delta^3({\bx}-{\bq})\right]\nn\,.
\eea
Note the action is quadratic in the fluctuations, since for the unperturbed background, $\langle I^L(\bq,\eta)\rangle_S = 0$, and the terms proportional to the traces turn into a total derivative.\vskip 4pt Since we keep off-shell modes of the metric (ignoring radiation) and couplings to the Ricci tensor, we may as well write the action in terms of the full multipole moments, including traces~\cite{thorne,bala},
\beq
Q^{i_1\cdots i_n} \equiv  \sum_{k=1}^{k=[n/2]} \frac{ n! (2n-4k+1)!!}{(n-2k)!(2n-2k+1)!!(2k)!!}~ \delta^{(i_1i_2}\cdots \delta^{i_{2k-1}i_{2k}} I^{i_{2k+1}\cdots i_n)\ell_1\ell_1 \cdots \ell_k\ell_k}\,.
\eeq
Then, the dynamics which derives from $S_{\rm LEFT}$ and the Einstein-Hilbert action can be equivalently represented by 
\bea\
&&\hspace{1cm}\frac{d^2 {\bz}^i_L ({\bq},\eta)}{d\eta^2} + {\cal H} \frac{d {\bz}^i_L ({\bq},\eta)}{d\eta}  = \ba^i_S(\bq,\eta) \label{eq:boxed1}\\
&&  - \partial_i \left[\Phi_L(\bx,\eta) + \sum_{\ell=2} \frac{1}{\ell!} Q^{i_1\ldots i_\ell}(\bq,\eta) \partial_{i_1\ldots i_\ell} \Phi_L (\bx,\eta)\right]_{\bx\to \bz_L(\bq,\eta)}\,,\nn
\eea
where
\beq
\ba^i_S \equiv -\left.\partial_i V_S(\bq,\bx,\eta)\frac{}{}\right|_{{\bx\to \bz_L({\bq},\eta)}} \ ,
\eeq
and for the Newtonian potential,
\beq
\label{eq:PoissonL}
\pd^2 \Phi_L(\bx,\eta) =  \frac{3}{2} {\cal H}^2\Omega_M  \delta_{M,L}(\bx,\eta)\,.
\eeq
The {\it mass}-density perturbation is given by
\beq
\label{eq:deltamL}
\delta_{M,L}(\bx,\eta) = \delta_{N,L}(\bx,\eta) + \sum_{\ell=2} \frac{1}{\ell!} \partial_{i_1\ldots i_\ell} {\cal Q}^{i_1\ldots i_\ell}(\bx,\eta)\, ,
\eeq
with  $\delta_{N,L}(\bx,\eta) \equiv \int d^3\bq \, \delta^3(\bx-\bz_L(\bq,\eta))$ the {\it number}-density. We also introduced
\beq
{\cal Q}^{i_1\ldots i_\ell}(\bx,\eta) \equiv \int d^3\bq \,Q^{i_1\ldots i_\ell}(\bq,\eta)\, \delta^3(\bx-\bz_L(\bq,\eta))\,. \label{eq:calmul}
\eeq
It is convenient to write these equations in $\bx$-Fourier space,
\bea
 \label{eq:displacement_fourier}
&&\hspace{3cm}\frac{d^2}{d\eta^2}  \bz_L(\bq_1,\eta) + {\cal H} \frac{d}{d\eta} z_L(\bq_1,\eta) =  \ba_S(\bz_L(\bq_1,\eta))+ \\ &&+ \frac{3}{2} {\cal H}^2 \Omega_M \int d^3 \bq_2  \int_k ~\frac{i\bk}{\bk^2}\exp\left[i \bk \cdot (\bz_L(\bq_1,\eta)-\bz_L(\bq_2,\eta))-\frac{1}{2}\bk^i \bk^j \big(Q_c^{ij}(\bq_1) + Q_c^{ij}(\bq_2)\big)+ \cdots \right]\nn\,,\\
&&\hspace{1cm}\Phi_L(\bk,\eta)= -\frac{3}{2} {\cal H}^2 \Omega_M~\frac{1}{\bk^2} \int d^3\bq~{\rm exp}\left[ -i \bk \cdot \bz_L(\bq,\eta) - \frac{1}{2}\bk^i \bk^j Q_c^{ij}(\bq) + \cdots \right] \ , \label{eq:displacement_fourier2}
\eea
where $Q_c^{i_1\ldots i_\ell}$ stands for the `connected' multipoles, see \cite{left}. The expectation value of the exponential of a quantity can be performed as the exponential of the connected cumulants of the same quantity, see e.g. \cite{CLPT}. This exponentiation may be useful to perform resummations, as we discuss briefly in sec.~\ref{sec:LEFTres}.\vskip 4pt

\subsubsection{Field Redefinitions}
\vskip 4pt

As we mentioned before, consistency requires we keep the Ricci tensor in the effective action through the traces of the multipole moments. This is one of the main differences with NRGR, where these terms can be removed by a field redefinition, up to `contact' interactions. In the continuum limit, however, these terms survive. Nevertheless, we can still perform field redefinitions, applying the leading order equations of motion for the long-wavelength fields into the higher derivatives interactions. The traces will not disappear, and instead they will be smoothed into terms proportional to the density.\vskip 4pt Consequently, the expression in \eqref{effnewt} transforms into (in the Newtonian limit)
\bea
 \label{effnewt2}
 S_{\rm LEFT} &\simeq& \int  d \eta d^3{\bx} d^3{\bq}\,a^4(\eta)  \Bigg\{-\rho_E({\bz}_L({\bq},\eta),{\bx})- \delta\rho_L({\bz}_L({\bq},\eta),{\bx})\,\tilde\Phi_L({\bx},\eta)\\ &+&\rho_L({\bz}_L({\bq},\eta),{\bx})\Bigg[ \frac{1}{2}\left(\frac{d{\bz}_L({\bq},\eta)}{d\eta}\right)^2 + \sum_{\ell=2} \frac{1}{\ell!} I^{i_1\ldots i_\ell}(\bq,\eta) \partial_{i_1\ldots i_{\ell-2}} E^{\tilde\Phi}_{i_{\ell-1}i_\ell}(\bx,\eta)\nn\\ &-&\sum_{\ell=0} \tilde C^{i_1\ldots i_\ell}(\bq,\eta) \partial_{i_1\ldots i_\ell} \delta_{N,L}(\bx,\eta)\Bigg]\Bigg\}+\cdots \nn\,,
\eea
with $\tilde C^L(\bq,\eta) = \frac{3}{2}{\cal H}^2\Omega_M\,C^L(\bq,\eta)$. Notice the action is now a function of a new potential, $\tilde \Phi(\bx,\eta)$. (The~full~redefinition also reshuffles some other terms which are absorbed into the Wilson coefficients.) The advantage is that the source of gravity is now only due to the trace-free part of the multipole moments. The equations of motion become
\bea
&&\hspace{1cm}\frac{d^2 {\bz}_L ({\bq},\eta)}{d\eta^2} + {\cal H} \frac{d {\bz}_L ({\bq},\eta)}{d\eta}  = \ba_S(\bq,\eta)\\
&&  - \partial_i \left[\tilde \Phi_L(\bx,\eta) + \sum_{\ell=0} \tilde C^{i_1\ldots i_{\ell}}  \partial_{i_1\ldots i_{\ell-2}} \delta_{N,L}(\bx,\eta)+ \sum_{\ell=2}\frac{1}{\ell!} I^{i_1\ldots i_{\ell}}(\bq,\eta) \partial_{i_1\ldots i_{\ell}} {\tilde\Phi}_L (\bx,\eta)\right]_{\bx\to \bz_L(\bq,\eta)}\,,\nn
\eea
together with
\bea
\label{eq:PoissonL2}
\pd^2 \tilde\Phi_L(\bx,\eta) &=&  \frac{3}{2} {\cal H}^2\Omega_M  \tilde\delta_{M,L}(\bx,\eta)\,,\\
\tilde\delta_{M,L}(\bx,\eta) &=& \delta_{N,L}(\bx,\eta) + \sum_{\ell=2} \frac{1}{\ell!} \partial_{i_1\ldots i_{\ell}} {\cal I}^{i_1\ldots i_{\ell}}(\bx,\eta)\, .
\eea
The new potential is sourced by a mass-density perturbation which only involves the trace-free multipole moments. (Here the ${\cal I}^L(\bx,\eta)$ are defined similarly to \eqref{eq:calmul}.) As long as we restrict ourselves to observables quantities in terms of the displacement fields, the effective action in \eqref{effnewt2} is an equally valid descriptions of the dynamics.
\subsection{Smoothing}\label{sec:smoothing}
\begin{figure}[t!]
\centering
\includegraphics[width=0.5\textwidth]{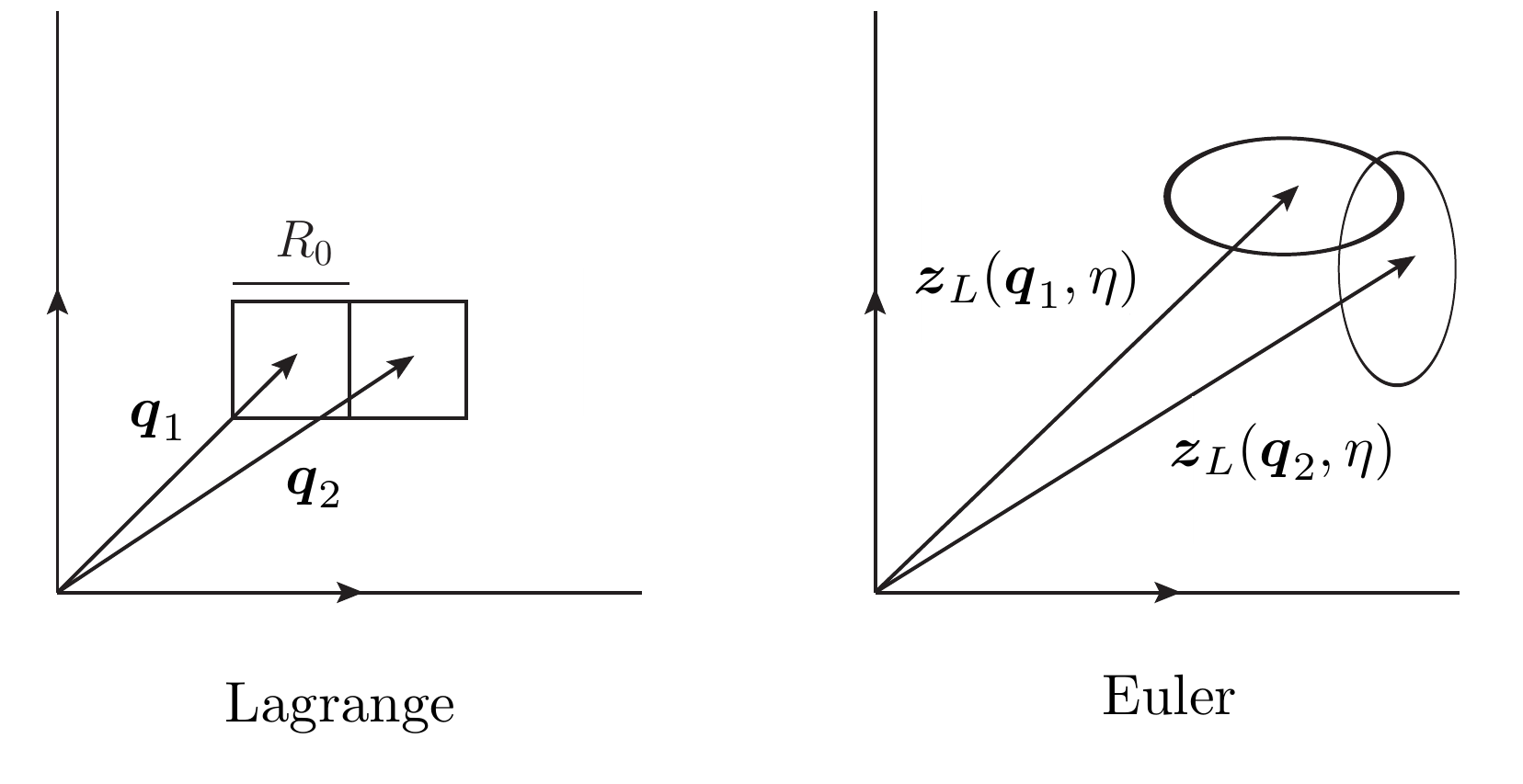}
\caption{Left panel: Finite regions of size $R_0$ in Lagrangian space.  Right panel: Evolution in Eulerian space. The vector $\bz_L(\bq,\eta)$ describes the motion of the center-of-mass for each Lagrangian region. Notice that, upon evolution, nearby regions may overlap.}
\label{figL1}
\end{figure}

\vskip 4pt
As we discussed in sec.~\ref{sec:effect}, the long-distance action is obtained by integrating out the short-distance degrees of freedom. The latter may be defined with respect to a smoothing scale, such that we only keep modes with momentum $k\lesssim R_0^{-1}$. As a result we obtain dynamical equations for finite regions of size $R_0$, centered at $\bz_L(\bq,\eta)$, defined as (see Fig.~\ref{figL1})
\beq
\bz_L(\bq,\eta) \equiv  \int d^3\bq_1 W_{R_0}(\bq,\bq_1) \bz(\bq_1,\eta),
\eeq
where $W_{R_0}(\bq,\bq_1)$ is a window function, obeying \beq \int d^3{\bq}_1\; W_{R_0}({\bq},{\bq}_1) = 1\,\label{window}.\eeq 
\vskip 4pt

In the Newtonian limit the point-particle effective action in \eqref{eq:actiondm} may be then written as
 \bea
&&S_{\rm pp}^{\rm DM}=\int a^4 d\eta \int d^3{\bx} \int d^3{\bq} d^3{\bq}_1~W_{R_0}({\bq},{\bq}_1) \ \times\\ \nn
&& \left\{\rho({\bz}({\bq}_1,\eta),{\bx})\left[-1+\frac{1}{2}\left(\frac{d{\bz}({\bq}_1,\eta)}{d\eta}\right)^2  \right]-\delta\rho({\bz}({\bq}_1,\eta),{\bx}) (\Phi_S({\bx},\eta)+\Phi_L({\bx},\eta))\right\}\ ,\nn  \label{actionW}
\eea 
after decomposing $\Phi$ into short- and long-distance contributions relative to the cutoff scale $R_0^{-1}$. We then split the displacement into motion of the center-of-mass and short-distance displacements, 
\beq
\label{expz2}
{\bz} ({\bq}_1,\eta) = {\bz}_L({\bq},\eta) + \delta {\bz}({\bq},{\bq}_1,\eta)\ , 
\eeq
with
\beq\label{eq:zshort}
\int d^3 \bq_1\; W_{R_0}(\bq,\bq_1)\; \delta{\bz}({\bq},{\bq}_1,\eta)=0\ .
\eeq
The $\bq$-label accounts for the given region whereas the $\bq_1$-label identifies the constituents.\vskip 4pt
The effective action follows by integrating out the hard modes ($\Phi_S,\delta\bz$). Let us concentrate first on the displacements. Then, Taylor expanding the expression for $\rho({\bz}({\bq}_1,\eta),{\bx})$ around $\bz_L(\bq,\eta)$, we obtain (after integrating by parts)
\beq
S_{\rm pp}^{\rm DM}\supset\frac{1}{\ell!}  \int d^3{\bx} \,d^3{\bq}\, d\eta\, a^4 \rho_L ({\bz}_L({\bq},\eta),{\bx}) \left(\int d^3{\bq}_1\;W_{R_0}({\bq},{\bq}_1) \delta {\bz}^{i_1}(\bq_1,\bq)\ldots \delta {\bz}^{i_\ell}(\bq_1,\bq)\right) \partial_{i_1\ldots i_\ell} \Phi_L ({\bx},\eta)\,.
\eeq
From here we identify (in the Newtonian limit)
\beq
\label{matchqij} 
Q^{i_1\cdots i_n}_{R_0}({\bq},\eta) = \int d^3{\bq}_2\; W_{R_0}({\bq},{\bq}_1)\; \delta { \bz}^{i_1}({\bq},{\bq}_1,\eta)\ldots \delta { \bz}^{i_n}({\bq},{\bq}_1,\eta)\,.
\eeq
For the other terms we find
\beq\label{kinetic} 
S_{\rm pp}^{\rm DM}\supset  \int d\eta d^3 \bx d^3 \bq~\bar \rho_m(\eta) a^4 \delta^3(\bx - \bz_L(\bq,\eta)) \left[\frac{1}{2}\left(\frac{d \bz_L(\bq,\eta)}{d\eta}\right)^2 + K(\bq,\eta)\right]\ ,  
\eeq
for the kinetic piece, with
\beq
K(\bq,\eta) = \frac{1}{2}\int d^3 \bq_1 W_{R_0}(\bq,\bq_1) \delta \dot \bz^i(\bq,\bq_1,\eta) \delta \dot \bz^i(\bq,\bq_1,\eta)\ ,
\eeq
and from the coupling to the short-distance potential,
\beq
S_{\rm pp}^{\rm DM}\supset\int d^3 \bx d^3 \bq \;\bar\rho_M(\eta)\,\delta^3 (\bx - \bz_L(\bq,\eta))\; V_S(\bq, \bx,\eta)\,,
\eeq
with 
\beq
V_S(\bq,\bk,\eta) =\int d^3\bx\; e^{- i\bk\cdot\bx} \;V_{S}(\bq,\bx,\eta)\equiv \int d^3 \bq_1 \; W_{R_0}(\bq,\bq_1)\; \Phi_S(\bk,\eta)\; e^{i\bk\cdot \delta \bz(\bq,\bq_1,\eta)}\ .
\eeq
As expected, the main contribution to $V_S$ comes from hard modes, with $k \sim (\delta z)^{-1}$.\vskip 4pt

Notice we could have directly obtained these expressions by using the results in sec.~\ref{sec:gravradM}, in the Newtonian limit. There we found, for instance for the multipole moments, e.g. \eqref{eq:lolij},
\beq
Q_{R_0}^{i_1\ldots i_\ell} (\bq,\eta)= \int d^3\bx \,{\cal T}_{R_0}^{00}(\bq,\bx,\eta) \bx^{i_1} \cdots \bx^{i_\ell}\,,
\eeq
at leading order in the PN expansion. In our case, the stress-energy tensor of a region of size $R_0$, and center-of-mass $\bq$ in Lagrangian space, is dominated by the mass-density, 
\beq
\label{eq:nobulk}
{\cal T}_{R_0}^{00}(\bq, \bx,\eta) =  \bar\rho_M (\eta) a^3(\eta) \int d^3\bq_1\, \delta^3\left(\bx - \delta\bz(\bq,\bq_1,\eta)\right) W_{R_0}(\bq,\bq_1)\,.
\eeq
From here we recover the expression in \eqref{matchqij}.\,\footnote{~Note we did not include the bulk motion in \eqref{eq:nobulk}, since the multipoles are defined in the rest frame of the `body.' This was not an issue in sec. \ref{sec:gravradM} where we assumed the binary system was at rest.} The overall factor of $\bar\rho_M$ is because of our definitions. In sec.~\ref{sec:radgrav} we introduced {\it mass} multipole moments, whereas in LEFT: $[Q^L] \sim$ {\it length}\,$^{L}$.

\subsection{Background, Response and Stochastic Terms}
\vskip 4pt

As we discussed in sec.~\ref{sec:effect}, we decompose the multipole moments into \cite{left}
\beq
\label{eq:qsr}
Q^{i\ldots i_n} = \underbrace{\left\langle Q^{i\ldots i_n}\right\rangle_S}_{\rm background} + \underbrace{\left(Q^{i\ldots i_n}\right)_{\cal R}}_{\rm response} + \underbrace{\left(Q^{i\ldots i_n}\right)_{\cal S}}_{\rm stochastic} \, ,
\eeq
and similarly for the binding potential, $V_S$.  Notice we added an stochastic term. The latter accounts for the fact that, due to the statistical nature of the initial conditions, in each realization there is a random departure from the expectation value. While the effective action in LEFT is written in terms of a local expansion in space and time, the (relatively long) Hubble time scale in the problem introduces an extra feature which we did not encountered before in NRGR. Namely, the possibility of time non-locality for the response functions.\vskip 4pt 

The symmetries of the problem constrain the type of terms which may appear in the expansion for the response, e.g. $\partial_i\partial_j\Phi_L,\; \partial_j\bs^i_L,$ etc., as dictated by rotational invariance and the equivalence~principle. For~instance, for the quadrupole in \eqref{eq:calmul} we may write,
\bea
\label{responseI}
\left({\cal Q}^{ij}(\bz_L({\bq},\eta),\eta)\right)_{{\cal R}}= \int d\eta^\prime\Big[\hspace{-0.6cm}&&\,G_1^{ij,lk}(\eta;\eta^\prime)\; \partial_l\partial_k \Phi_L(\bz_L({\bq},\eta^\prime)) \\ && + \, G_2^{ij,lk}(\eta;\eta^\prime)\; \partial_l \bs_{L,k}({\bq},\eta^\prime)+\ldots \Big]\ ,\nn
\eea
where $\bs_L = \bz_L -\bq$, and $G_a^{ij,lk}(\eta;\eta^\prime)$ are retarded Green's function which depend on the dynamics at short distances.
The non-locality in time is inherited by the typical variation of the retarded Green's functions compared with the Hubble scale. This complicates the construction of the EFT, which now requires a more elaborate structure. However, this non-locality first appears at two-loop order and furthermore one can show it does not produce large effects \cite{Carrasco:2013mua,Zaldarriaga:2015jrj}.\vskip 4pt At one-loop order, on the other hand, we can proceed as follows. At linear order the displacement scales as the gradient of the potential, and is given by \cite{Zeldovich}
\beq
\bs_L^{(1)}(\bk,\eta) =  i \frac{\bk}{\bk^2} D(\eta)\delta_0(\bk)\ .
\eeq
Then, the response of the quadrupole may be written solely in terms of $\partial^i \bs_L^j$, as 
\bea
\left(Q^{ij}({\bq},\eta)\right)_{{\cal R}} &=& \left[\int d\eta^\prime~\tilde G_1^{ij,kl}(\eta;\eta^\prime)\; \frac{D(\eta')}{D(\eta)}\right]\partial^k \bs^{(1)l}_L(\bq,\eta)+\cdots   \\ &=& l^2_{ij,kl}(\eta) \, \partial^k \bs^{(1)l}_L(\bq,\eta) + \cdots \ ,\nn
\eea
where $l_{ijkl}^2(\eta)$ is defined in terms of the integral in the brackets and has units of {\it length}$\,^2$. We can decompose it further into irreducible representations and write (the minus signs are for convenience)
\beq 
\left(Q_{ij}({\bq},\eta)\right)_{\cal R} = - \frac{1}{3} l^2_T(\eta) \delta_{ij}~\partial_k  s_k(\bq,\eta) - l_{TF}^2(\eta) \left(\frac{1}{2}(\partial_i  s_j({\bq},\eta) + \partial_j  s_i({\bq},\eta)) - \frac{1}{3} \delta_{ij} \partial_k  s_k({\bq},\eta)\right)\,.
\label{eq:quadr_response}
\eeq
As usual, counter-terms for the $\ell$-loop divergences are evaluated at $(\ell-1)$-loop order. Hence, the expression in \eqref{eq:quadr_response} is sufficient to one-loop order. In addition, for the background piece we introduce another length scale,
\beq
\label{eq:qijS1}
\langle Q^{ij}({\bq},\eta)\rangle_S \equiv l_S^2(\eta) \frac{1}{3} \delta_{ij}\ .
\eeq
Similarly, the dependence on long-wavelength perturbations for the binding potential, $V_S$, induces a response function for the acceleration which also has a time-dependent coefficient \cite{left}
\beq
\label{ashortresp}
\ba^i_{S}(\bz_L({\bq},\eta)) = \frac{3}{2} {\cal H}^2 \Omega_M ~l_{\Phi_S}^2(\eta)  \partial_\bq^i ( \partial_\bq \cdot \bs_L(\bq,\eta))\,.
\eeq
Notice that the symmetries precluded a term proportional to $\partial_\bq^2 \bs^i_L$ in \eqref{ashortresp}. See \cite{Carrasco:2013mua} for an implementation of this procedure in the Eulearian approach~to~two-loops.

\section{Renormalization \& Composite Operators}\label{sec:LEFTren}

\subsection{Quadrupole Moment}\label{sec:LEFTrQ}
\vskip 4pt

\begin{figure}[t!]
\centering
\includegraphics[width=0.5\textwidth]{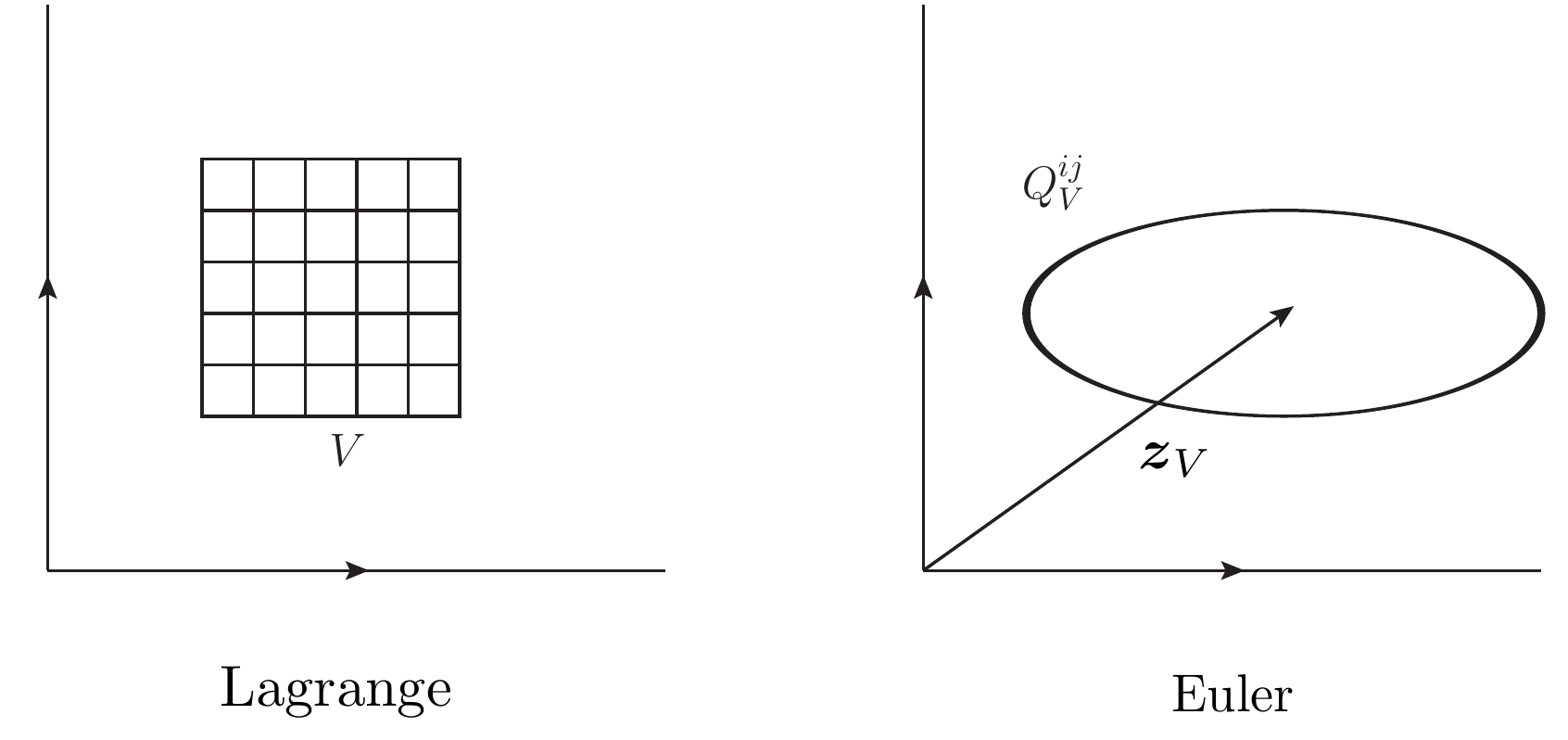}
\caption{ Left panel: a region of size $V$ in Lagrangian space containing several cells of size $R_0$. Each cell evolves with its own quadrupole moment $Q_{R_0}^{ij}(\bq,\eta)$, as shown in Fig. \ref{figL1}. Right panel: The same region in Eulerian space.} 
\label{figL2}
\end{figure}

Let us consider the quadrupole moment of a volume $V$ in Lagrangian space, determined by a set of smoothed regions, see Fig. \ref{figL2},
\beq
\label{eq:qijv}
Q_{V}^{ij}(\eta) = \frac{1}{V} \int_V d^3 \bq~(\bz_L^i(\bq,\eta)-\bz^i_V)(\bz_L^j(\bq,\eta)-\bz^j_V) + Q_{R_0}^{ij}(\bq,\eta) \ . 
\eeq
The center-of-mass is defined as, 
\beq
\bz_V(\eta) = \frac{1}{V} \int_V d^3 \bq~\bz_L(\bq,\eta)\ .
\eeq
While the $Q_{R_0}^{ij}$ depend on a smoothing scale, we expect the $Q^{ij}_V$ to be a well defined measurable quantity. That is the case provided we remove the dependence on $R_0$. From \eqref{eq:qijv} we obtain
\beq
\label{qvij}
Q_V^{ij} + \bz_V^i \bz_V^j  =\frac{1}{V} \int_V d^3 \bq ~\left[\left(\bs^i_L (\bq,\eta) \bs^j_L(\bq,\eta) + Q_{R_0}^{ij}(\bq,\eta)\right) +  \bq^i \bs_L^j(\bq,\eta)+ \bq^j \bs_L^i(\bq,\eta) + \bq^i \bq^j\right]\ .
\eeq
The left-hand side is cutoff independent and therefore the same must be true for the right-hand side. This means the expectation value, and correlation with a long-wavelength displacement perturbation, must be independent of~$R_0$.\vskip 4pt Let us start by taking the expectation value on the background of the short modes,
\beq
\langle Q_V^{ij} + \bz_V^i \bz_V^j \rangle_S = \frac{1}{V} \int_V d^3 \bq \left\langle \bs^i_L (\bq,\eta) \bs^j_L(\bq,\eta) + Q_{R_0}^{ij}(\bq,\eta)\right\rangle_S+l_V^2{}_{ij} \, ,
\eeq
where  $l_V^2{}_{ij}=\int_V d^3 \bq\; \bq^i \bq^j$ is a geometric factor associated to the Lagrangian volume $V$. Using \eqref{eq:qijS1}, we require
\beq
\left\langle \bs^i_L (\bq,\eta) \bs^j_L(\bq,\eta)\right\rangle_S + l^2_S \frac{\delta_{ij}}{3} \,,
\eeq
to be cutoff/smoothing independent. For example, for a scaling universe 
\beq
\left\langle \bs^i_L (\bq,\eta) \bs^j_L(\bq,\eta)\right\rangle_S =  \frac{1}{3} \delta_{ij} A \int^\Lambda_0 dk \frac{4\pi^2}{(2\pi)^3} k^n= \frac{1}{3} \delta_{ij} \int^\Lambda_0 dk \frac{k^n}{k_{\rm NL}^{n+3}(\eta)}=\frac{1}{3} \delta_{ij} l_\Lambda^2(\eta) \ ,
\eeq
where $l_\Lambda$ is define in \eqref{eq:lLam} with $\Lambda =R_0^{-1}$. Then, splitting $l_S = l_{S,\rm ren}(\eta) + l_{S,\rm ct}(\eta)$, we fix the counter-term
\beq
\label{lsct}
l_{S,\rm ct}^2(\eta) = - l_\Lambda^2(\eta)\ .
\eeq
Furthermore, the response must also be cutoff independent when correlated with the displacement, 
\bea\label{eq:quad_res}
&&\hspace{2cm}\left\langle \bs_L^k(\bq,\eta ) \left( Q_V^{ij}(\eta) + \bz_V^i(\eta) \bz_V^j(\eta)\right)\right\rangle =\\ \nonumber
&& \left\langle \bs_L^k(\bq,\eta) \;\times\;\frac{1}{V} \int_V d^3 \tilde\bq \left[ \bs^i_L (\tilde \bq,\eta) \bs^j_L(\tilde\bq,\eta) + \left(Q_{R_0}^{ij} (\tilde\bq,\eta)\right)_{\cal R} \right]\right\rangle\\ \nonumber
&& +  \frac{1}{V} \int_V d^3 \tilde \bq\left[\tilde\bq^i \left\langle \bs^k_L(\bq,\eta)\bs^j_L (\tilde\bq,\eta)\right\rangle +\tilde\bq^j \left\langle \bs^k_L(\bq,\eta)\bs^i_L (\tilde\bq,\eta)\right\rangle\right]\, .
\eea
The second line entails the renormalization of the displacement field, which we review in sec.~\ref{sec:dispL}. The first line involves the product of the displacement at {\it the same point}, or a `composite operator' \cite{left},
\beq
\label{compositeqij}
\left\langle s_L^k(\bq,\eta) \left[s^i_L (\tilde\bq,\eta) s^j_L(\tilde\bq,\eta) + \left(Q_{R_0}^{ij}(\tilde\bq,\eta)\right)_{\cal R} \right]\right\rangle\ .
\eeq
We require this combination to be cutoff independent. From the results in sec.~\ref{sec:pitfalls} we find
\beq
\left\langle \bs^l(\bk,\eta) \int_\bp s_i( \bp- \bk,\eta) s_j(-\bp,\eta)\right\rangle'\,  \toUV ~ i  \bk^2 P_L(k)  \frac{4}{35} l_\Lambda^2(\eta) \left( -\frac{\bk^i \bk^j \bk^l}{\bk^6} +  2\delta_{ij}\frac{\bk^l}{\bk^4}\right).
\eeq
(We removed the overall delta function, denoted with the prime.) The divergence must be absorbed into the correlation with the response \eqref{eq:quadr_response}, 
 \beq
\label{ctqij}
\left\langle \left(Q^{ij}_{R_0}\right)_{\cal R}(\bk,\eta) \bs^l(-\bk,\eta)\right\rangle' = i P_L(k) \bk^2 \left( \frac{\bk^l}{\bk^4} \frac{1}{3} (l_{T}^2(\eta)-l_{TF}^2(\eta)) \delta_{ij} + l_{TF}^2(\eta) \frac{\bk^i \bk^j \bk^l}{\bk^6}\right),
\eeq
with suitable counter-terms. The following choices remove the cutoff dependence: 
\beq\label{eq:counter_sol}
l_{TF,{\rm ct}}^2(\eta) = \frac{4}{35} l_\Lambda^2 (\eta)\,,~ l_{T,{\rm ct}}^2(\eta) = -\frac{4}{7} l_\Lambda^2(\eta)\, .
\eeq
The final (finite) answers still depends on renormalized parameters, $l_{S,T,TF,\rm ren}(\eta)$. As we discussed in~sec.~\ref{smatch} these are obtained via a matching procedure at a given $|\bk| < k_{\rm NL}$, either by comparison with numerical simulations or observations. From the universality of the Wilson coefficients we can thus predict the value of correlation functions at another scale. After the smoothing/cutoff is removed ($R_0 \to 0$) the matching wavelength, $|\bk|$, determines the typical length-scale, i.e. $V^{1/3} \sim |\bk|^{-1}$. In~this case, in order to keep the lowest order contributions from the multipoles we must work in the $|\bk|$-soft limit. Alternatively we could fix a finite smoothing size, $R_0 \simeq V^{1/3} \gtrsim k_{\rm NL}^{-1}$, which then serves as the physical matching scale, and measure directly the quadrupole moment say from numerical simulations. One can then use renormalization group techniques to study the scale dependence on the smoothing scale toward longer distances.\,\footnote{~This is equivalent to obtaining the renormalization group equations by studying the (in-)dependence of physical quantities on the cutoff/smoothing scale, see e.g. \cite{zee}.} 

\subsection{Displacement}\label{sec:dispL}
\vskip 4pt
For the displacement field it is useful to introduce a {\it source} term, ${\cal F}_{\rm LEFT}(\bq,\eta)$, as~\cite{left}
\beq
\label{source}
\ddot  \bs_L^i(\bq_1,\eta) + {\cal H} \dot \bs_L^i(\bq_1,\eta) + \frac{3}{2} {\cal H}^2 \Omega_M \int_\bk  {i \bk^i \over \bk^2 } \int d^3 \bq_2 \;e^{ i \bk \cdot (\bq_1 -\bq_2)}\;  i\,\bk^j \left(\bs^j_L(\bq_2,\eta)-\bs^j_L(\bq_1,\eta)\right) = {\cal F}_{\rm LEFT}^i(\bq_1,\eta)\ ,
\eeq
which we then use compute the $(n)$-order displacement in LEFT, i.e.
\beq
\bs^{(n)}_L(\bq,\eta) = \int d\eta^\prime G(\eta,\eta^\prime) {\cal F}_{\rm LEFT}^{(n)}(\bq,\eta^\prime)\ ,
\eeq
in terms of the Green's function associated with the linear theory. The source includes the non-linearities in \eqref{boxed0}-\eqref{boxed02} from LPT, together with the multipole moments and binding potential in LEFT, see \eqref{eq:boxed1}-\eqref{eq:PoissonL}. The latter are split into pieces, as in \eqref{eq:qsr}, which are incorporated into the source term,
\beq
\label{sourceLEFT}
{\cal F}_{\rm LEFT}(\bq,\eta) = {\cal F}_{\rm LPT}(\bq,\eta) + {\cal F}_S (\bq,\eta)+ {\cal F}_{\cal R}(\bq,\eta) +{\cal F}_{\cal S}(\bq,\eta)\,.
\eeq
The displacement is renormalized provided ${\cal F}_{\rm LEFT}(\bq,\eta)$ has a finite expectation value and correlation function with other displacement fields, order by order in perturbation theory. We discuss the power counting rules at the end of the section. We start with the renormalization due to the background and response and later include the stochastic term. For the sake of notation we drop most of the `LEFT' labels  in what follows, as well as the explicit time dependence, unless otherwise noted.

\subsubsection{Background and Response Terms}
\vskip 4pt

\begin{figure}[t!]
\centering
\includegraphics[width=0.2\textwidth]{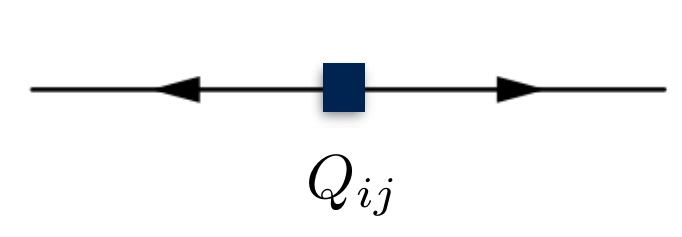}
\caption{Diagrammatic representation of a quadrupole insertion in LEFT, either response or background. To one-loop order the counter-terms are evaluated at tree-level.}
\label{CQ}
\end{figure}

Let us perform an explicit computation to one-loop order. We start by expanding the source \cite{left},
\bea
&&{\cal F}^l(\bq_1) = \ba^l_S(\bq_1) +  \frac{3}{2} {\cal H}^2 \Omega_M  \int_\bk   {i \bk^l \over \bk^2 } \int d^3 \bq_2 \exp[ i \bk \cdot (\bq_1 -\bq_2)] \times \\ 
&&\qquad\qquad\times \left\{  \tfrac{1}{2}(i\bk_i)(i\bk_j)\big[ s_i(\bq_1) s_j(\bq_1) +Q_{ij}(\bq_1)+s_i(\bq_2) s_j(\bq_2)+ Q_{ij} (\bq_2)- 2 s_i(\bq_1) s_j(\bq_2)\big]  \right.   \nn \\ 
&&\qquad\qquad\qquad +   \frac{1}{6}(i \bk_i)(i \bk_j) (i \bk_r) \left[ 3 s_r(\bq_1) (s_i(\bq_2) s_j(\bq_2)+Q_{ij}(\bq_2))- 3 s_r(\bq_2) (s_i(\bq_1) s_j(\bq_1)+Q_ {ij}(\bq_1))\right.\nn\\ 
&&\qquad\qquad\qquad+ \left.\left.  \left(s_i(\bq_1) s_j(\bq_1) +3 Q_{ij}(\bq_1)\right) s_r(\bq_1) - \left(s_i(\bq_2) s_j(\bq_2) +3 Q_{ij}(\bq_2)\right) s_r(\bq_2) \right]\right\} + \cdots \,.\nn \label{forcep}
\eea
We immediately recognize the factors of $(\bs_i(\bq_a) \bs_j(\bq_a)+Q_{ij} (\bq_a))$ from sec.~\ref{sec:LEFTrQ}. This combination produces finite results after the counter-terms are chosen as in \eqref{lsct} and \eqref{eq:counter_sol} \cite{left}, see fig.~\ref{CQ}. There are, however, remaining terms which are potentially UV sensitive. One such term is in the second line of \eqref{forcep}, which after correlating with the displacement reads (in Fourier space)
\beq
\label{newint}
\langle {\cal F}_l \bs_k\rangle' (\bk) \supset -\frac{3}{2}{\cal H}^2 \Omega_M \int_\bp  {i \bp_l \over \bp^2 } (i\bp_i)(i\bp_j)\; i\, C_{ijk}(\bp,-\bk,\bk-\bp)\ .
\eeq
Notice that the integral in \eqref{newint} is slightly different from what we found for the quadrupole renormalization. It is easy to show the above term is equivalent to a correlation, in $\bq$-space, with
\beq 
{\cal F}_l(\bq)\supset \calo_l (\bq) \equiv  \bs_j(\bq) { \partial_l \partial_j \partial_k\over \partial^2} \bs_k(\bq)\ . 
\eeq
Then, the divergent part in \eqref{newint} becomes \cite{left}
\bea
\langle {\cal O}_l\bs_j\rangle'(\bk) =  \frac{3}{2}{\cal H}^2 \Omega_M \bk^2 C_{lj}^{(11)}(\bk,-\bk) \left( \frac{2}{7} l_\Lambda^2\right).
\eea
There is another source of divergences which arises from the cubic terms in the third line of~\eqref{forcep}. These~`crossed' terms, for instance
\beq
\int_\bk   {i \bk^l \over \bk^2 } \int d^3 \bq_2 \exp[ i \bk \cdot (\bq_1 -\bq_2)](i \bk_i)(i \bk_j) ( i\bk_r)\langle \bs_m (\bq) \bs_i (\bq_2)\rangle \langle \bs_r (\bq_1) \bs_j(\bq_2)\rangle\ ,
\eeq
are not renormalized by the quadrupole moment, which instead gives
\beq
\langle \bs_l (\bq) \bs_k (\bq_1)\rangle \langle \bs_i (\bq_2) \bs_j(\bq_2) + Q_{ij}(\bq_2)\rangle \to~{\rm finite}\ .
\eeq
The divergent contribution becomes
\beq
\langle {\cal F}_l\bs_j\rangle' (\bk) \supset  \frac{3}{2}{\cal H}^2 \Omega_M  C^{(11)}_{ij}(\bk,-\bk) \int_\bp\left[  {(\bp_l-\bk_l)(\bp_r-\bk_r)(\bp_m-\bk_m)(\bp_k-\bk_k) \over |\bp-\bk|^2 } - {\bp_l \bp_m \bp_k \bp_r \over \bp^2 }  \right] C_{kr}^{(11)}(\bp,-\bp)\ ,
\eeq
which, after isolating the UV part, reduces to \cite{left}
\beq
\label{eq:div_3}
\langle {\cal F}_l \bs_j\rangle' (\bk) \supset  \frac{3}{2}{\cal H}^2 \Omega_M  \bk^2 C^{(11)}_{jl}(\bk,-\bk) \left(\frac{13}{15} l_\Lambda^2\right) \, .
\eeq
Similarly to what we discussed in sec.~\ref{sec:renCVCR}, the above analysis suggests we need a new counter-term. The latter is due to the acceleration induced by the response of the potential $V_S$ on short scales, see \eqref{ashortresp},
\beq
\langle \ba^i_{S,\cal R}\bs^j\rangle(\bk) = -\frac{3}{2} {\cal H}^2 \Omega_M  l_{\Phi_S,\rm ct}^2 \bk^2 C^{(11)}_{ij}(\bk,-\bk)\ ,
\eeq
and we find
\beq
 l_{\Phi_S,\rm ct}^2= \left(\frac{13}{15}+\frac{2}{7}\right)~ l_\Lambda^2  = \frac{121}{105}l_\Lambda^2\, .
 \eeq
This choice guarantees the displacement is renormalized and remains finite as we remove the cutoff.\vskip 4pt Notice we could have started directly with the displacement field, in which case only a combination of counter-terms are fixed by renormalization. This combination may then be determined by matching for the displacement alone. The fact that we are demanding our theory to produce finite results also for the multipole moments, which we can measure/observe, led us to the new counter-term. This was readily at hand in LEFT, yet an explicit computation forces it upon us as a consistency requirement in the long-wavelength effective theory. 
 
 \subsubsection{Stochastic Term}
\vskip 4pt
There is another type of divergence which is not removed by a counter-term from the response. It is due to the contribution from $C^{(22)}_{ij} (\bk,-\bk)$, see sec.~\ref{sec:pitfalls} and Fig.~\ref{lss1loop}. The reason this requires an extra counter-term is simply because, unlike the response, it is not proportional to $P_L(k)$. As we see in diagrammatic form, it is equivalent to a `self-energy' contribution to the two-point function. By inspection of the integral in \eqref{eqQ12} it enters as  $k^2 \Lambda^{2n-1}$ in the displacement. The renormalization procedure is straightforward. The equation for the displacement includes already a contribution from ${\cal F}_{\cal S}(\bq,\eta)$, see \eqref{sourceLEFT}. The divergence in $C^{(22)}(\bk,-\bk)$ is then removed by a (contact) counter-term
\beq
\label{eq:stoca}
\left\langle \bs^{i}_{\cal S,\rm ct} \bs^j_{\cal S,\rm ct}\right\rangle \propto \bk^i \bk^j \Lambda^{2n-1}\,.
\eeq
The stochastic term has zero expectation value, but accounts for departures from the average in the correlation functions, see e.g. \cite{disfluid,eftinfdis,eftinfdis2}. At the end of the day we find
\beq
\bk^i \bk^j C_{ij}^{\cal S}(\bk,-\bk) \propto  k^4\,,\label{eq:Sk4}
\eeq
for the mass-density power spectrum (see below). This term can be obtained by matching, either into numerical simulations or directly by observation, reproducing the above scaling behavior. Notice it may be neglected in the soft limit $k \ll k_{\rm NL}$. However, as we move toward the non-linear scale the stochastic terms may be important and compete with the contributions from the response, scaling as~$k^2 P_L(k)$, e.g.~\cite{Baldauf:2015tla}.

\subsection{Mass-Density}\label{sec:MassD}
\vskip 4pt

We have succeeded in renormalizing the theory in Lagrangian space, including multipole moments and displacements. The mass-density field in \eqref{eq:deltamL}, on the other hand, is a derived quantity. It features not only the number-density, defined through \eqref{mapqx} for the long-wavelength perturbations, but also contributions from the multipole moments. This means the correlation functions for the mass-density must be finite, to one-loop order, with the choices of counter-terms already made.\vskip 4pt 

The mass-density correlation function can be written as, see \eqref{eq:PoissonL} and \eqref{eq:displacement_fourier2},
\beq
\label{densityre}
\langle\delta_{M}(\bk,\eta_1)\delta_{M}(\bk,\eta_2)\rangle'=\int d^3 \bq\; e^{-i\bk\cdot \bq}  \left\langle e^{-i\, \bk\cdot (\bs(\bq,\eta_2)-\bs({\bf 0},\eta_1))-\frac{1}{2}\bk^i \bk^j (Q^{ij}_c(\bq,\eta_2) + Q^{ij}_c({\bf 0},\eta_1))+\cdots}\right\rangle\ ,
\eeq
with $\bq = \bq_2 - \bq_1$ and $\br=|\bx_2-\bx_1|$. By Taylor expanding the exponential and keeping terms which contribute to one-loop order, we find that the necessary counter-terms combine into \cite{left}
\be
P_{1\ell} (k)\supset - \frac{1}{3}l^2_{\delta_M,\rm ct} \bk^2 P_L(k)\,,
\ee
with
\beq
\label{eq:l2dm}
l^2_{\delta_M} \equiv  l^2_{\theta} +  l_S^2+l_T^2 +2 l^2_{TF}\ .
\eeq 
The $l^2_{S,T,TF,{\rm ct}}$ counter-terms precisely cancel the divergences from composite operators which appear in the Taylor expansion, combining once again into factors of $(\bs^i\bs^j+ Q^{ij})$. There is another term in \eqref{eq:l2dm},~$l^2_{\theta}$, defined as
\beq
\label{eq:thetact}
l^2_{\theta} \equiv \frac{1}{6}\left( l_T^2 + 2 l_{TF}^2 +2l_S^2+ 6l_{\Phi_S}^2\right),
\eeq
which appears from the relationship between density and displacement, 
\beq \delta_M(\bk,\eta) = -\partial^i_\bq \bs_i(\bq,\eta) \equiv - \theta(\bk,\eta)+ \cdots\,.\label{delft}\eeq
The equation of motion for $\theta$ reads \cite{left}
\beq
\ddot \theta(\bq,\eta) + {\cal H}\dot \theta(\bq,\eta) -\frac{3}{2}{\cal H}^2\Omega_M \theta(\bq,\eta) = \label{thetaeq}
 \frac{3}{2}{\cal H}^2\Omega_M \, l^2_{\theta}\,\partial_\bq^2 \theta(\bq,\eta) +\cdots\, .
\eeq
Therefore, in order to renormalize the mass-density our previous choices of counter-terms must combine in $l^2_{\theta,\rm ct}$, as in \eqref{eq:thetact}, to cancel the cutoff dependence we find at one-loop order,
\beq\label{eq:31lpt}
\langle \theta^{(1)}\theta^{(3)}\rangle' (\bk)= \bk^i \bk^j C^{(13)}_{ij}(\bk,-\bk) ~\toUV~\frac{8}{63} l_\Lambda^2 \bk^2 P_L(k)\, .
\eeq
To show that is the case we first solve \eqref{thetaeq} using the Green's function for an EdS universe,
\beq
G(a,a')=-\frac{2}{5}\frac{1}{{\cal H}^2 a} \left[\frac{a}{a'}-\left(\frac{a}{a'}\right)^{-3/2}\right]\ .
\eeq
Since the counter-term is evaluated at linear order, $l_{\theta,\rm ct}(\eta) \propto l_\Lambda(\eta) \propto D^2(\eta) = a^2(\eta)$. Hence, using $\theta^{(1)} \propto a(\eta)$, we have
\beq
\theta_{\rm ct}(\bq,\eta) = \frac{1}{6} l^2_{\theta,\rm ct}(\eta_0) \,a^2(\eta)\,\partial^2 \theta^{(1)}(\bq,\eta)\ .
\eeq
From here we obtain
\beq
\langle \theta^{(1)} \theta_{\rm ct}\rangle' (\bk) =- \frac{1}{6}l^2_{\theta,\rm ct}(\eta) \; \bk^2 P_L(k) \ ,
\eeq
and using the values for the counter-terms we found before, we arrive at
\beq
\frac{1}{6} l^2_{\theta,\rm ct} = \frac{1}{36}\left(-2-\frac{4}{7} + \frac{8}{35} -6\times \frac{121}{105} \right) l_\Lambda^2 =\frac{8}{63} l_\Lambda^2\ ,
\eeq
which precisely cancels the divergence in \eqref{eq:31lpt}, as advertised.\vskip 4pt

In summary, we have shown that all the parameters in LEFT can be adjusted to renormalize the theory. As we notice for the case of the mass-density, or the displacement, the coefficients in LEFT may enter in different linear combinations. That is the case provided the fields are not connected by a conservation law.\,\footnote{~In Euler space up to two-loops~\cite{Carrasco:2013mua} only one parameter is necessary for the mass-density field, i.e. the `sound speed' \cite{eftfluid,eftlss}. The same parameter also enters in the computation of correlations of the momentum, since the two are connected by matter conservation.  However, if one is interested instead in the velocity, new parameters are required. That is the case because $\bv \equiv \boldsymbol{\pi}/\rho$, which is a composite operator \cite{left,Mercolli:2013bsa,Carrasco:2013mua}.} Therefore, as long as we limit ourselves to a subset of possible observables, only partial information may be needed. Nevertheless, the consistency of the theory imposes also finite results for the multipole moments, which require an independent set of measurements. 

\subsection{Power Counting}\label{sec:powerL}
\vskip 4pt

After the theory is renormalized the remaining step in order to make predictions is to establish how many coefficients are necessary to a given order, namely the power counting rules. The first terms in the derivative expansion in LEFT scale as $l_{A, {\rm ren}}^2 k^2 P_L$ (for $A=\{S, T,TF,\Phi_S\}$). For instance, for the mass-density power spectrum in a scaling universe we find,
\beq
\Delta^2_{l_{A,{\rm ren}}}(k) = \gamma_{A,{\rm ren}} \left(\frac{k}{k_{\rm NL}}\right)^{n+5}\,,
\eeq
where $\gamma_A \equiv k_{\rm NL}^2l_A^2$ is a dimensionless parameter, and we introduced the notation
\beq
\label{eq:powerD2}
\Delta^2(k) \equiv \frac{k^3}{2\pi^2} P_L(k) = \left(\frac{k}{k_{\rm NL}}\right)^{n+3}\,.
\eeq 
On the other hand, the (renormalized) $N$-loop order in perturbation theory scales as~\cite{Pajer:2013jj,Carrasco:2013sva} 
\beq
\Delta^2_{(N)}(k) \propto \left(\frac{k}{k_{\rm NL}}\right)^{(n+3)(N+1)} \quad \to\quad \Delta^2_{(1)} \propto  \left(\frac{k}{k_{\rm NL}}\right)^{2n+6},~~{\rm etc}.
\eeq
It is then straightforward to show higher order terms in LEFT enter as
\beq\label{eq:counter_contrib}
\Delta^2_{Q,a_S} (k) \propto  \left(\frac{k}{k_{\rm NL}}\right)^{(n+3)(N+1)+2p}, 
\eeq
to $N$-loop order in perturbation theory. Here $p$ is related to the number of derivatives involved, either in the coupling to the multipole moments or through response functions. (Since we have interactions which are local each additional derivative contributes a factor of $(k/k_{\rm NL})^{2p}$ in the power spectrum.) The counting is complete with the stochastic term, see \eqref{eq:Sk4}, which scales as \cite{eftfluid,eftlss,Pajer:2013jj,Carrasco:2013mua} 
\beq
\label{noise}
\Delta^2_{\cal S} \propto \left(\frac{k}{k_{\rm NL}}\right)^7 \,. 
\eeq
These rules then determine which terms contribute to the computation of the power spectrum to a given order, provided $k < k_{\rm NL}$. Notice that, depending on the value of $n$, different terms are more important than others. For instance, when $n> -1$ the correction from $\Delta^2_{l_A}(k)$ dominates over the one-loop result and are comparable for $n=-1$, assuming $\gamma_{A,\rm ren} \simeq {\calo (1)}$.\vskip 4pt

Since a cutoff introduces a scale, the power counting applies {\it only} in terms of renormalized parameters. That is after counter-terms have been added and cutoff (or smoothing scale) is safely removed. This~is an important point. When a cutoff is used to regularize the loop integrals the (power-law) divergent piece and counter-term enter at the same order, as it is required to produce finite results. However, this should not mislead us into thinking that both, the renormalized loop contribution and LEFT parameter(s), wound up at the same order. In fact, for a scaling universe we could have used dim. reg., as we discussed at length in part \ref{sec:part2}.  In that case power-law divergences are discarded, while incorporating instead the non-analytic behavior. The power counting rules apply in dim. reg., but the renormalized parameter may not depend on scale, only when logarithms are present. That is when both contributions become of the same order.\vskip 4pt 

For example we find a logarithmic divergence at $n=-1$, when both the one-loop and renormalized contributions from LEFT are comparable \cite{left}. In that case we have, 
\beq
\Delta_{(1)}^2(k) = k^3 (\bk^i  \bk^j) C_{(1\ell)}^{ij}(\bk,-\bk) = \frac{k^2}{k_{\rm NL}^2} \left\{1 +  \frac{k^2}{k_{\rm NL}^2} \left( {\rm finite} - \frac{1}{6} l^2_{s,\rm ren}(\mu)\; k_{\rm NL}^2 -\frac{4}{63} \log (k/\mu)\right)\right\}\label{cij}\ ,
\eeq
after an $1/\epsilon$ pole is removed by a counter-term. The $l_{s,\rm ren}^2$ is a (combined) renormalized parameter which depends on an arbitrary renormalization scale $\mu$ (see \cite{left} for more details). The renormalization group equation in this case is somewhat trivial, 
\beq
\mu \frac{d}{d\mu} l_{s,\rm ren}^2 (\mu) =  -\frac{24}{63} k_{\rm NL}^{-2}\, ,
\eeq
but it tells us nonetheless the expected typical size for the renormalized parameter, i.e. \beq l_{s,\rm ren}^2(\mu \simeq k_{\rm NL}) \simeq k_{\rm NL}^{-2}\,.\eeq 
On the other hand, when $n> - 1$, at one-loop order we find power-law divergences. If a cutoff is implemented, the highest power of $\Lambda$ scales as $l_\Lambda^2 \propto \Lambda^{n+1}$, see \eqref{eq:lLam}. As we discussed, a series of counter-terms are needed to remove the dependence on $l_\Lambda^2$. Nevertheless, from the previous scaling rules we notice that --once the divergences are removed-- the contribution from $l^2_{s,\rm ren}$ dominates over the one-loop result. However, the $l^2_{s,\rm ren}$ coefficient becomes scale independent, due to the missing $k^2 \log k$ term for $n>-1$, unlike the $n=-1$ case. In general, logarithmic divergences at one-loop order appear when $n=-1+2m$, with $m$ a positive integer. (Similarly for the stochastic term when $n=1/2 + m$, e.g. \eqref{eq:stoca} \cite{left,Pajer:2013jj}.) As we notice from \eqref{eq:counter_contrib}, these are renormalized into higher multipole moments, and higher powers of derivatives in response functions. For example, for $n=1$, we get a correction scaling as $k^4P_L \log k/\mu$ (together with a $1/\epsilon$ pole removed by a counter-term). The renormalized one-loop result would be subleading with respect to the leading order term(s) in LEFT, scaling as~$k^2P_L$, but of the same order as the $k^4 P_L$ contributions.\vskip 4pt 

This discussion suggests that, unlike the cutoff regularization which mixes up different orders in the $k/k_{\rm NL}$ expansion, using dim. reg. --at least in a scaling universe-- neatly isolates the relevant pieces form the loop diagrams. Unfortunately one cannot simply implement dim. reg. for the case of our universe, which presents many different relevant scales. Hence, a cutoff remains the chosen regularization scheme. Therefore, only after counter-terms are judiciously added one obtains the power counting rules described above. It~is possible, nonetheless, to approximate relatively well the real universe --modulo the BAO oscillations-- with a piece-wise power spectrum of the form \cite{Carrasco:2013mua}
\beq
P_L(k) = (2\pi)^3 \Bigg\{ \begin{array}{c} \frac{1}{k_{\rm NL}^3} \left(\frac{k}{k_{\rm NL}}\right)^{-2.1}~{\rm for}\, k > k_{\rm tr} \\  \frac{1}{\tilde k_{\rm NL}^3} \left(\frac{k}{\tilde k_{\rm NL}}\right)^{-1.7}~{\rm for}\, k < k_{\rm tr}\,,
\end{array}
\eeq
in the range $k \simeq 0.1 - 1\, h$ Mpc$^{-1}$, where $k_{\rm NL} \simeq 4.6 \, h$ Mpc$^{-1}$, $\tilde k_{\rm NL} \simeq 1.8 \, h$ Mpc$^{-1}$ and $k_{\rm tr} \simeq 0.25 \, h$ Mpc$^{-1}$.  This helps understand different contributions at two-loop order in the Eulerian approach. See the discussion in \cite{Carrasco:2013mua} for more details.

\subsection{A Non-Renormalization Theorem}

There has been considerable interest in back-reaction effects from short-distance inhomogeneities in the evolution of the long-distance universe. In particular, whether the former could be responsible for the late time cosmic acceleration, see e.g. \cite{Green:2014aga, Buchert:2015iva,Green:2015bma} and references therein. In \cite{eftfluid} a rather simple argument for the negative was presented. Let us consider a single virialized region of physical size $\ell_{\rm vir}$ centered at $\bx=0$. Let us work in a locally free-falling frame and introduce the quadrupole moment of a smoothed region, 
\beq
\label{eq:El0}
Q_{\Lambda}^{ij}(t) \equiv \int d^3{\bx}' W_{\Lambda}(\bx')\,{\cal T}^{00} (\bx',t){\bx}^\prime_i {\bx}^\prime_j\,,
\eeq
with $W_{\Lambda}(\bx')$ a window function in Euler space obeying $\Lambda^{-1} \gg \ell_{\rm vir}$. The Ward identity then implies (see \eqref{eq:magg}) 
\beq
\left[{\cal T}^{ij}\right]_{\Lambda}(t) \equiv \int d^3 {\bx}~W_{\Lambda}(\bx'){\cal T}^{ij} (\bx',t) = \frac{1}{2} \frac{d^2}{dt^2} Q_{\Lambda}^{ij}(t)\, ,
\eeq
up to corrections of order $\ell_{\rm vir}\Lambda \ll 1$. We now average over a Hubble time, and since $\ell_{\rm vir} \ll v H^{-1}$, we~find
\beq
\label{eq:El1}
\left\langle\left[{\cal T}^{ij}\right]_{\Lambda}(t)\right\rangle = \frac{1}{ H^{-1}} \left(\frac{d}{dt} Q^{ij}_\Lambda(H^{-1})-\frac{d}{dt} Q^{ij}_\Lambda(0)\right) \sim \frac{1}{H^{-1}} \,\rho\, v \, \ell_{\rm vir} \sim \rho v^2 \left(\frac{\ell_{\rm vir}}{v H^{-1}}\right) \ll \rho v^2 \,.
\eeq
Therefore, virialized structure `decouples,' since its contribution is suppressed compared to non-virialized sources, for which $p_{\rm ren} \sim \rho v^2 >0$ \cite{eftfluid}.\vskip 4pt

A similar argument can be made in LEFT and, moreover, within full fledged general relativity. Let us take a region in Lagrangian space of size $R_0$ with center-of-mass $\bz_L(\bq,\eta)$, which evolves into a virialized structure of size $\ell_{\rm vir}$ in Euler space.  Moreover, for an isolated region and momenta $|\bk|^{-1} \gg \ell_{\rm vir}$, we can furthermore remove the traces from the effective action in \eqref{eq:LEFTac}. Therefore, we obtain
\beq
\label{eq:TL}
\left[{\cal T}^{jk}\right]_L(\bk,\eta)=\left[{\cal T}^{jk}_{\rm pp} \right]_L(\bk,\eta) + e^{i \bk\cdot \bz_L(\bq,\eta)}\left[ \sum_\ell \frac{(-i)^{\ell-2}}{\ell!} \bk^{i_1} \ldots \bk^{i_{\ell-2}} \frac{d^2}{d\eta^2}I^{jk\,i_1\ldots i_{\ell-2}}(\bq,\eta)\right] +\cdots\,,
\eeq
where the $I^{i_1\cdots i_\ell}$ are the STF mass multipole moments and the ellipsis include similar contributions from the $J^{i_1\cdots i_\ell}$. The point-particle term, ${\cal T}^{jk}_{\rm pp}$, follows from \eqref{eq:actiondmL} and includes a renormalized density and pressure, with $p_{\rm ren} \simeq \rho_{\rm ren} v^2 > 0$, due to the kinetic and potential energy on short scales. 
\vskip 4pt
It is clear that, in the limit $|\bk| \ell_{\rm vir} \ll 1$, the expression in \eqref{eq:TL} is dominated by the lowest orders in the multipole expansion. We are left then with two extra terms incorporating the finite size of the region, at leading order in derivatives,
\beq
 \left[{\cal T}^{ij}\right]_L(\bk,\eta) = \left[{\cal T}^{ij}_{\rm pp} \right]_L(\bk,\eta) + e^{i \bk \bz_L(\bq,\eta)}\left(\frac{d^2}{d\eta^2} I^{ij}(\bq,\eta) + \frac{4}{3}  \bk^l \epsilon^{ikl} \frac{d}{d\eta} J^{jk}(\bq,\eta)\right)+ {\cal O} (|\bk|\ell_{\rm vir}) \,.
\eeq
The current term, however, is suppressed in the Post-Newtonian limit, $v \ll 1$. Hence, by taking an average over a Hubble time we recover the same result as before. There are nonetheless two minor differences. First of all the STF parts of the multipoles encode tidal effects, and thus are further suppressed since these are induced by curvature. Secondly, the dependence of $I^{ij}(\bq,\eta)$ on moments of ${\cal T}^{\mu\nu}(x)$ is slightly more complicated than in \eqref{eq:El0}, also for $J^{ij}(\bq,\eta)$, see sec. \ref{sec:gravradM}. Nevertheless, all these extra terms are subleading in the PN expansion, and may be neglected in the above analysis.\vskip 4pt

These arguments are naturally generalized when more than one region is incorporated, provided they do not overlap upon evolution. As we discussed in sec.~\ref{sec:longdistance}, when regions overlap we cannot longer remove the traces. In that case we may use a field redefinition, obtaining a Poisson equation which is sourced only by the STF multipoles, see \eqref{eq:PoissonL2}.\footnote{The transformation can be applied in full general relativity, using $R_{\mu\nu} \propto T_{\mu\nu}- \frac{1}{2} g_{\mu\nu} T$, to remove the Ricci tensor.} The physics remains unchanged, provided we use observable quantities to measure/define the expansion of the universe. (If the reader is not satisfied with this argument, it is straightforward to show the contribution from the traces is highly suppressed for virialized structures.)\vskip 4pt A~similar decoupling occurs also for correlation functions, such as the power spectrum. First of all, the new terms from LEFT are clearly suppressed in the soft limit, scaling as $l^2_{\rm ren} k^2 P_L(k)$ at leading order \cite{left}.  Furthermore, one can show that the renormalized coefficients themselves receive most of their contribution from modes near the non-linear scale, with small corrections from virialized structures~\cite{Garny}.

\section{Resummation}\label{sec:LEFTres}
In sec.~\ref{sec:powerL} we set up power counting rules in LEFT for a scaling universe. We did so in terms of the ratio $k/k_{\rm NL}$, which is equivalent to using the linear displacement (or mass-density) power spectrum. However, this is not satisfactory in our cosmology, mainly due to the various scales involved. While~this does not have major consequences in the way we handle the impact of non-linear scales, identifying the relevant parameters becomes essential to properly incorporate the imprint of soft(er) modes in the dynamics.

\subsection{Expansion Parameters}
\begin{figure}[t!]
\centering
\includegraphics[width=0.55\textwidth]{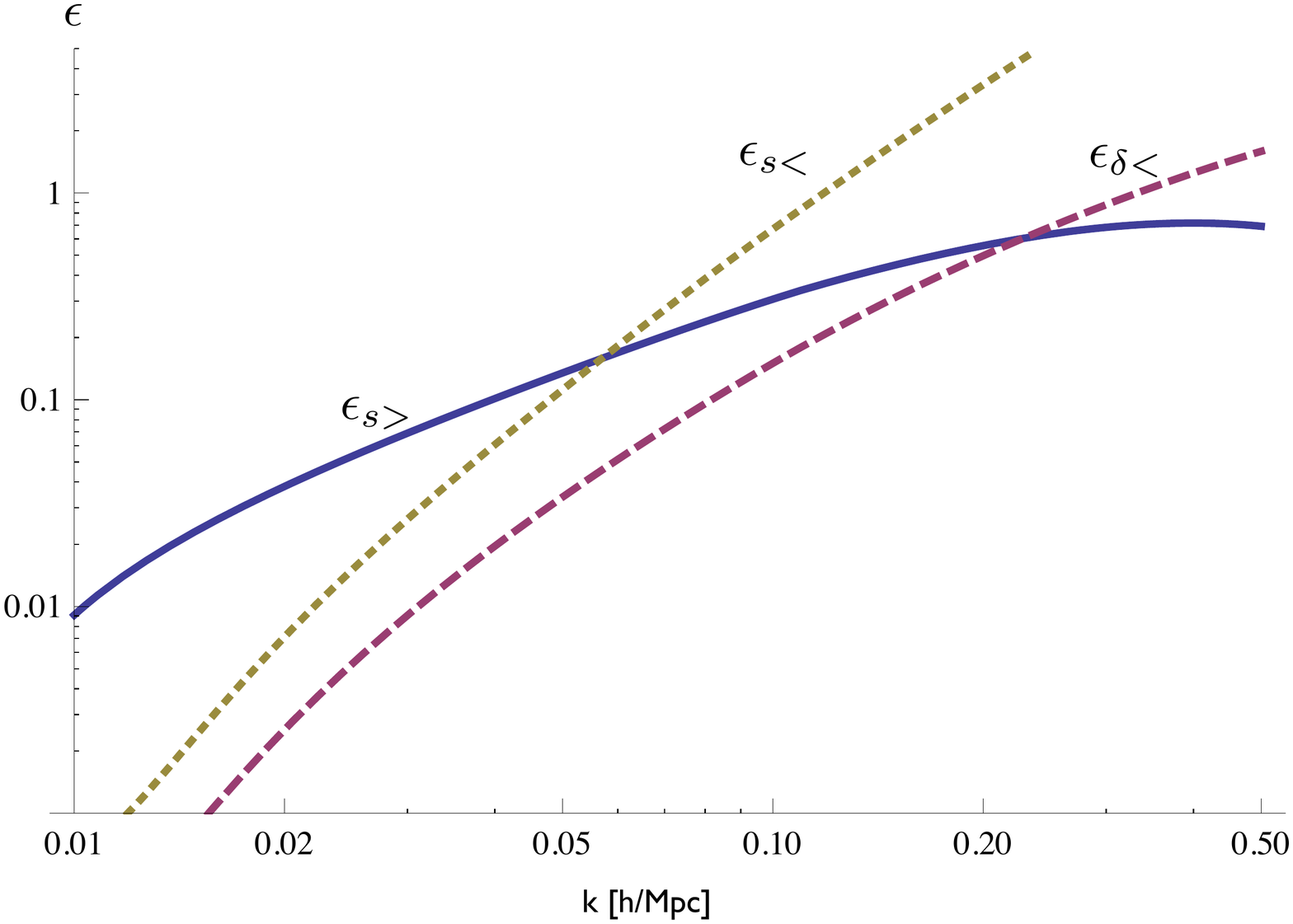}
\caption{\small  Parameters quantifying the relevance of modes longer ($\epsilon_{s<}(k)$) and shorter ($\epsilon_{s>}(k)$) than a given~$k$, and tidal effects ($\epsilon_{\delta <}(k)$), plotted for the real universe. (See the text.)}
\label{epsilons}
\end{figure}

\vskip 4pt For a given mode, $\bk$, different scales affect the dynamics in different ways. It is then convenient to split the various contributions to the mass-density power spectrum into three categories: 
\bea 
\label{eq:epsX}
\epsilon_{s>}(k) &=& \bk^i\bk^j \int_\bp  C_{ij}(\bp,-\bp,\eta)\,\theta(|{\bp}|-|{\bk}|)\,,\\
\epsilon_{s<}(k) &=& \bk^i\bk^j \int_\bp  C_{ij}(\bp,-\bp,\eta) \, \theta(|{\bk}|-|{\bp}|)\,,\nn \\
\epsilon_{\delta <} (k) &=& \int_\bp  P(p,\eta) \, \theta(|{\bk}|-|{\bp}|)\, . \nn
\eea 
\bit \im $\epsilon_{s>}(k):$ Variance of the displacements produced by modes with $|\bp|>|\bk|$.  
\im $\epsilon_{s<}(k):$ Variance of the displacements produced by modes with $|\bp|<|\bk|$. 
\im $\epsilon_{\delta <}(k):$ Variance of the density fluctuations produced by modes with $|\bp|<|\bk|$. 
\eit
There is in principle another parameter, $\epsilon_{\delta>}(k)$, which we could introduce. However, while $\epsilon_{\delta<}(k)$ accounts for tidal effects through the long-wavelength gravitational potential, the $\epsilon_{\delta>}(k)$ is a fictitious parameters. That is the case because mass conservation guarantees there is no dependence on such parameter in the final expressions. For~example, in perturbation theory to one-loop order, we find for the mass-density power spectrum
\bea
&&P_{1\ell}(k) \propto P_L(k) \, \epsilon^{L}_{\delta<}(k)\,~~~~ (\,k_{\ell} \ll k\,)\,, \\
&&P_{1\ell}(k) \propto P_L(k)\, \epsilon^{L}_{s>}(k)\,~~~~ (\,k_{\ell} \gg k\,)\,, \nn
\eea
where $k_{\ell}$ is the loop momenta and the $\epsilon^L$-parameters are defined with the linear power spectrum in~\eqref{eq:epsX}.
The $\epsilon_{s>}(k)$ is clearly responsible for the imprint of the UV modes on the long-distance universe, and is incorporated into the derivative expansion in LEFT, after renormalization. On the other hand, the displacements in $\epsilon_{s<}(k)$ do not enter directly in correlation functions at equal times.\,\footnote{~This is ultimately related to the equivalence principle, see e.g.~\cite{Carrasco:2013sva,Ben-Dayan:2014hsa,Garny}.} (In fact, in Euler space there is a cancelation between $P^{(22)}(k)$ and $P^{(13)}(k)$.) However, they change the final location for the short-distance scales. The $\epsilon_{s<}(k)$ is ultimately responsible for the broadening of the BAO peak.\vskip 4pt We plot the $\epsilon$-parameters for our cosmology in Fig.~\ref{epsilons}. Notice, for $k > 0.1\, h$ Mpc$^{-1}$, $\epsilon_{s<}(k)$ becomes rapidly the largest. Therefore, it must be properly included in order to achieve accurate results. 

\subsection{Soft Displacements to all Orders} 
\vskip 4pt
In LEFT we do not expand in  $\epsilon_{s<}(k)$, which is then naturally resummed in the final expressions. This has two main advantages. 
Firstly, we do not encounter IR divergences. That is not the case in the Eulerian approach, where intermediate IR divergences cancel out after all terms are added up, requiring a careful manipulation of the integrals, see e.g. \cite{Carrasco:2013sva,Blas:2013bpa}. Secondly, the resummation of $\epsilon_{s<}(k)$ incorporates the shift in the mass-density field, which as we mentioned is behind the smearing of the BAO peak.\vskip 4pt

The resummation is easier to identify in Euler space. The perturbative solution reads 
\bea
\dl(\bx) &=& \dl^{(1)}(\bx)+ \dl^{(2)}(\bx) + \dl^{(3)}(\bx) + \cdots \ ,
\eea
where, for instance to second order, 
\bea\label{delta2}
\dl^{(2)}(\bx)&=&d_i^{(1)} (\bx)\partial_i \dl^{(1)}(\bx)+ {17\over 21} \left(\dl^{(1)}(\bx)\right)^2 + \frac{2}{7} K_{ij}^{(1)}(\bx) K_{ij}^{(1)}(\bx)\ ,  \\
d_i^{(1)}(\bx)& =& - \int {d^3 k \over (2\pi)^3} {i k_i \over k^2} \, \dl^{(1)}(\bk)\, e^{i \bk\cdot\bx} \label{eq:d1L}\ , \\
K_{ij}^{(1)}(\bx)& =&  \int {d^3 k \over (2\pi)^3} \left({ k_i k_j\over k^2}-{1 \over 3}\delta_{ij}\right)  \dl^{(1)}(\bk) \, e^{i \bk\cdot\bx}\ \,.
\eea
Notice there are two types of terms.  The fact that the Eulerian perturbative approach expresses all quantities as a function of $\delta^{(1)}$ hides the fact that the last two terms in  $\delta^{(2)}$ entail two factors of $\delta^{(1)}$, whereas the first has a displacement, $d^{(1)}_i$, instead. The latter are contributions which are resummed in a Lagrangian approach. In other words, if we keep terms from displacements only, we have
\beq
\label{shifted}
\underbrace{\delta(\bx)}_{\rm disp.}=\delta^{(1)}+d_k^{(1)} \partial_k \delta^{(1)}+ {1\over 2} d_k^{(1)} d_l^{(1)} \partial_k\partial_l\delta^{(1)} + \cdots = \delta^{(1)}(\bx +  \boldsymbol{d} )\ ,
\eeq
where $\bd$ is given in \eqref{eq:d1L}. (This is also related to the eikonal approximation, e.g. \cite{Ben-Dayan:2014hsa}.) \vskip 4pt In Lagrangian space the resummation can be seen as a coordinate transformation, which is implicit in the relationship between density and displacement,
\beq
\label{eq:mapqx2}
1+\delta({\bx})=\left.\Big[{\rm det} \Big(1+\tfrac{\partial \bs_i}{\partial \bq_j}\Big)\Big]^{-1}\right|_{\bq= \bx - \bs(\bq)}\,.
\eeq
There are several terms which are non-linear in the leading order displacement. Either from expanding the determinant in powers of ${\partial \bs^{(1)}_i}/{\partial \bq_j}$, or from the $(n)$-order in LPT, ${\partial \bs^{(n)}_i}/{\partial \bq_j}$. Finally we have the coordinate transformation. These are the terms resummed in LEFT,
\beq {\partial \bs^i}/{\partial [\bq^j}(\bx -\bs)] =  {\partial \bs^i}/{\partial \bq^j} + \bs^l {\partial^2 \bs^i}/{\partial \bq^j\partial \bq^l} + \cdots\,.\eeq  
The displacement terms may dominate in various situations of interest, for instance, in the presence of a long-wavelength gravitational potential, $\Phi_L$. As a consequence of the equivalence principle such potential may be removed by a coordinate transformation, up to first order in derivatives. This is the type of scenarios where the resummation becomes important.\,\footnote{~The resummation is also relevant for scaling universes with $n<-1$. There is an IR enhancement, which also makes the integrals UV finite. As we emphasized this does not mean a renormalization procedure is not required.}\vskip 4pt As an example, the presence of the BAO peak in the correlation function, $\xi(x)$, makes these terms relevant. This can be seen by comparing the terms with a displacement, $d_L^2\, \xi^{\prime\prime}(x)$, relative to those containing higher powers of the density, $\delta_L^2 \xi(x)$, where $\delta_L$ and $d_L$ are produced by the long-wavelength mode. Because of the sharpness of the BAO feature, $x^2 \xi^{\prime\prime}(x)/\xi \sim 10^2$, the terms with a displacement are enhanced with respect to the dynamical ones. See appendix C in \cite{left}, also e.g. \cite{Bernardeau:2011vy,Baldauf:2011bh,Sherwin:2012nh,Kehagias:2013yd,Peloso:2013zw,Creminelli:2013mca,Creminelli:2013poa,Creminelli:2013nua,Schmidt:2013gwa,Valageas:2013zda,Kehagias:2013paa,Senatore:2014via,Ben-Dayan:2014hsa,Mirbabayi:2014gda,Baldauf:2015xfa} for more details on the treatment of IR modes.

\subsection{Exponentiation}
\vskip 4pt
The expression in LEFT for the correlation function in the long-wavelength limit reads
\beq
\label{kcumuleft}
1+\xi_L(\br) =  \int d^3 \bq~ \int_\bk e^{-i\bk\cdot (\bq - \br)} \,K_L(\bq,\bk)\,,
\eeq
where, see \eqref{densityre},
\beq
K_L(\bq,\bk,\eta)\equiv \left\langle e^{i\, \bk\cdot (\Delta \bs(\bq,\eta))-\frac{1}{2}\bk^i \bk^j (Q^{ij}_c(\bq_1,\eta) + Q^{ij}_c(\bq_2,\eta))+\cdots}\right\rangle\,,
\eeq
and $\Delta\bs(\bq,\eta) \equiv \bs_L(\bq,\eta)-\bs_L({\bf 0},\eta)$. Retaining up to the third moment, we write
\beq
\label{logkL}
\log K_L (\bq,\bk,\eta) = -\frac{1}{2} A_L^{ij}(\bq,\eta) \bk_i \bk_j - \frac{i}{6} W_L^{ijl}(\bq,\eta) \bk_i \bk_j\bk_l +  \ldots\ ,
\eeq
with the following correlation functions
\bea
A_L^{ij}(\bq,\eta) &\equiv& \left\langle \Delta \bs^i(\bq,\eta) \Delta \bs^j(\bq,\eta) \right\rangle_c + 2 \left\langle Q_c^{ij}(\bq,\eta)\right\rangle_S\ , \\
W_L^{ijl}(\bq,\eta) &\equiv& \left\langle \Delta \bs^i(\bq,\eta) \Delta \bs^j(\bq,\eta) \Delta \bs^l(\bq,\eta)\right \rangle_c +3 \, \Delta\left \langle  \bs^{(i} Q^{jl)}_{\cal R} \right\rangle (\bq,\eta)\ .
\eea
The last term is defined as: \beq \Delta\left \langle  \bs^i Q^{jk}_{\cal R}\right \rangle (\bq,\eta)  \equiv\left \langle  \bs^i(\bq,\eta) Q^{jk}_{\cal R}({\bf 0},\eta)\right\rangle -  \left\langle  \bs^i({\bf 0},\eta) Q^{jk}_{\cal R}(\bq,\eta)\right \rangle\,.\eeq
Formulas such as (\ref{kcumuleft}) depend on the exponential of connected moments. Keeping terms in the exponential consistently sums all $\epsilon_{s<}$ effects, and thus provides a more accurate correlation function \cite{left}. This was incorporated for instance in \cite{Senatore:2014via}, where a residual oscillatory behavior in the comparison between the EFT computations in \cite{Carrasco:2013mua} and numerical simulations (due to the BAO features) was addressed.\vskip 4pt 

Notice that, in sec.~\ref{sec:MassD}, in order to renormalized the theory we expanded the exponential. In this approximation we ended up with a combined parameter, $l^2_{\delta M,{\rm ren}}$.\,\footnote{~The combined parameter then becomes analogous to the (dimensionless) sound-speed term, $c^2_{s}$, in the Eulerian approach, e.g.~\cite{eftlss,Carrasco:2013mua}. For example in the notation of ~\cite{Carrasco:2013mua}: $l_{\delta_M}^2= 2\pi\; c_{s}^2/k_{\rm NL}^2$.} However, had we kept the finite pieces from the multipole moments upstairs in \eqref{kcumuleft}, the parameters would not add up in the same manner. This is nonetheless consistent with our previous observations regarding resummation. The point of keeping terms in the exponent is to resum the soft displacements in powers of $\epsilon_{s<}$\,, whereas the role of the renormalized parameters, and counter-terms, is to incorporate short-distance physics encoded in $\epsilon_{s>}$\,. Therefore we may proceed as follows. First, the UV contribution from LPT and the counter-terms --which must cancel each other-- may be kept in the exponential without lose of accuracy.  Afterwards, when the results are cutoff-independent, we may bring downstairs the terms which involve the renormalized (finite) contributions from the displacement and multipoles. At this point the renormalized coefficients combine into one, $l^2_{\delta_M,\rm ren}$, which integrates the physical effects from the UV modes. The remaining terms are kept in the exponent, thus incorporating IR effects \cite{left,Senatore:2014via}. Hence, the advantage of LEFT versus the Eulerian counter-part is not on the regularization/renormalization procedure (which, nonetheless, it is perhaps more intuitively implemented in LEFT), but in the way the soft physics is naturally incorporated. 
\newpage
\section{Summary of Part~\ref{sec:part3}}\label{sec:sum3}
\vskip 12pt
\begin{figure}[h!]
\centering
\includegraphics[width=0.8\textwidth]{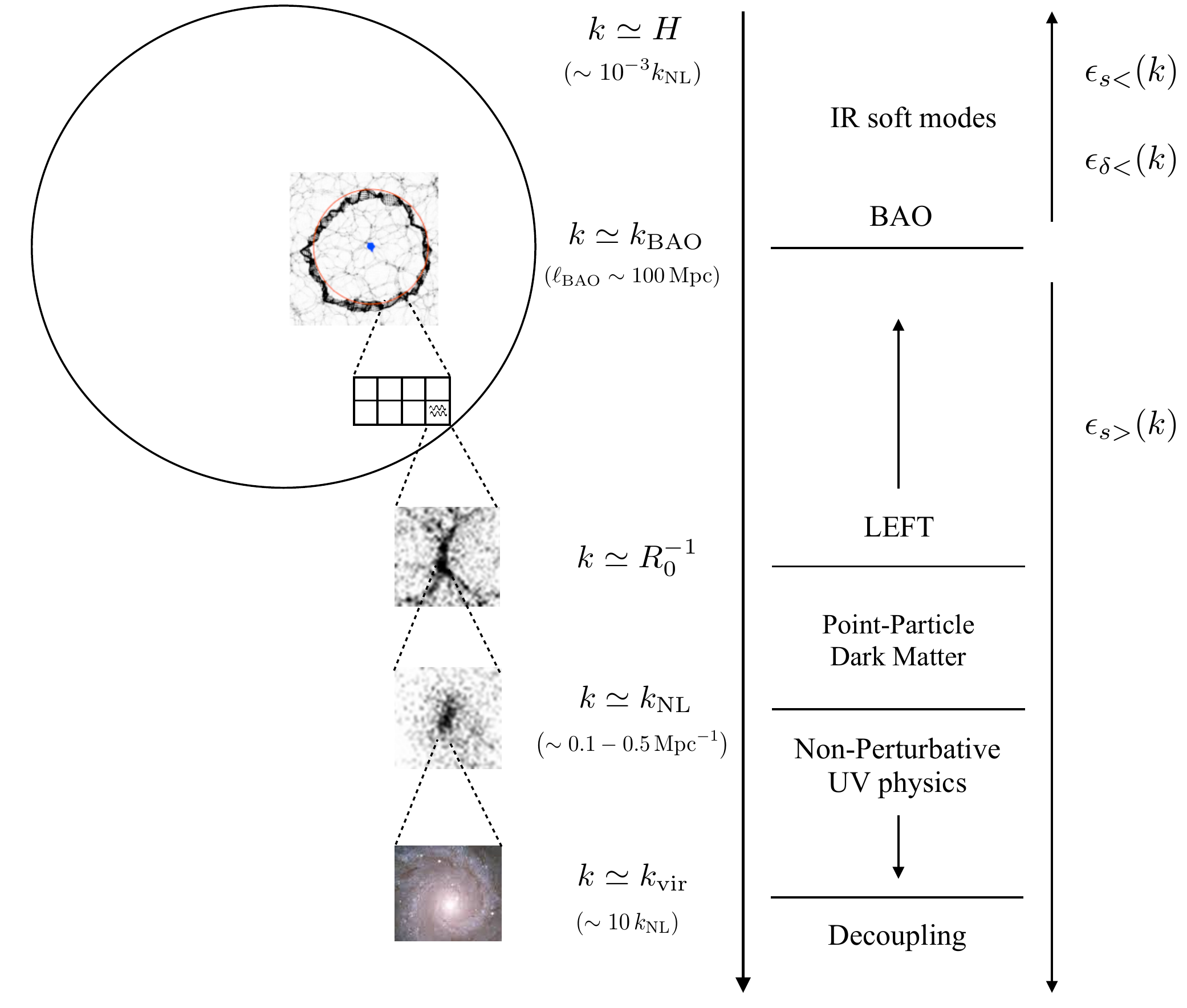}
\caption*{The (L)EFT approach to Large Scale Structures.}
\end{figure}
Cosmological large scale structures may be described in terms of the continuum limit of an effective theory for extended objects in Lagrangian space: LEFT. There is a separation of scales in space (shown above) which allows for the implementation of a multipole expansion, similar to what we discussed for the binary case. While the effective action is local in space {\it and} time, the response functions for the multipoles may display time non-locality. The would-be divergences for the displacement and composite operators in LPT are renormalized by counter-terms, including background, response and stochastic terms.\vskip 4pt At a given $k$ the perturbative contributions are organized in terms of three parameters. The standard expansion parameter describing tidal effects $(\epsilon_{\delta<})$, the imprint of hard modes through renormalized coefficients $(\epsilon_{s>})$, and the effects of soft displacements $(\epsilon_{s<}$). The latter are resummed to all orders in Lagrangian space. LEFT is thus uniquely suited to incorporate the resummation of  long-wavelength displacements and the impact of short-distance physics on long-distance observables, paving the way to more accurate constraints on dark matter, dark energy, and primordial physics. As a fully relativistic theory, LEFT can also be applied to analyze data from surveys approaching the Hubble volume, testing gravity in the deep infrared. 
\newpage
 
 \addcontentsline{toc}{part}{Concluding Remarks \& Outlook}
  
\part*{Concluding Remarks \& Outlook}\label{sec:conclusion}
 \vspace{-0.1cm}
{\it  `I will, therefore, take occasion to assert that the higher powers of the reflective intellect are more decidedly and more usefully tasked by the unostentatious game of draughts than by all the elaborate frivolity of chess. In the latter, where the pieces have different and bizarre motions, with various and variable values, what is only complex is mistaken (a not unusual error) for what is profound.'} -- Edgar A. Poe \vskip 2pt \noindent `The murders in the rue Morgue,'  Graham's Magazine (1841).\vskip 4pt

Throughout this review we have shown how EFT methods can be successfully applied across different length-scales: from black holes to cosmological structures. In all the cases the main element is the existence of a separation of scales in the problem which allows for a perturbative treatment. For instance, in the binary inspiral problem with comparable masses, most of the analytic computations have been performed within the realm of the PN expansion. In this context, NRGR provides a systematic framework, with textbook regularization and renormalization techniques, and a set of Feynman rules which can be automatized entirely using {\it Mathematica}~code. The EFT approach has thus proven to be very powerful, having a major impact in the modeling of spinning binary compact objects, and providing key ingredients to construct accurate GW templates in the forthcoming era of multi-messenger astronomy \cite{ligo16,ligodetect}.\vskip 4pt  
While NRGR has led to remarkable achievements, recent developments in gravity suggest that the Feynman technology --which features prominently in the calculations-- may be somewhat less efficient, and other type of (on-shell) methods may be more suitable to compute certain observables, e.g. \cite{BCF,BCFW,zvi1,zvi2,zvi3,onshell1,onshelltree,onshell2,zvi4,JJTASI}. The main reason is that Feynman diagrams derive from an action where locality and global symmetries are manifest, e.g. Lorentz invariance, at the expenses of introducing redundancies such as local gauge invariance. For example, the photon has two on-shell helicity states, however the action for electrodynamics requires a four-vector potential $A^\mu(x)$. The existence of redundancies increases the complexity of the calculations, with a number of diagrams needed to enforce the gauge invariance of the results, i.e. the~Ward~identity. The same gauge invariance is also responsible for the equivalence we found between different computations of the total radiated power.  Namely, either directly in terms of the square of the amplitude; from the conservation of the pseudo stress-energy tensor in the radiation zone; or through the study of radiation-reaction effects. Even though in principle on-shell methods may not seem to be directly applicable to the calculation of the binding potential, a naturally off-shell quantity, we may nonetheless wonder whether a more efficient computational approach, or more ambitiously non-perturbative methods, may be incorporated into the EFT framework. Initial momentum in this direction appeared in~\cite{iraduff,eftvaidya}.\vskip 4pt

Another question, arising from the study of finite size effects, is the nature of the black hole state. In particular, the origin of the results we encountered for the response functions to long-wavelength~perturbations, and the vanishing of the electric- and magnetic-type Love numbers for black holes in four dimensions. As we argue, this is a fine tuning in the EFT which demands an explanation. However,~such an understanding can only be found in the short-distance theory. In view of the recent discussions on the `black hole information paradox' and the `black hole firewall,' e.g. \cite{Almheiri:2013hfa,Giddings:2013jra,Maldacena:2013xja,hooft,Dvali:2015cwa,Hawking:2015qqa}, it is thus tantalizing to ponder whether properties of black holes on long-distance scales may teach us something about the quantum theory of gravity, and ultimately the degrees of freedom responsible for many peculiar features. We conclude this review by adding a few remarks and pointing out some future directions.

\section*{Beyond Perturbation Theory: A Deeper Structure}
\phantomsection
\addcontentsline{toc}{section}{~~~~Beyond Perturbation Theory: A Deeper Structure}

\begin{figure}[t!]
\centering
\includegraphics[width=0.6\textwidth]{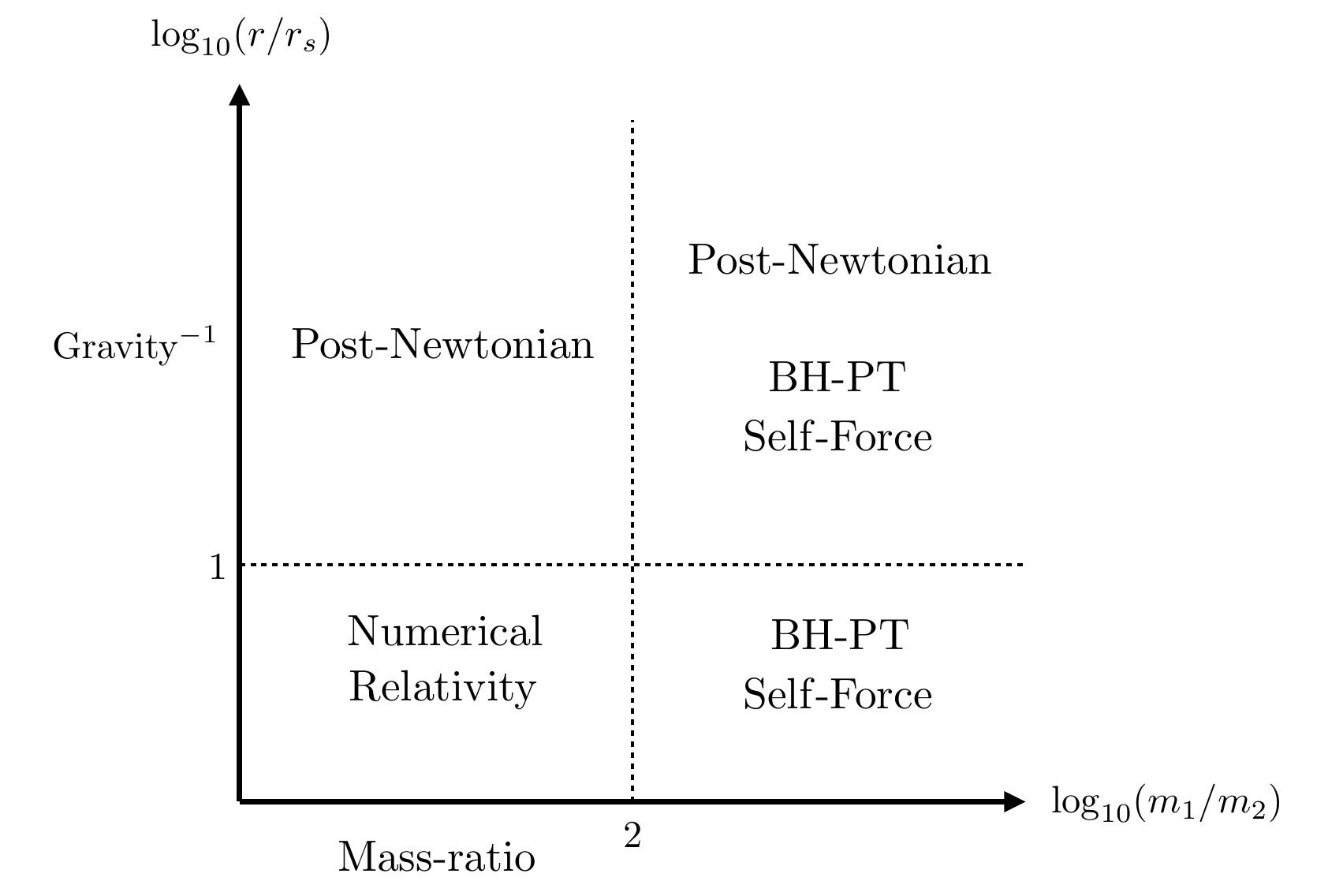}
\caption{Various approaches to the two-body problem, depending on the strength of the gravitational interaction and mass ratio.}
\label{bhpt}
\end{figure}

We have concentrated on binary systems in the inspiral phase, with separation $r/r_s \sim 1/v^2 \gg 1$. 
In~such case, the PN expansion applies for any value of the mass ratio $q_m \equiv m_2/m_1$. There are, however, other methods available when $q_m \ll 1$, for example black hole perturbation theory (BH-PT) and the gravitational self-force program. Altogether these approaches complement numerical techniques to cover all of the parameter space, see Fig.~\ref{bhpt}. For more details on standard methods and a complete set of references see e.g. \cite{Blanchet,Buoreview,ALT}. For a review of the self-force problem on the EFT side see~\cite{chadreview}.\vskip 4pt

An attempt at describing --analytically-- the entire number of cycles during the inspiral, merger and ringdown phases, has been originally put forward in \cite{eob1} and dubbed the Effective One Body (EOB) approach. As it is suggested by the name-tag, the main idea is to map the gravitational two-body problem into the dynamics of a single body with an effective mass and evolving in an effective geometry. For non-rotating objects, the latter is obtained as a deformation of Schwarzschild's metric in powers of the (symmetric) mass ratio. A somewhat related approach is the (semi-analytic) gravitational self-force program for extreme-mass-ratio inspirals, e.g. \cite{poissongsf,chadreview}. In both cases the symmetric mass ratio, $\nu_m$, plays a crucial role \cite{Blanchet,Buoreview,ALT}.\vskip 4pt
Originally the EOB approach was inspired by work in electrodynamics \cite{eobqed}, which is essentially the eikonal approximation discussed in sec.~\ref{sec:eik}. For a linear theory the effective action can be computed to all orders in the PN expansion by summing the series shown in Fig.~\ref{fig1}, and simply corresponds to the one-graviton exchange. The resulting Hamiltonian may be reduced to a one-body problem by going to the center-of-mass frame. To the extent that electromagnetism is described by a linear theory, this does not provide much of a hint about how to incorporate a one-body approach in gravity. On the contrary, as we see in NRGR the relevant diagrams involve higher $n$-point functions, see Fig~\ref{fig:4pn} \cite{riccardocqg}, which are not easily mapped into motion in an effective metric. Moreover, it is not clear either how this effective metric is carried over to the non-perturbative regime. For instance, it does not solve Einstein's equation in vacuum, and instead requires a non-trivial (and plausibly non-local) stress-energy tensor.\vskip 4pt  Despite these shortcomings the EOB approach has attracted significant attention in the last years, see~\cite{Buoreview} and references therein. In particular because of its ability to model remarkably well the binary's dynamics, after a `calibration' procedure, when compared against numerical simulations and perturbative solutions. Therefore, even at a phenomenological level, the success of the EOB may be a hint of a deeper structure in gravity. This is further supported by equally impressive results from semi-analytic self-force computations at leading order in the mass ratio. For instance, the comparison/prediction of high PN order coefficients in the expansion of the binding energy, including logarithmic terms associated with tail (of-tails) effects up to 7.5PN \cite{Blanchet:2014bza,Blanchet:2015iaa}. (Half-integer contributions may enter in the conservative sector through hereditary effects \cite{Blanchet:2015iaa}, see sec.~\ref{sec:RenQM}.)\vskip 4pt

The self-force computation involves a non-trivial comparison. In the PN regime we write the binding energy as an expansion in the small PN parameter $x$,
\beq
\label{eq:EpnA}
e_{\rm pn} (x,\nu_m) = \left( 1+ \sum_n \left(\sum_i^n e^{(n,i)}_{\rm pn} \nu^i_m \right) x^n\right)+ \delta e_{\rm pn}(\log x)\,,\eeq
where $e(x,\nu_m) \equiv E(x,\nu_m)/E_0$ and $E_0=-\mu_m x/2$. From self-force calculations we have instead
\beq
e_{\rm gsf} (\nu_m,x) = \left( 1 + \sum_k e^{(k)}_{\rm gsf}(x) \nu_m^k\right)\,,
\eeq
as an expansion in the mass ratio.  For instance, for the analytic terms this expression can be expanded for small $x$,
\beq
e^{(k)}_{\rm gsf}(x) \simeq \sum_n e_{\rm gsf}^{(k,n)} x^n\,,
\eeq
and then used to compare both results. One finds an outstanding agreement between the two at leading order in $\nu_m$, 
\beq
e_{\rm gsf}^{(1,n)} \simeq e_{\rm pn}^{(n,1)}\,
\eeq
to high $n$PN order, and similarly for the logarithmic corrections. For example, see e.g. \cite{binigsf,LeTiec:2011ab},
\bea
e_{\rm gsf}^{(1,4)} &\simeq& 153.8803\,,\\ 
e_{\rm pn}^{(4,1)} &=& \frac{9037}{1536}\pi^2+\frac{1792}{15}\log 2 + \frac{896}{15} \gamma_E -\frac{123671}{5760} \simeq  153.8838  \,.
\eea
Conversely, one can use the numerical self-force results for $e^{(1)}_{\rm gsf} (x)$ to predict the value of the PN coefficients at leading order in $\nu_m$. Not surprisingly, since in the self-force problem one does not expand in a small velocity parameter \cite{poissongsf}, this program has been extended to (very-)high PN order (21.5PN) in \cite{gsf1,gsf2}, including logarithmic corrections (13PN). These have not yet been reproduced in the PN framework.\,\footnote{~The second-order self-force is known \cite{gsf2d1,gsf2d2}, however a similar program has not been yet developed to ${\cal O}(\nu_m^2)$. In \cite{largeN} a different expansion (the large-$N$ limit) was introduced, and used to calculate the self-force to high accuracy in the ultra-relativistic regime, $\gamma \gg 1$. This may be useful to perform similar comparisons with numerical methods, e.g.~\cite{highnum,cardoso,loustohigh}.} \vskip 4pt

What is remarkable about these manipulations is the following. While the comparison for certain gauge invariant observables, e.g. \cite{Detweiler:2008ft}, can be safely made in the extreme-mass-ratio regime, $q_m \ll 1$, the self-force computations are used to predict PN coefficients at leading order in $\nu_m$, for {\it all} mass ratios, which is somewhat puzzling. If $e_{\rm GR}(\nu_m,x)$ denotes the exact result in general relativity, then we have the approximations
\beq
 e_{\rm gsf}(\nu_m,x) \simeq e_{\rm GR}(\nu_m \ll 1,x)\,,~ e_{\rm pn}(x,\nu_m) \simeq e_{\rm GR}(\nu_m, x \ll 1) \,.
\eeq
Hence, in the upper-right corner of Fig.~\ref{bhpt}, $e_{\rm gsf} \simeq  e_{\rm pn}$. However,  in principle this cannot be extrapolated to the upper-left, and let alone the lower-left, corner. And yet agreement is found at linear order in~$\nu_m$.\vskip 4pt In the self-force program one originally computes $e_{\rm gsf}(q_m,x)$ as an expansion in $q_m \ll 1$. The extrapolation in $q_m$ for comparable masses, however, produces incorrect results. This is not surprising, the $q_m$ parameter is not invariant under~$1\leftrightarrow 2$, since there is a clear asymmetry in the extreme-mass-ratio case. On the other hand, by construction, the PN expansion is symmetric. The~trick is, before extrapolating, to write the self-force result in a manifestly symmetric manner, using $q_m  = \nu_m + {\cal O}(\nu_m^2)$, such that $e_{\rm gsf}(\nu_m,x) \simeq e_{\rm gsf}(q_m \simeq \nu_m,x)$ at leading order in the mass ratio. However, for comparable masses we have $\nu_m \simeq 1/4$, which is not a particularly small expansion parameter. This is even more puzzling if one extends the self-force computations into the numerical regime, where one also finds remarkable agreement~\cite{ALT}.\vskip 4pt 

There are a few (heuristic) comments one can insert here. For example, if we look at the terms in the PN series expansion of \eqref{eq:EpnA}, we notice that the ${\cal O}(\nu_m^2)$ correction starts also at ${\cal O}(x^2)$, and moreover it is further suppressed by a factor of $1/24$, e.g. \cite{Blanchet}.\,\footnote{~This appears to be a generic property of the PN expansion. Namely, at ${\cal O}(x^n)$ the coefficients of the $\nu^n_m$ terms are highly suppressed. Amusingly, up to $n=6$, one finds the pre-factors $\left(\tfrac{1}{12},\tfrac{1}{24}, \tfrac{35}{5184}, \tfrac{77}{31104}, \frac{1}{512},\frac{2717}{6718464}\right)$ respectively, e.g.~\cite{binigsf2}.} While this argument suggests we may be able to extrapolate to the PN regime, we do not expect the PN expansion to converge toward merger. Therefore, this does not explain the agreement with numerical relativity. On the other hand, the EOB program appears to push the PN framework beyond the inspiral phase, which may then lead us to speculate that perhaps all these features are somewhat connected to a deeper non-perturbative structure. This, in turn, may be manifest also in the cancelations observed in on-shell methods for scattering amplitudes, and invalidate the intuition one may get about the PN expansion from staring at the diagrams in Fig.~\ref{fig:4pn}.\vskip 4pt Amusingly, a source of inspiration behind the EOB approach is also the work of \cite{todorov1,todorov2,todorov3}, which is in spirit similar to the implementation of on-shell methods in \cite{iraduff,eftvaidya}. The idea is based on the Lippmann-Schwinger equation \cite{Lippmann}, relating the (on-shell) scattering amplitude with the (off-shell) binding potential. The main difference is that in the work of \cite{iraduff,eftvaidya} one does not resort to Feynman's technology, and instead the gravitational scattering amplitude (for spin-$0$ scalar particles) is obtained in a {\it bootstrap} manner, see e.g.~\cite{BCF,BCFW,zvi1,zvi2,zvi3,onshell1,onshelltree,onshell2,zvi4,JJTASI}. This approach still relies somewhat on a loop expansion, however, it surpasses the need of complicated $n$-graviton vertices. More recently an intriguing duality has been discovered between Yang-Mills and gravity~\cite{zvi1,zvi2,zvi3,zvi4}. This duality was used in \cite{donal1,donal2} to relate stationary and maximally symmetric spacetimes, to solutions in a non-abelian gauge theory. Unfortunately, while the results in \cite{zvi1,zvi2,zvi3,zvi4,donal1,donal2} are promising, in its current incarnation the duality discussed in \cite{donal1} relies on having found the gravitational solution {\it first}, prior to implementing the transformation. Perhaps similar techniques may be uncovered one day for the binary system, shedding light on the aforementioned properties and turning the two-body problem analytically solvable.

\section*{The Black Hole (Quantum) State}
\phantomsection
\addcontentsline{toc}{section} {~~~~The Black Hole (Quantum) State}

In sec.~\ref{sec:eff} we mentioned that --for black holes in four dimensions-- the real part of the response function goes to zero as $\omega \to 0$. This led to the vanishing of the (renormalized) $C_{E(B)}$ coefficients. In fact, this turns out to be the case for all the electric- and magnetic-type $\ell$-pole moments \cite{smolkinlove,poissonlove,damourlove}. As we alluded before this is somewhat puzzling, in particular since in sec.~\ref{sec:abs} we discovered non-trivial absorptive properties for black holes. It is then plausible this feature may be ultimately linked to the --rather peculiar-- nature of the black hole (quantum) state.\,\footnote{~Incidentally, unexpected cancelations (computed through the fluid/gravity correspondency) have also been found in \cite{flugrav}, albeit for dissipative coefficients at higher order in the derivative expansion. In~light of the `membrane paradigm'~\cite{membrane} this may be more than just a coincidence.}\vskip 4pt
Let us look at the quadrupole term. The retarded Green's function in \eqref{responseret2} may~be~written~as, 
\beq
\label{eq:Jan2}
\frac{2G_N}{r_s^5} f_{\rm bh}(\omega) = f_0(\mu) + \beta_0 \log \omega^2/\mu^2 + i \frac{1}{45} (\omega r_s) + (\omega r_s)^2\left(f_2(\mu) + \beta_2 \log\omega^2/\mu^2\right)  + \cdots \,,
\eeq
where we included the contribution from \eqref{eq:univImf} to the imaginary part (with the correct boundary condition). Here the $f_{0,2}(\mu)$   renormalize the coefficients of the $E_{\mu\nu} E^{\mu\nu}$ and $\dot E_{\mu\nu} \dot E^{\mu\nu}$ terms in the effective action, respectively (see \eqref{Se2}). The $\mu$-independence leads to a renormalization group trajectory.\vskip 4pt 

Following sec.~\ref{smatch}, an explicit matching calculation yields $(f_0,\beta_0)_{\rm bh}=0$ and $(f_2(r_s^{-1}),\beta_2)_{\rm bh}\neq 0$, after removing a logarithmic divergence in dim. reg. \cite{steinchak2}.\,\footnote{~The retarded Green's function is analytic in the upper-half plane. Therefore, its real part cannot be zero for all frequencies (below the cutoff of the EFT) while having a non-zero imaginary part. Hence, the non-trivial absorptive properties unavoidably imply new terms --beyond minimal coupling-- in the point-particle effective action.} The vanishing of $f_0$ in \eqref{eq:Jan2}, plus the lack of logarithmic contribution at order $\omega^0$, implies $C^{\rm ren}_E(\mu)=0$  for black holes, at {\it all} scales. Although this is not inconsistent, it clashes against some basic expectations. While the logarithms may not be present, we could still have power-law divergences requiring a counter-term, $C_E^{\rm ct}(\Lambda)$. These are set to zero in dim. reg. mainly because all possible terms are already present in the effective action. Furthermore, we expect the coefficients to be determined by the short-distance scale, $r_s$, as in e.g. \eqref{rs5}. (Similarly to the $C_{ES^2}$ coefficient in sec.~\ref{sec:finiteSS}.) Since there is no apparent enhanced symmetry when $C_E=0$, nothing prevents it from receiving (plausible large) corrections. Hence, the fact that all of the $C_{E(B)}^{(\ell-2)}$ coefficients vanish --unprotected by symmetries-- implies a `fine tuning' from the EFT point of view.\,\footnote{~Likewise for the hierarchy problem in particle physics, e.g. \cite{iraeft}. Since power-law divergences may be set to zero in dim. reg., one may claim there is no issue. However, one still cannot explain why $m_{\rm Higgs}$ is much {\it lighter} than a UV scale which in principle could be as high as the Planck mass (supported thus far by the LHC). The problem worsens for the cosmological constant, e.g. \cite{WLambda1,WLambda2}. (For an attempt to ameliorate the hierarchy for the latter see e.g.~\cite{nonlocalgia,zee1,zee2}.)} This unusual property constrains the black hole (quantum) state. For instance, for the spectral decomposition of the retarded Green's function, one finds~\cite{walterunp,iragrg} 
\beq
\label{eq:spectral}
{\rm Re}\, f_{\rm bh}(\omega_0\to 0) = {\rm PV} \sum_n \frac{\big| \big\langle {\rm bh}\big| \hat Q^{ij}\big| n\big\rangle\big|^2}{\big(E_{\rm bh}-E_n\big)} \to 0\,.
\eeq
Here $\hat Q^{ij}$ is the quadrupole operator, and $| {\rm bh}\rangle$ the black~hole's state, with $\langle{\rm bh}|\hat H |{\rm bh}\rangle =E_{\rm bh}$, and $|n\rangle$ the eigenvectors with $E_n$ eigenvalues. Then, provided $\hat Q^{ij}|E_0\rangle \neq 0$, we would conclude a black~hole cannot be simply described in terms of a (pure) vacuum state, with $E_{\rm bh}=E_0$.\,\footnote{Notice similar arguments could have been used in electrodynamics for the $J_\mu A^\mu$ coupling. However, in such case the real part vanishes because of a $U(1)$ symmetry, which implies the associated charge annihilates the vacuum, thus preventing a mass for the photon, yet allowing a non-zero imaginary part. (A photon mass also leads to a discontinuity in the number of degrees of freedom \cite{ArkaniSM}.) A non-zero mass is generated in the broken phase, as we conclude from~\eqref{eq:spectral}.} 
Inspired by Bekenstein-Hawking we may then think instead of a thermal state, in which case we would have \cite{walterunp}
\beq
\sum_{n,m} e^{-E_m/T_{\rm bh}}\,\, \frac{\big|\big\langle m \big| \hat Q^{ij}\big| n\big\rangle \big|^2}{\big( E_m - E_n\big)} \to 0\,.
\eeq
Furthermore, since all the $C_{E(B)}^{(\ell-2)}$ vanish, this would have to apply for all multipole moments, imposing a rather non-trivial set of conditions.\vskip 4pt 

At this point, and following this line of argument, one may wonder whether one can write down a dispersion relation involving the $C_{E(B)}$ coefficients. Let us consider the retarded Green's function,
\beq
\label{eq:GTbp}
iG^{\rm ret}_{ij,kl}(t-t',\bk-\bk') \equiv i\left\langle \left[{\cal T}^{ij}(t,\bk),{\cal T}^{kl}(t',\bk')\right] \right\rangle  \theta(t-t')\,,
\eeq
in terms of the stress-energy tensor of the `source,' in this case an isolated black hole. Let us also introduce $g_{\rm bh}(\omega,\bp)$, as in \eqref{responseret2}, by factoring out the tensor structure. In the low frequency limit, the absorption cross section is thus given by
\beq
\label{eq:absom3}
\sigma^{\rm bh}_{\rm abs}(\omega) \propto \frac{1}{\omega} \text{Im}\, g_{\rm bh}(\omega,\bp)\,,
\eeq
after expanding in powers of $(\omega,\bp)$ and evaluating at forward scattering, with $|\bp|=\omega$. For example, at lowest order in $\omega$ we have ${\cal T}^{ij}(\omega,0) \simeq \tfrac{1}{2} \omega^2 Q_E^{ij}(\omega)$, therefore
\beq
\label{eq:gvsf}
g_{\rm bh}(\omega \to 0 , \bp \to 0) \simeq \omega^4 f_{\rm bh}(\omega \to 0) + \cdots\,.
\eeq 
Causality implies that $g_{\rm bh}(\omega,\bp)$ is analytic in the upper-half $\omega$-plane, e.g. \cite{sean}, hence it obeys a dispersion relation. Following the scaling in \eqref{eq:gvsf} let us assume the UV behavior of the retarded Green's function is such that,
\beq
\label{eq:KKr}
{\rm Re}\, g^{(4)}_{\rm bh}(\omega_0\to 0,\bp\to 0) \propto ~ {\rm PV}\,\int_{0}^{+\infty} \frac{d\omega}{\pi}\frac{{\rm Im}\, g_{\rm bh}(\omega,\bp)}{\omega^5}\,,
\eeq
for the fourth-order derivative, thus isolating $C_E$ on the left-hand side. This expression may be naively justified as follows.  As it is known, the cross section approaches a geometric value at high frequencies, $\sigma_{\rm abs}(\omega \to \infty) \propto r_s^2$ \cite{matzner,Mashhoon,sanchez}. Therefore, if a relationship like \eqref{eq:absom3} were to hold in the UV, the expression in \eqref{eq:KKr} would be valid. However, the positivity condition of the imaginary part, e.g. \cite{sean}, implies the right-hand side cannot vanish, whereas the left-hand side is zero for black holes in $d=4$. Hence, we conclude \eqref{eq:KKr} cannot be true. (It may still be valid for neutron stars, and black holes in $d>4$, imposing a lower bound on the Love number.) In turn, this means $g_{\rm bh}(\omega,\bp)$ does not have the claimed UV properties, and subtractions are needed.\,\footnote{~Something similar occurs in the study of dissipative effects within an EFT approach to fluid dynamics~\cite{disfluid,nicofluid1,nicofluid2}.} Therefore, while the absorption cross section approaches a constant value, the retarded Green's function in \eqref{eq:GTbp} continues to grow. \vskip 4pt

Even though this is not necessarily an issue, after all the EFT expansion modifies the UV behavior above the cutoff scale, it is nonetheless somewhat peculiar and signals the imprint of non-analytic behavior, e.g. \cite{qsnm,Leaver}. To press on this issue let us then consider the gravitational scattering amplitude off of a black hole, ${\cal A}_{\rm bh}(\omega,\theta)$.  Assuming ${\cal A}_{\rm bh}(\omega,\theta)$ is polynomially bounded, we can write down a dispersion relation in the forward limit, involving the $n$-th order derivative at zero frequency on the left-hand side, and an integral over the imaginary part on the right-hand side. As it is well known, the amplitude presents a series of so called `quasi-normal' modes \cite{qsnm}, which manifest themselves as resonances and branch cuts in the complex $\omega$-plane, e.g. \cite{Leaver}. Therefore, the exact form of the dispersion relation requires a careful study of the analytic properties of the scattering amplitude. At low frequencies, however, the latter may be computed within the EFT framework. This entails contributions from the mass term as well as the $C_{E(B)}$ coefficients. The $t$-channel exchange and seagull-type diagrams only contribute low powers of $\omega$, and therefore they cancel out for $n \geq 4$. (The singularity in the forward direction is thus avoided by acting with the derivatives prior to sending $\theta\to 0$.) As we discussed in sec.~\ref{smatch}, taking $n=4$ we then isolate the contribution from $C_{E(B)}$, in the $\omega \to 0$ limit. This~leads to a relationship between the Wilson coefficients and an integral of the imaginary part(s) (related to the total cross section), plausibly also contributions from isolated poles. This is often referred as a {\it sum rule}, e.g. \cite{Distler,Baumann:2015nta}. The right-hand side must vanish for black holes, because the left-hand side does, but it remains finite for neutron stars. Hence, for the latter it is also a specific representation for the Love numbers. We leave this open for further exploration.\vskip 4pt

All these results hint at a non-trivial black hole state. In particular the apparent lack of {\it hair}, even in gravitational backgrounds such as the one induced by a companion \cite{nohair}. This resulted in the vanishing of {\it all} the electric- and magnetic-type Love numbers in four dimensions \cite{smolkinlove,poissonlove,damourlove,Poissontidal,Panitidal}. At the same time, black holes {\it deform} when rotation is present, e.g. \cite{HartleS}, and furthermore also allow for {\it polarizability}, e.g. \cite{Dampol}, as well as absorptive properties, e.g. \cite{page}. It~would be then interesting to contrast all of these features against putative theories that aim at incorporating short-distance gravitational degrees of freedom.\,\footnote{~One way the black hole state may be probed is through high energy gravitational scattering \cite{sgrav,erice,gia2}. An EFT approach \cite{largeN,steveprog} could also be useful at large impact parameters, to study the onset of black hole formation, perhaps shedding light on the analytic properties of the gravitational $S$-matrix \cite{sgrav,erice}. Another interesting regime to explore the manifestation of all these properties is the large-$D$ limit of general relativity~\cite{larged1,larged2}.} For instance, the Kerr/CFT correspondency \cite{kcft1,kcft2,gkcft,kcft3,kcft4,kcft5} and more generally the role of conformal symmetry \cite{hooft}, (non-violent) non-locality \cite{gidNL,gidNL2,gidNL3,gidring}, or the `corpuscular' framework \cite{gia1,gia01,gia12,gia3}. Moreover, the recent results associated with the Bondi-Metzner-Sachs (or BMS) group may not be unrelated, e.g. \cite{Strominger:2013jfa, gia3,Hawking:2015qqa}, due to the --conjectured-- presence of `soft hair', e.g. \cite{Hawking:2016msc,giahair}. In~light of all of these developments, it is not unthinkable we may be on the brink of discovering a deeper structure in gravity, which may be probed with a new generation of gravitational wave detectors, e.g. \cite{Giddings:2016tla}. Hopefully this will be reviewed before the $2\times 10^2$ anniversary of general relativity.\vskip 4pt

Through direct implementation, or parameterizing its own demise, the EFT framework offers a remarkable set of tools to systematically study a familiar --though complex-- structure, or to unravel unexpected behavior. Concluding this review we hope the reader will consider incorporating the EFT formalism to the toolbox of skills to be mastered by a modern theoretical physicist.\,\footnote{~\url{http://www.ictp-saifr.org/?page_id=9163}} 

\begin{center}
{\bf \huge Acknowledgments}
\end{center}
\vskip 0.5cm
The work reviewed here was conducted during the last several years. My input has been supported at different stages~by: The Department of Energy grants DOE-ER-40682-143 and DEAC02-6CH03000 (CMU); The Foundational Questions Institute grant RFPI-06-18 (UCSB); The National Science Foundation grant No.04-56556 (KITP); NASA grant NNX10AH14G (Columbia~U.);  The~National Science Foundation grant AST-0807444 and Department of Energy grant DE-FG02-90ER40542 (IAS); The German Science Foundation within the Collaborative Research Center SFB 676 (DESY); The Simons Foundation and S\~ao Paulo Research Foundation Young Investigator Awards, grants 2014/25212-3 and 2014/10748-5 (ICTP-SAIFR).\vskip 4pt 

I am indebted to my collaborators, colleagues and friends for sharing their wisdom throughout these years, and for innumerable illuminating discussions which helped me shape the perspective on the material presented here. In particular: Daniel Baumann, Raphael Flauger, Chad Galley, Rodolfo Gambini, Steve Giddings, Walter Goldberger, Dan Green, Ben Grinstein, Sean Hartnoll, Lam Hui, Thomas Konstandin,  Adam Leibovich, Alex Le Tiec, Diana Lopez-Nacir, Alberto Nicolis, Joe Polchinski, Jorge Pullin, Andi Ross, Ira Rothstein, Leonardo Senatore, Ari Socrates, Matias Zaldarriaga, Tony Zee and Jure Zupan. I am also grateful to Daniel Baumann, Dan Green and Sean Hartnoll for helpful comments on the draft. I thank Marc Kamionkowski for inviting me to put this review together and patiently waiting for its completion; and last --but not least-- to Patricia Porto, Imme Roewer and Emiliano A. Porto for the immense and unconditional support. 
\vskip 12pt Dedicated to Isabel (Perla), `{\it \underline{No te olvides...}}' 
\newpage
\renewcommand*{\appendixname}{}

\appendix
\addcontentsline{toc}{part}{Appendix}
\noindent
{\huge\bf Appendix}

\section{Field Redefinitions}\label{app:field-redef}

Let us go back to the contribution from the $C_{R(V)}$ coefficients to the effective action in~\eqref{acrcv}, 
\beq
\label{app:crcv}
C_R \int d^4 x~ \delta^4(x^\alpha-x^\alpha(\tau))~ R(x)~d\tau  + C_V \int d^4 x~\delta^4(x^\alpha-x^\alpha(\tau)) R_{\mu\nu}(x) v^\mu(\tau) v^\nu(\tau) d\tau\, .
\eeq
According to Einstein's equations the Ricci tensor, $R_{\mu\nu}(x) \propto \delta^4(x^\alpha-x^\alpha(\tau))$, vanishes everywhere except when it is evaluated on the worldline of the particles, where instead blows up. This means that, on-shell, these terms are either zero or infinity. We show next they can be consistently set to zero everywhere by means of a field redefinition.\vskip 4pt

In general we are not allowed to use the equations of motion in the action. (For example, the action for a free scalar field vanishes on-shell.) However, we are allowed to make a change of field-coordinates in the path integral. To illustrate the procedure let us study an instructive example involving a single real scalar field in five dimensions. The action is \cite{DGP}
\beq
S= M_5^3 \int d^4xdy \partial_A\phi(x,y)\partial^A\phi(x,y) +
\Mp^2\int d^4xdy\delta(y)
\partial_{\mu}\phi(x,y)\partial^{\mu}\phi(x,y)\,,\label{acdgp}
\eeq
with $M_5$ a mass scale. Here  $A= 0\cdots 4$ and the metric is $\mbox{diag} (+,-,-,-,-)$. In addition to the Einstein-Hilbert action, we have an extra term which is localized in a $d=4$ `brane.' This is a simplified version of the so called `DGP model' \cite{DGP}. Notice it resembles our case, where we have fields localized on the worldline of a point-particle. The Green's function obeys \cite{DGP}
\beq
(M_5^3\partial_A\partial^A+
\delta(y)\Mp^2\partial_{\mu}\partial^{\mu})G(x,y,0,0)=\delta^4(x)\delta(y) \label{ret},
\eeq 
whose solution, with retarded boundary condition $G_R(x,y,0,0)=0$ for $x^0<0$, reads 
\beq
G_R(p,y)=\frac{1}{\Mp^2p^2+2M_5^3p}e^{-p|y|}\label{gret}\,,
\eeq
in mixed Euclidean Fourier space (where $(p,y) \equiv (p_1, p_2, p_3, p_4,y)$ and $p\equiv \sqrt{p_1^2+p_2^2+p_3^2+p_4^2}$).\vskip 4pt

Let us perform now the following field redefinition in the action of \eqref{acdgp},
\beq
\phi (x,y) \to \phi(x,y) - \frac{\Mp^2}{M_5^3}\delta(y) \phi (x,y)\,.
\eeq
It is straightforward to show that the variation of the bulk term cancels the one on the brane, and the {\it new} action becomes
\begin{eqnarray}
{\tilde S}&= M_5^3 \int d^4xdy \partial_A\phi(x,y)\partial^A\phi(x,y)
+\Mp^2\int d^4xdy~\delta(y)
\phi(x,y)\partial^2_{y}\phi(x,y) \nonumber \\
&- \frac{\Mp^4}{M_5^3}\delta(0)\int
d^4xdy~\delta(y)\partial_{\mu}\phi(x,y)\partial^{\mu}\phi(x,y).
 \label{acdgp2}
\end{eqnarray}
Notice, however, the term on the brane is `regenerated,' up to an overall factor of $\delta(0)$, together with a term proportional to $\partial_y^2$, also localized.\vskip 4pt We can easily show the new action in \eqref{acdgp2} encodes the same physics as the original one in \eqref{acdgp}. The equation for the Green's function, ${\tilde G} (p,y)$, is given by
\beq
\left(M_5^3(p^2-\partial_y^2)+\Mp^2\delta(y)\partial_y^2-\delta(0)\delta(y)\frac{\Mp^4}{M_5^3}p^2\right)\tilde G_R(p,y) = \delta(y)\, .\label{newgre}
\eeq
After some manipulations we conclude that the retarded Green's functions, $\tilde G_R$ and $G_R$, are related by:
\beq
{\tilde G}_R=\frac{G_R}{\left(1+\frac{\Mp^2}{M_5^3}\delta(0)\right)} \equiv Z^{-1}~G_R\,.
\eeq
As expected, we recover our original expression up to an overall normalization factor, and the physics remains unaltered. The $Z$ factor precisely accounts for the divergences that we encounter dealing with these localized terms. (Moreover, using dim. reg. we would set $\delta(0) \to 0$, and $Z_{\rm dim. \,reg.} \to 1$.)\vskip 4pt

In our case, we can follow these manipulations to remove the terms in \eqref{app:crcv}. We perform a field redefinition of the metric,
\beq
\label{eq:dgapp}
\delta g^{\mu\nu}(x) =  c\left(\frac{1}{2\Mp^2} \int d\tau \frac{\delta^4(x^\alpha-x^\alpha(\tau))}{\sqrt{g}(x)}\right)g^{\mu\nu}(x),
\eeq
with $c$ an arbitrary coefficient. Hence, up to surface terms which do not alter the conclusions,
\beq
\delta S_{\rm EH} = -2\Mp^2 \int d^4x~G_{\mu\nu}(x) \delta g^{\mu\nu}(x) = c \int R_{\alpha\beta}(x^\alpha(\tau))g^{\alpha\beta}(x^\alpha(\tau)) d\tau\, ,
\eeq
with $G_{\mu\nu} - R_{\mu\nu} - \frac{1}{2} g_{\mu\nu} R$, Einstein's tensor. This shows that the coefficient of terms proportional to the Ricci scalar are arbitrary. Hence, choosing $c = -C_R$ we may remove it from the effective action altogether. All coordinate invariant observables are then independent of $C_R$. A similar procedure eliminates $C_V$ \cite{nrgr}.\vskip 4pt

Let us make a final remark regarding these manipulations. After applying \eqref{eq:dgapp} there is actually a left-over piece, proportional to $\delta^3(\bx_1(\tau)-\bx_2(\tau))$, in the effective action. For the two-body system this delta-like potential is similar to the so called `Darwin term,' which accounts for the interaction between the wave-functions of the electron and positron, and contributes to the fine structure of the energy levels of the Hydrogen atom.  Therefore, in a quantum mechanical world, we cannot simply set this --distributional-- term to zero, because wave-functions may not vanish at $r=0$. (The expectation value of the Darwin term is non-zero for $\ell=0$ states.) On the other hand, in the classical theory, this term {\it always} vanishes except when the constituents of the binary overlap. In such case we find a pure divergence, which can be removed as we just explained. This is not possible, however, in the continuum limit --and the evolution of large scale structures-- as we discussed in part~\ref{sec:part3}. For the latter the Ricci-type contributions may be replaced by terms proportional to the (smooth) mass-density, and derivatives thereof. As we showed, these terms end up playing an important role in the consistency of the theory.

\section{Toolkit}\label{app:Fey}
\subsection*{Gravitational Couplings}
\phantomsection
\vskip 4pt
The relevant couplings that are needed to compute all spin effects to NLO order are summarized here:
\bea
L_{mh} &=&- \frac{m}{2\Mp}h_{00}\,,\\
L_{mhv} &=& -\frac{m}{\Mp} h_{0i}v^i\,,\\
L_{mhv^2} &=& -\frac{m}{2\Mp}\left(h_{ij}v^iv^j + \frac{1}{2} h_{00} v^2\right)\,,\label{Ap:lv2m}\\
L_{mhv^3} &=& -\frac{m}{2\Mp} h_{0i} v^i {\bv}^2\,, \\
L_{mh^2} &=& \frac{m}{8\Mp^2}h_{00}h_{00}\label{Ap:h2m}\,,\\
L_{mh^2v} &=& \frac{m}{2\Mp^2} h_{00}h_{0i}v^i\label{Ap:h2mv}\,,\\
L_{Sh}&=& \frac{1}{2\Mp}h_{i0,k}S^{ik}\,,\label{Ap:sgnr1}\\
L_{Shv} &=& \frac{1}{2\Mp}\left(h_{ij,k}S^{ik}v^j + h_{00,k}S^{0k}\right),\label{Ap:sgnr15}\\
L_{Shv^2} &=& \frac{1}{2\Mp}\left(h_{0j,k}S^{0k}v^j +
h_{i0,0}S^{i0}\right)\,,\\
L_{Shv^3} &=& \frac{1}{2\Mp} h_{ik,0}S^{k0} v^i
\label{Ap:sgnr2}\,, \\
L_{Sh^2} &=& \frac{1}{4\Mp^2}S^{ij} \big({h^{\mu}}_j h_{0\mu,i} + h_{kj} h_{0i,k}\big) \label{Ap:sgh2v4}\\
L_{Sh^2v} &=& \frac{1}{4\Mp^2}\Big[S^{ij} {h^l}_j \left(h_{kl,i} -
h_{ki,l}\right)v^k - S^{ij}h_{0j} h_{0i,0} + S^{i0}\left(h_{00}h_{00,i}+{h^l}_ih_{00,l}\right)\Big]\label{Ap:sgh2v5}\\
L_{hS^2} &=& -\frac{C_{ES^2}}{4m\Mp}h_{00,ij}S^{ik}S^{jk}\,,\label{Ap:fsr1}\\
L_{hS^2v} &=& -\frac{C_{ES^2}}{2m \Mp} h_{0l,ij}v^l S^{ik}S^{jk}\,,\label{Ap:fsr2}\\
L_{hS^2v^2} &=& \frac{C_{ES^2}}{2m
\Mp}\Big[\frac{1}{2}h_{00,ij}S^{i0}S^{j0}+   S^{0k} S^{jk}h_{00,lj}v^l +\\
&&  S^{ik} S^{jk}\big(h_{il,0j}v^l-\frac{1}{2}h_{lr,ij}v^rv^l + h_{li,jr}v^lv^r - 
\frac{{\bv}^2}{4}h_{00,ij}\big)\Big]\,,\nn \\
\label{Ap:fsr3}
L_{h^2S^2} &=& \frac{C_{ES^2}}{8m
\Mp^2} S^{ik} S^{jk}\Big[h_{00,i}h_{00,j}+h_{0i,l}h_{0l,j}-h_{0l,j}h_{0l,i}
 + h_{00,l}(h_{ij,l}-h_{il,j}-h_{jl,i})\\ && + h_{0l,i}h_{0j,l}-h_{i0,l}h_{j0,l} + h_{00}h_{00,ij} - 2 h_{li}h_{00,lj}\Big]\nn\,.
 \eea
 \newpage
\subsection*{Integrals}
\phantomsection
\vskip 4pt
The integrals we find throughout our calculations in the PN framework are regularized in dim. reg. by analytic continuation in $d$, the number of spacetime dimensions. The most common master integral we find is given by
\beq
\int \frac{d^{d-1}\bk}{(2\pi)^{d-1}}\frac{e^{i\bk\cdot\br}}{({\bk}^2)^\alpha}=\frac{1}{(4\pi)^{\frac{d-1}{2}}}\frac{\Gamma\left(\frac{d-1}{2}-\alpha\right)}{\Gamma(\alpha)}\left(\frac{{\br}^2}{4}\right)^{\alpha-\frac{d-1}{2}}\,,
\eeq
which allows us to transform expressions in momentum space, from the Feynman diagrams, into coordinate space. By taking derivatives with respect to $\br^i$ we can construct tensor integrals, for example,
\begin{align}
\int\frac{d^{d-1}\bk}{(2\pi)^{d-1}}\frac{\bk^i\bk^j\bk^l\bk^m e^{i\bk\cdot\br}}{({\bk}^2)^\alpha}&=\frac{1}{(4\pi)^{\frac{d-1}{2}}}\frac{\Gamma\left(\frac{d-1}{2}-\alpha\right)}{\Gamma(\alpha)}\left(\frac{{\br}^2}{4}\right)^{\alpha-\frac{d-1}{2}-2}\Bigg(\frac{1}{4}\left(\delta^{il}\delta^{jm}+\delta^{im}\delta^{jl}+\delta^{ij}\delta^{lm}\right) \nn\\
&+\left(\frac{2\alpha-d-5}{4}\right)\left(\delta^{ij}{\hat \br}^l{\hat \br}^m+\delta^{il}{\hat \br}^j{\hat \br}^m+\delta^{im}{\hat \br}^j{\hat \br}^l+\delta^{jl}{\hat \br}^i{\hat \br}^m+\delta^{jm}{\hat \br}^i{\hat \br}^l+\delta^{lm}{\hat \br}^i{\hat \br}^j\right)\nn \\
&+\left(\alpha-\frac{d-1}{2}-2\right)\left(\alpha-\frac{d-1}{2}-3\right){\hat \br}^i{\hat \br}^j{\hat \br}^l{\hat \br}^m\Bigg)\,,
\end{align}
and so on and so forth.\vskip 4pt  

In general, in the conservative sector, at $n$-order in the PN expansion we expect ${\bf 3}$-momentum integrals, equivalent to $n$-loop diagrams in quantum field theory. In many instances integrals can be computed as nested $\ell$-loop integrations, with $\ell < n$. Often, using several tricks, e.g. \cite{smirnov}, these reduce to products with $\ell = 1$. In such case, a useful master (one-loop) integral is the following
\beq
\int \frac{d^{d-1}\bk}{(2\pi)^{d-1}}\frac{1}{\left({\bk}^2\right)^\alpha\left(({\bk+\bp})^2\right)^\beta}=  \frac{1}{(4\pi)^{\frac{d-1}{2}}}\frac{\Gamma(\alpha+\beta-\frac{d-1}{2})}{\Gamma(\alpha)\Gamma(\beta)}\frac{\Gamma(\frac{d-1}{2}-\alpha)\Gamma(\frac{d-1}{2}-\beta)}{\Gamma(d-1-\alpha-\beta)}\left(\bp^2\right)^{\frac{d-1}{2}-\alpha-\beta}\,,
\eeq
which enters for example in the tail effect (see appendix A of \cite{andirad}). These manipulations simplify the number of loop integrals required, leaving a handful of irreducible ones at a given order. See \cite{riccardocqg} for a more detailed discussion.\vskip 4pt 

In the radiation theory we also encounter similar integrals after multipole expanding the potential-radiation vertices. However, as we notice  diagrammatically, for instance in Fig.~\ref{fig:tmn}, the higher order source multipoles are obtained in terms of lower order loop integrals with potential modes. This is the case because in the computation of the gravitational amplitude (and pseudo stress-energy tensor) we remove a propagator from the worldline and put it on-shell (and soft). This means, to compute the total radiated power to $n$PN order we require at most $(n-1)$-loop integrals from the conservative sector. 

  \newpage
\section*{References}
\phantomsection
  \addcontentsline{toc}{part}{References}

\bibliographystyle{elsarticle-num}
 \bibliography{Refs}
\end{document}